\documentclass[aas_macros]{fpfslut}

\usepackage[T1]{fontenc}
\usepackage{lmodern}
\usepackage[czech,english]{babel}

\usepackage{pdfpages}
\usepackage{dirtytalk}
\usepackage{mathptmx}
\usepackage{esvect}
\usepackage{staff}
\usepackage{tcolorbox}
\usepackage{tabularx}
\usepackage{array}
\usepackage{colortbl}
\tcbuselibrary{skins}
\usepackage{wrapfig}
\usepackage{adjustbox}
\usepackage{makecell}
\usepackage{longtable}

\usepackage{pdfpages}
\hypersetup{
        colorlinks=true,
        allcolors=blue,
        }
\tcbset{tab2/.style={enhanced,fonttitle=\bfseries,fontupper=\normalsize\sffamily,
		colback=black!2!white,colframe=white!20!black,colbacktitle=black!40!white,
		coltitle=black,center title}}

\newcommand{\apj}{Astrophys. J.}   
\newcommand{\apjl}{Astrophys. J. Lett.}   
\newcommand{\aap}{Astron. Astrophys.}   
\newcommand{\mnras}{Mon. Not. R. Astron. Soc.}   
\newcommand{\prd}{Phys. Rev. D}   

\newcommand{\mdot}{\dot{m}}
\newcommand{\Mdot}{\dot{M}}
\newcommand{\pgas}{p_{\mathrm{gas}}}
\newcommand{\uint}{u_{\mathrm{int}}}
\newcommand{\der}{\mathrm{d}}
\newcommand{\msol}{{\mathrm{M}}_\odot}
\newcommand{\koral}{\texttt{KORAL}}
\newcommand{\schw}{Schwarzschild}
\newcommand{\risco}{r_{\mathrm{ISCO}}}
\newcommand{\rg}{r_{\mathrm{g}}}
\newcommand{\tg}{t_{\mathrm{g}}}
\newcommand{\Ledd}{L_{\mathrm{Edd}}}
\newcommand{\Medd}{\dot{M}_{\mathrm{Edd}}}

\definecolor{OliveGreen}{HTML}{3C8031}
\definecolor{BrickRed}{HTML}{B6321C}

\begin{document}

\author{Debora Lančová}
\title{Computer modelling of accretion processes in binary systems with black holes and neutron stars}
\thesiskind{Dissertation}
\supervisor{doc. RNDr. Gabriel Török, Ph.D.}
\advisor{Prof. Marek Abramowicz, Ph.D.}
\yearofsubmission{2023}
\maketitle

\def\@acknowledgementname{jedna}

\begin{acknowledgement}
	\vspace{1cm}
	
I would like to express my heartfelt gratitude to my advisors, colleagues, family, and friends for their unwavering support and guidance throughout my doctoral journey.

\vspace{1cm}

\textit{\say{So long and thanks for all the fish.}}	 	

\textit{Douglas Noël Adams}

\end{acknowledgement}

\tableofcontents

\mainmatter

\part*{Preface}

This dissertation is written as an annotated collection of selected papers presenting the results I achieved during my doctoral studies at the Institute of Physics of the Silesian University in Opava. I was a co-author of six papers published in prestigious international journals and four proceedings papers. In one paper and two proceeding papers, I was the leading author. One additional paper is submitted for peer review. For the annotation, I  selected a total of 6 papers\footnote{Papers selected for the annotation are denoted by an asterisk symbol in the list of my publications.}.

The first part of the thesis  is devoted to an overview of the topics addressed within the selected papers. Chapter \ref{chap:intro} summarizes the astrophysical picture of accreting systems, focusing on the behaviour of X-ray binaries that exhibit effects of the general relativity in combination with radiation magnetohydrodynamic in a complex interplay. In Chapter \ref{chap:AD}, a~more detailed introduction to the modelling of accretion disks in general relativity is given. Chapter \ref{chap:mhd} introduces general relativistic radiative magnetohydrodynamics, a~method that enables the self-consistent modelling of magnetized fluids in general relativity, including the dynamical interaction with radiation and the exchange of energy and momentum between matter and radiation. Then, in Chapter \ref{chap:puffy}, the puffy disk is introduced, an accretion disk model resembling an ultraluminous state of microquasars derived from the results of numerical simulations. Finally, the \ref{chap:QPOs} chapter focuses on modelling X-ray variability and quasi-periodic oscillations using analytical models of accretion disks but employing an advanced description of the oscillations and outlining the implications of the complex modelling for observational data.

\section*{Full list of papers published in international journals}

\begin{itemize} 
	\setlength\itemsep{1em}

\item Wielgus, M., Lančová, D., Straub, O., Kluźniak, W., Narayan, R., Abarca, D., Różańska, A., Vincent, F., Török, G., \& Abramowicz, M.; Observational properties of puffy discs: radiative GRMHD spectra of mildly sub-Eddington accretion (2022), \mnras, 514, 780, DOI: \url{10.1093/mnras/stac1317}$^*$

\item Török, G., Kotrlová, A., Matuszková, M., Klimovičová, K., Lančová, D., Urbancová, G., \& Šrámková, E.; Simple Analytic Formula Relating the Mass and Spin of Accreting Compact Objects to Their Rapid X-Ray Variability (2022), \apj, 929, 28, DOI: \url{10.3847/1538-4357/ac5ab6}$^*$

\item Kotrlová, A., Šrámková, E., Török, G., Goluchová, K., Horák, J., Straub, O., Lančová, D., Stuchlík, Z., \& Abramowicz, M. A.; Models of high-frequency quasi-periodic oscillations and black hole spin estimates in Galactic microquasars (2020), \aap, 643, A31, DOI: \url{10.1051/0004-6361/201937097}$^*$

\item Bakala, P., De Falco, V., Battista, E., Goluchová, K., Lančová, D., Falanga, M., \& Stella, L.; Three-dimensional general relativistic Poynting-Robertson effect. II. Radiation field from a rigidly rotating spherical source (2019), \prd, 100, 104053, DOI: \url{10.1103/PhysRevD.100.104053}. 

\item Lančová, D., Abarca, D., Kluźniak, W., Wielgus, M., Sądowski, A., Narayan, R., Schee, J., Török, G., \& Abramowicz, M.; Puffy Accretion Disks: Sub-Eddington, Optically Thick, and Stable (2019), \apjl, 884, L37, DOI: \url{10.3847/2041-8213/ab48f5}$^*$

\item De Falco, V., Bakala, P., Battista, E., Lančová, D., Falanga, M., \& Stella, L.; Three-dimensional general relativistic Poynting-Robertson effect: Radial radiation field (2019), \prd, 99, 023014, DOI: \url{10.1103/PhysRevD.99.023014}.

\end{itemize}
\vspace{0.2cm}

\section*{List of proceedings and submitted papers}

\begin{itemize} 
	\setlength\itemsep{1em}

\item Lančová, D., Yilmaz, A., Wielgus, M., Dovčiak, M., Straub, O., \& Török, G.; Spectra of puffy accretion discs: the kynbb fit (2023), Astronomische Nachrichten, 344, e20230023, DOI: \url{10.1002/asna.20230023}$^*$ 

\item Šrámková, E., Matuszková, M., Klimovičová, K., Horák, J., Straub, O., Urbancová, G., Urbanec, M., Karas, V., Török, G., \& Lančová, D.; Oscillations of fluid tori around neutron stars (2023), Astronomische Nachrichten, 344, e20220114, DOI: \url{10.1002/asna.20220114}. 

\item Matuszková, M., Klimovičová, K., Urbancová, G., Lančová, D., Šrámková, E., \& Török, G.; Oscillations of non-slender tori in the external Hartle-Thorne geometry (2022), arXiv e-prints, arXiv:2203.10653, DOI: \url{10.48550/arXiv.2203.10653}$^*$

\item Lančová, D., Bakala, P., Goluchová, K., Falanga, M., De Falco, V., \& Stella, L.; The study on behaviour of thin accretion disc affected by Poynting-Robertson effect (2017), RAGtime 17-19: Workshops on Black Holes and Neutron Stars, 127. 

\item Karas, V., Klimovi{\v{c}}ov{\'a}, K., Lan{\v{c}}ov{\'a}, D., Svoboda, J., T{\"o}r{\"o}k, G., Matuszkov{\'a}, M., {\v{S}}r{\'a}mkov{\'a}, E., {\v{S}}pr\v{n}a, R.,  \& Urbanec, M.; Timing of accreting neutron stars with future X-ray instruments: towards new constraints on dense matter equation of state (2023), submitted to Contrib. Astron. Obs. Skalnat\'{e} Pleso 

\end{itemize}

\section*{Presentations at international conferences and invited seminars}

\begin{itemize} 
	\setlength\itemsep{0.2em}
\item RAGtime 20, Opava, Czech Republic; October 2018; \textit{GRRMHD simulation of thin accretion disk stabilized by magnetic field}, talk

\item Heraeus Seminar on Accretion in Strong Gravity, Wilhelm und Else Heraeus Stiftung, Bad Honnef, Germany; February 2019; \textit{Global GRRMHD simulation of thin accretion disk}, talk

\item 12th Integral Conference, Geneva, Switzerland; February 2019; \textit{Global GRRMHD simulation of thin accretion disk}, poster

\item RAGtime 21, Opava, Czech Republic; September 2019; \textit{Puffy accretion disks: sub-Eddington, optically thick, and stable}, talk

\item Future of X-ray Timing 2019, Amsterdam, Netherlands; October 2019; \textit{Glo\-bal GRRMHD simulation of thin accretion disk}, talk

\item BHI colloquium, Harvard University, USA; December 2019; \textit{Puffy accretion disks: sub-Eddington, optically thick, and stable}

\item RAGtime 23,  Opava, Czech Republic; September 2021; \textit{Puffy accretion disk: observational properties and inner structure}, talk

\item 9th Microquasar Workshop, Cagliari, Italy;  September 2021; \textit{Puffy accretion disk: sub-Eddington, optically thick, and stable}, poster

\item Seminar, NORDITA, Stockholm, Sweden; November 2021; \textit{Puffy accretion disks: new results and plans}

\item Seminar, Institute of Astronomy, Cambridge University, Cambridge, United Kingdom; November 2021; \textit{Global GRRMHD simulation of stable, optically thick sub-Eddington accretion disk – the Puffy disk}

\item Seminar, INAF Sicily, Palermo, Italy; March 2022; \textit{Puffy accretion disk: sub-Eddington, optically thick, and stable}

\item Pharos conference, Rome, Italy; May 2022; \textit{Broadened iron lines exhibited by accreting relativistic compact object}, poster

\item XMM-Newton meeting, Black hole accretion under the x-ray microscope, Madrid, Spain; June 2022; \textit{Observational properties of puffy accretion disk}, poster

\item Ten Years of High-Energy Universe in Focus - NuSTAR 2022, Cagliari, Italy; June 2022, \textit{Numerical model of a stable sub-Eddington accretion disk – the puffy disk}, talk

\item COSPAR 2022, Athens, Greece; July 2022; \textit{Observational properties of puffy accretion disk}, talk

\item 31st Texas Symposium on Relativistic Astrophysics, Prague, Czech Republic; September 2022; \textit{GRRMHD simulation of sub-Eddington accretion onto stellar mass black hole}, talk

\item RAGtime 23, Opava, Czech Republic; October 2022; \textit{GRRMHD simulation of sub-Eddington accretion onto stellar mass black hole}, talk

\item Third Athena Scientific Conference, Barcelona, Spain; November 2022; \textit{Observational properties of puffy accretion disk}, poster

\item Timescales in Astrophysics Conference, Abu Dhabi, UAE; January 2023; \textit{Timescales of BH accretion disks}, talk/poster

\item HEAD meeting, Waikoloa Village, Hawaii, USA, March 2023; \textit{Puffy accretion disks: sub-Eddington, 
optically thick, and stable}, poster 

\item IBWS 2023, Karlovy Vary, Czech Republic; May 2023; \textit{Puffy accretion disks: sub-Eddington, 
optically thick, and stable}, poster 

\end{itemize}

\vspace{0.2cm}

\part{Annotation}

\chapter{Introduction: Accretion onto compact objects} 
\label{chap:intro}

\begin{figure}
    \centering
    \includegraphics[width=\linewidth]{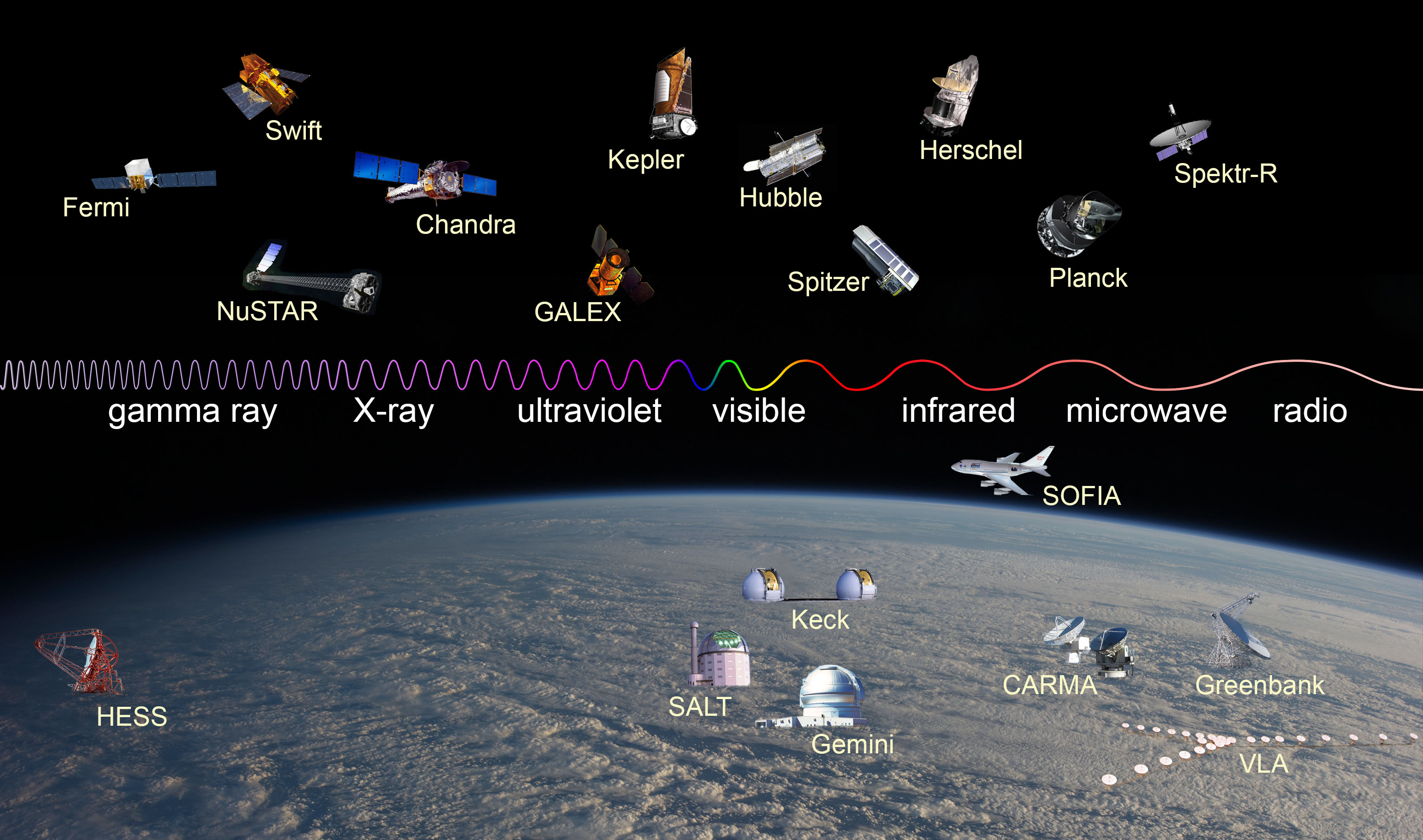}
    \caption{Space and land-based observatories from the NASA fleet, covering almost the whole electromagnetic spectrum (NASA).}
    \label{fig:observatories}
\end{figure}

Compact objects such as black holes (BHs) or neutron stars (NSs) power some of the most energetic sources observed in the Universe. By studying these sources across different wavelengths, various parameters of their central engines have been determined through different methods and techniques, including multi-messenger astronomy. The emission of electromagnetic radiation from these objects spans across a~wide range of energies, and modern observatories, both on Earth and in space, cover nearly the entire electromagnetic spectrum (see Figure \ref{fig:observatories}).

A broad spectrum of accreting compact object sources can be observed and studied thanks to these advanced facilities, even though they are often spatially unresolved and appear as point sources. These multi-wavelength observations provide valuable insights into the composition, variability, and spectral properties of these sources. To date, over 500 Galactic X-ray binaries (XRBs) containing a~compact object have been discovered  \citep{Avakyan2023,Neumann2023}, and even more extragalactic sources, such as BHs in the centres of other galaxies, isolated NSs wandering across the Galaxy \citep{Treves2000}, (extra)galactic NSs or intermediate mass BHs, e.g., in the ultra-luminous X-ray sources \citep[ULXs,][]{ULXs}, and isolated BHs only observed while consuming a~random passing-by object as a~tidal disruption event \citep[TDE,][]{Hills1975}, and many others. 

\begin{figure}[b]
    \centering
    \includegraphics[width=1\linewidth]{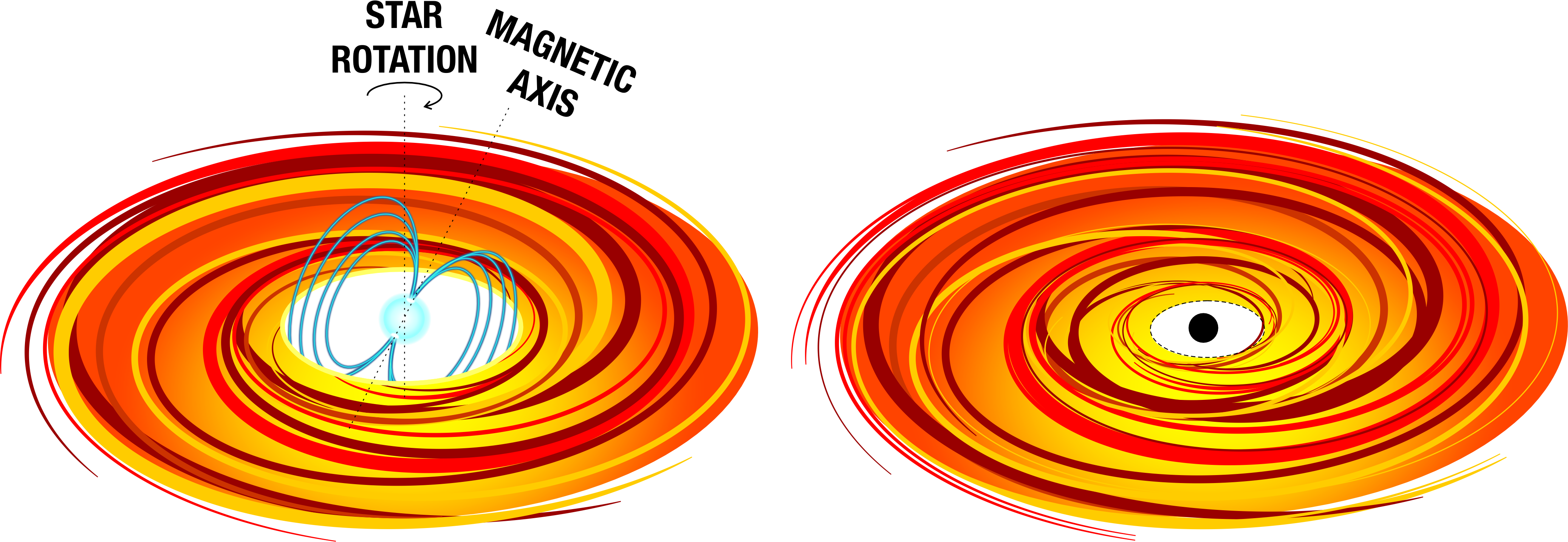}
    \caption{Accretion disk in a NS system (\textit{left}) with the disk truncated at the magnetospheric radius due to the presence of a~strong magnetic field, and in a BH system (\textit{right}) extending all the way to the innermost stable circular orbit (dashed line).}
    \label{fig:chap1:NS}
\end{figure}

In our Galaxy, the brightest sources are expected to contain compact objects. Among them, the NS systems are usually easily distinguishable due to the observation of periodic pulsations or by the occasional occurrence of thermonuclear bursts, which are not observed in BH sources. The pulsations arise from the rotation of the NS and the misalignment between its rotational and magnetic axes (see Figure \ref{fig:chap1:NS}). NS systems undergo a~complex life cycle, from being bright in the radio band to bright in the X-ray band, dimming down and being reborn, and occasionally emitting fast and intense X-ray bursts. Fast time variability, often manifested as high-frequency quasi-periodic oscillations (QPOs), is also more prominent in NS sources compared to BH sources, see Chapter \ref{chap:QPOs}, \cite{tor-etal:2022}, and references therein. 

\section{Black holes}

Since the discovery of the first BH systems in the 1970s \citep{Webster1972}, they have become the focus of numerous observation campaigns across the electromagnetic (EM) spectrum. BHs have served as laboratories for testing fundamental theories and expanding our understanding of the Universe. Explorations of the BH systems have also revealed their crucial role in galaxy formation and evolution. They have helped to understand processes that govern the behaviour of matter and energy in the most extreme environments.

BHs exist across a~wide range of masses in the Universe, spanning from very light stellar-mass BHs with mass $M\sim 10\,\msol$ (where $\msol$ is the solar mass) to supermassive BHs in the centres of galaxies with mass up to $10^{10}\,\msol$ \citep{Dullo2021}. Despite the enormous range of masses, many properties of BH systems are remarkably similar. They are observed due to the accretion of matter from their surroundings, often appearing as highly luminous sources, particularly in the X-ray band. BH systems can also produce jets, collimated fast outflows originating in the polar regions, which are bright in radio wavelengths. In specific orientations, the jets may be Doppler-boosted in the observer's direction, which makes the system even more luminous.

\begin{figure}[b]
    \centering
    \includegraphics[width=1\linewidth]{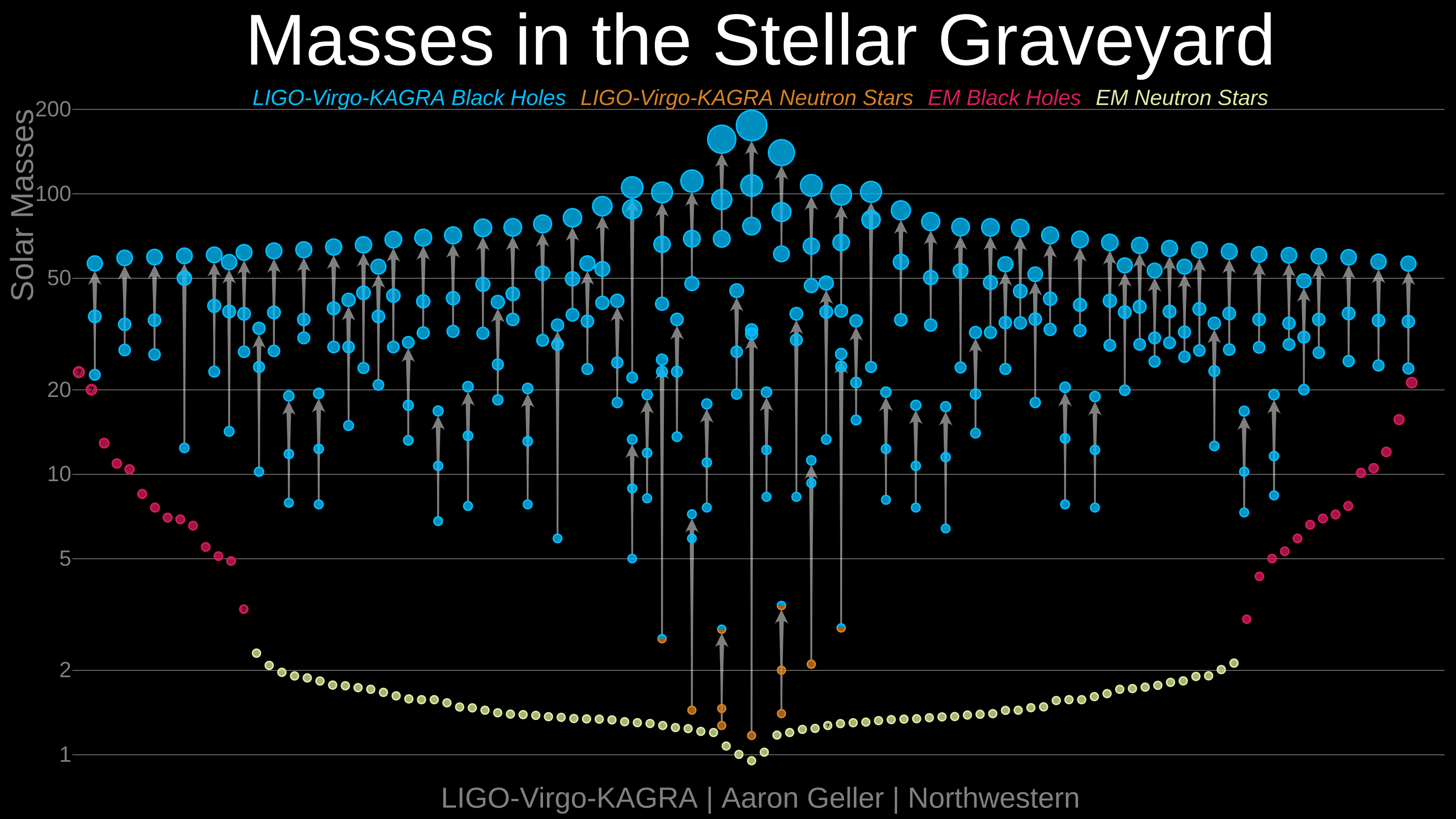}
    \caption{Range of masses of known BHs and NSs observed in the EM spectrum or in the GW. Some sources are still to be confirmed,  denoted with question marks. \textit{Visualization: LIGO-Virgo-KAGRA / Aaron Geller / Northwestern}.}
    \label{fig:1:graveyard}
\end{figure}

In the case of BHs, the source of the radiation is not the compact object itself but rather its immediate surroundings. When the material with a~non-zero angular momentum falls onto a~BH, it forms an accretion disk that can extract and radiate away a~significant portion, up to 42 \%, of the rest-mass energy from the infalling matter.  This process is one of the most efficient mechanisms for converting matter into radiation, second only to annihilation. The presence of a~compact object and the extreme gravitational curvature of spacetime in its vicinity profoundly impact the behaviour of matter, leading to observable properties that differ significantly from those expected in a~flat spacetime. As a~result, the effects of general relativity (GR) are often crucial for an accurate description of accretion and other processes near compact objects \citep{Salpeter1964,Lynden-Bell1969,Bardeen1972,AccretionPower}.

\subsection{Supermassive black holes}

Supermassive BHs are commonly found at the centers of galaxies, a~fraction of which (about 1 \%) accrete matter at very high rates from their surroundings environment. Depending on the density of material accumulated in their vicinity, some of these BHs can release a~tremendous amount of energy in the form of radiation from the accretion disk and powerful jets from the polar regions. These objects are often referred to as active galactic nuclei \citep[AGN, ][]{Ambartsumian:1958qna}. 

The discovery of AGN was closely linked to the observation of quasars (quasi-stellar objects). Quasars were initially unidentified sources with extremely high luminosity and peculiar spectral properties that did not contain the usual spectral lines of known elements \citep{StrangeSpectra}. This posed a~significant puzzle for astronomers and physicists for quite some time until it was realized that these lines corresponded to the conventional spectral lines of hydrogen, helium, and other elements but heavily redshifted to much lower frequencies. This redshift indicated that the sources were extremely distant and located in an expanding Universe \citep{1963Natur.197.1040S}. 

\begin{figure}[bh!]
    \centering
    \includegraphics[width=1\linewidth]{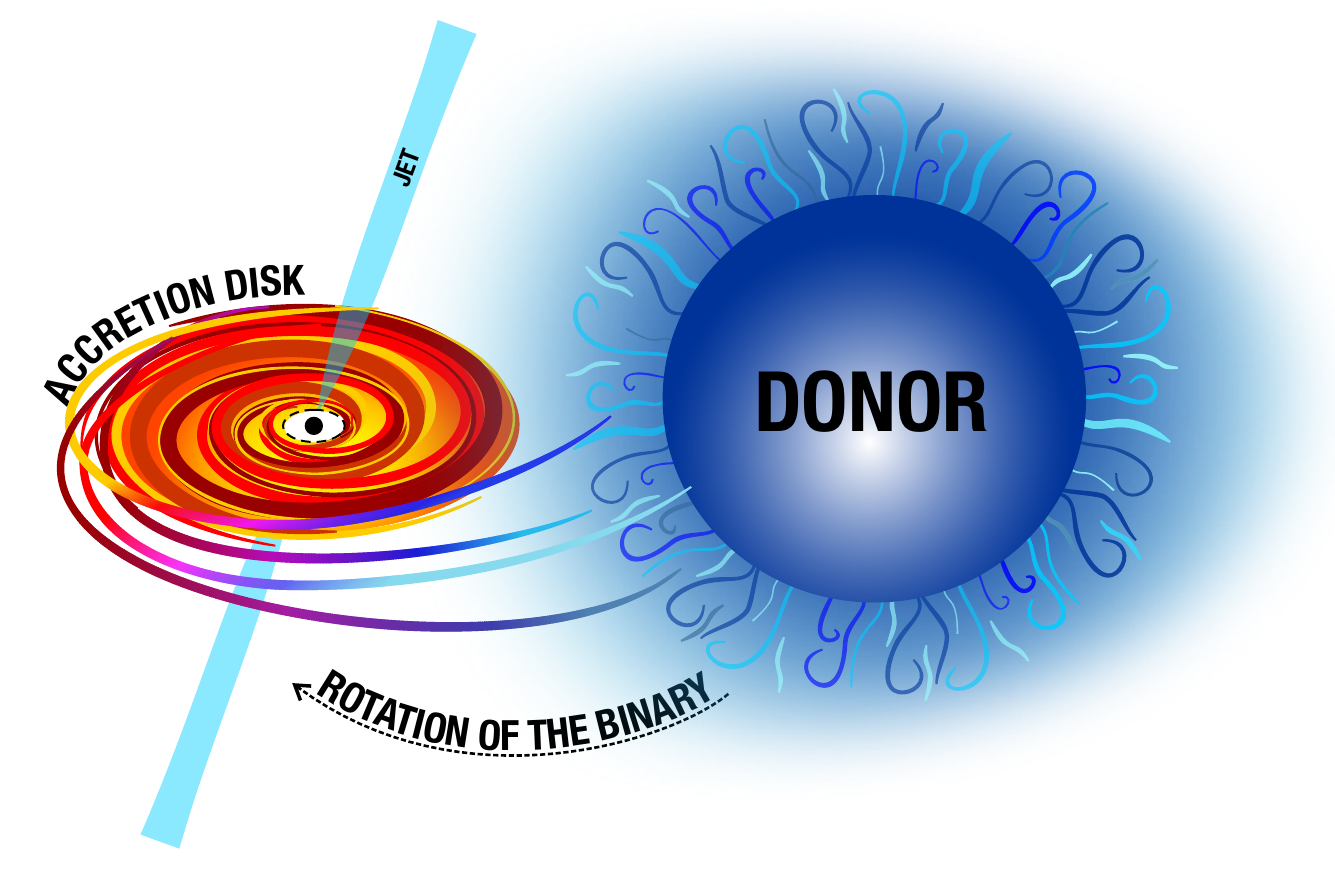}
    \caption{Illustration of typical  HMXB  in a~microquasar source with radio loud jet.}
    \label{fig:1:HMXB}
\end{figure}

The presence of standard spectral lines raised a~new question: How can a~source be so luminous when also exhibiting fast variability that suggests a~relatively compact size? The answer came in the form of accretion onto a~compact object in an AGN. No other process could steadily release such an enormous amount of energy within such a~small volume. At the same time, more luminous sources were found -- blazars and Seyfert galaxies, with similar properties to quasars. Blazars are believed to be AGN observed on-axis, with their jets pointing directly toward Earth \citep{1995PASP..107..803U}.

\subsection{Stellar mass black holes}

Another family of BHs are the stellar-mass (or low-mass) BHs, with mass ranging from a~few to tens of $\msol$. These BHs are formed during the final stage of the life of a very massive star ($M\gtrsim 40\, \msol$) in supernova explosions. If a~stellar-mass BH is a part of a~binary system and accretes matter from its companion star, the system can be easily observed, particularly within our own Galaxy, since it is highly luminous.

Stellar-mass BH systems serve as excellent laboratories for testing the effects of GR in strong gravitational fields, studying the behaviour of matter under extreme conditions, and other aspects of modern astrophysics, including alternative theories of gravity. As such, they have the potential to advance our understanding of fundamental physics and contribute to the development of new theories that can explain some of the most enigmatic phenomena in the Universe.

The mass range of the stellar mass BHs is theoretically determined by the Tolman–Oppenheimer–Volkoff limit on one side, which sets the maximum mass on a~NS at approximately $3\,\msol$ \citep{maxNSmass,Rhoades1974}, which is considered to be the minimal mass of a~BH of stellar origin. The upper mass limit for BHs created through supernova explosions is proportional to the maximum mass of the progenitor star, yielding a~mass limit of around $80\,\msol$  \citep[see, e.g.,][]{Belczynski2010} and references therein). The upper limit existence follows from the fact that supernova explosions of very massive stars are more sensitive to the pair instability, and no remnant is left behind. However, the gravitational wave (GW) observations \citep{Abbott2016} showed that the mass range of BHs (and NSs as well) is much broader than obtained through the EM observations since BHs can also be formed through mergers of less massive objects, see Figure \ref{fig:1:graveyard}.

\section{Binaries with compact objects}

XRBs are one of the brightest sources in our Galaxy. A~typical XRB consists of a~compact object and a~companion star that transfers material onto the compact object (\cite{Remillard2006} and references therein). This material forms an accretion disk, and as it spirals inward, it liberates gravitational energy, leading to the emission of the X-rays in the innermost regions closest to the compact objects.

\begin{figure}[b]
    \centering
    \includegraphics[width=1\linewidth]{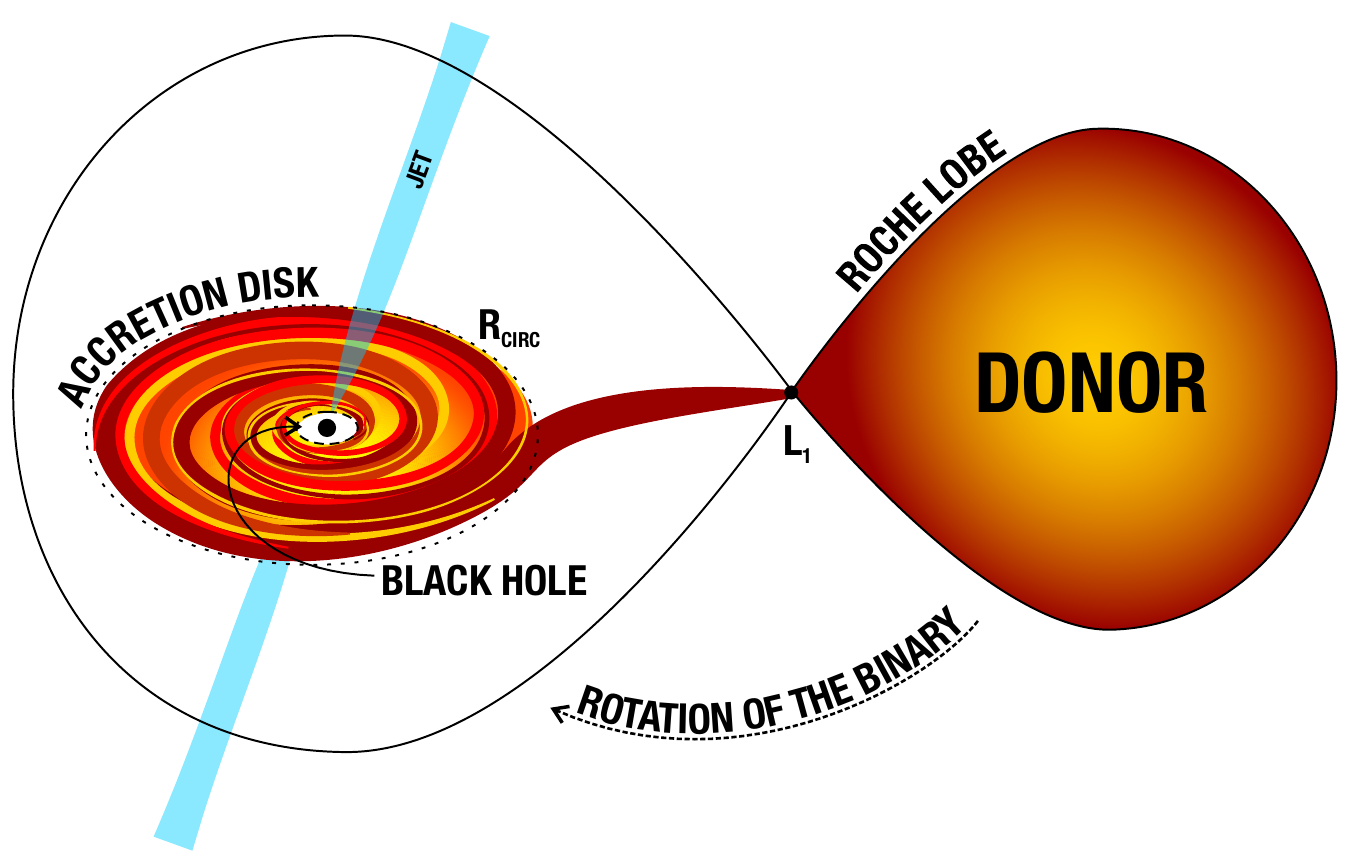}
    \caption{Illustration of typical LMXB in a~microquasar source with a radio-loud jet.}
    \label{fig:1:LMXB}
\end{figure}

XRBs can be divided into two categories based on the mass of the companion and the mechanism of mass transfer: High mass and low mass XRBs \citep[HMXBs and LMXBs,][]{Bradt1983}. The essential properties of both are summarized in Table \ref{tab:1:xbs}, and see also illustration in Figures \ref{fig:1:HMXB} and \ref{fig:1:LMXB}. A~key parameter in these systems is the mass of the \textit{donor} ($M_D$), from which the material is transported onto the compact object, known as the \textit{accretor} of mass $M_A$.

\begin{table}[ht]
\label{tab:1:xbs}
\centering
\begin{adjustbox}{width=1\textwidth}
\renewcommand{\arraystretch}{1.8}
\begin{tabular}{c c c c c c c c} 
        Type & Companion & \makecell{Mass ratio \\ ($M_D/M_A$) }& \makecell{Orbital \\ period}  &  \makecell{Mass transfer\\  mechanism} & Example & Observed in \\ \hline \hline 
        \textbf{HMXB} & Supergiant (O-B) & $>1$ & days & Wind accretion & \texttt{Cygnus X-1} & \makecell{Optical, UV, \\ X-rays} \\ \hline
        \textbf{LMXB} & Main sequence star & $<1$ & hours & \makecell{Roche lobe \\ overflow} & \makecell{\texttt{GRS 1915+105} (BH) \\ \texttt{4U 0614+091} (NS)} &  X-rays \\
        
\end{tabular}
\end{adjustbox}
\caption{Typical properties of LMXBs and HMXBs.} 
\end{table}

In LMXBs, a~low-mass donor star fills its Roche lobe and transfers material onto a~companion compact object through the $\mathrm{L_1}$ point. 
Suppose the LMXB originates from the evolution of a~binary star (as opposed to dynamical interactions of two or three bodies in dense clusters). In that case, the angular momentum of the BH or NS and the accretion disk will likely be parallel to that of the binary system since it was accreted from the disk matter over millions of years of co-evolution of both components.

LMXBs predominantly emit radiation in the X-ray band and are relatively faint in other wavelengths. Due to the source of X-rays lying deep in the strong gravitational field of the compact object, studying LMXBs provides an opportunity to obtain valuable insights into the properties and behaviour of matter under extreme gravitational conditions.


\section{Microquasars and their phenomenology}

Jets from the central regions are frequently observed in the XRBs, which resemble scaled-down quasars. As a~result, these systems are often referred to as microquasars. \citep{MIrabel1992}. 

\begin{figure}[b]
    \centering
    \includegraphics[width=1\linewidth]{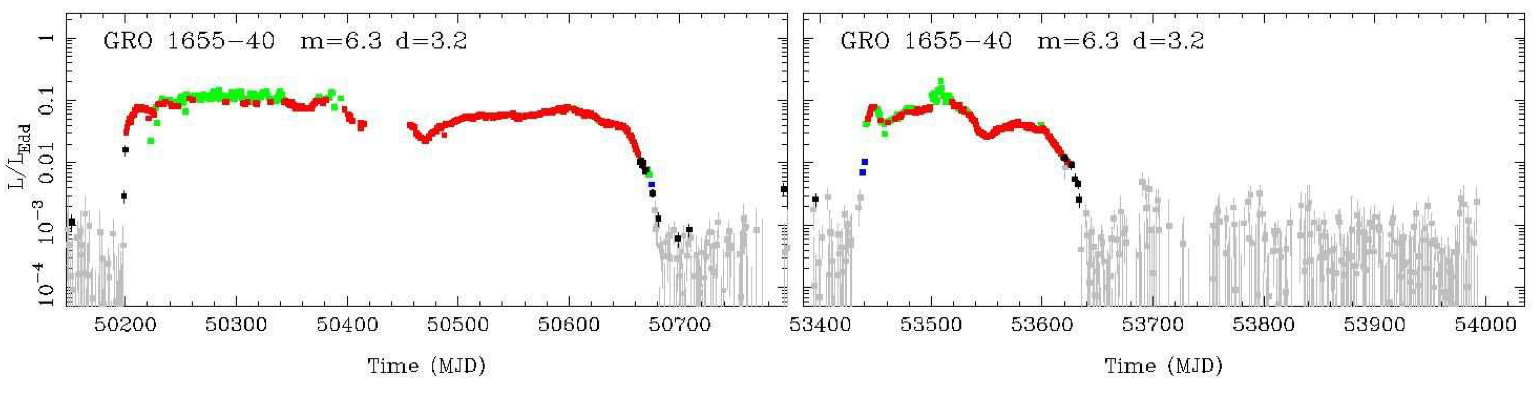}
    \caption{Observed light-curve of typical Galactic microquasar \texttt{GRO 1655-40} showing two episodes of luminous outburst (colours) and quiescent (grey) periods, adopted from \cite{Done2007}.}
    \label{fig:1:LC}
\end{figure}

Microquasars are primarily observed in X-rays, and when jets are present, they can also be detected in the radio wavelengths. The comparison of observational data obtained from these two bands, which originates from very different physical processes,  provides valuable insights into the extreme environment surrounding compact objects. It makes the microquasars a~popular target of observations and a~hot research topic, see \cite{vsechnosecsim}.

BHs microquasars are known to undergo irregular cycles of outbursts and quiescent periods (see Figure \ref{fig:1:outburst}), characterized by distinct spectral and timing properties of the observed signal \citep{Done2007,Tetarenko2016}. During the quiescent phase, which can last for years, the luminosity of the source is extremely low, making some sources undetectable. However, during an outburst, an active period of very high luminosity lasting on average for 100 days \citep{Tetarenko2016}, the spectral and timing properties change rapidly. The evolution of the outburst is often represented by a~\textit{q}-shaped curve on a~hardness-intensity diagram (HID), where the X-ray intensity is plotted against the observed X-rays hardness ratio (typically $(4-10)/(2-4)\,\mathrm{keV}$). The HID traces the X-ray luminosity and energy spectrum variations throughout the outburst; see Figure \ref{fig:1:outburst}. 

This behaviour is characteristic for all known microquasars (most of which are BH LMXBs), regardless of the donor's mass or accretion regime, even in systems with a~NS instead of a~BH at the centre \citep{2014MNRAS.443.3270M}. NS systems are generally more persistent, with relatively high X-ray luminosity, and do not exhibit the transient behaviour typical for BH microquasars. However, several transient NS systems have been observed. 

The evolution of transient NS LMXBs during an outburst appears to be more complex than that of BHs systems, likely due to the presence of a~solid surface and an anchored magnetic field. Some of the spectral states the NS systems exhibit are similar to those of  BH systems \citep{vdKlis2006}, although the curve on the HID differs. In the case of NS systems, a colour-colour diagram (CCD) is often used instead of the HID, which shows the ratio of harder and softer counts against each other.

\begin{figure}[t]
    \centering
    \includegraphics[width=1\linewidth]{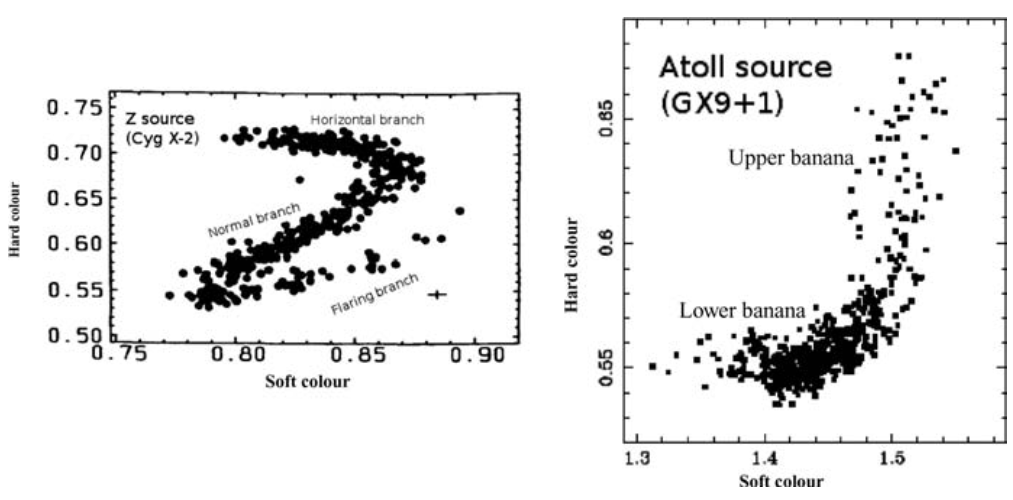}
    \caption{CCD for a~Z source \texttt{Cygnus X-2} (left) and atoll source \texttt{GX9+1}. The different spectral states recognized in both kinds of sources are labelled, adapted from \cite{XrayUniverse}.}
    \label{fig:1:banana}
\end{figure}

NS systems can be classified as Z sources or Atoll sources  \citep{1989A&A...225...79H} based on their luminosity and shape on the CCD, and a~few sources have been observed to transition between these two states  \citep{2009ApJ...696.1257L,XrayUniverse}, see Figure \ref{fig:1:banana}. 

The outburst behaviour is driven by fundamental physical processes occurring in the accretion disk, leading to state transitions regardless of environmental factors. Therefore, modelling the outbursts and the accretion disk during state transitions is crucial for understanding the physics of accretion.

\subsection{Spectral states of black hole microquasars}

\begin{figure}[b]
    \centering
    \includegraphics[width=1\linewidth]{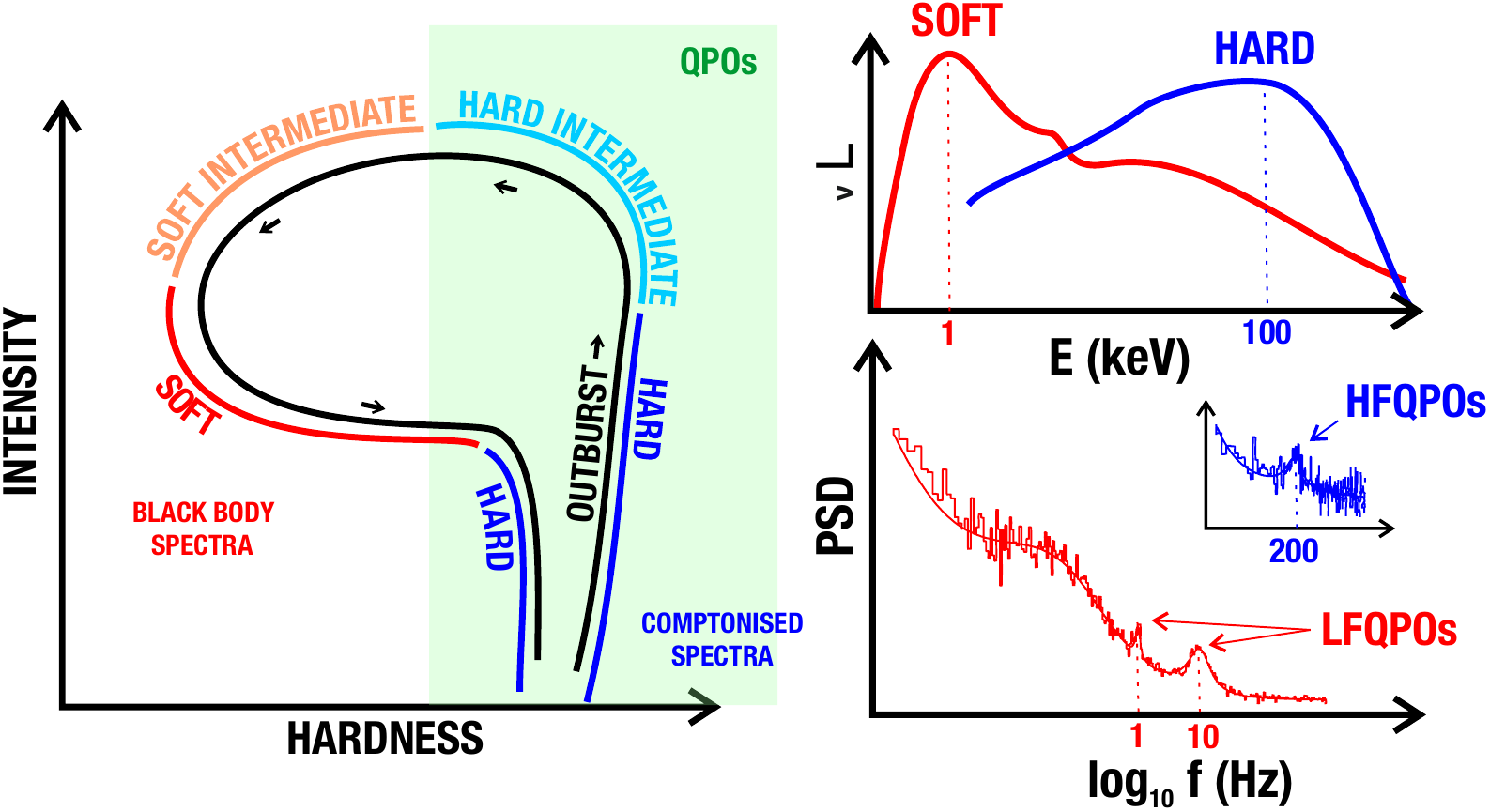}
    \caption{The $q$ curve in the HID of a~BH microquasar during an outburst (left), spectral and timing properties (right). Data for the spectra and PSD from \cite{Gierlinski1999} and \cite{Motta2022}}
    \label{fig:1:outburst}
\end{figure}

Several spectral states can be distinguished during an outburst in BH systems, corresponding to different accretion regimes. The state transitions between these states strictly follow the order shown in Figure \ref{fig:1:outburst}, meaning that the system must go through the hard and intermediate states before reaching the soft state \citep{1995ApJ...442L..13M}. These spectral states are commonly classified into five categories as follows:

\begin{itemize}
    \item \textbf{Hard state}: In the hard state, the majority of emitted photons are in the hard X-ray band, and spectral shape corresponds to a~combination of low-temperature thermal emission and strong non-thermal processes. This state can be modelled by a~truncated thin disk, with a~hot geometrically thick optically thin disk formed in the inner region. The photon energy distribution follows a~power law with a~photon index typically ranging from $1.5$ to $2$ \citep{Zdziarski1996,Zycki1999}, which is usually modelled as upscattering of the thermal seed photons from the truncated disk in the inner optically thin disk or a~corona. The system can remain in the hard state for weeks. Strong variability is observed in this state, especially as a~red noise component plus low frequency (LF) QPOs.

    \item \textbf{Intermediate states}: From the hard state, the system rapidly transitions into intermediate hard and intermediate soft states, distinguished by their timing properties \citep{vdKlis2006}.  The frequency of QPOs typically shifts to higher values as the system moves from the intermediate hard state to the intermediate soft state \citep{Wang},  suggesting that the inner edge of the accretion disk moves inward.

    \item \textbf{Ultra-luminous state}: In some sources, an extremely luminous state is observed \citep{2021MNRAS.503..152M}, characterized by a~thermal spectrum with a~strong, high-energy tail. This state can be modelled, e.g., using the puffy accretion disk  \citep{Lancova2019,Wielgus2022}.
    
    \item\textbf{Soft state}: In the soft state, the peak of the spectral energy distribution (SED) is in the soft X-ray band, and the shape corresponds to a~multi-colour thermal spectrum with a~high-energy tail.  The luminosity is high, particularly in the soft X-ray band \citep{1995ApJ...442L..13M}. This state is modelled as the thermal emission from a~geometrically thin optically thick disk extending to the innermost stable circular orbit (ISCO), with a~Comptonizing corona responsible for the high-energy tail. The time variability is significantly reduced in the soft state.
    
\end{itemize}

An integral aspect of spectral state modelling is a~hot corona, responsible for the up-scattering of photons to very high energies. The geometry of the corona has been the subject of extensive discussion in microquasar astrophysics, with preferred models including the lamp-post, sandwich, or spherical corona models \citep{2017bhlt.book.....B}.  However, recent measurements of polarization angles from XRB observations have placed additional constraints on the corona's geometry \citep{Podgorny2023}.

During a~complete outburst, the source undergoes a~sequence of spectral states, starting from the hard state and transitioning through the intermediate hard and intermediate soft states before reaching the soft state. Then it transitions back to a~less luminous hard state and eventually returns to quiescence\footnote{Originally, these states were labelled as soft/high and hard/low, but this originated in the early stages of the X-ray astronomy when the observatories were unable to detect the hard X-rays emitted during the hard state. It is now understood that the luminosity in both states can be similar.}. 
In some cases, an outburst may not progress beyond the hard or intermediate hard state, as was observed, e.g., in 2019, in a~newly discovered XRB \texttt{MAXI J1348-630} \citep{Alabarta2021}. 

\subsection{Timing properties of XRBs}
\label{sec:1:timing}

\begin{figure}
    \centering
    \includegraphics[width=1\linewidth]{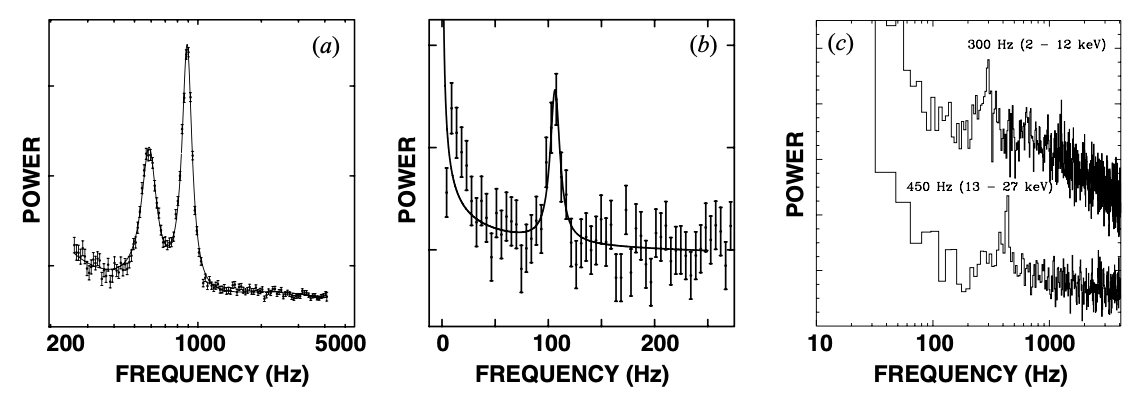}
    \caption{A typical PSD diagram of a~NS source, showing the kHz QPOs (\textit{left}), and LF QPOs (\textit{middle}), and the HF QPOs from a~BH source (\textit{right}). The peaks in the high frequencies are typically weaker in BHs XRBs than those of NS sources. From \cite{vdKlis2006}.}
    \label{fig:1:QPOs}
\end{figure}

XRBs exhibit variability across a~wide range of timescales in all wavelengths, with the fastest variation reaching frequencies of approximately 1 kHz in X-rays. This subsection focuses on rapid variability, particularly on the QPOs, which are observed as strong, broad peaks in the power spectral density (PSD) diagram, obtained as a result of Fourier analysis of observed X-ray light curves (see Figures \ref{fig:1:QPOs} and \ref{fig:1:outburst}). The properties of QPOs evolve in conjunction with spectral state transitions in the accretion disk.

QPOs originate in the innermost and hottest regions of the accretion disk. However, the exact physical mechanism behind QPOs still needs to be fully understood, although several theoretical models have been proposed (a short overview of selected models can be found in Chapter \ref{chap:QPOs}). QPOs provide valuable insight into the accretion processes in binary systems. For instance, the frequency and amplitude of the QPOs can be used to estimate the mass and spin of a~compact object \citep{Motta2014,tor-etal:2022,Kotrlova2020,Gol-etal:2019} or, in the case of a~NS, it can put constraints on the dense matter equation of state, \citep{Urbancova2019,2013MNRAS.433.1903U}, as well as the size and structure of the accretion disk, and the properties of the surrounding gas. Thus, QPOs can serve as a~valuable tool for studying the physics of extreme environments.

QPOs are observed across a~wide range of frequencies, from mHz to kHz. QPOs are typically classified based on their characteristic frequencies and the type of source in which they are observed. The most common classification distinguishes low-frequency (LF) and high-frequency (HF) QPOs. LF QPOs in BH systems are usually labelled as Type A, B, or C,  while in NS systems, they are referred to as HBO (horizontal branch oscillations), NBO (normal branch oscillations), and FBO (flaring branch oscillations) \citep{vdKlis2004,Casella2005}. HF QPOs are primarily associated with BH sources, whereas in NS sources, they are commonly known as kHz QPOs, or the twin-peak QPOs. Observed frequencies of HF QPOs can be directly linked to the characteristic frequencies of orbital motion in the innermost areas of an accretion disk surrounding a~compact object, where the GR effects govern the dynamics

Another intriguing phenomenon is the burst oscillations observed in bursting NS systems. These oscillations have frequencies that closely match the rotational frequency of the NS itself. They are observed in a~few sources exhibiting coherent pulsations due to magnetospheric accretion. Moreover, QPOs with mHz frequencies were also discovered, likely linked to unstable nuclear reactions on the surface of NS \citep{Revnivtsev2001}.

Figure \ref{fig:1:QPOs} demonstrates the significant differences between BH and NS HF QPO signals. However, there are also many similarities. The kHz QPOs in NS sources are typically observed as two strong simultaneous peaks with strongly correlated lower and higher frequency variations. These peaks are distinct and prominent, clearly rising above the noise level.  In contrast, HF QPOs in BH systems are more challenging to observe and have been detected as simultaneous pairs in only a~few cases. In the case of the NS sources, the properties of observed QPOs can be explained by the source of oscillations connected to the boundary layer on a~NS surface and modulation of its luminosity by oscillations of accreting structures in the innermost region close to the NS affected by the quadrupole momentum of the NS \citep{2022arXiv220310653M}. In  BH sources, the absence of a~boundary layer results in matter from the innermost regions of the disk drifting toward the event horizon, making it challenging for emitted radiation to escape the powerful gravitational pull.

%
%
%

\chapter{Physics of accretion disks}
\label{chap:AD}

Apart from annihilation, accretion is the most energetic process in the universe, capable of releasing up to 42\% of the binding rest-mass energy from the accreting material in the form of EM radiation. Due to the extremely high temperatures in the innermost area of the disk,  which can reach up to $10^8$ K, it is mainly in the X-ray band. Consequently, the study of accretion is a~hot topic in high-energy astrophysics, particularly when it involves sources with compact objects, especially BHs. 

Accretion plays a~crucial role in understanding the mechanisms driving the growth of celestial objects and is an essential process in studying the formation and behaviour of BHs. However, accretion physics is a~complex research field requiring various physical effects from different disciplines. These include plasma physics, hydrodynamics, electrodynamics, and GR. Furthermore, implementing advanced computational and numerical methods is essential for the realistic modelling of accreting systems.

In an accretion disk, the material experiences strong gravitational forces from the central object, and the presence of a~strong magnetic field can significantly influence the disk's structure, dynamics, and the generation of powerful jets observable in radio wavelengths.  Moreover, nuclear physics is crucial for understanding the behaviour of matter at high densities and temperatures. Finally, GR is necessary to accurately describe the behaviour of matter in the strong gravitational field near the central compact object, accounting for relativistic effects such as frame dragging and gravitational lensing.

Incorporating these effects into a~realistic model of an accretion disk presents a~major challenge that requires a~multi-disciplinary approach and advanced computational techniques. However, the potential rewards are significant, as a~deeper understanding of accretion disks can provide valuable insights into the formation and evolution of BHs and the behaviour of matter under extreme conditions. This chapter introduces analytical models of accretion within the framework of GR,  with a~specific focus on accretion disks in galactic XRBs.

\section{Spherical accretion and the Eddington limit}

The simplest accretion case is the spherical one, also known as the Bondi accretion, first described by \cite{Bondi}. It calculates the spherically symmetric accretion of a~homogeneous gas cloud surrounding a~gravitating body. Further studies by \cite{Parker1958}  investigated the spherically symmetric outflow, describing phenomena such as solar winds. 

The description of spherical accretion allows the definition of several fundamental characteristics that can also describe general properties of accretion flows in different regimes. One such characteristic is the mass accretion rate, denoted as $\Mdot$, representing the rate at which material falls onto the central object. The mass accretion rate is proportional to the radial velocity of the gas and its density, given by $\Mdot \propto -v^r\rho$, where the negative sign  indicates that a~positive value of $\Mdot$ represents accretion, while a~negative value represents outflow.

Assuming a~steady spherical accretion onto a~radiating body, a~maximal value of the luminosity of the central object can be established, at which the radiation pressure balances the gravitational pull on the infalling matter. Beyond this critical luminosity, spherical accretion cannot take place. This limit is the Eddington luminosity, one of the most essential concepts in astrophysics. The exact value can be derived considering a~radiating body of mass $M$ and luminosity $L$, surrounded by a~cloud of fully ionized hydrogen gas. Then the effective cross-section is taken as the Thomson cross-section $\sigma_\mathrm{T}$, and the proton mass $m_{\mathrm{p^+}}$  is used as the mass of the hydrogen atoms. By balancing the radiation force on each electron at a~radius $r$,
\begin{equation}
    F_r = \frac{\sigma_\mathrm{T}L}{4\pi c r^2}
\end{equation}
\noindent with the gravitational pull on a~hydrogen core at the same radius,
\begin{equation}
    F_g = \frac{GMm_{\mathrm{p^+}}}{r^2},
\end{equation}
\noindent where $c$ is the speed of light and $G$ the gravitational constant,  we can derive the value of the Eddington luminosity $L_{\mathrm{Edd}}$ using the electron scattering opacity $\kappa_{es}=\sigma_\mathrm{T}/m_{\mathrm{p^+}}$, as
\begin{equation}
    L_{\mathrm{Edd}} = \frac{4\pi c GM}{\kappa_{es}}=1.3\times 10^{38}\left(\frac{M}{\msol}\right)\,\mathrm{erg\cdot s^{-1}}.
\end{equation}

\noindent From this value, a~critical $\dot{M}_{\mathrm{Edd}}$ can be derived,
\begin{equation}
    \dot{M}_{\mathrm{Edd}} = \frac{L_{\mathrm{Edd}}}{\eta c^2},
\end{equation}

\noindent where $\eta$ is the energy conversion efficiency. In the case of the BH accretion, $\eta$ can be derived from the specific angular momentum of the BH, reaching values up to $\eta=0.42$ for Keplerian rotation of the disk and maximally rotating BH. It is common to use $\eta=1$, which is also used in this thesis.

However, the GR corrections are neglected in these calculations. The strong gravity influences the photon trajectories, which leads to the formation of an Eddington capture sphere, studied in \cite{2012A&A...546A..54S}. Effects of the radiation from a~compact star on the motion of a~test particle were also studied in \cite{2019PhRvD..99b3014D} and for a~rotating star in \cite{2019PhRvD.100j4053B}\footnote{I was a~co-author of these two papers; however, they are not a~part of the collections of papers commented in this thesis, since their topic is far from the main subject of this thesis.}.

Although the $\Ledd$ derivation is based on simple assumptions, the value is also valid for disk accretion. Nevertheless, the accretion can also become super-Eddington in the case of geometrically thick accreting tori \citep{Jarosz1980,Abramowicz+1978}, as discussed in Section \ref{sec:2:overwiev}. 

\section{Accretion in strong gravity}

BHs are very simple objects, as Chandrasekhar said in his famous monograph \cite{chandra}. They can be described by only three parameters: mass $M$, specific angular momentum (spin) $a$, and electric charge $Q$. As a~result, simple units can be used to describe BH spacetime across the whole range of masses, such as the gravitational radius, $\rg = GM/c^2$, for length and $\tg = \rg/c$ for time. These geometrized units will be used from here forward, laying $G=c=1$.

Parameters describing the accretion flow can be written in simple forms using this units, 

\begin{align*}
    m&=\frac{M}{\msol} & \dot{m} &= \frac{\dot{M}}{\Medd}  & r &= \frac{r_*}{\rg} \\
\end{align*}

\noindent where $M$ is the mass of the compact object, $\dot{M}$ is the mass-accretion rate, and $r_*$ is the radius in cm. 

\begin{figure}
    \centering
    \includegraphics[width=0.9\linewidth]{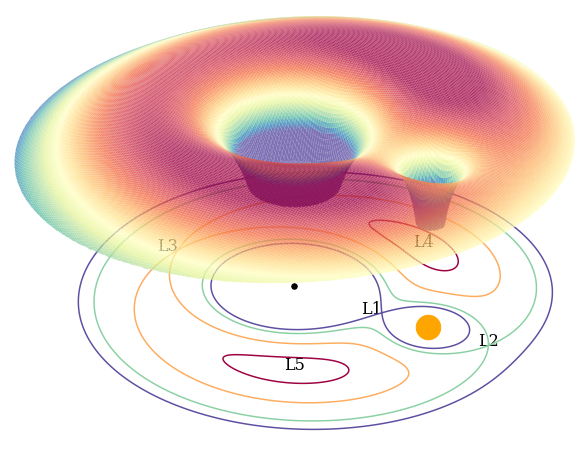}
    \caption{Roche potential of two orbiting bodies with a~mass ratio $q=0.1$, representing a~BH and a~main sequence star in LMXBs. The bottom part of the figure shows the contours of the Roche potential, with the Lagrange points $\mathrm{L_1}$-$\mathrm{L_5}$ highlighted. The potential wells are cut off to enhance the bottom part's visibility.}
    \label{fig:2:roche}
\end{figure}

In a~LMXB, an accretion disk forms when a~companion star fills up its Roche lobe (see Figure \ref{fig:2:roche}), and its material starts to flow onto the compact object through the $\mathrm{L_1}$ point. A~Roche lobe is the largest closed equipotential surface of the Roche potential, which combines the effects of gravitational forces acting on components of a~binary system, the gravitational pull and centrifugal force. The companion can fill the Roche lobe during the final stage of life of the companion star as it depletes its hydrogen reserves and begins burning heavier elements like helium, or if the orbit shrinks due to the binary evolution.

The matter from the companion object has a~non-zero angular momentum, and the conservation laws prevent it from radially falling into the centre. Instead, it forms a~disk-like structure in the rotational plane of the binary system. Within this accretion disk, thermal and viscous processes lead to the angular momentum transport outward, gradually causing the material to lower its orbit until it reaches the inner edge of the disk. Nevertheless, the situation is more complex in the case of a~NS disk and will be discussed below.

At the $\mathrm{L_1}$ point, the material from the companion star is injected into the potential well of the accretor with high angular momentum and settles gradually into the lowest energy state, which corresponds to a~Keplerian circular orbit with the same angular momentum as the gas had when it passed through the $\mathrm{L_1}$ point.The radius of this orbit, known as the circularization radius $R_\mathrm{circ}$, can be easily derived from the mass ratio $q$ and the separation $s$ of the binary system as

\begin{equation}
    R_\mathrm{circ}/s = \left(1+q\right)\left(0.5 - 0.227 \log q\right)^4, 
\end{equation}

\noindent using a~fitted formula for the location of the $\mathrm{L_1}$ point. The $R_\mathrm{circ}$ corresponds to the outer edge of an accretion disk \citep{AccretionPower}, see also Figure \ref{fig:1:LMXB}.

\subsection{Inner edge of an accretion disk}

Accretion disks around standard stars or white dwarfs are expected to extend all the way to the surface of the central object. In the case of compact objects, the accretion disk is strongly influenced by strong gravitational and magnetic fields and fast rotation. In NS systems, the disk can reach the surface of the NS under certain conditions, and it is typically truncated further away due to interactions with the NS's magnetic field.

\begin{figure}[t]
    \centering
    \includegraphics[width=1\linewidth]{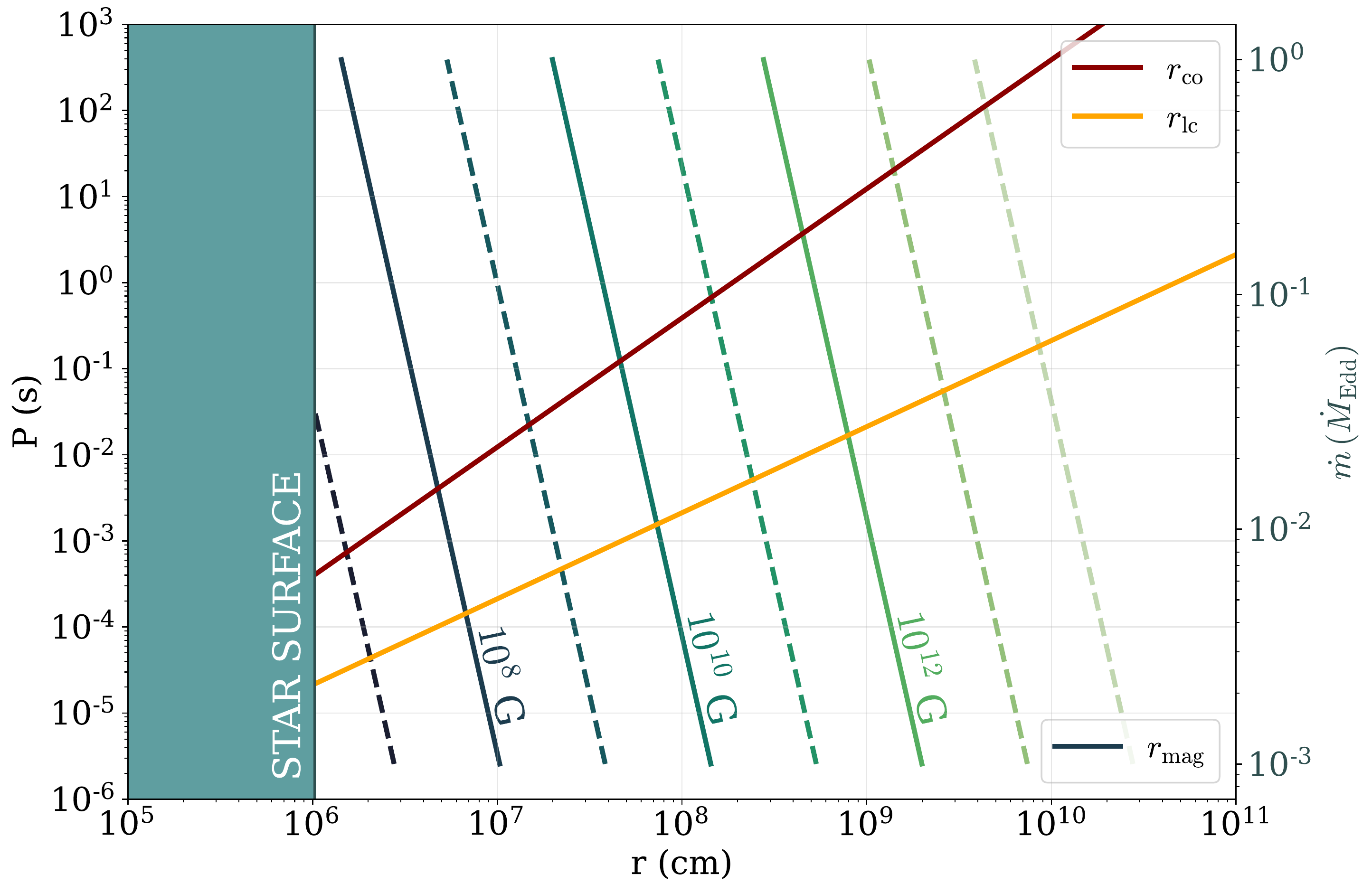}
    \caption{The three important radii in a~NS system: The $r_{\mathrm{co}}$ and $r_{\mathrm{lc}}$ are plotted as a~function of the period on the left axis, while the $r_{\mathrm{mag}}$ is shown as a~function of luminosity on the right axis for  magnetic field strength $B_*$ ranging from $10^{7}$ to $10^{14}$ G. The calculations are performed for a~NS with $M=1.4\,\msol$ and  $R_* = 10^6\,\mathrm{cm}$ in the Newtonian gravity.}
    \label{fig:2:NSradii}
\end{figure}

Three important radii in NS systems determine the position of the inner edge of the accretion disk and the accretion regime. The position of these radii depends on the parameter of the NS itself and the accretion flow.  These radii are the light cylinder radius $r_{\mathrm{lc}}$, the corotation radius $r_{\mathrm{co}}$, and the magnetospheric radius $r_{\mathrm{mag}}$. Accretion can only occur if the following conditions are met: $r_{\mathrm{mag}} < r_{\mathrm{co}} < r_{\mathrm{lc}}$, see Figure \ref{fig:2:NSradii}.

The three radii can be derived as \cite{AccretionPower}

\begin{align}
    &r_{\mathrm{lc}} = \left(4.709\times10^{10}\,\mathrm{cm}\right) P, \\
    &r_{\mathrm{co}} = \left(1.5\times10^{8}\,\mathrm{cm}\right) m^{1/3} P^{2/3}, \\
    &r_{\mathrm{mag}} = \left(2.9\times10^{8}\,\mathrm{cm}\right) m^{-1/7} \mu_{30}^{4/7} \mdot^{-2/7}, \label{eq:2:rmag}
\end{align}

\noindent where $P$ is the rotation period of the NS with mass $m$. The expression for the $r_{\mathrm{mag}}$ is given for a~dipole magnetic field on the NS surface and other assumptions. $\mu_{30}$ is there the magnetic moment $\mu=\frac{B_*}{R_*^3}$ in the units of $10^{30}\,\mathrm{G\cdot cm^3}$, $B_*$ is the magnetic field strength at the surface of NS, and $R_*$ is the star's radius. The $r_{\mathrm{mag}}$ corresponds to the location where the ram pressure in the accretion disk balances the magnetic pressure of the NS's magnetic field, and it indicates the inner edge of the accretion disk, see also Figure \ref{fig:chap1:NS}.

NS systems can exhibit three regimes. In the accreting regime, where $r_{\mathrm{mag}} < r_{\mathrm{co}}$, the angular velocity of the disk is lower than that of the NS,  allowing the material to accrete onto the NS's surface. In the propeller regime, where $r_{\mathrm{co}} < r_{\mathrm{mag}} < r_{\mathrm{lc}}$, the disk's angular velocity is higher than that of the NS, and the material is ejected from the system or accumulated on the $r_{\mathrm{mag}}$. Lastly, in the radio pulsar regime, where $r_{\mathrm{mag}} > r_{\mathrm{lc}}$, no accretion can take place. 

In the BH systems,  where there is no hard surface or anchored magnetic field of the central object, the position of the inner edge of the accretion disk is solely determined by the system's geometry, which is governed by the GR effects. An important effect of the GR is the existence of a~last stable orbit -- a~specific radius below which no stable circular orbit can exist. This orbit is usually labelled as ISCO, the innermost stable circular orbit. In the analytical model of a~thin accretion disk, the ISCO is commonly considered as the inner edge of the accretion disk, based on the assumption that the shear of the fluid drops rapidly on ISCO and the torque vanishes. These models assume that the material is too hot in the plunging region (between ISCO and the horizon), the flow is there laminar, and cannot contribute to the observed thermal spectra. The position of ISCO depends purely on the properties of the BH spacetime, such as its mass or angular momentum, see equation (\ref{eq:2:kerrisco}).

\section{Black hole spacetimes}
\label{sec:2:metrics}

The framework of GR is necessary to describe the vicinity of BHs or other compact objects due to the strong curvature of spacetime caused by their extreme compactness. GR describes curved spacetimes using the metric tensor $g_{\mu\nu}$  \citep{1973grav.book.....M}. The spacetime interval between two infinitesimally close events with spacetime coordinates $(x)^A$ and $(x)^B$ can be calculated using the metric tensor $g_{\mu\nu}$ as

\begin{equation}
\der s^2 = g_{\mu\nu} \der x^\mu \der x^\nu,
\end{equation}

\noindent where $\der x=(x)^A - (x)^B$ and using the Einstein notation
\begin{equation}
    x_\mu x^\mu = \sum_{\mu=0}^{N-1} x_\mu x^\mu  = x_0 x^0 + x_1 x^1 + ... +x_N x^N,
\end{equation}

\noindent where N is the dimension of the vectors. 

The spacetime interval is independent of the choice of coordinates, and the metric tensor provides us with complete information about the curvature of spacetime and how to measure distances at a~given point.

\subsection{Static spacetime - Schwarzschild metric}

The \schw{} metric describes a~non-rotating BH with zero charge \citep{1916AbhKP1916..189S}. The metric is static, spherically symmetric, and in the \schw{} coordinates $\left(t,r,\theta,\phi\right)$, using the  $\left(-,+,+,+\right)$ (space-like) signature\footnote{which is used everywhere within this thesis}, the spacetime element is given by

\begin{equation}
    \der s^2 = -\left(1-\frac{2M}{r}\right) \der t^2 + \left(1-\frac{2M}{r}\right)^{-1}\der r^2 + r^2 \der \theta^2 + \left(r^2  \sin^2\theta\right)\der \phi^2.
    \label{eq:2:schw}
\end{equation}

In these coordinates, the metric becomes singular at the gravitational radius $r_{Schw}=2\,\rg$, where $g_{tt} = 0$, which also corresponds to the surface of infinite redshift. The spacetime is within this surface causally disconnected, so this surface is referred to as the event horizon. However, this singularity is only a~coordinate one and can be eliminated by choosing a~different coordinate system, unlike the singularity at $r=0\,\rg$ corresponding to $g_{rr}=0$, which is physical.  

Several significant surfaces can be found for the geodesical motion of a~test particle around a~\schw{} BH. Apart from the ISCO at $\risco = 6\,\rg$, there is the marginally bound orbit (circular but unstable) at $r_{\mathrm{mb}} = 4\,\rg$, and the photon orbit $r_{\mathrm{ph}} = 3\,\rg$, an unstable circular orbit for mass-less particles like photons.

\subsection{Rotating black holes: the Kerr metric}

Both EM and GW observations suggest that most BHs, especially stellar-mass BHs, are rapidly rotating  \citep{Reynolds2021}. A~rotating BH can be described using the Kerr metric, based on a~solution of the Einstein equations for a~rotating BH with zero charge in vacuum \citep{kerr:1963}. The rotation is parameterized with a~dimensionless spin parameter $a = J/M^2$, where $J$ is the BH angular momentum. The most convenient coordinate system for the Kerr metric is the Boyer-Lindquist (BL) coordinates $\left(t,r,\theta,\phi\right)$, the spacetime element is then given by

\begin{equation}
    \der s^2 = - \left(1 - \frac{2 M r}{\Sigma}\right) \der t^2 - \frac{4 M a r}{\Sigma}\sin^2\theta\der t \der \phi + \frac{\Sigma}{\Delta} \der r^2 + \Sigma\der \theta^2 + \frac{A}{\Sigma}\sin^2\theta \der \phi,
    \label{eq:2:kerr}
\end{equation}

\noindent where
\begin{align}
    \Sigma &= r^2 + a^2\cos^2\theta, \\
    \Delta & = r^2 - 2 Mr + a^2,\\
    A & =  \left(r^2 + a^2\right)^2 - a^2\Delta\sin^2\theta,
\end{align}

\noindent which reduces to the \schw{} metric for $a = 0$ and BL coordinates reduces to the \schw{} coordinates. 

Two event horizons (surfaces where $g^{rr}=0$) exist for the Kerr metric, $r_h^\pm$, where ($+$)  denotes the outer and ($-$) the inner horizon. These horizons do not coincide with the surfaces of infinite redshift, the ergosphere $r_{e}^\pm$, where $g_{tt} = 0$:

\begin{align}
    r_{\mathrm{e}}^\pm &= M \pm \sqrt{M^2 - a^2\cos^2\theta},  \\
    r_{\mathrm{h}}^\pm & = M \pm \sqrt{M^2 - a^2}.
\end{align}

These surfaces, specifically $r_{e}^+$ and $r_{h}^+$, define the ergoregion, an area where no static observers can exist and the spacetime corotates with the BH. This effect extends the boundary of the ergoregion, causing the spacetime to be dragged into corotation with the BH. This phenomenon is known as frame dragging. Consequently, a~frame of an observer with zero angular momentum (ZAMO) can be found, which appears to orbit the BH for an observer at infinity, even though its angular momentum is locally zero.

The existence of the ergosphere allows for the extraction of rotational energy from the BH \citep{Penrose1971, Blandford1977}, powering jets in AGN and microquasars, which has also been confirmed in numerical simulations \citep{Tchekhovskoy2011}.

\begin{figure}
    \centering
    \includegraphics[width=\linewidth]{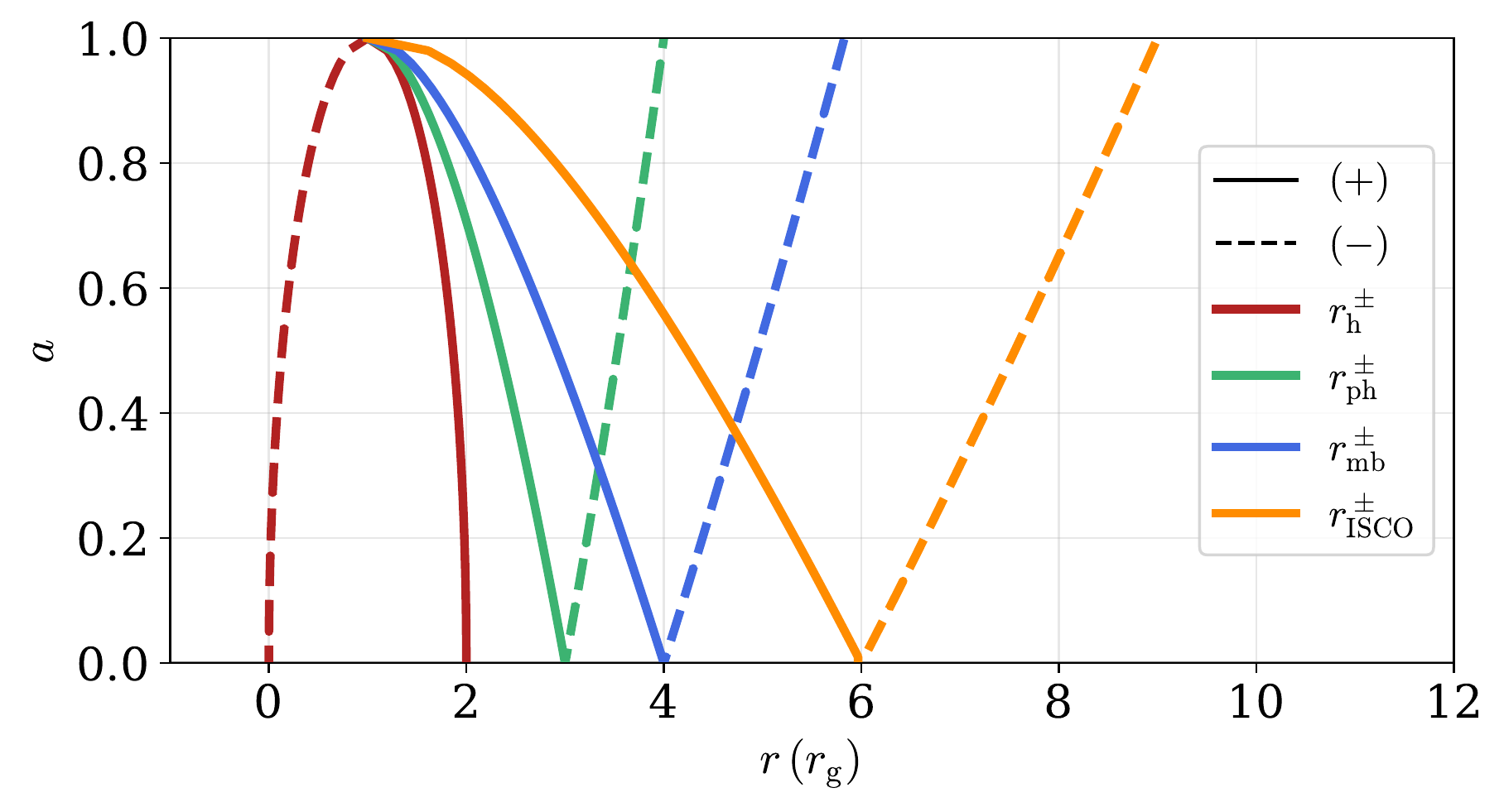}
    \caption{The inner $r_{\mathrm{h}}^-$ and outer $r_{\mathrm{h}}^+$ event horizons and important radii of the Kerr metric at the equatorial plane as a~function of $a$. The full lines represent the prograde motion, while the dashed lines represent the retrograde motion (or the inner and outer event horizon). The values transition to the Schwarzschild metric ones at $a=0$.}
    \label{fig:2:impr}
\end{figure}

Describing the geodesic motion in the vicinity of a~rotating BH becomes significantly more complex than the static one, particularly for non-equatorial motion. It is important to distinguish between prograde ($+$) and retrograde ($-$) motion with respect to the rotation of the BH. The radii of important orbits as a~function of $a$ are shown in Figure \ref{fig:2:impr}, where for $a=0$, the values transition to those corresponding to a~non-rotating BH. The expressions on the equatorial plane are given by

\begin{align}
    &r_{\mathrm{ph}}^\pm  = 2M\left\{1+\cos\left[\frac{2}{3}\cos^{-1}\left(\mp\frac{a}{M}\right)\right]\right\}  ,\\
    &r_{\mathrm{mb}}^\pm  = 2M\mp a + 2\sqrt{M\left(M\mp a\right)},\\
    &\risco^\pm  = M\left[3 + Z_2 \mp \sqrt{\left(3-Z_1\right)\left(3+Z_1+2Z_2\right)}\right], \label{eq:2:kerrisco}
\end{align}

\noindent where

\begin{align}
    & Z_1 = 1 + \left(1-a^2\right)^\frac{1}{3} \left[\left(1+a\right)^\frac{1}{3} + \left(1-a\right)^\frac{1}{3}\right], \\
    &Z_2 = \sqrt{3a^2 + Z_1^2}.
\end{align}

These radii are important for modelling the structure and temporal variability of accretion disks. The photon orbit radius provides insights into the properties of the compact object that can be used to test the predictions of GR, which may become possible in the era of BH imaging within the Event Horizon Telescope (EHT) collaboration, \citep{Paugnat2022,Wielgus2020}. The ISCO, on the other hand, is often associated with the inner edge of the accretion disk and can be inferred from observations using the spectral-based fitting methods \citep{CF,reflection}, enabling measurements of the mass and spin of a~compact object. Finally, the location of the marginally bound orbit determines the maximal radial extension of the marginally overflowing accretion tori. In principle, comparison of the expected frequencies of oscillations of these tori with frequencies of observed HF QPOs allows for estimation of the mass and spin of the central object as well \citep{tor-etal:2022}.

In most cases, the Kerr or even the \schw{} metric is sufficient to describe accretion disks' structure and observational pictures in the case of the BH system. For the NS case, the \schw{} metric is generally acceptable, although the Hartle-Thorne metric \citep{HartleThorne1968} is used when considering the effects of the star's quadrupole moment. The Hartle-Thorne metric describes the spacetime around a~compact, rigidly rotating, axially symmetric body accurately to the second order of angular momentum of the star and the first order of its quadrupole moment.

\section{Analytical models of accretion disks}

Analytical solutions for a~flow of plasma in an accretion disk are derived based on the energy, mass, and momentum conservation principles. However, in order to obtain these solutions, several assumptions are typically made. These assumptions include:
\begin{itemize}
    \item Neglecting self-gravitation effects, assuming that the gravitational influence of the disk on itself is negligible.
    \item Assuming that the disk midplane lies in the equatorial plane of the central object.
    \item Using the ideal gas approximation.
    \item Assuming a~steady and axisymmetric disk ($\partial/\partial t = 0$, $\partial/\partial \phi = 0$).
    \item Neglecting the effects of a~large-scale magnetic field.
    \item Assuming hydrostatic equilibrium in the vertical direction.
    \item Assuming Keplerian motion of the fluid, setting $\Omega = \Omega_K = \sqrt{\frac{GM}{r^3}}$,
\end{itemize}

\noindent These assumptions lead to analytical solutions for the disk structure. However, these solutions may not fully capture the complexities of real-world accretion processes, and more sophisticated numerical simulations are often needed for a~more accurate description.

The first analytical model of accretion disk was introduced by \cite{Shakura+Sunyaev1973}, who described a~geometrically thin, optically thick disk. In the same year, \cite{Novikov+Thorne1973} and then \cite{Page1974} extended the solution to include the effects of GR, providing corrections for a~rotating Kerr BH.

The thin disk model assumes a~small vertical extension. It is radiatively efficient, and the inner edge is located at $\risco$, where the boundary condition is such that the stress is zero. The thin disk quickly gained popularity and has been widely used to fit observational data from sources with compact objects \citep{kerrbb}. In this model, the viscosity responsible for the transport of angular momentum is parameterized by the viscous coefficient $\alpha$, the viscous stress $t_{r\phi}$ in the disk is assumed to be proportional to the pressure $p$,
\begin{equation}
    t_{r\phi} = \alpha p.
\end{equation}

The $\alpha$ parametrization used in the thin disk model operates well even though the specific source of viscosity remained unknown for quite a~long time. For many years, various possibilities were considered to explain the origin of viscosity in astrophysical plasma. It was not until almost 20 years after the thin disk model was constructed that \cite{Balbus1991,Balbus1998} introduced the concept of the magnetorotational instability (MRI) as a~potential source of viscosity. The observations of accretion in astrophysical sources showed that the value of $\alpha$ is typically in the range between $\alpha=0.01-0.1$ \citep[e.g.][]{2011ApJ...742...85G}. Similar values were obtained in numerical simulations \citep{2010MNRAS.408..752P}. However, the simulations showed that the value of $\alpha$ is far from constant in the disk, with a~distinct peak in the inner parts of the disk and the plunging region \citep{2013MNRAS.428.2255P}.

\subsection{Magnetorotational instability}

The MRI is an instability occurring in magnetized rotating systems, such as accretion disks. It is caused by the differential rotation of the ionized gas carrying a~low-scale magnetic field. The stretching of magnetic field lines gives rise to turbulences in the accretion flow and carries the angular momentum outward, even in the case of a~very weak magnetic field. 

\begin{figure}[t]
    \centering\includegraphics[width=1\linewidth]{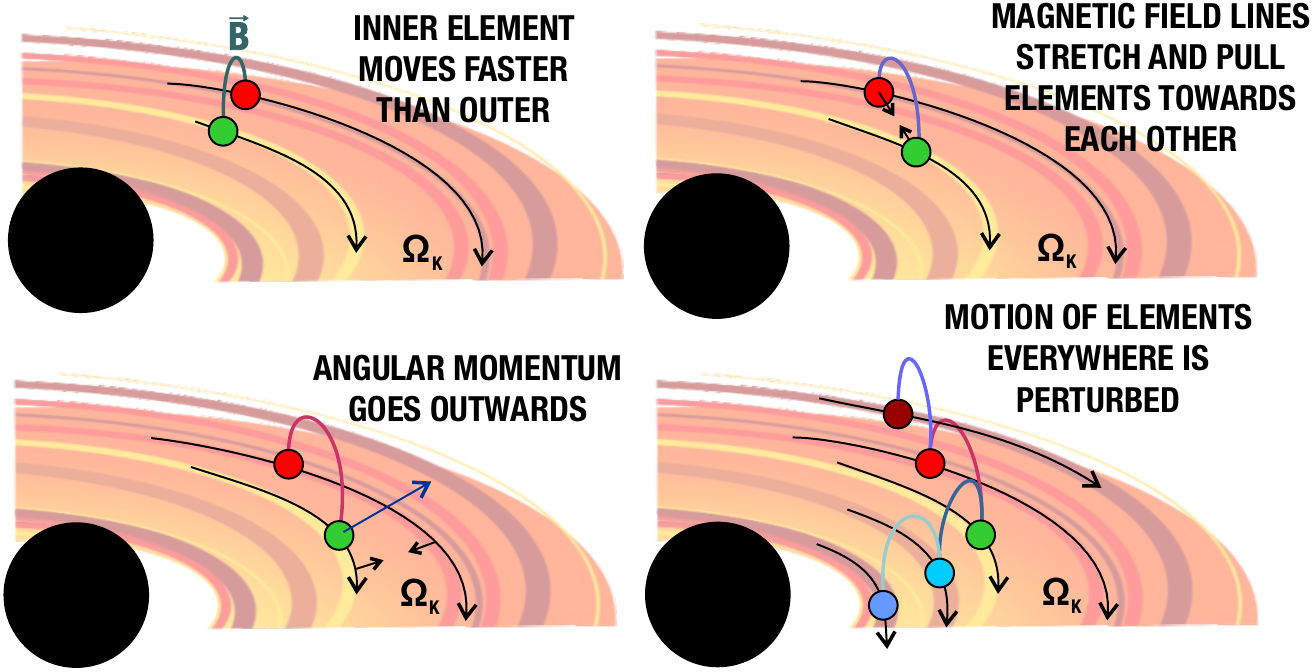}
    \caption{Schematic explanation of the MRI}
    \label{fig:2:MRI}
\end{figure}

In the ideal MHD approximation, where the fluid is considered a~perfect conductor, the electric field vanishes in the frame comoving with the fluid, and the magnetic field lines are frozen into the fluid. As a~result, the field lines act as springs connecting two fluid elements, one on a~larger orbit with lower angular velocity and the other on a~lower orbit with higher angular velocity. As the fluid orbits, the inner element moves away from the outer one due to the differential rotation, causing the magnetic spring to stretch and pull the inner element back. Consequently, the inner element loses angular momentum and slows down, while the outer element gains this angular momentum and speeds up. This process occurs continuously throughout the disk, gradually transferring angular momentum outward and rising turbulences throughout the fluid. Figure \ref{fig:2:MRI} shows a~schematic picture of the MRI.

The MRI mechanism is often employed in MHD simulation to trigger turbulences and force the matter to flow towards the central object without implementing an artificial viscosity term into the source terms of the flow equations. Simulations showed that the $\alpha$ disk assumptions agree with the simulations results, e.g. in \cite{2012MNRAS.420..684P} or \cite{2013MNRAS.428.2255P}. Even though  MRI is not the only magnetic instability present in a~disk, it matches the $\alpha$ viscosity in magnitude. The value of $\alpha$ derived from simulations is lower in the disk body by order of magnitude than fitted from observations using an analytical model, which, however, assumes the value of $\alpha$ constant throughout the whole disk. 

The presence of a~magnetic field in the disk is also likely to play a~crucial role in stabilizing the luminous regimes of the thin disk model, as extensively discussed and analytically explored \citep{Begelman2007,Oda2009}, and confirmed in simulations \citep{Sadowski2016,Lancova2019,Mishra2022,Huang2023}.

\section{Overview of analytical models}
\label{sec:2:overwiev}

\begin{figure}
    \centering
    \includegraphics[width=\linewidth]{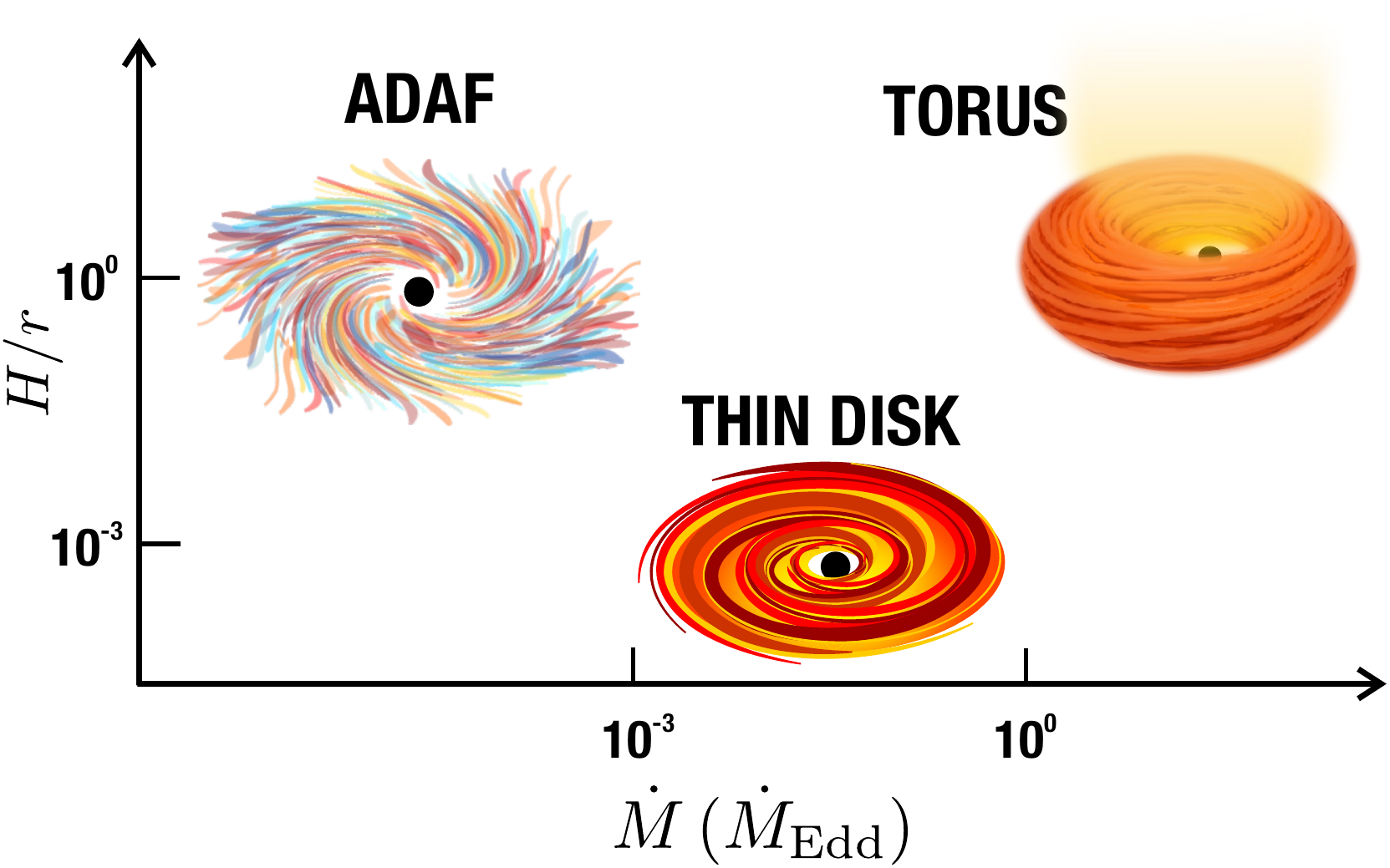}
    \caption{Illustration of three accretion disk models for various mass accretion rates and disk thicknesses.}
    \label{fig:2:disks}
\end{figure}

Many analytical models of accretion disks have been proposed over the years, corresponding to different accretion regimes, typically characterized by $\Mdot$. These models differ in terms of the optical depth $\tau$, half-thickness ratio to radius $H/r$, radiative efficiency, or temperature $T$ (see Figure \ref{fig:2:disks}). 

For a~very low mass accretion rate, such as in the case of the BH in the centre of the Galaxy, \texttt{Sgr $\mathrm{A^*}$}, the accretion disk is optically thin, and the gas density is so low that Coulomb interactions are no longer effective. As a~result, the electrons and ions have significantly different temperatures, with the electrons being much colder than the ions \citep{NarayanMcClintock2008}.  In this regime, radiative cooling is insufficient, and the dominant cooling mechanism is advection; These types of disks are often referred to as Advection-Dominated Accretion Flows (ADAFs) and emit a~power-law spectrum with a~strong Compton component.

As the mass accretion rate increases, the gas density also increases, Coulomb coupling becomes efficient, and the ions and electrons temperatures become equal. The disk becomes optically thick, and the radiation cooling becomes efficient. In this regime, the disk flattens into a~thin disk of \cite{Shakura1976}. However, once the radiation pressure starts to dominate, the thin disk model is known to be thermally and viscously unstable \citep{Shakura1976,Lightman1974}. This regime corresponds to mass accretion rates down to approximately $\mdot \approx 0.01$ for a~stellar mass compact object. 

\begin{figure}
    \centering
    \includegraphics[width=\linewidth]{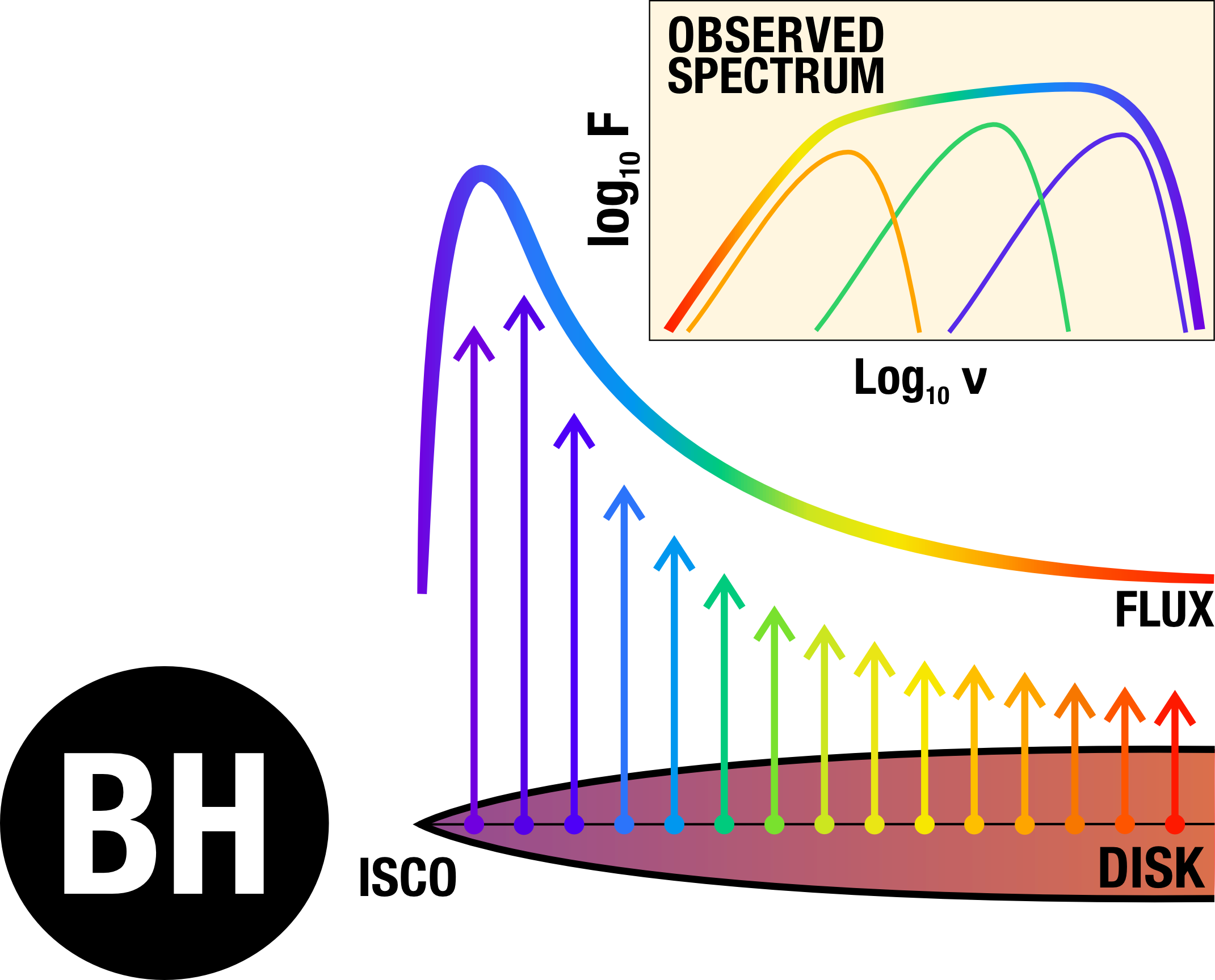}
    \caption{Illustration of a~thin disk and its local flux. The inset shows an observable multi-black body spectrum as a~sum of black-body spectra corresponding to the temperature on certain radii.}
    \label{fig:2:thin}
\end{figure}

Nevertheless, observations have shown that thin disks can still be observed at luminosities corresponding to much higher mass accretion rates during outbursts  \citep{Li2005,McClintock2014} when the disk is in a~soft spectral state. The thin disk emits a~multi-temperature black body spectrum, illustrated in Figure \ref{fig:2:thin}.

\begin{figure}
    \centering
    \includegraphics[width=\linewidth]{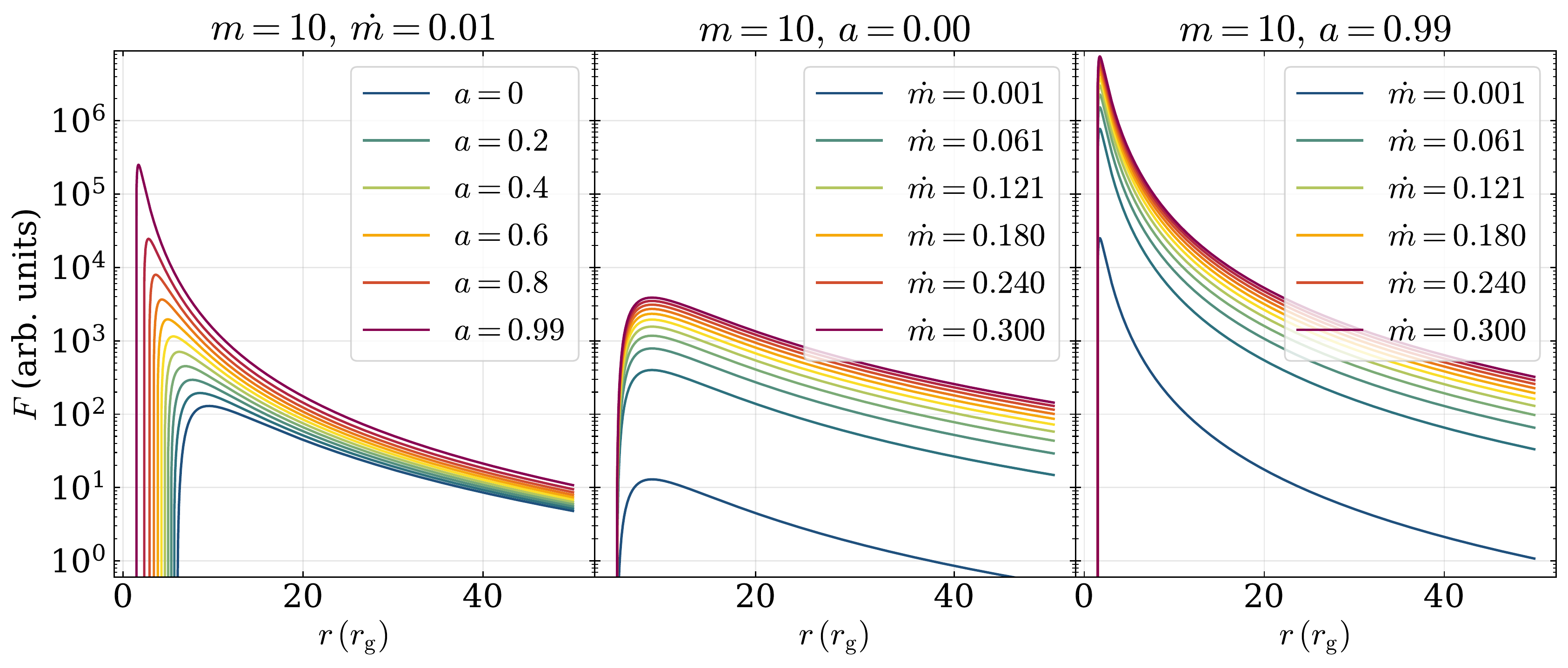}
    \caption{Radial profile of the local flux $F$ of a~thin disk as a~function of spin $a$ (\textit{left}), the $\mdot$ for a~non-rotating BH (\textit{middle}), and maximally rotating BH (\textit{right}) panel, for $M=10\,\msol$. }
    \label{fig:2:NTflux}
\end{figure}

The local flux of a~thin disk is given as \citep{Page1974}
\begin{equation}
    F = \frac{\mathcal{Q}}{\mathcal{B}\sqrt{\mathcal{C}}}\frac{3 G M \Mdot}{8 \pi r^3},
\end{equation}

\noindent where $\mathcal{Q},\mathcal{B}$ and $\mathcal{Q}$ are the relativistic correction, which can be found, e.g. in Appendix~A of \cite{Kato2008}. In the non-relativistic limit, it becomes $(1-\sqrt{{r_{in}}/{r}})$. The local flux is zero under the inner edge and maximal close to the inner edge, e.g., for a~non-rotating BH at $\sim 10\,\rg$. The flux strongly depends on the position of the inner edge, which is linked to the spin of the central BH, while the dependence on the $\mdot$ is not so strong, as shown in Figure \ref{fig:2:NTflux}. This feature of the thin disk model is used to establish the spin and mass of a~BH from observations.

It is evident that the thin disk model lacks an important part that would stabilize it even at higher mass accretion rates. Several stabilizing mechanisms have been proposed so far, \citep{Rozanska1999,Oda2009,Ciesielski2012,Zhu2013}; however, the magnetic field and magnetic pressure seem to be most plausible \citep{Begelman2007,Li2014}, even though the disk then turns to be geometrically thick, as was also shown in numerical simulations \citep{Sadowski2016,Lancova2019,Mishra2019,Huang2023}. 

\begin{figure}
    \centering
    \includegraphics[width=0.9\linewidth]{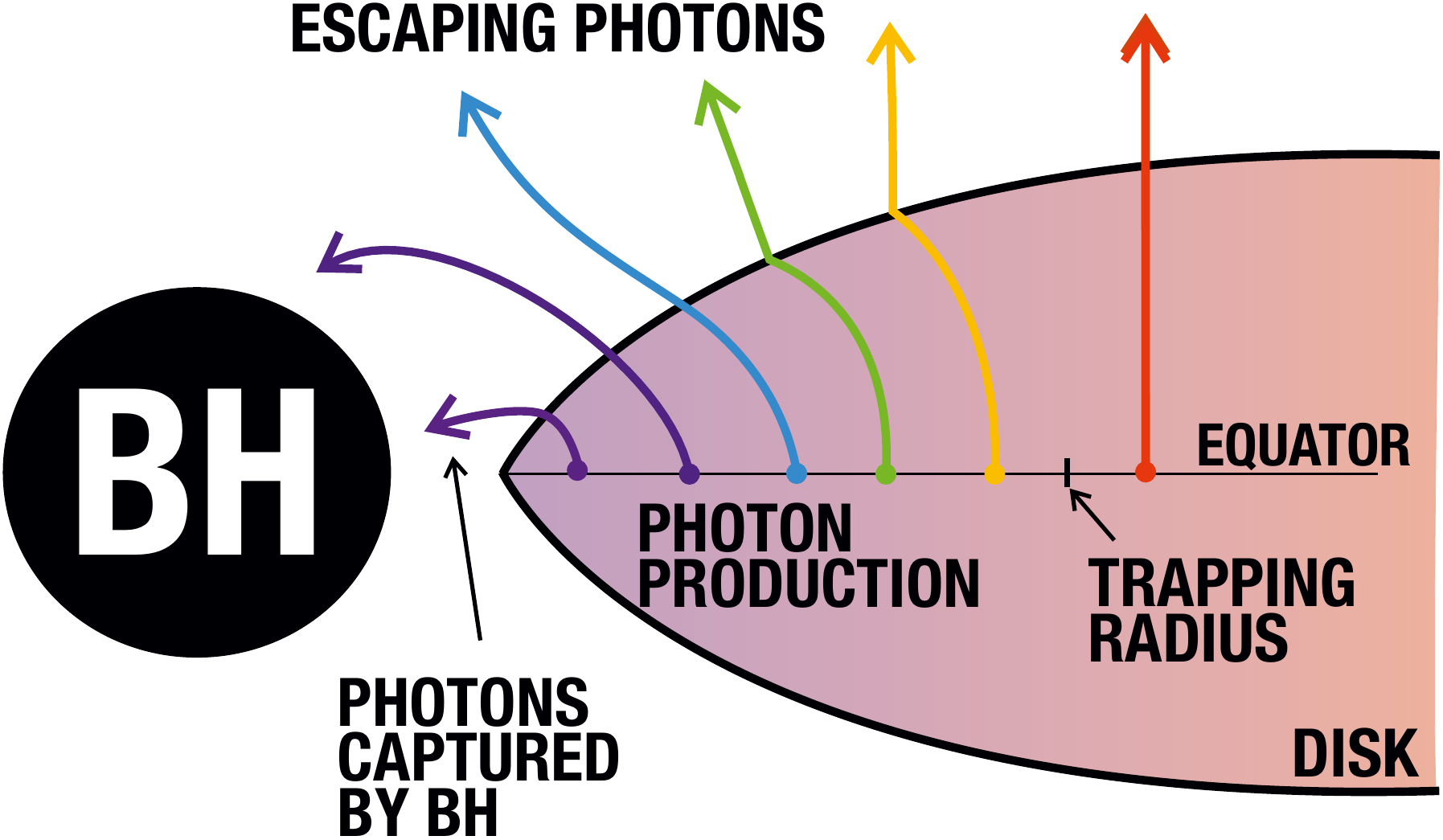}
    \includegraphics[width=1\linewidth]{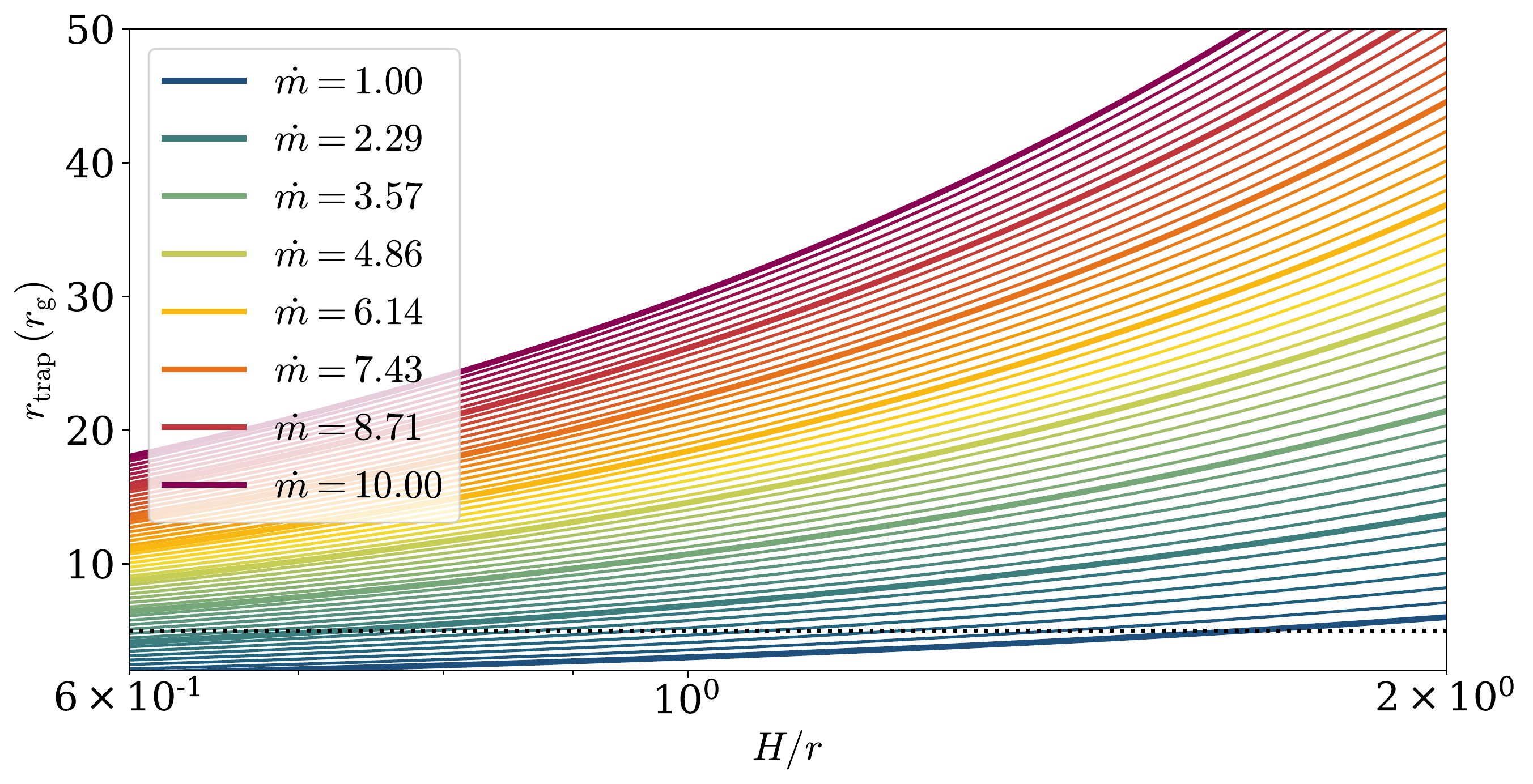}
    \caption{\textit{Top:} Illustration of photon trapping in a~thick accretion disk.\\ 
    \textit{Bottom:} The photon trapping radius as a~function of disk thickness for selected values of $\mdot$. The dotted line represents the $\risco$ for a~non-rotating BH.}
    \label{fig:2:adv}
\end{figure}

For $\mdot\sim 1$, the advection of radiation (photon trapping) becomes an important cooling mechanism, and this state corresponds to the slim disk model, an (essentially one-dimensional) analytic model of an optically thick advective disk, with $H/r\approx 1$ \citep{AbramSlim}. Photon trapping occurs when the disk opacity is so high that the photons generated in the plasma are effectively trapped and cannot reach the surface. Instead, they are carried away with the matter of the disk, as shown in Figure \ref{fig:2:adv}. Photon trapping can be quantitatively described by comparing the photon diffusion time, $t_{diff}=3 H \tau/c$ \citep{MihalasMihalas}, and the accretion time, $t_{acc} = r/|v^r|$. The radius at which photon trapping becomes important can be determined using the continuity equation for disk accretion and electron scattering opacity \citep{Kato2008}:
\begin{equation}
    r_\mathrm{trap} = 3 \left(\frac{H}{r}\right) \mdot \rg,
\end{equation}
\noindent as a~function of $\mdot$ and the disk thickness to radius ratio, ${H}/{r}$. For high $\mdot$ and ${H}/{r}$, the trapping radius is larger than $\risco$, and the advection of radiation becomes an important cooling mechanism in the slim disk model.  

The slim disk is often used to model the luminous state of XRBs and ULXs \citep{Foschini2006}. It is less radiatively efficient than the thin disk since a~fraction of the photons are advected into the BH. The inner edge of the slim disk is located inside the ISCO. However, the slim disk spectral model, \texttt{slimbb}, \citep{2009ApJS..183..171S}, when applied on a~data from microquasar in outburst, does not retrieve consistent values of BH spin for different observed luminosity when it increases from $L\approx 0.01\, L_\mathrm{Edd}$ up to $L\approx 0.5\, L_\mathrm{Edd}$, corresponding to the radiation pressure-dominated regime of accretion. This was demonstrated in the case of \texttt{LMC X-3}, in \cite{Straub2011}. When fitting the observed spectra at different luminosity, the derived value of the BH spin was not constant, and the same behaviour was seen for the thin disk model.

The Eddington limit is interpreted as the maximal luminosity a~disk can reach, and consequently, the $\dot{M}_\mathrm{Edd}$ the highest possible mass accretion rate before radiation pressure prevents accretion. However, this value is derived under many assumptions, such as spherical accretion. Higher luminosities can be reached by breaking them, such as considering a~thick accretion disk (or torus) with $H/r > 1$. In such a~case, a~funnel forms around the rotational axis, allowing radiation to escape and reduce radiation pressure acting on the accreting material. The escaping radiation is also highly collimated towards observers close to the rotational axis and obscured by the thick torus body for equatorial observers.  Super-Eddington accretion is actually frequently observed in the Universe, as seen in ULXs, AGN, and microquasars.

The most popular model of the super-Eddington accretion disk is the \say{Polish doughnut} \citep{Jarosz1980,Abramowicz+1978}, describing a~radiation pressure-supported accretion torus. In the Polish doughnut model, the matter distribution is assumed to be stationary and axially symmetric and the angular momentum constant within the disk. The distribution of matter in the disk is limited by the equipressure surfaces resulting from the \say{von Zeipel theorem} \citep{1924MNRAS..84..665V,1975ApJ...197..745S}, which states that the surfaces of constant angular velocity and angular momentum coincide for a~barotropic fluid.

\begin{figure}[b]
    \centering
    \includegraphics[width=0.7\linewidth]{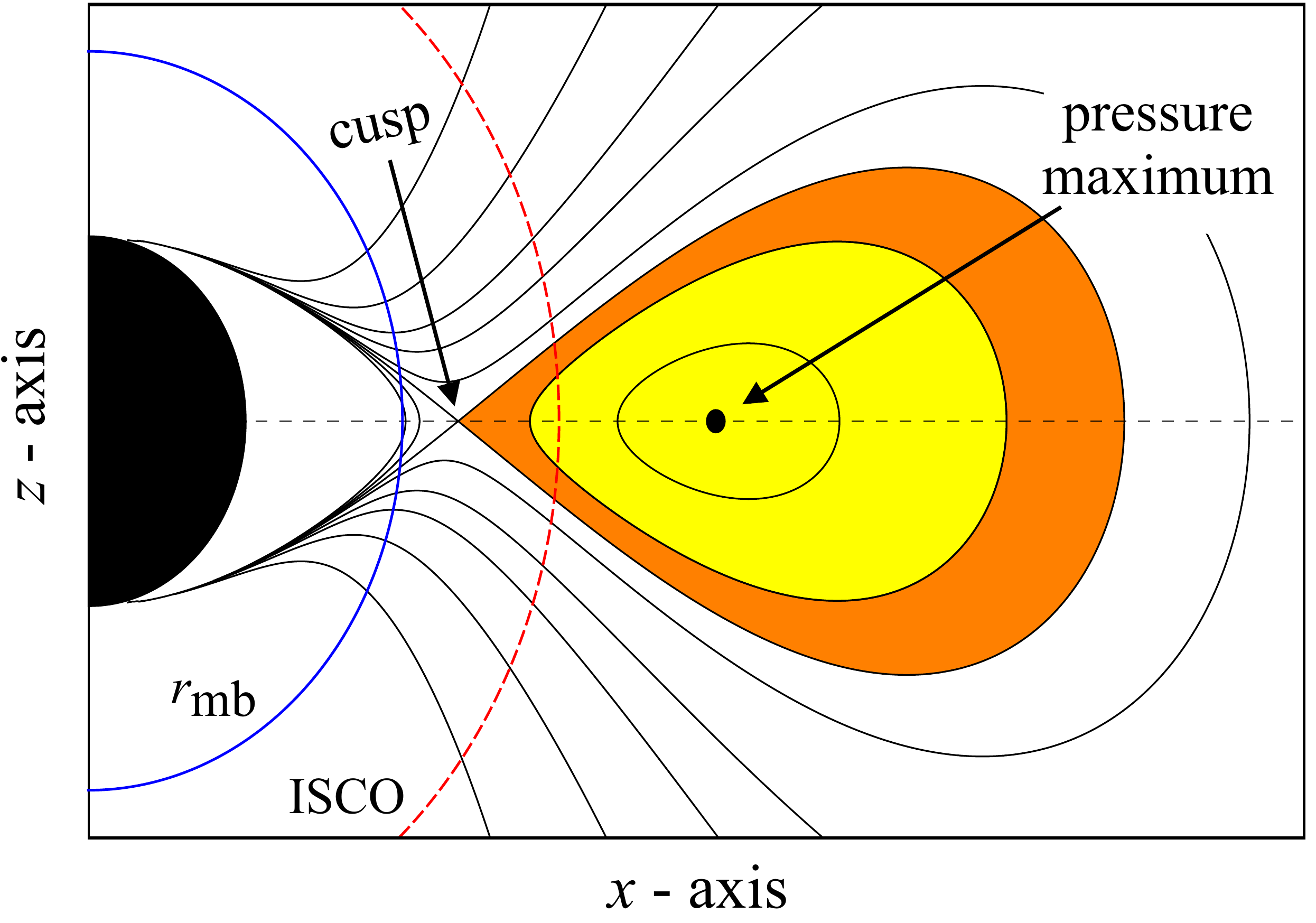}
    \caption{Illustration of the cusp torus configuration. Adopted from \cite{Kotrlova2020}, based on \cite{Jarosz1980}.}
    \label{fig:2:cusp}
\end{figure}

A particular case of the Polish doughnut model is the cusp torus, which is the largest physical configuration of the accretion tori, where the fluid of the disk fills the largest closed equipressure surface, forming a~cusp on its inner edge, between the $\risco$ and $r_\mathrm{mb}$, see Figure \ref{fig:2:cusp}. The cusp torus is implemented in various QPOs models, which will be further elaborated in Chapter \ref{chap:QPOs}  \citep{torok2016mnras,tor-etal:2022,Blaes2006,AbramowiczKluzniak2001, KluzniakAbramowicz2001}.

\section{Modelling the accretion disk in microquasars}

Accretion disks in microquasars are commonly modelled using a~simple yet powerful thin disk model based on a~simple and convenient description of the viscosity in the fluid using the $\alpha$ parametrization, where a~single constant $\alpha$ relates the fluid stress and pressure. 

The alternating periods of luminous outbursts and quiescent fades are often attributed to state transitions of the accretion disk caused by unidentified instabilities \citep{Done2004,Vincentelli2023}. However, the time scale of  fundamental instabilities in the thin disk model is much shorter than the observed variability. Yet, the emitted spectrum and general behaviour are consistent with the predictions of analytical models in many aspects.

The stability of the sub-Eddington accretion disk plays an important role in modelling the entire life cycle of XRBs. However, accurately modelling the whole outburst and state transitions of a~XRB in complex numerical simulations remains a~challenge.  In our work, \citep{Wielgus2022,Lancova2023},  we successfully modelled the observable signal from simulations of a~mildly sub-Eddington radiation pressure-dominated yet stable puffy disk  \citep{Lancova2019} and compared it to the expected signal from XRB. We found that the simulations can be compared to observed data and that the estimation of source parameters, such as BH spin, is strongly influenced by the specific topology of the puffy disk system compared to that obtained using standard analytical models. 

The oversimplified one-dimensional $\alpha$-disk cannot capture the dynamics and three-dimensional structure of turbulent accretion flows in the vicinity of a~BH. A~self-consistent accretion model must address the three-dimensional interplay between gravity, fluid dynamics, magnetic fields, and radiation. It should also capture the turbulent nature of the flow driven by the MRI, which provides the crucial mechanism for angular momentum transport. Such a~model can be obtained using advanced numerical methods for modelling plasma, such as in the framework of global general relativistic radiative magnetohydrodynamic (GRRMHD) simulations.  However, simulating optically thick, geometrically thin accretion structures is computationally expensive, requiring high resolution of the three-dimensional numerical grid \citep{Balbus1998}, short integration time steps, and the use of implicit solvers to capture the evolution of radiation field  \citep{Sadowski2013}. Consequently, many published works sacrifice either radiation \citep{Liska2019}, magnetic field \citep{Mishra2019}, or general relativity \citep{Jiang2019}.

The topic of the sub-Eddington accretion has been a~major part of my work during my studies, resulting in the development of the puffy disk model \citep{Lancova2019}, where the results of  advanced simulations were presented as one of the first self-consistent simulations of a~stable mildly sub-Eddington accretion.

\chapter{GRRMHD simulation of sub-Eddington accretion}
\label{chap:mhd}



The global general relativistic (GR) radiative (R) magneto (M) hydrodynamics (HD) method is used to simulate the flow of magnetized plasma, including the exchange of energy and momentum with radiation in the framework of GR. Astrophysical fluids in accretion disks, primarily composed of hydrogen, usually have high enough temperatures to be ionized ($T\gtrsim 10^4$ K). The presence of free electrons in the plasma leads to the generation of a~magnetic field through electric currents, which significantly influences the properties of the flow.

In global simulations, the computational domain contains the entire accretion disk and thus captures the complex dynamics and the interactions between the disk, corona, outflows, and jets. On the other hand, a~shearing box (or shearing sheet in 2D) simulates the plasma flow within a~relatively small box placed at a~specific location in the disk, such as the center, surface, or base of a~jet. Shearing box simulations help study the structure and small-scale dynamics of the flow. 

The development of the GRMHD method began in the late 1980s \citep{Anile1989}, followed by the development of complementary numerical methods in the 1990s \citep{Komissarov1999}. In the early 2000s, several codes were developed, taking advantage of the rapid advancements in high-performance and supercomputing, which led to a~broader implementation of MHD and GRMHD in accretion disk research.

One of the most widely used GRMHD codes is the \texttt{HARM} \citep{Gammie2003},  which has since been extended and modified in various variants and derivatives, such as the \texttt{HARMPI} and \texttt{H-AMR} codes \citep{HARMPI,HAMR} heavily parallelized and/or including advanced treatment of radiation, or the \texttt{HARM\_COOL} code \citep{Janiuk23}, which includes cooling by neutrinos. The \koral{} code, described below, was developed alongside \texttt{HARM} and shares many methods and algorithms. Other notable codes include the \texttt{BHAC}, implementing advanced nuclear equations of state for the matter, neutrino leakage, and the possibility to use numerical metrics for non-BH spacetimes \citep{BHAC}. The \texttt{IllinoisGRMHD} \citep{Etienne2015} is designed for dynamical metric simulations of BH and NS mergers within the \textit{Einstein Toolkit} framework \citep{EinsteinToolkit}. Many other MHD codes are available for studying different accretion regimes, such as the \texttt{Athena++} \citep{Athena}, \texttt{Cosmos++} \citep{Cosmos} or non-relativistic but resistive code \texttt{PLUTO} \citep{Mignone2007}, among others. This chapter introduces the equations and numerical methods as implemented in the \koral{} code. However, it represents only one road that can be taken. To describe other existing methods would be beyond the scope of this work.

This chapter aims to summarize the methods and features of the \koral{} code in the version used in the puffy disk simulations published in \cite{Lancova2019}. Since its first publication in \cite{Sadowski2013}, new features have been added to the code, and no comprehensive article summarizing these updates exists. The \koral{} code was tested against other codes in the \textit{The Event Horizon General Relativistic Magnetohydrodynamic Code Comparison Project} \citep{Porth2019}.

The following list provides a~selection of articles related to the \koral{} code, although some more might need to be included. \koral{} was first introduced as a~GRRHD code in \cite{Sadowski2013}. The MHD was added in \cite{Sadowski2014}, followed by the photon-conserving Comptonization and radiation viscosity in \cite{Sadowski2015_compton}. The mean-field dynamo, which allows for long 2D simulations without losing the azimuthal component of the magnetic field and suppressing the MRI, was introduced in \cite{Sadowski2015_dynamo}. Two-temperature heating and cooling of electrons and ions, enabling simulations of low-density flows where the ion and electron gases have different temperatures due to inefficient Coulomb interaction, were implemented in \cite{Sadowski2017}, together with self-consistent calculation of the adiabatic index and advanced treatment of opacities. The evolution of relativistic non-thermal electrons was incorporated in \cite{Chael2017}.


\section{GRRMHD equations}

\koral{} solves the conservation equations of mass, momentum, and energy on a~fixed grid in an arbitrary metric. The code adopts geometrized units where $G=c=1$, the gravitational radius $\rg = GM/c^2$ as the unit of length, and $\tg = \rg/c$ as the unit of time. Greek indices range from 0 to 3, while Latin indices range from 1 to 3. 

The conservation equations for a~fluid with a~rest mass density $\rho$ and lab frame 4-velocity $u^\mu$ are:

\begin{align}
&\left(\rho u^\mu\right)_{;\mu} = 0, \label{eq:3:momentum_conv}\\
&\left( T^\mu\phantom{}_\nu\right)_{;\mu} = G_\nu,\label{eq:3:matter_conv}\\
&\left( R^\mu\phantom{}_\nu\right)_{;\mu} = -G_\nu,\label{eq:3:rad_conv}\\
&\left( F^{*\mu\nu}\right)_{;\mu} = 0,
\label{eq:3:magflux_conv}
\end{align}

\noindent where $R^\mu\phantom{}_\nu$ is the radiation stress-energy tensor, $G_\nu$ the radiative 4-force density describing the exchange of energy and momentum between the gas and radiation (see section \ref{sec:3:rad}), and $F^{*\mu\nu}$ the dual tensor of the electromagnetic field.

The ideal MHD approximation is used in the \koral{} code, assuming infinite conductivity of the fluid,  which leads to the vanishing of the electric field in the fluid frame. Under this assumption, $F^{*\mu\nu}$ can be given in a~simple form, 

\begin{equation}
    F^{*\mu\nu} = b^\mu u^\nu - u^\mu b^\nu,
    \label{eq:3:Ftensor}
\end{equation}

\noindent where $b^\mu$ is related to the magnetic field $B^i$ as \citep{Komissarov1999} 

\begin{equation}
    B^i = F^{*i0} = F^{*it} = b^i u^t - u^i b^t,
\end{equation}

\noindent and vice versa, 

\begin{align}
b^t &= B^i u_i, \\
b^i &= \dfrac{B^i + b^t u^i}{u^t}.
\end{align}

The gas stress-energy tensor, $T^{\mu\nu}$, is given as

\begin{equation}
T^{\mu\nu}= \left(\rho + \uint + \pgas + b^2\right)u^\mu u^\nu + \left(\pgas+ \frac{b^2}{2}\right)g^{\mu\nu} - b^\mu b^\nu, 
\end{equation}

\noindent where $\uint$ is the internal energy density, $\pgas$ is the gas pressure, and $g^{\mu\nu}$ represents the metric tensor. 

For the adiabatic equation of state,  the gas internal energy $\uint$ in the comoving frame and the gas pressure $\pgas$ are related by
\begin{equation}
    \pgas = (\Gamma-1) \uint
\end{equation}

\noindent where  $\Gamma$ is the adiabatic index. Typically, $\Gamma$ is assumed to be $5/3$ for gas pressure and $4/3$ for optically thick radiation pressure-dominated or relativistic fluid.

From (\ref{eq:3:Ftensor}) is evident that $F^{*tt}=0$, therefore the time component of equation (\ref{eq:3:magflux_conv}) is not an evolution equation. However, the $\nabla\cdot \mathbf{B}=0$ constraint can be directly delivered from there. It can also be shown that the magnetic 4-vector is perpendicular to the 4-velocity in the fluid frame,  $b^\mu u_\mu=0$ \citep{Komissarov1999}.

The Maxwell equation in the lab frame then consists of the induction equation
\begin{equation}
\partial _t \left(\sqrt{-g}B^i\right) = -\partial_j\left(\sqrt{-g} b^i u^j - u^j b^i \right),
\end{equation}
\noindent where $\sqrt{-g}$ is the metric determinant. This equation governs the evolution of the magnetic field and the divergence-free constraint,
\begin{equation}
\frac{1}{\sqrt{-g}}\partial_i \left(\sqrt{-g}B^i\right) = 0,
\label{eq:3:divB}
\end{equation}

\noindent which ensures $\nabla \cdot \mathbf{B} = 0$ constrain. However, equation (\ref{eq:3:divB}) is not directly evolved to prevent the production of magnetic monopoles due to numerical errors. Instead, the flux-CT method of \cite{Toth2000} is implemented, which solves the flux-interpolated constrained transport. 

\subsection{Radiation}
\label{sec:3:rad}

The hot material in accretion disks produces EM radiation, which interacts with the gas and can significantly influence the fluid's dynamics and temperature by emission and absorption. If the disk is optically thin, the photons escape freely and can be up-scattered by the inverse Compton process. However, this process has little dynamical influence on the fluid, and the consequences to the observed spectrum can be calculated separately during post-processing. On the other hand, if the fluid is optically thick, the photons are effectively trapped and influence the flow dynamics by generating radiation pressure. Photon production in the fluid is also an essential cooling mechanism that must be considered when modelling radiatively efficient accretion regimes. 

Hence, the radiation should be an inseparable part of sub-Eddington accretion disk simulations and be evolved simultaneously with the fluid in radiatively efficient flows. However, capturing the complicated dynamics of gas and radiation interaction is challenging, and some sacrifices for the sake of computational simplicity have to be made. 

The ideal way of following each photon trajectory is not feasible in an optically thick medium where the photon's mean path is extremely short, and an enormous number of individual photos would be needed to capture the dynamics correctly. Based on the various moments of the radiation field, the method used in the \koral{} code can adequately solve the problem in both optically thin and thick media by handling the radiation as an additional \say{fluid} in the domain. The radiation tensor $R_{\mu\nu}$ exchanges momentum and energy with the matter through the radiation 4-force density $G_\nu$, and the exchange is conservative,  $\left(T^\mu\phantom{}_\nu + R^\mu\phantom{}_\nu\right)_{;\mu} = 0$ (see equations (\ref{eq:3:matter_conv}) and (\ref{eq:3:momentum_conv})).

$R_{\mu\nu}$ consists of components corresponding to various moments of the fre\-quency-integrated specific intensity $ I = \int_\nu I_\nu \der \nu$. In the fluid frame\footnote{the hats are denoting the fluid frame throughout this work} the radiative energy density $\widehat{E}=\int \widehat{I}_\nu \der\nu \der\Omega$, flux $\widehat{F}^i=\int \widehat{I}_\nu \der \nu \der\Omega N^i$ and the pressure tensor $\widehat{P}^{ij} = \int \widehat{I}_\nu \der \nu \der \Omega N^iN^j$, where $\nu$ is frequency, $\Omega$ the solid angle and $N^i$ the unit vector in direction $x^i$, together form \citep{MihalasMihalas,Sadowski2013}

\begin{equation}
   \widehat{\mathbf{R}}=
   \begin{pmatrix}
        \widehat{E} & \widehat{F}^i\\
        \widehat{F}^j & \widehat{P}^{ij}
    \end{pmatrix}
    .
\end{equation}

This relatively simple formulation is a~function of all components of $R^{\mu\nu}$ in an arbitrary frame; thus, to close the set of conservation equations (\ref{eq:3:rad_conv}), all components of $R^{\mu\nu}$ are needed.
However, only the radiative energy density $R^{tt}$ and fluxes $R^{ti}$ are known in an arbitrary frame. In the \koral{} code,  the \textbf{M1} closure scheme \citep{Levermore1984,Sadowski2013} is implemented to find the full form of $R^{\mu\nu}$.

The scheme assumes the existence of a~\say{radiation rest frame}, in which the radiation is isotropic\footnote{this frame is denoted by tildes}, and where radiation tensor acquires a~simple form of
\begin{equation}
   \tilde{R}^{\mu\nu}=
   \begin{pmatrix}
        \tilde{E} & \tilde{E}/3\\
        \tilde{E}/3 & 0
    \end{pmatrix}
    = \frac{4}{3}\tilde{E}\tilde{u}_R^\mu\tilde{u}_R^\nu + \frac{1}{3}\tilde{E}\eta^{\mu\nu},
\end{equation}

\noindent where $\tilde{u}^\mu_R=\left(1,0,0,0\right)$ is the radiation rest frame 4-velocity, and the metric $\eta^{\mu\nu}$ is locally Minkowski. 

Since this equation is in covariant form, an arbitrary frame expression can be found: 

\begin{equation}
    R^{\mu\nu} = \frac{4}{3}\tilde{E} u^\mu_R u^\nu_R + \dfrac{1}{3}
    \tilde{E} g^{\mu \nu}.
\end{equation}

\noindent However, $\tilde{E}$ should always be interpreted as the radiation energy density in the radiation rest frame.

The $M1$ closure works sufficiently well in both optically thin and thick regi\-mes, as long as there is only one source of radiation. Since the radiation is handled as a~fluid, in a~multiple beams intersection, radiation from one source would be interacting with another as a~fluid would do. Fortunately, in astrophysical simulations, such a~scenario is usual (see the extensive testing in \cite{Sadowski2013} for reference). 

\subsection{Interactions between the fluid and radiation}

The radiation 4-force density $G^\mu$, describing the conservative interactions between radiation and the fluid, is given by \citep{MihalasMihalas, Sadowski2013, Sadowski2015_compton, Sadowski2015_dynamo}

\begin{equation}
    \widehat{G}^\mu = \int \left(\chi_\nu I_\nu - \eta_\nu \right) \der \nu \der \Omega N^i,
\end{equation}

\noindent which  takes a~relatively simple form for the frequency-integrated quantities:
\begin{equation}
    \widehat{G}^\mu = 
    \begin{pmatrix}
    \kappa_a\rho\left(\widehat{E} - 4\pi\widehat{B}\right)\\
    \left(\kappa_a + \kappa_{es}\right)\rho\widehat{F}^i
    \end{pmatrix},
\end{equation}

\noindent where $\widehat{B}$ is the frequency-integrated Planck function given by $\widehat{B} = a T_g^4/4\pi$, where $a$ is the radiation constant and  $T_g$ the gas temperature. $\chi = \kappa_a + \kappa_{es}$ is the total opacity coefficient consisting of the absorption and electron scattering opacity, respectively\footnote{
In the simulations of the puffy disk, we adopted $\kappa_a = 6.4\times10^{22} \rho T^{-7/2}\,\mathrm{cm}^2\cdot\mathrm{g}^{-1}$ and $\kappa_{es} = 0.34\,\mathrm{cm}^2\cdot\mathrm{g}^{-1}$.}.

In \cite{Sadowski2014} the covariant form of $G^\mu$ was introduced, 
\begin{equation}
    G^\mu = -\rho \left[\chi R^{\mu\nu}u_\nu + \left(\kappa_{es}R^{\alpha\beta}u_\alpha u_\beta + 4\pi\kappa_a B\right)u^\mu\right].
\end{equation}

However, this approach neglects the exchange of momentum \textit{and} energy during the particle-photos interactions and is appropriate only to model the elastic scattering and thermal absorption. The $G^0$ term corresponds to the gain of fluid energy density due to absorption ($\kappa_a\rho\tilde{E}$) and loss due to emission ($\kappa_a\rho a T_g^4$). For the black body radiation, the effective radiation temperature is defined as $T_R^4 = {\tilde{E}}/{a}$ in the fluid frame. The $\widehat{G}^0$ in the fluid frame is then 
\begin{equation}
    \widehat{G}^0 = \kappa_a \rho a \left(T_R^4 - T_g^4\right),
    \label{eq:3:G0mu}
\end{equation}
\noindent which drives the system towards local thermal equilibrium (LTE).

The space components, $G^i$, on the other hand, describe a~symmetric absorption and re-emission of photons by the fluid, which drives the system to a~state of zero radiation flux in the fluid frame. 

To enhance this exchange and to include the energy changes during scattering, Comptonization is implemented in \koral{}, and it comes in two options. The more advanced one is the photon conserving scheme introduced in \cite{Sadowski2015_compton}, where the radiation is treated as a~Bose-Einstein fluid, and the photon number density $n_i$ is evolved throughout the simulation. However, this significantly increases the computational cost of the simulations, and for many applications, a~more straightforward black body Comptonization is sufficient.

The radiation 4-force density in the fluid frame $\widehat{G}^\mu$ can be written as 

\begin{equation}
    \widehat{G}^\mu = \widehat{G}_{\mathrm{BB}}^\mu - \widehat{G}^\mu_\mathrm{Compt},
\end{equation}

\noindent where $\widehat{G}_{\mathrm{BB}}^\mu$ is given by (\ref{eq:3:G0mu}) and the term corresponding to the black-body Comptonization is given by
\begin{equation}
    \widehat{G}^0_\mathrm{Compt} =\rho \widehat{E}\kappa_{es} \left[\frac{4k_B\left(T_g-T_R\right)}{m_e}\right]\left(1+3.683\frac{k_B T_g}{m_e} + \frac{4k_B T_g}{m_e}\right)\left(1+\frac{4k_B T_g}{m_e}\right)^{-1},
\end{equation}
where $k_B$ is the Boltzmann constant and $m_e$ the electron mass. The last two expressions in parentheses are fitted formulas corresponding to corrections for relativistic electrons; see \cite{Sadowski2015_compton}\footnote{There is a~typo in \cite{Sadowski2015_compton}, missing 4 in the last term.}. 

\subsection{Radiative viscosity}

The \textbf{M1} closure tends to overestimate the angular momentum of photons close to the rotational axis, which leads to the creation of unphysical shocks, which significantly alters the behaviour of both gas and radiation in the axis regions and can result in misinterpretation of the results. In \koral{}, it is addressed by introducing an artificial viscous term in the radiation, following \cite{Coughlin2014}, to diffuse the shocks, as presented in Appendix B of \cite{Sadowski2015_dynamo}.

The viscous term is given by:
\begin{equation}
    R^{\mu\nu}_{\mathrm{visc}} = -2\nu_{\mathrm{rad}}\tilde{E}\sigma_R^{\mu\nu},
\end{equation}

\noindent where $\nu_{\mathrm{rad}}= \alpha_{\mathrm{rad}}\lambda$ is the radiative viscosity coefficient, defined as the product of a~constant $\alpha_{\mathrm{rad}}$ coefficient\footnote{typical value is $\alpha_{\mathrm{rad}} = 0.1$ based on calculations and simulations in \cite{Sadowski2015_dynamo}} and photon's mean free path $\lambda$. $\sigma_R^{\mu\nu}$ is the shear tensor,

\begin{equation}
    \sigma_R^{\mu\nu} = \frac{1}{2}\left(u^\mu_{R;\alpha}h_R^{\alpha\nu}+ u^\nu_{R;\alpha}h_R^{\alpha\mu}\right) - \frac{1}{3}h_R^{\mu\nu},
\end{equation}

\noindent where $h^{\mu\nu}_R=g^{\mu\nu}+u_R^\mu u_R^\nu$ is the projection tensor. 

The viscosity term is suppressed in the optically thick regions by setting  $\lambda = \mathrm{min}\left(r,\frac{1}{\rho\chi}\right)$. The maximal value of $\nu_{\mathrm{rad}}$ is also limited to ensure that the CFL condition is not violated. In the end, the viscous term is added to the radiative fluxes in the \textbf{M1} scheme as

\begin{equation}
    R^{i\nu} = R^{i\nu} +  R^{i\nu}_{\mathrm{visc}}.
\end{equation}
 
\subsection{Mean field dynamo}
\label{sec:3:dynamo}

The main engine for angular momentum transport in the accretion disk is the MRI, yet it also leads to the dissipation of magnetic energy and fading of the magnetic field. However, the magnetic field is simultaneously replenished due to the mean-field dynamo effect. This process naturally occurs in 3D simulations, where all three components of $\mathbf{B}$ are evolved. In 2D simulations, on the other hand, the axisymmetry prevents the generation of the poloidal component of the magnetic field, as implied by Cowling's anti dynamo theorem \citep{Cowling1933,Brandenburg1995,Sadowski2015_dynamo}. 

Therefore, a~2D simulation cannot last for a~prolonged time due to the decaying magnetic field (estimations maximal duration is about 5000 $\tg$), but the decay can still influence the results even in earlier stages. In \cite{SadowskiNarayan2016}, a~set of very long 3D simulations of super-Eddington accretion was published, and one of them was compared to a~2D simulation of the same parameters. The results confirmed that the artificial mean-field dynamo implemented in the \koral{} code enables extremely long 2D simulations with time-averaged properties almost identical to the 3D runs; however, the time variability and radiative luminosity were notably higher in the dynamo simulations. Even though the dynamical properties of the flow are captured well, one should keep in mind that only a~full 3D treatment of the problem can lead to reliable results.  

The saturated state, when the dynamo term counterbalances the magnetic dissipation, corresponds to a~state when the magnetization $\beta$ and the magnetic tilt angle $\xi$ are

\begin{align}
    &\beta = \frac{p_\mathrm{mag}}{\pgas} \approx 0.1\\
    &\xi = \frac{b^r b^\phi}{b^2}\approx\frac{\mathbf{B}_{\widehat{p}}}{B_{\widehat{\phi}}} \approx 0.25,
\end{align}
\noindent where $B_{\widehat{p}}$ and $B_{\widehat{\phi}}$ are the orthonormal poloidal and azimuthal components of the magnetic field \citep{Sorathia2012}. The mean magnetic field evolution equation \citep{Brandenburg2001} can be written as 

\begin{equation}
    \frac{\partial}{\partial t}\mathbf{B} = \alpha_{dyn}\nabla\times\mathbf{B}+\eta_{md}\Delta\mathbf{B},
    \label{eq:3:dynamo_evolution}
\end{equation}
\noindent where $\alpha_{dyn}$ is the dynamo coefficient and $\eta_{md}$ the magnetic diffusivity coefficient. To balance the magnetic diffusion by the generation of a~poloidal magnetic field, the (\ref{eq:3:dynamo_evolution}) can be rewritten as:
\begin{equation}
        \frac{\partial}{\partial t}\mathbf{B}_{\widehat{p}} = \alpha_{dyn}\left(\nabla\times\mathbf{B}\right)_{\widehat{p}}
\end{equation}
\noindent  Using the assumption of Keplerian motion (the magnetic field is generated on the dynamical timescale $1/\Omega_K$) and the magnitude of poloidal current $|\mathbf{J}_{\widehat{p}}| \approx |B_{\widehat{\phi}}|/H$, where H is the half-thickness of the disk, the increment of the azimuthal vector potential is

\begin{equation}
    \frac{\partial}{\partial t}A_{\widehat{\phi}} = \alpha_{dyn}\xi \Omega_K B_{\widehat{\phi}}.
\end{equation}

To prevent over-saturation of the magnetic field, errors in the plunging regions, and excessively high $\xi$ values, the \koral{} code introduces factors that control the intensity of the dynamo process, as described in Appendix A of \cite{Sadowski2015_dynamo}.

After generating the magnetic field through the dynamo process, the new magnetic field increment $\der\mathbf{B}_{{dyn}}$ is added to the existing magnetic field by computing the curl of $\der \mathbf{A}_{dyn}$: $\der\mathbf{B}_{dyn}=\nabla\times \der \mathbf{A}_{dyn}$. Subsequently, the azimuthal component $B^\phi$ is damped to maintain saturated values of $\xi$. This damping process ensures that the dynamo-generated magnetic field does not become too dominant, leading to unrealistic behaviour or numerical issues.

\section{Numerical methods}

The conservation laws (\ref{eq:3:momentum_conv})-(\ref{eq:3:matter_conv})  in the coordinate basis of $\left(t,x^i\right)$ are given by \citep{Gammie2003}
\begin{align}
    &\frac{\partial}{\partial t}\left(\sqrt{-g}\rho u^t\right) + \frac{\partial}{\partial x^i}\left(\sqrt{-g}\rho u^i\right) = 0,\\
    &\frac{\partial}{\partial t}\left(\sqrt{-g}T^{t}\phantom{}_\nu\right) + \frac{\partial}{\partial x^i}\left(\sqrt{-g}T^{i}\phantom{}_\nu\right) = \sqrt{-g}T^{\alpha}\phantom{}_\beta\Gamma^{\beta}\phantom{}_{\nu\alpha} + \sqrt{-g}G_\nu ,\\
    &\frac{\partial}{\partial t}\left(\sqrt{-g}R^{t}\phantom{}_\nu\right) + \frac{\partial}{\partial x^i}\left(\sqrt{-g}R^{i}\phantom{}_\nu\right) = \sqrt{-g}R^{\alpha}\phantom{}_\beta\Gamma^{\beta}\phantom{}_{\nu\alpha} - \sqrt{-g}G_\nu,
\end{align}
\noindent where $\Gamma^{\alpha}\phantom{}_{\beta\gamma}$ are the Christoffel symbols describing the spacetime curvature. 

However, this form is not stable in numerical schemes because it takes the values of the right-hand side as exact from the cell centres when the left-hand sides spacial derivatives are not exact results of the numerical differential. Thus, in the \koral{} code, a~static metric is expected, and the equations can be written as 

\begin{align}
    &\frac{\partial}{\partial t}\left(\rho u^t\right) + \frac{\partial}{\partial x^i}\left(\rho u^i\right) = \frac{-\rho u^i}{\sqrt{-g}}\frac{\partial}{\partial x^i}{\sqrt{-g}}, \label{eq:3:coordbas1}\\
     &\frac{\partial}{\partial t}\left(T^{t}\phantom{}_\nu\right) + \frac{\partial}{\partial x^i}\left(T^{i}\phantom{}_\nu\right) = T^{\alpha}\phantom{}_\beta\Gamma^{\beta}\phantom{}_{\nu\alpha} - \frac{T^{i}\phantom{}_\nu}{\sqrt{-g}}\frac{\partial}{\partial x^i}{\sqrt{-g}}  +G_\nu ,\label{eq:3:coordbas2}\\
     &\frac{\partial}{\partial t}\left(R^{t}\phantom{}_\nu\right) + \frac{\partial}{\partial x^i}\left(R^{i}\phantom{}_\nu\right) = T^{\alpha}\phantom{}_\beta\Gamma^{\beta}\phantom{}_{\nu\alpha} - \frac{R^{i}\phantom{}_\nu}{\sqrt{-g}}\frac{\partial}{\partial x^i}{\sqrt{-g}}  -G_\nu .\label{eq:3:coordbas3}
\end{align}

\noindent These modified equations ensure numerical stability in the simulations.

The \koral{} code solves a~system of differential equations (\ref{eq:3:coordbas1})-(\ref{eq:3:coordbas3})  using the finite volume Godunov scheme, which in general solves 
\begin{equation}
    \frac{\partial}{\partial t} \mathbf{Q} + \nabla \cdot \mathbf{F}\left(\mathbf{Q}\right) = \mathbf{S}\left(\mathbf{Q}\right),
    \label{eq:3:godunov}
\end{equation}

\noindent in other words, the time change of some vector quantity  $\mathbf{Q}$ plus the divergence of the flux of $\mathbf{Q}$ is equal to the source of $\mathbf{Q}$. The finite-volume method means that (\ref{eq:3:godunov}) is calculated in a~finite-volume element.

In a~first-order Godunov scheme, the state at the time ($t+\Delta t$) is computed using the averaged value of $\mathbf{Q}$ in the whole cell, minus fluxes through all cell faces and then adding the source term. In a~second-order Godunov scheme,  values at $(t+\Delta t)/2$ are used to advance the whole system in time. The second-order scheme generally provides more accuracy compared to the first-order scheme, but it also requires more computational resources. 

\subsection{The Riemann problem}

\begin{figure}
    \centering
    \includegraphics[width=1\linewidth]{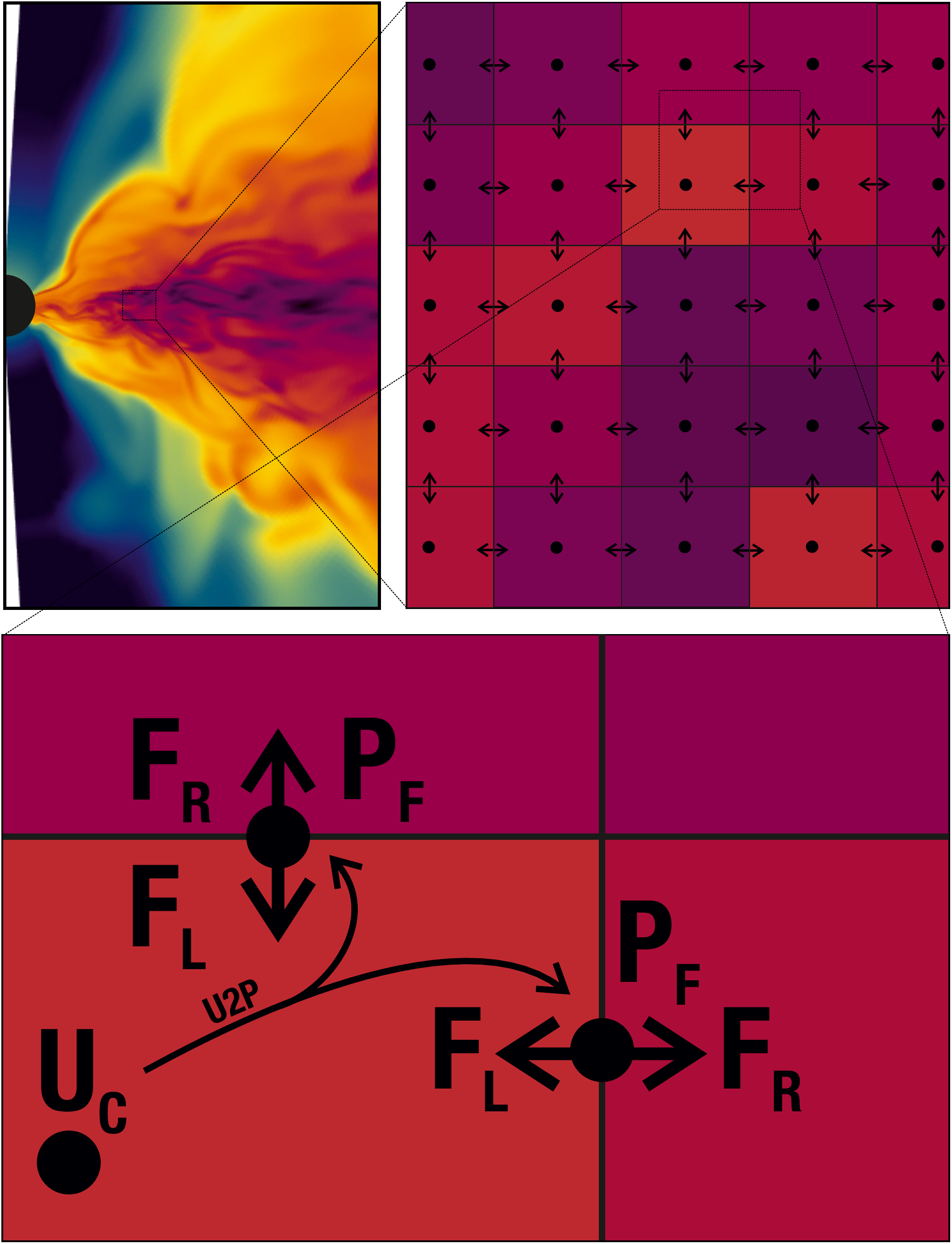}
    \caption{Illustration of the MHD algorithm in 2D. \textit{Top left}: Snapshot of a~density map from accretion disk simulation. \textit{Top right}: Zoom in onto the grid, which consists of homogeneous cells. The cell centres and fluxes are indicated. \textit{Bottom}: Zoom-in on a~cell boundary, showing the conserved quantities stored at the cell centres and left and right fluxes through the cells' faces, which are calculated from the primitives reconstructed at the faces. To solve the total flux through each cell face, the Riemann problem is solved.}
    \label{fig:3:scheme}
\end{figure}

The complementary problem involves solving the flux through neighbouring cells. In a~discrete code, the fluxes at cells' faces are reconstructed from the values at the centres since all cells are considered  homogeneous. At each interface between two cells, a~discontinuity exists, resulting in two values of flux: the left $\mathbf{F}^{\mathrm{L}}$ and right $\mathbf{F}^{\mathrm{R}}$ flux (see Figure \ref{fig:3:scheme}). To find the total flux through the cell face, a~Riemann problem, a~well-known problem in hydrodynamics, has to be solved.

A one-dimensional Riemann problem can be represented by a~tube containing two fluids in different states separated by a~thin membrane. Once the membrane is removed, the fluids mix and the time evolution of this state is the Riemann problem. A~semi-analytical solution can be found in specific cases, such as the Sod tube problem \cite{1978JCoPh..27....1S}.

However, in 2D simulation, the Riemann solver needs to be called four times for each cell at each time step, and in 3D simulation, it needs to be called six times for each cell at each time step. Furthermore, in the case of radiative MHD, the problem has to be solved separately for the gas and radiation. Implementing a~precise yet complex solver can thus significantly influence the computational load.  For this reason, an approximate solver is usually implemented in MHD codes. One such solver is the HLL (Harten–Lax–van Leer) Rieman solver \citep{Harten1983}, which is also implemented in \koral{}. 

The HLL solver approximates the solution by the superposition of left and right moving waves, corresponding to the minimal and maximal velocity of the signal in the fluid. If the left and right maximal speeds, $S^{\mathrm{L}}$ and $S^{\mathrm{R}}$, are derived, the flux through a~cell face is given by

\begin{equation}
    \mathbf{F}= \begin{cases}\mathbf{F}^{\mathrm{L}} & 0<S^{\mathrm{L}} \\ \frac{S^{\mathrm{R}} \mathbf{F}^{\mathrm{L}}-S^{\mathrm{L}} \mathbf{F}^{\mathrm{R}}+S^{\mathrm{L}} S^{\mathrm{R}}\left(\mathbf{Q}^{\mathrm{R}}-\mathbf{Q}^{\mathrm{L}}\right)}{S^{\mathrm{R}}-S^{\mathrm{L}}} & S^{\mathrm{L}} \leq 0 \leq S^{\mathrm{R}} \\ \mathbf{F}^{\mathrm{R}} & S^{\mathrm{R}}<0\end{cases},
\end{equation}

\noindent where $\mathbf{Q}^R$ and $\mathbf{Q}^L$ are the right and left state respectively \citep{Toro}.  

This method also imposes a~constrain on the size of the time step $\Delta t$, since the maximal signal speed $S^{\mathrm{max}}$  in a~cell  cannot exceed $\Delta x/\Delta t$, where $\Delta x$ is the cell size. Otherwise, the signal would traverse the entire cell and influence fluxes on both faces within a~single time step. This constraint on the maximum $\Delta t$ size is known as the Courant–Friedrichs–Lewy (CFL) condition and can be expressed as

\begin{equation}
    \frac{S^{\mathrm{max}} \Delta t}{\Delta x} \leq C,
    \label{eq:4:CFL}
\end{equation}

\noindent where $C < 1$. Choosing a low value of $C$ leads to more precise results at the price of significantly increasing the computational load.

\subsection{Implicit-explicit time-stepping}

In the optically thick regime, the $G^\nu$ may become much larger than the conserved quantities, and the evolution equation stiff,  requiring an implicit solver to evolve it. In the \koral{} code, the second order Runge-Kutta implicit-explicit (IMEX)  stepper is implemented, following \cite{Pareschi2005} and \citep{Sadowski2015_dynamo}.

The implicit scheme in the \koral{} code evolves the time and spatial derivatives in (\ref{eq:3:coordbas3}) and (\ref{eq:3:coordbas2}) separately, as the advection term is stable in the explicit step due to the speed of light limit, which restricts the time step size $\Delta t$. The time derivative is then found implicitly as 

\begin{align}
    & T^t\phantom{}_\nu\left(t + \Delta t\right) - T^t\phantom{}_\nu\left(t\right) = \Delta t G_\nu\left(t + \Delta t\right)\\
    & R^t\phantom{}_\nu\left(t + \Delta t\right) - R^t\phantom{}_\nu\left(t\right) = -\Delta t G_\nu\left(t + \Delta t\right),
\end{align}

\noindent which leads to a~system of four equations due to the conservation laws. This system is solved for the primitive quantities. The IMEX scheme calculates the explicit and implicit terms together, enabling an efficient and accurate evolution of the optically thick fluid  \citep[for detailed description, see, e.g.,][]{McKinney2014}. 

In addition to the IMEX scheme, the \koral{} code offers other time-stepping solvers, such as the standard explicit Runge-Kutta in the first or second order or the Heun solver. However, the IMEX scheme is considered the best choice for problems involving optically thick fluids, as it efficiently handles the stiff behaviour of the equations.

\subsection{Conserved and primitives variables and reconstruction scheme}

In equation (\ref{eq:3:godunov}), the fluxes and source terms on the left-hand side can be expressed as simple functions of a~well-defined set of primitive quantities that naturally describe the flow, such as the density, velocity, and internal energy. However, these quantities are not conserved during evolution. Fortunately, it is possible to find another set of  quantities  such that they are also functions of the primitives and are conserved during evolution. Equation (\ref{eq:3:godunov}) can be then rewritten as:
\begin{equation}
    \frac{\partial}{\partial t} \mathbf{U}\left(\mathbf{P}\right) + \nabla \cdot \mathbf{F}\left(\mathbf{P}\right) = \mathbf{S}\left(\mathbf{P}\right),
    \label{eq:3:godunov_withP}
\end{equation}

\noindent where $\mathbf{U}$ and $\mathbf{P}$ are the vectors of conserved and primitive quantities, respectively.

The number of equations to solve in MHD codes depends on the included physics; for the case of the puffy disk simulations, using 3D GRMHD and radiation, there are 12 equations and 13 primitives. Specifically, six for the hydrodynamics part,  three for the magnetic field and 4 for the radiation. The vectors of primitive and conserved quantities are:

\begin{align}
    &\mathbf{P} = \left[\rho,\uint,u^\mu/u^t,S,B^i,\tilde{E},\tilde{F}^i\right], \label{eq:3:primitives}\\
    &\mathbf{U} = \left[\rho u^t,T^t_t+\rho u^t,T_i^t,B^i,R^t_t,R^t_i\right],
\end{align}

\noindent where the one extra primitive, $S$, is the gas entropy per unit volume. 

The entropy is used in the case of a~cold gas when $\uint\ll\rho$ and the numerical scheme can lead to negative values of $\uint$. At the end of each time step, the entropy is calculated and stored in the vector of primitives \citep{Sadowski2013}.  The entropy is given by 

\begin{equation}
    S = \frac{\rho}{\Gamma - 1}\ln{\frac{p}{\rho^\Gamma}}
\end{equation}

\noindent and evolved as 

\begin{equation}
    \left(Su^\mu\right)_{;\mu} = 0.
\end{equation}

The order of the primitives in (\ref{eq:3:primitives}) is kept the same as in the code to distinguish between the HD ($\rho,\uint,u^\mu/u^t,S$), magnetic ($B^i$) and radiation ($\tilde{E},\tilde{F}^i$) part. In the code, the number of primitives changes depending on the set-up; e.g., if the photon number conserving Comptonization is also included, the number of primitives is 14; for two-temperature plasma or relativistic electrons, it is much more.

The conversion from primitives to conserved quantities (p2u) is relatively straightforward. However, the inversion from conserved to primitives (u2p) is much more complicated, especially in the case of GRMHD (and even in the special relativistic MHD) because the u2p relations consist of an open set of nonlinear equations that must be solved numerically. u2p has to be performed at least once at each time step and, in the case of the second-order precision in \koral{}, even twice. Therefore, a~fast and precise method for the inversion is crucial for running the simulations effectively. 

The u2p inversion scheme in GRMHD involves solving an open set of algebraic equations, for which the Newton-Rhapson method is often used. Reducing the number of equations to solve is convenient to optimise the calculations. The algorithm implemented in the \koral{} code is well described in \cite{Noble2006}.

Several backup mechanisms are also implemented in the code in case the inversion fails, such as applying entropy inversion if the flow is too cold, as described above.  If all backup mechanisms fail, the code can also interpolate the faulty cell values from surrounding cells. In this case, the cell is flagged, the number of these fix-ups is summed over the whole domain, and the user is notified in the log. These cells also cannot be used for interpolation in another cell. If more of such cells are next to each other,  interpolation cannot be performed, the values of the primitives are set to the floor values, and the code will produce a~warning to the user that there is a~zone with a~potential problem, where the u2p inversion is failing, and the results can be unphysical.

The inversion is done separately for the MHD and radiation primitives, but the inversion scheme is almost identical. After each inversion, the floors are applied if obtained values reach them. 

A related problem is calculating values of primitive quantities in different parts of a~cell. The values are needed at the cell faces to calculate the fluxes. However, the values stored are at the cell centre, so the values of primitives at each face have to be \textit{reconstructed} to be consistent with the state of the fluid and avoid producing artificial shocks while not smoothing out large physical shocks.  Therefore, these numerical schemes are usually referred to as shock-capturing schemes.

Several reconstruction schemes are implemented in the \koral{} code, including the \textit{minmod} limiter. This scheme is controlled by only one diffusivity parameter $\theta_{MM}$ \citep{Sadowski2013}. Here, $\theta_{MM} = 1$ corresponds to the diffusive van Leer scheme, and $\theta_{MM} = 2$ to the monotonized central scheme. In simulations, $\theta_{MM} = 1.5$ is usually the best choice (see \cite{vLeer1979} and references therein).

\section{Running \koral{} }

\koral{} is written in the \texttt{C} programming language, employing numerical methods from the \texttt{GSL} library  \citep{galassi_gnu_2009} and using the \texttt{VisIt} library for visualization \citep{HPV:VisIt}. The structure of the code consists of 5 header files, 19 \texttt{C} files, and a~folder \texttt{OPACITIES}, where the tabular opacities introduced in \cite{Sadowski2017} are stored; this forms the core of the code. The user-defined problems are saved in the \texttt{PROBLEMS} folder. Anything that is problem-specific should be implemented in  a separate folder stored within the \texttt{PROBLEMS} folder. Each problem folder must include an \texttt{init.c} file, where the initial state is defined. Another mandatory file for each problem is the \texttt{define.h} header, which includes a~setup for the code, parameters of the spacetime and grid, or modifiers to the standard options or constants. 

Other optional files can be included, such as \texttt{prepinit.c} or \texttt{postinit.c}, which are called before and after initialization, respectively, and can include specific tools or modifiers for the initial state. The \texttt{kappa.c} and \texttt{kappaes.c} files allows the user to define their prescriptions for the absorption and electron scattering opacity. A \texttt{finger.c} file can be called at the beginning of each time step, e.g., to extract specific data or modify something. Additionally,  \texttt{bc.c} can be included, in which problem-specific boundary conditions are defined. The particular problem to be solved using the \koral{} core is specified in file \texttt{problem.h} in the root directory.

\begin{figure}
    \centering
    \includegraphics[width=1\linewidth]{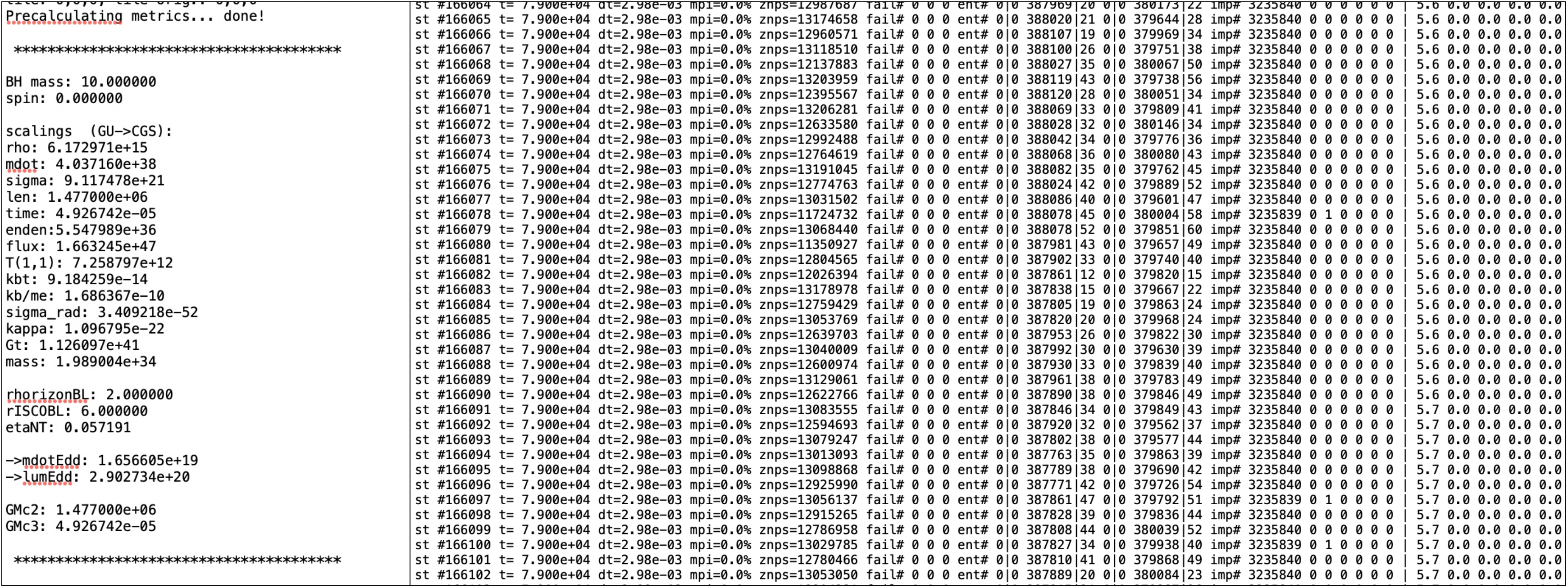}
    \caption{Example of \koral{} log output. \textit{Left:} The log printed right after the start shows general information about the set-up, including the BH mass and spin in the case of BH simulation, as well as scalings from the code units to the cgs units. \textit{Right:} Example of the log output during the run, including the step number and time in the simulation, the current time step size, MPI load, number of zones calculated per second, number of fails (i.e., failed u2p conversions in MHD or radiation part), and information about the implicit solver.} 
    \label{fig:3:log}
\end{figure}

The code also includes separate tools to process or modify the output data, such as tools for slicing and averaging over some plane, exporting outputs in more convenient \texttt{hdf5} format, averaging the results over specific time windows, or changing the resolution. Additionally, there is a~tool that enables turning on the radiation or other options of the code. Since the primitives are saved in the binary output files as a~1D array, if the number of primitives changes (e.g., by including radiation or photon number conserving), the entire structure of the binary file needs to be changed. 

The code is heavily parallelized using the message-passing interface (MPI), enabling simulations to run on supercomputers with a~large number of nodes and individual CPUs. Each node then process-specific part of the grid, and the values at the boundary are passed between them through the MPI. Simulations of the puffy disk were performed thanks to obtaining several grants at supercomputer \texttt{Prometheus} within the \texttt{PLGRID} infrastructure.

\subsection{\koral{} algorithm}

Having described important procedures required for GRRMHD in the previous sections, the algorithm of \koral{} can be summarized as follows:

\begin{enumerate}
    \item First, the code searches for an existing binary file to restart the simulation. If none is found, the \texttt{init.c} initializes a~new simulation. The initial state is always saved to a~separate array for the boundary condition calculations or other operations if needed.
    \item Ghost cells are filled using values corresponding to the boundary conditions. The number of ghost cells depends on the order of the reconstruction scheme, and typically, it is between 3 and 5 in each direction. They are required for  the flux calculations through the outer edges of the domain. Ghost cells are not evolved in time.
    \item Re-scaling or perturbation of the initial state is performed if defined by the user. For instance, when initializing an accretion disk with a~stable torus, the azimuthal fluid velocity  may be perturbed to trigger the MRI.
    \item The main time loop begins with the initial time step size $\Delta t$ estimated from characteristic wave speeds in the smallest cell. In the following steps, $\Delta t$ is calculated during the explicit step from the CFL condition. 
    \item User-defined \texttt{finger.c} is applied, and data distribution through MPI is performed when running on a~cluster.  
    \item The first implicit and explicit steps are performed to obtain the intermediate values, followed by the first u2p inversion. Data are exchanged over MPI, ghost cells' values are determined from the boundary conditions, and the dynamo term is applied in a~2D simulation.
    \item The first IMEX step is performed to obtain the intermediate values. 
    \item Second implicit and explicit steps are done in the same manner as before, but using the intermediate values. 
    \item The second IMEX step advances the whole solution in time.
    \item Second u2p inversion is performed, and the entropy is calculated and stored in primitives. 
    \item At the end of each time step, the total number of failed u2p inversions is counted, and the number of iterations of the implicit solver and  performance marks are calculated. All relevant information is printed into the log for the user to control the simulation status.
\end{enumerate}

The explicit and implicit steps include all manipulations with the primitives, reconstruction scheme, and Riemann problem-solving. The explicit step includes
\begin{enumerate}
    \item Primitives are reconstructed on the cell borders, and left and right fluxes are calculated.
    \item Riemann problem is solved using the HLL solver to calculate fluxes through all cell faces. The magnetic fluxes are determined using the CT to enforce $\nabla\cdot \mathbf{B} = 0$ constraint.
    \item The time step $\Delta t$ is calculated using the wave speed in each cell. The minimal value across the whole domain is then used everywhere. 
    \item Conserved quantities in $(t+\Delta t)$ are calculated using (\ref{eq:3:godunov_withP}).
\end{enumerate}

In the implicit step, the energy of the radiation and the fluid is compared in each cell. Then, the one which stores less energy is evolved implicitly, using the original values of primitives as the first guess. If a~solution is not found in some cell after a~predefined number of iterations, the values are interpolated from neighbouring cells, flagged, and printed in the log for the user. 

\koral{} saves its outputs into binary files as often as the user defines. The code can read these binary files, restart the simulations from a~certain point, or process them via several complementary scripts for different purposes. If it runs on a~serial machine, the code can directly create a~\texttt{VisIt}  \texttt{.silo} files, where most of the essential quantities are saved, which is helpful for a~fast check of the run or tuning of the initial setup. 

\subsection{Computational grid and resolution}

A correct choice of computational grid parameters is fundamental in MHD simulations, as it must have a~sufficient resolution to resolve the low-scale turbulences while considering computational demands. The smaller cell size leads to a~smaller $\Delta t$, as evident from the CFL condition (equation \ref{eq:4:CFL}). The issue becomes even more complex in the case of the GR due to singularities in the commonly used coordinate systems. However, the coordinate singularities can be removed by adopting the Kerr-Shild (KS)  coordinates \citep{Kerr2009}. 

To achieve an ideal resolution, the cells must be smaller in the inner region and the midplane of the accretion disk, usually at the equatorial plane. This can be easily achieved using the modified Kerr-Schild (MKS) coordinates \citep{McKinney2004}. Nevertheless, the simulation outputs are most conveniently stored in Boyer–Lindquist (BL) coordinates \citep{Boyer1967}. Transformation relations between all three in the Kerr metric for a~rotating BH are described below.

The BL coordinates are $\left(t,r,\theta,\phi\right)$, and the Kerr metric in BL coordinates is given by (\ref{eq:2:kerr}). The determinant of the metric is $g=-\Sigma^2\sin^2\theta$.

In the KS coordinates, $\left(t,r,\theta,\phi\right)$, the Kerr metric is expressed as:

\begin{align} 
\der s^2= & -\left(1-\frac{2 r}{\Sigma}\right) \der t^2+\left(\frac{4 r}{\Sigma}\right) \der r \der t+\left(1+\frac{2 r}{\Sigma}\right) \der r^2+ \Sigma \der \theta^2 \nonumber \\ 
& +\sin ^2 \theta\left[\Sigma+a^2\left(1+\frac{2 r}{\Sigma}\right) \sin ^2 \theta\right] \der \phi^2 \\ 
& -\left(\frac{4 a r \sin ^2 \theta}{\Sigma}\right) \der \phi \der t-2 a\left(1+\frac{2 r}{\Sigma}\right) \sin ^2 \theta \der \phi \der r, \nonumber 
\end{align}

\noindent and the determinant is $g=\Sigma^2\sin^2\theta$. The $r$ and $\theta$ coordinates are the same as in the BL coordinates, $r^{\mathrm{KS}} = r^{\mathrm{BL}}$ and $\theta^{\mathrm{KS}} = \theta^{\mathrm{BL}}$. 

Implementing the MKS coordinates provides the advantage of using a~logarithmic grid, which concentrates the resolution in $r$ and $\theta$ direction. The transformations for $t^{\mathrm{MKS}}$ and $\phi^{\mathrm{MKS}}$ remain the same as in the KS coordinates, and the other coordinates transform as

\begin{align}
    & r^{\mathrm{KS}} = R_0 + \exp \left( r^{\mathrm{MKS}}\right),\\
    & \theta^{\mathrm{KS}} =\left\{ \frac{\tan \left[H_0 \pi \left(\theta^\mathrm{MKS} - \frac{1}{2}\right)\right]}{\tan\left(\frac{\pi}{2}H_0 \right)} + 1 \right\}\frac{\pi}{2},
\end{align}

\noindent where $R_0$ and $H_0$ are parameters controlling how much are the cells  concentrated in $r$ and $\theta$ directions. The expression for $\theta$ transformation is a~slightly modified version of the MKS coordinates presented in \cite{McKinney2004}. 

Other coordinate systems are implemented in the \koral{} code, e.g., cylindrified coordinates covering precisely the polar regions to resolve the jets, similar to those introduced in \cite{Ressler2017}. 

However, internally the code considered all cells as cubes, and the coordinate transformation is applied only when necessary. 
 
\subsection{Resolving the MRI}

One challenge of the GRMHD is the scalability of turbulences, which makes it difficult to achieve fully converged solutions that resolve turbulences on all scales. When changing the resolution, convergence in an accretion disk GRMHD simulation is typically determined by the stability of important parameters (e.g., mass accretion rate, density scale height, radiative luminosity). However, increasing resolution to reach convergence is impractical and computationally expensive. 

In studies by \cite{Hawley2011,Hawley2013}, quality parameters were derived to assess simulation convergence based on the assumption that the characteristic wavelength of the MRI must be well resolved so the MRI-induced turbulences are maintained. Although satisfying these conditions is important for obtaining reliable and physical results, limitations and approximations of the GRMHD must still be considered.

The quality parameters, $Q_\theta$ and $Q_\phi$, correspond to the ratio of the cell size to the characteristic MRI wavelength 
\begin{equation}
    \lambda_\mathrm{MRI} = \frac{2\pi\nu^A}{\Omega}, 
\end{equation}
\noindent where $\nu^A$ is the Alfvén speed and $\Omega$ is orbital frequency. The quality parameters are then defined as 
\begin{align}
&Q_\theta=\frac{\lambda_{\mathrm{MRI}}}{\Delta \theta}=\frac{2 \pi\left|\nu^A_{\theta}\right|}{\Omega \Delta \theta},\\
&Q_\phi=\frac{\lambda_{\mathrm{MRI}}}{r \Delta \phi}=\frac{2 \pi\left|v^A_{\phi}\right|}{\Omega \Delta \phi}.
\end{align}

\noindent To maintain the turbulences, the non-linear growth of MRI has to be resolved, which couples both parameters. If $\langle Q_\phi \rangle\geq20$, then $\langle Q_\theta\rangle\geq10$ is needed. However, if $\langle Q_\phi \rangle$ is lower, $\langle Q_\theta \rangle$ can compensate if it is high enough \citep{Hawley2011}.

\chapter{Puffy accretion disk}
\label{chap:puffy}

The puffy disk is a numerical model of an accretion disk based on the results of the GRRMHD simulation,
targeted to study properties and behaviour of sub-Ed\-dington accretion onto a stellar-mass BH. Unlike the traditional thin disk model, which assumes a geometrically thin and optically thick structure, the puffy disk consists of a vertically extended and optically thick layer surrounding a geometrically thin core. The puffy disk is stabilized by the magnetic pressure against the thermal and viscous instabilities, even in the radiation pressure-dominated regime. The puffy disk simulations provide a detailed model of the dynamics, radiation properties, and stability of sub-Eddington accretion in XRBs by employing advanced numerical techniques and radiative transfer methods. 

Simulation of the sub-Eddington accretion helps to interpret the data obtained from observations of accretion systems, such as microquasars containing a BH. By comparing the simulated observational signatures of the puffy disk with the actual astrophysical data, we can refine our models and gain insights into the physical processes occurring within these systems. Additionally, commonly used tools and methods for analyzing data from observations can be tested against the puffy disk simulations \citep{Wielgus2020,Lancova2023}.   

\section{Simulating a sub-Eddington accretion disk}

Two apparent approaches exist to initiate a simulation of sub-Eddington accretion onto a compact object, such as a BH. The first method involves setting up a thin disk by adopting existing solutions like those proposed by \cite{Novikov+Thorne1973} with a Gaussian distribution of quantities in the vertical direction. Alternatively, one can employ the steady thin disk solution by \cite{Kluzniak2000}. Then, thread the initial state with a weak magnetic field to induce turbulence that drives the matter towards the central BH. 

However, in this particular setup, in the regime when the radiation pressure dominates over the gas pressure (e.g., for $\mdot \gtrsim 0.1$ for a stellar-mass BH), the thermal and viscous instabilities lead to the collapse of the disk \citep{Lightman1974,Shakura1976,Fragile2018,Jiang2013}. 

\begin{figure}
    \centering
    \includegraphics[width=1\textwidth]{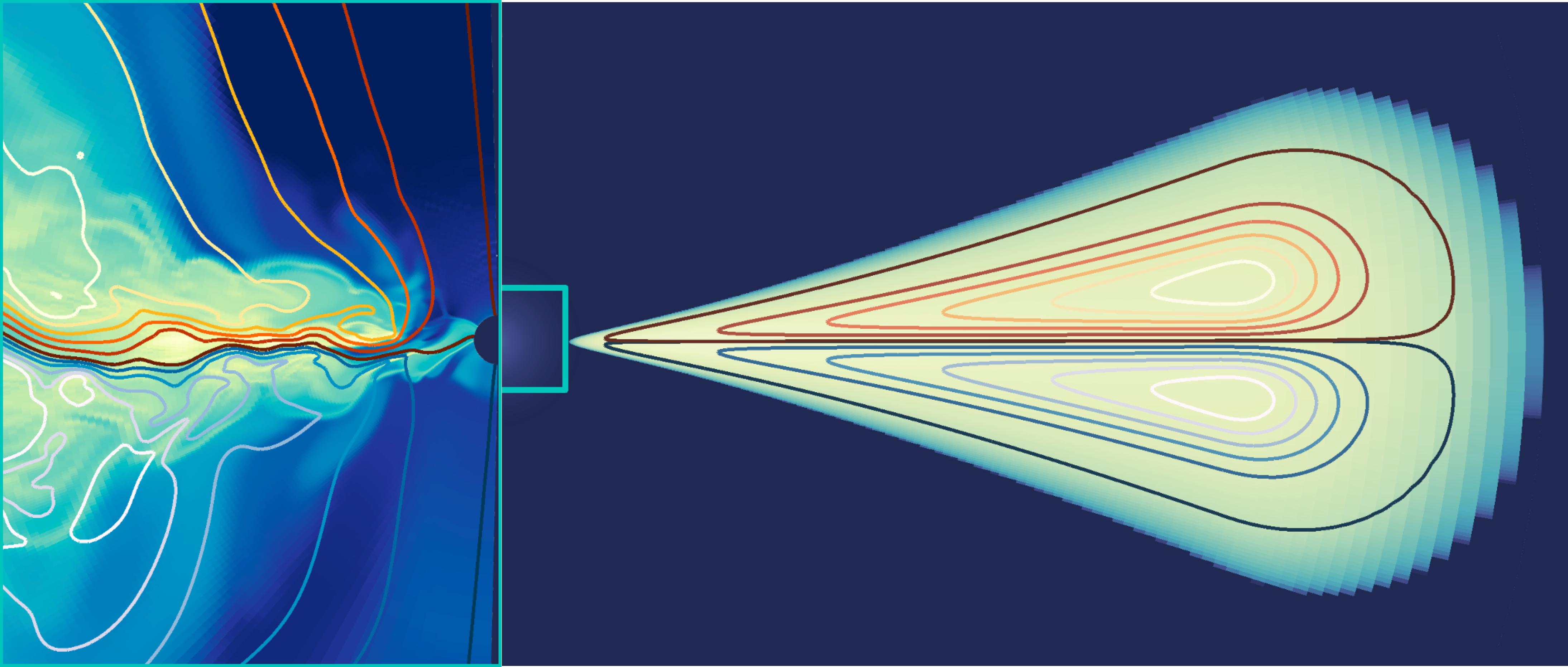}
    \caption{Logarithmic density plots of the initial torus with magnetic field loops (\textit{right}) and the initial state for the puffy disk simulation (\textit{left}) in the inner regions denoted by the green rectangle. The magnetic field lines are shown in both parts.}
    \label{fig:init}
\end{figure}

An effective strategy to maintain a stable sub-Eddington disk involves introducing an ad hoc cooling function as an additional source term in the MHD equations (\ref{eq:3:godunov}). This function removes excess or injects sparse energy generated within the flow during each time step, thereby regulating the disk's thickness to radius ratio $H/r$ \citep[e.g.][]{Penna2013, Shafee2008}. However, while this approach can enforce stability, it is obviously artificial and does not fully address the underlying issue of disk stability.

On the other hand, accurately resolving the MRI  in the central region of an extremely thin disk with a very low $\mdot$ presents its own challenges.  Achieving this requires an exceptionally high resolution throughout the entire simulation domain or implementing methods that dynamically increase resolution where needed. One such approach is the adaptive mesh refinement, as employed in simulations of tilted accretion disks by \cite{Liska2019}. This method allows for targeted refinement of regions of interest, enabling a more precise representation of the MRI and its effects in razor-thin disk cores.

\begin{figure}
    \centering
    \includegraphics[width=1\linewidth]{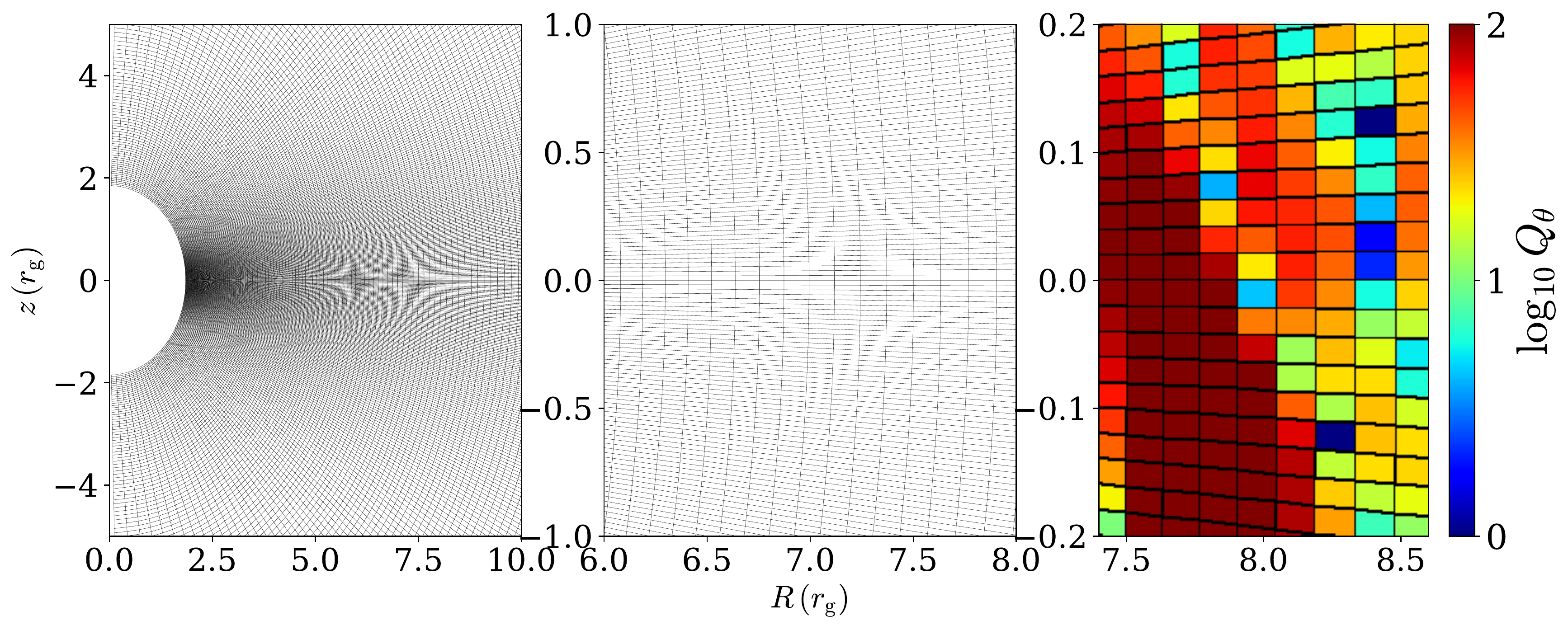}
    \caption{Computational grid for the puffy disk simulations in the $r,\theta$ plane. \textit{Left and middle:} Detailed view of the computational grid. \textit{Right}: The quality parameter in the $\theta$ direction, $Q_{\theta}$. }
    \label{fig:4:grid}
\end{figure}

Another approach to simulate sub-Eddington accretion involves modelling the behaviour of matter falling onto a BH from a distant reservoir on a Keplerian orbit. However, such simulations can have excessively long duration due to their dependency on the position of the external source of matter. As a result, simulating the entire system of a BH and a star as a matter source, resembling a real XRB, becomes unfeasible within the GRMHD framework.

An alternative approach is considering a stable matter reservoir closer to the central object with on a Keplerian orbit. Then, a more manageable and realistic scenario can be obtained by focusing the analysis solely on the innermost regions where the disk is self-consistently formed. 

This approach was adopted in \cite{Sadowski2016}, focused on the disk stability and comparison of different magnetic field topologies. In the puffy disks project, we used the same set of simulations, expanding them to lower $\mdot$, and conducted a comprehensive analysis of the disk's general properties in relation to standard analytical models. 

Consequently, while the results described the disk as a "thin disk" in \cite{Sadowski2016}, our findings in \cite{Lancova2019} revealed significant differences when compared to the analytical models, leading to the introduction of a novel accretion disk model, the puffy disk.



\section{Initial configuration for the Puffy disk simulations}

The puffy disk simulations were initiated using a thick accretion torus of \cite{Penna2013_torus}, as illustrated in the right panel of Figure \ref{fig:init}. The inner edge of the torus is positioned at $r_{in} = 42\,\rg$. The torus is threaded by a quadrupole magnetic field, with the counterclockwise loops above and clockwise loops below the equator. Initially, the magnetic field strength is set not to exceed $\beta = \pgas/p_{\mathrm{mag}}=20$.

The simulation is carried out in 2D for approximately $\sim60\, 000\, \tg$ using the mean-field dynamo (see section \ref{sec:3:dynamo}) to sustain the MRI in axisymmetric simulation. This duration allows sufficient time for the matter to form a disk in the inner region (left panel of Figure \ref{fig:init}). Once the $\mdot$ stabilizes, the data is extended into a 3D grid, and the simulation continues for $\sim 2000\,t_g$ to achieve a relaxed solution. This final state is considered the actual initial state of the simulations.

\begin{figure}
    \centering
    \includegraphics[width=\linewidth]{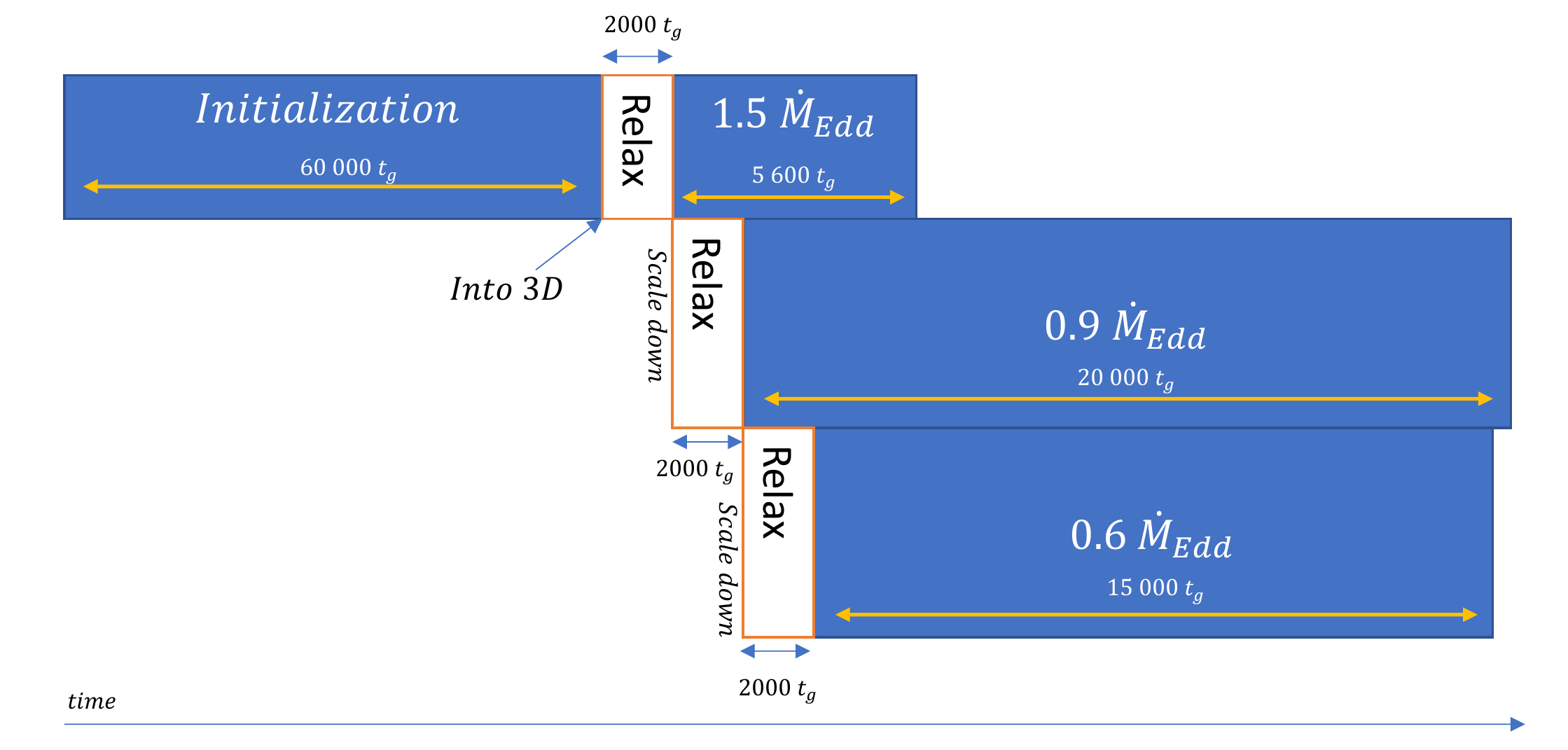}
    \caption{The timeline of the puffy disk simulations showing the initial state in 2D and the time needed for the solution to relax.}
    \label{fig:4:timeline}
\end{figure}

The first run, labelled as \texttt{S15}, is conducted with a super-Eddington $\mdot$ but is relatively short. To obtain a lower $\mdot$, the density and magnetic field strength are scaled down by a small constant factor in the whole domain, conserving the gas temperature and maintaining a constant magnetic-to-gas pressure ratio. This rescaled simulation is run for an extended period and results in $\mdot$ dropping to $\sim 0.9$.  Following another rescaling, the $\mdot$ decreases further to $\sim 0.6$, leading to the puffy disk as described in \cite{Lancova2019} (simulation \texttt{S06}).

Attempting another rescaling with the original resolution becomes problematic, as the inner region close to the equatorial plane becomes unresolved, leading to the rapid collapse of the disk. Table \ref{tab:sims} provides an overview of all three simulations and selected parameters, and Figure \ref{fig:4:timeline} the timeline of the simulations.

\begin{table}
    \caption{Puffy disk simulations and selected parameters, including the density scale-height $h_\rho$, photosphere height $h_\tau$ to radius ratio, total duration $T_{run}$ of each simulation in $\tg$, number of orbits at $r=10\,\rg$, and seconds.}
    \centering
    \begin{tabular}{c c c c c c c}
        \hline
        Simulation & $\mdot$ & $h_\rho/r$ & $h_\tau/r$ & $T_{run}\, (\tg)$ & $T_{run}\, (\mathrm{orbits})$ & $T_{run}\, (\mathrm{s})$  \\\hline \hline
        \texttt{S15} & $1.51$ & $0.20$ & $1.0$ & $5600$ & $28$ & $0.27$\\ \hline
         \texttt{S09} & $0.89$ & $0.15$ & $1.0$ & 20000 & 100  & 0.98 \\ \hline
         \texttt{S06} & $0.58$ & $0.10$ & $1.0$&  17400 & 87 & 0.86
\\ \hline
    \end{tabular}
    \label{tab:sims}
\end{table}

\subsection{Computational grid and resolution}

The computation grid resolution is ($320,320,32$) in $N_r, N_\theta$, and $N_\phi$ respectively. In the $\phi$ direction, the domain covers a $\pi/2$ wedge rather than the full $2\pi$ to optimize computational resources. For verification, a full 3D  simulation was presented in \cite{Sadowski2016}, demonstrating that the results quantitatively match those obtained from the $\pi/2$-wedge simulation. 

The calculations are performed on the MKS coordinates, employing logarithmic spacing in $r$ and $\theta$, concentrating cells in the innermost regions, achieving a significant effective resolution. The computational grid can resolve the MRI even in the dense disk core while satisfying the minimal requirements for the quality parameters throughout the disk, see Figure \ref{fig:4:grid}.

\subsection{Stability}

In the analytical thin disk model, the balance between the viscous heating  $Q^+_\mathrm{visc}$ and radiative cooling  $Q^-_\mathrm{rad}$ is maintained when the disk is primarily governed by gas pressure. However, when the radiation pressure exceeds the gas pressure, this balance is swiftly disrupted, resulting in disk collapse. This phenomenon was also demonstrated in GRMHD simulations, such as those conducted by \cite{Mishra2016}. 

In order to study the stability of an accretion disk, an ideal simulation would need to span a duration corresponding to the viscous timescale at the outer edge of the disk, which is approximately $t_{\mathrm{acc}}\sim 10^{17}\,\tg$ ($10^{13} \,\mathrm{s}$ for $10\,\msol$ BH), for the outer edge located at $100\, \rg$. However, within the framework of GRMHD, the longest simulations have only recently reached a duration of up to $10^5\,\tg$ \citep{Liska2019}. Conducting a radiative simulation of such an extended duration would require immense computational power. Nonetheless, the stability can still be investigated in shorter simulations, e.g.,  by examining if the simulation remains stationary on the $T-\Sigma$ diagram, where $\Sigma$ is the surface density. The balance in the azimuthal ($\phi$) direction can be demonstrated through the radial profile of $\mdot$, which should remain constant if the mass inflow is stable.

\begin{figure}[b]
    \centering
    \includegraphics[width=1\linewidth]{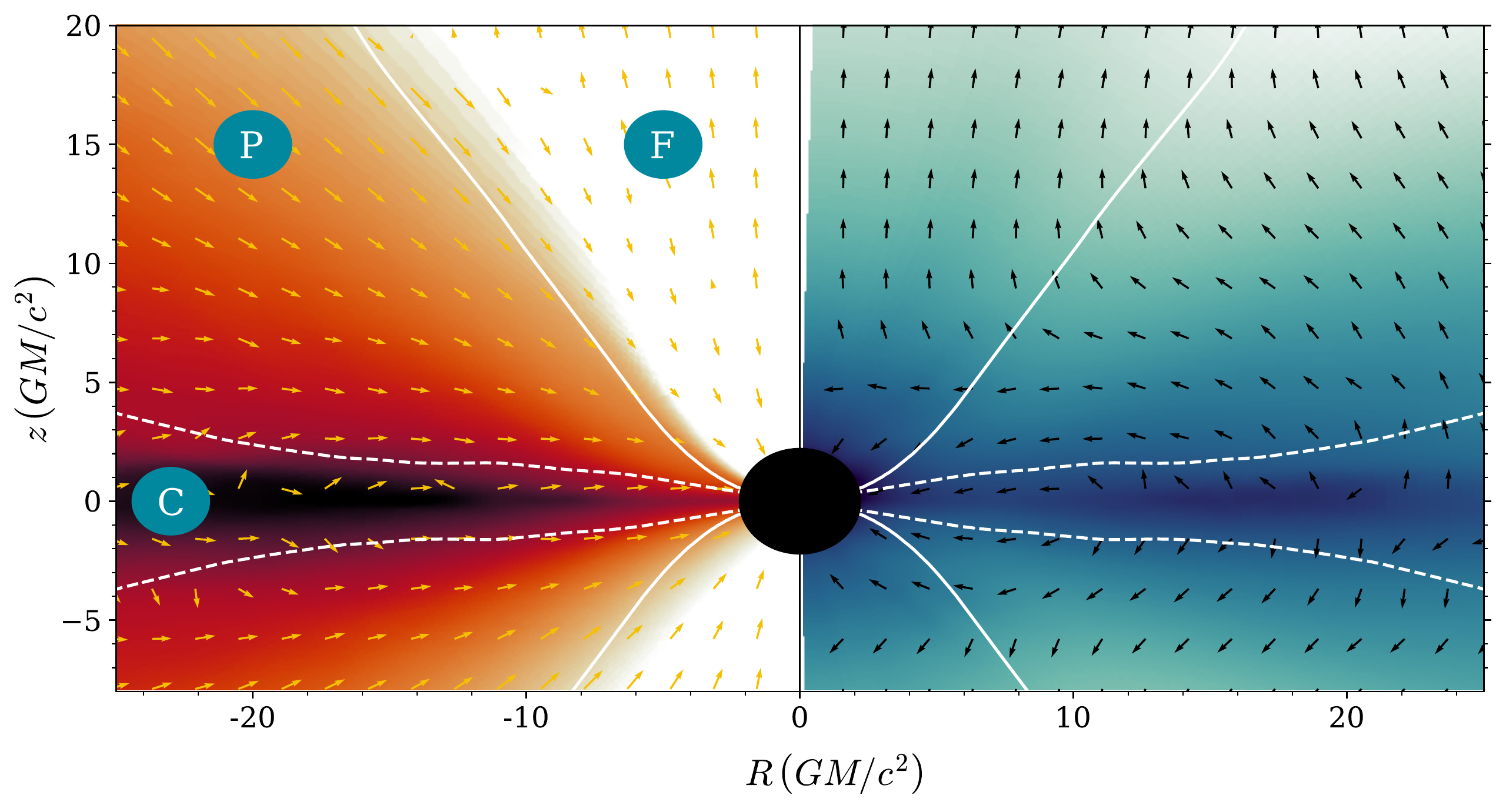}
    \caption{The structure of the puffy disk: Gas and radiation in the $r-\theta$ plane using the time-averaged data from \texttt{S06}. The full white line represents the photosphere, while the dashed line indicates the density scale height. 
    \textit{Left:} Logarithm of density with denoted regions (C - core, P - puffy, F - funnel). Quivers show the gas momentum flux. \textit{Right:} Logarithmic colour map displaying radiation energy density, accompanied by arrows indicating the radiation flux.}
    \label{fig:4:moments}
\end{figure}

\begin{figure}
    \centering
    \includegraphics[width=\linewidth]{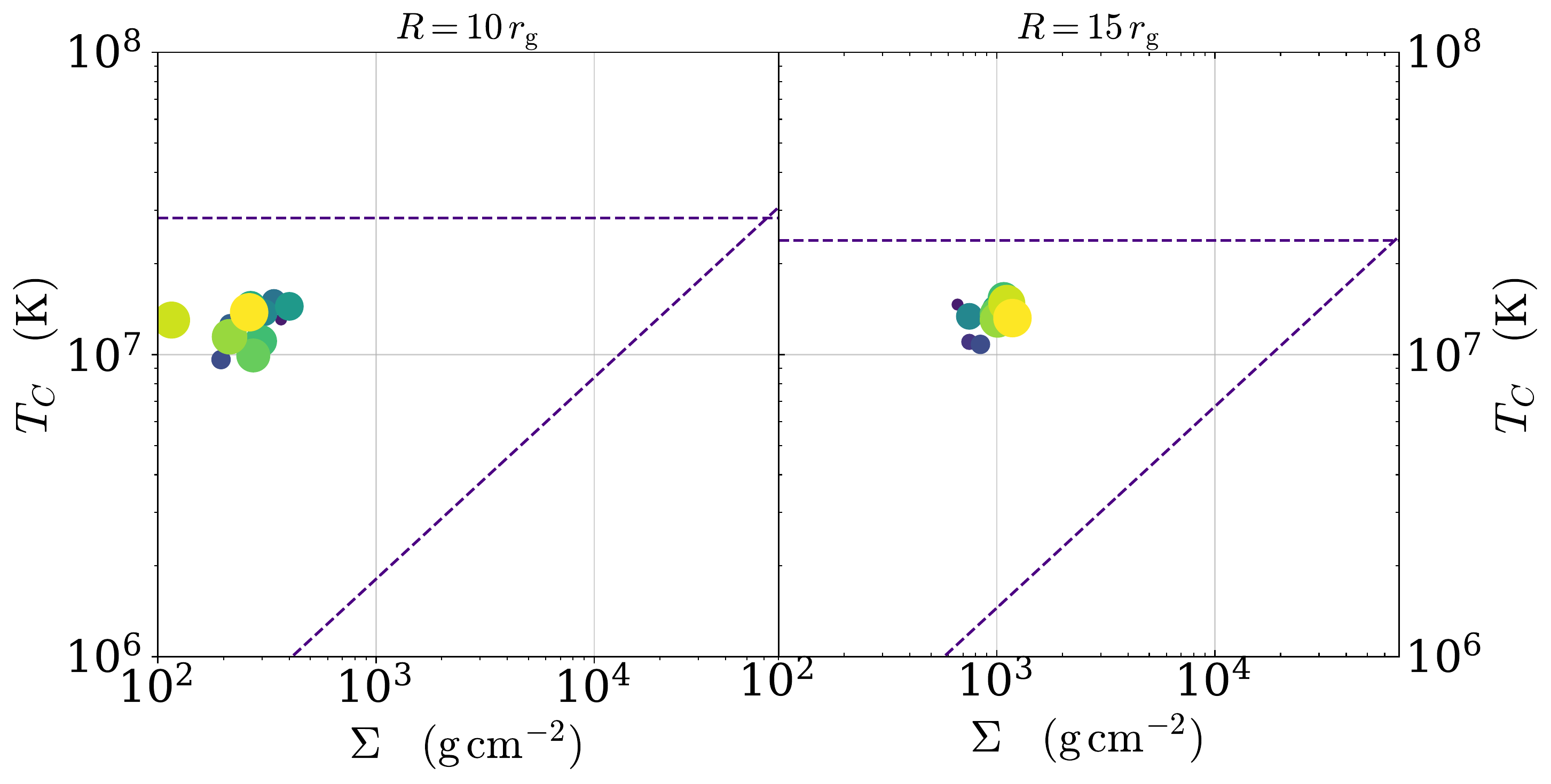}
    \caption{Thermal equilibrium ($T-\Sigma$) diagram for a standard thin disk at $r=10\,\rg$ (\textit{left}) and $r=15\,\rg$ (\textit{right}), using $\alpha = 0.1$, $m=10$ and non-rotating BH. The dots represent the time evolution of the puffy disk during the \texttt{S06}; the colour of the dots changes from blue to yellow, and their size grows over time. }
    \label{fig:4:T2Sigma}
\end{figure}

In the puffy disk simulation, the heating and cooling processes are balanced throughout the entire duration, thanks to the vertical net flux of the magnetic field, as is illustrated in Figure 4 of \cite{Sadowski2016}. Figure \ref{fig:4:T2Sigma} shows that the puffy disk maintains thermal equilibrium throughout the simulation. Additionally, Figure \ref{fig:4:mdot} indicates that the total $\mdot$ remains stable up to relatively large radii, providing further evidence of a relaxed and stable simulation.

\section{Properties of the puffy disk}

The puffy disk represents a sub-Eddington accretion disk stabilized by magnetic pressure. The disk is geometrically and optically thick. It can be divided into three distinct regions, each with unique properties:
\begin{itemize}
    \item Core (C on Figure \ref{fig:4:moments}): This region resembles a standard thin disk, geometrically thin and dense.
    \item Puffy region (P on Figure \ref{fig:4:moments}): A vertically extended optically thick area of hot magnetized gas. It resembles a warm corona, as discussed in \cite{Gronki2020} and \cite{Rozanska2015}.
    \item Funnel (F on Figure \ref{fig:4:moments}): Located above the disk, this optically thin region exhibits strong outflows.
\end{itemize}

The disk core is defined by the density scale height given by

\begin{equation}
h_\rho =\sqrt{\frac{\int \rho z^2 \der z}{\int \rho \der z}}.
\end{equation}

\noindent The photosphere of the disk is located where the gas becomes optically thin, corresponding to a surface denoted by $h_\tau$. The criterion for optically thin gas is when the vertically integrated optical depth, $\tau$, satisfies $\tau \leq 2/3$. The optical depth is calculated as
\begin{equation}
    \tau(z)= \int_{z_{\mathrm{out}}}^{z} \kappa_{es}\rho \der z.
\end{equation}
\noindent where the integration is performed from the outer boundary at $z_{\mathrm{out}}$ towards the equator.

\begin{figure}
    \centering
    \includegraphics[width=0.8\linewidth]{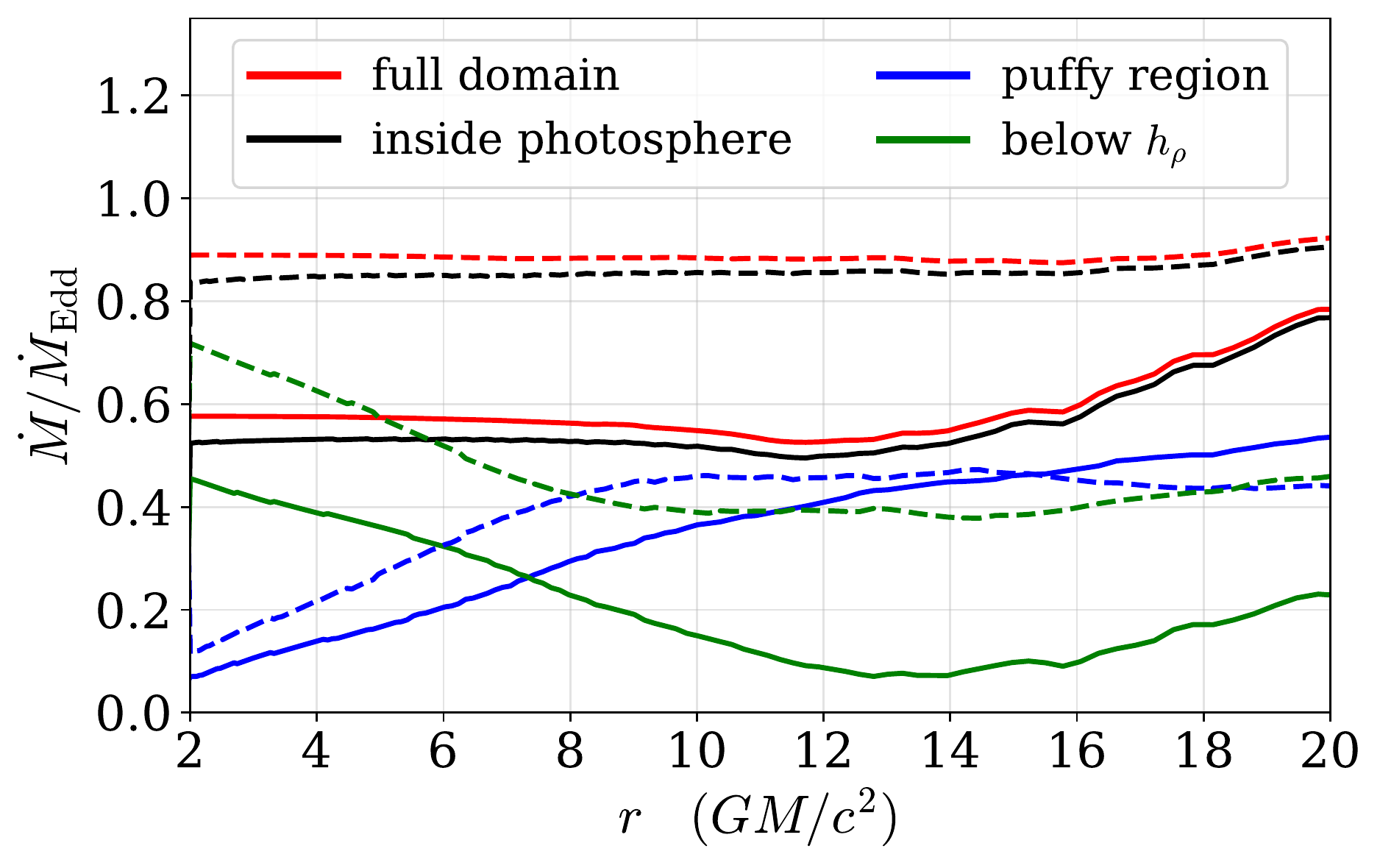}
    \caption{Mass accretion rate $\dot{m}$ in different regions for all three runs. Dashed lines correspond to \texttt{S09}, while the solid lines to the \texttt{S06}.}
    \label{fig:4:mdot}
\end{figure}

The core resembles a geometrically thin Keplerian disk of high density, gas temperatures around $10^7$ K, and subsonic accretion velocities. The mass accretion rate in this region is very low, and it even approaches zero in some areas (as shown in Figure \ref{fig:4:mdot}). Most of the accretion occurs in the puffy region, particularly at larger radii. 

\begin{figure}
    \centering
    \includegraphics[width=\linewidth]{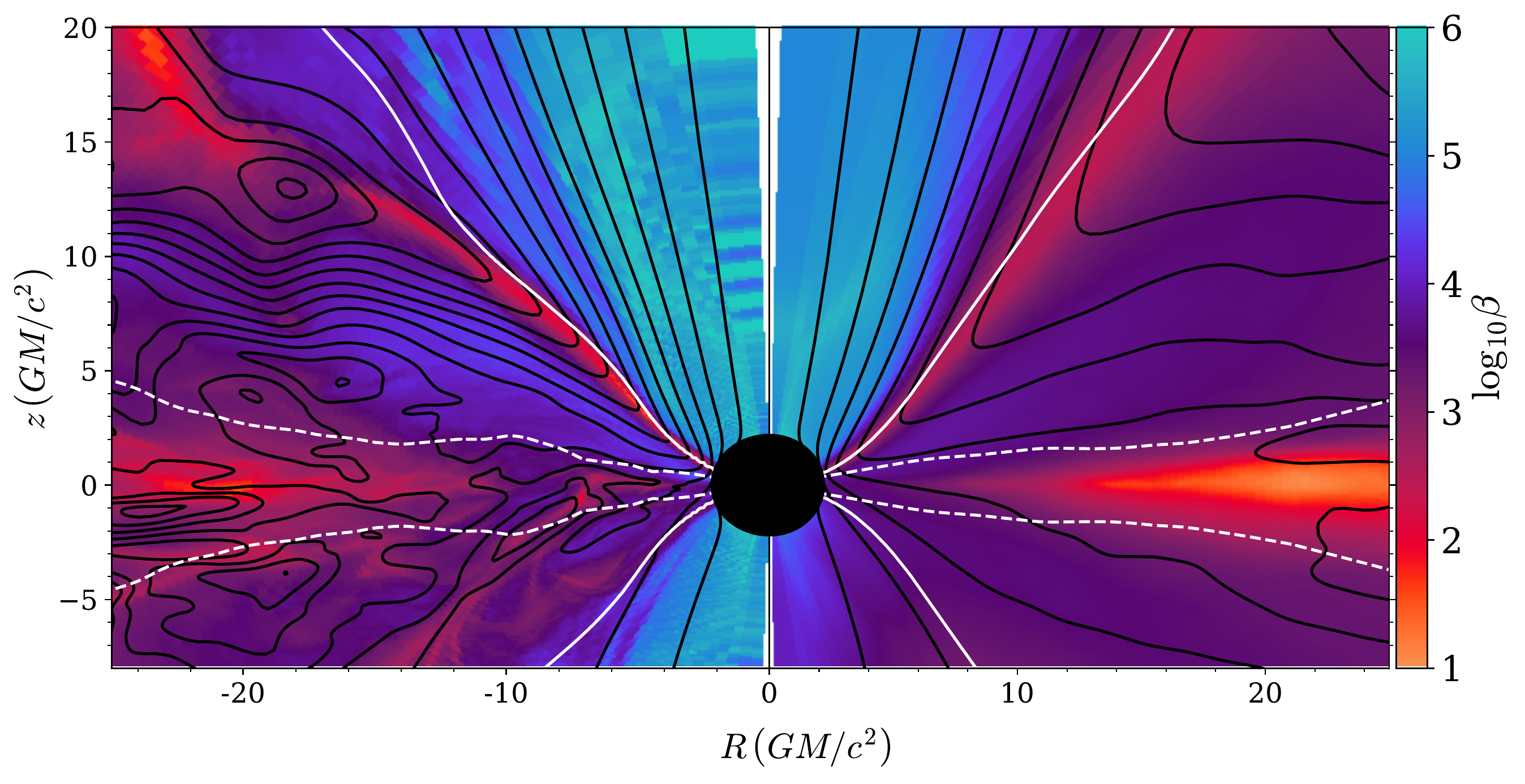}
    \caption{Magnetic field lines (black contours) and magnetization $\beta$ shown in colours for simulation \texttt{S06}. The white solid line represents the photosphere, while the dashed lines indicate the density scale height. The \textit{right} panel displays data averaged over both azimuthal angle and time, providing an overall view. In contrast, the \textit{left} panel shows a~snapshot from the simulation, averaged only over the azimuthal angle (close to the axis, the snapshot data shows numerical error due to lower resolution, these errors are being corrected during the simulation run and do not impact the overall results.).}
    \label{fig:4:magfield}
\end{figure}

The fluid within the disk shows a high level of magnetization, particularly in the funnel region, but is also higher in the puffy region than in the core. The magnetic field is turbulent in the disk; however, these turbulent fluctuations are averaged out over the entire simulation run to establish a magnetic field topology specific to this solution. In this topology, the azimuthal component dominates within the disk, while the radial component dominates in the funnel region, as shown in Figure \ref{fig:4:magfield}.

On the photosphere, the magnetic field  rapidly changes as the field lines are carried away by the gas. This process leads to a decrease in the magnetic pressure and potential formation of a current sheet and magnetic reconnection, which cannot be resolved in our simulations under the ideal MHD approximation.

In the optically thick regions of the puffy disk, the radiation approximately follows the density gradient, escaping through the funnel region and becoming collimated along the axis. In the innermost regions of the disk, a significant portion of the radiation is advected into the BH, contrary to what analytical models predict. A fraction of the radiation is also released, but closer to the centre than anticipated in a standard thin disk. As a result, the spectrum is shifted towards lower radii compared to the predictions of a standard thin disk model. 

Regarding the gas dynamics, a strong inflow is present in the puffy region of the disk. On the other hand, the gas motion in the core region of the disk is slower and more turbulent. This behaviour arises from the complex interplay between magnetic forces, radiation pressure, and gravitational effects within the disk. The slower and more turbulent motion in the core allows for energy dissipation and redistribution of angular momentum, leading to the formation and maintenance of the geometrically thin disk structure observed there.

The core temperature is slightly higher than the puffy region, as illustrated in Figure \ref{fig:4:temps}. The core temperature rapidly increases in the plunging region, but the temperature on the photosphere follows almost a  power-law distribution with no significant change under the $\risco$. 

\begin{figure}[t]
    \centering
    \includegraphics[width=\linewidth]{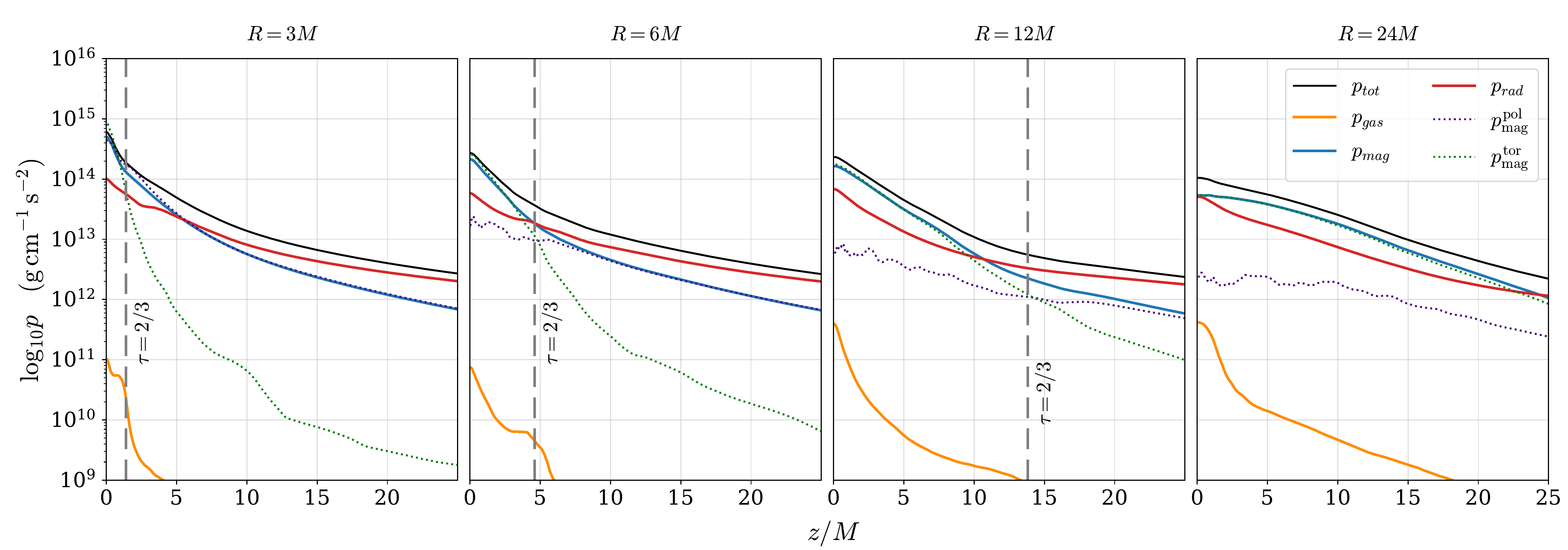}
    \caption{Vertical distribution on pressure components on selected radii for the \texttt{S06} simulation.}
    \label{fig:4:pres_vert}
\end{figure}

\begin{figure}[b]
    \centering
    \includegraphics[width=\linewidth]{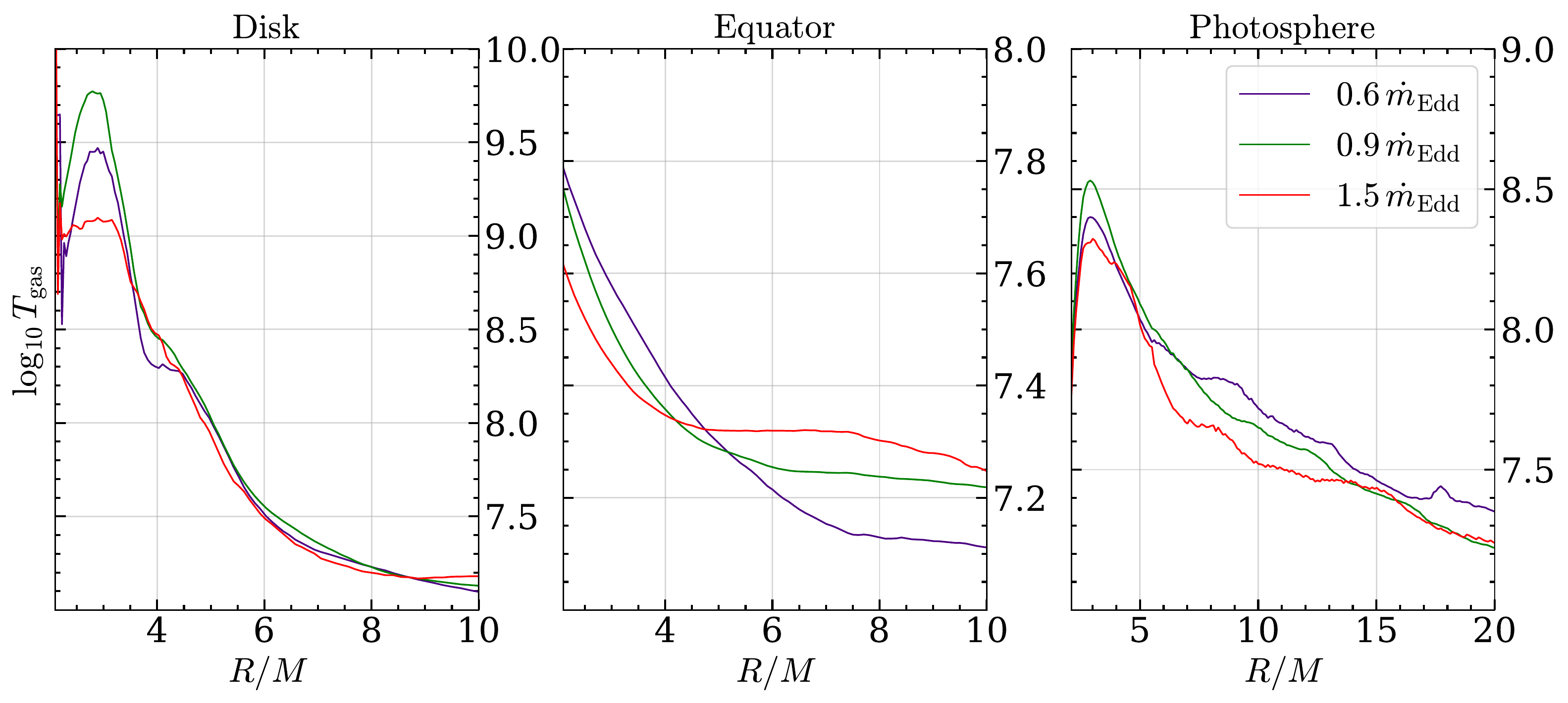}
    \caption{Radial profiles of the gas temperatures for all three simulations on different surfaces, at the core surface (\textit{left}), at the equator (\textit{middle}), and on the photosphere (\textit{right}).}
    \label{fig:4:temps}
\end{figure}

In both the core and puffy regions of the disk, the radiation and magnetic pressures surpass the gas pressure by several orders of magnitude, as shown in Figure \ref{fig:4:pres_vert}. The presence of magnetic pressure and its interplay with other physical processes contribute to the stability of the disk, despite the significant deviation from the traditional expectations of a sub-Eddington thin disk structure.

\section{Observational picture}

The analysis of these observational properties of the puffy disk and comparisons with analytical models were presented in \cite{Wielgus2022}, and in \cite{Lancova2023} it was extended to include also the \texttt{kynbb} spectral model in \textsc{xspec} and compare it with the \texttt{kerrbb}.

The puffy disk spectra and images were computed using the relativistic radiative post-processing code \texttt{HEROIC} \citep{2015MNRAS.451.1661Z,2016MNRAS.457..608N}. This code solves the complete radiative transfer problem on the time-averaged output from the converged simulations for three different mass-accretion rates. By accounting for various physical processes such as scattering effects, Comptonization, bremsstrahlung, and synchrotron radiation, \texttt{HEROIC} enables to study of observational properties and a direct comparison with standard analytical models.  Additionally, it provides an opportunity to validate commonly used tools by simulating observations.

\begin{figure}
    \centering
    \includegraphics[width=\linewidth]{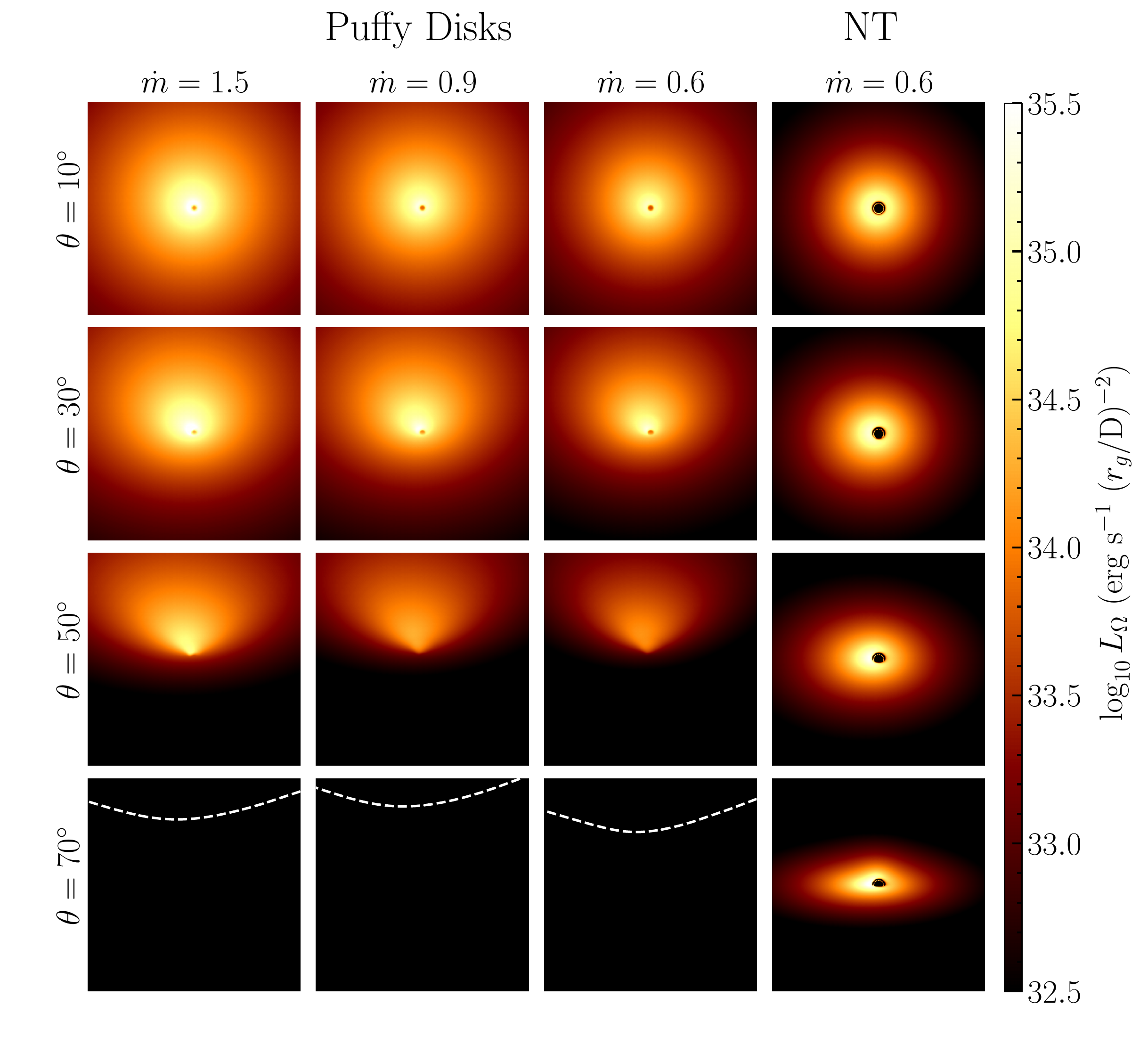}
    \caption{Three columns on the left show the images of the puffy disk for three different mass-accretion rates, while the right column is the image of a standard thin disk with $\mdot = 0.6$. The images are arranged in increasing inclination angles from top to bottom. The colour map scale is consistent across all images. However, in the last row, the puffy disk's luminosity is significantly lower, making its visibility less prominent than the other images. From \cite{Wielgus2022}}
    \label{fig:DO_images}
\end{figure}

One feature of the puffy disk is the beaming of radiation towards the rotational axis, accompanied by significant obscuration for viewing angles close to the equatorial plane. This behaviour stands in contrast to the standard thin disk model. The inner parts of the disk and the central BH are completely obscured for observers with high inclinations, see Figure \ref{fig:DO_images}. Conversely, for observers with low inclinations, the disk radiation extends all the way to the event horizon.

\begin{figure}
    \centering
    \includegraphics[width=\linewidth]{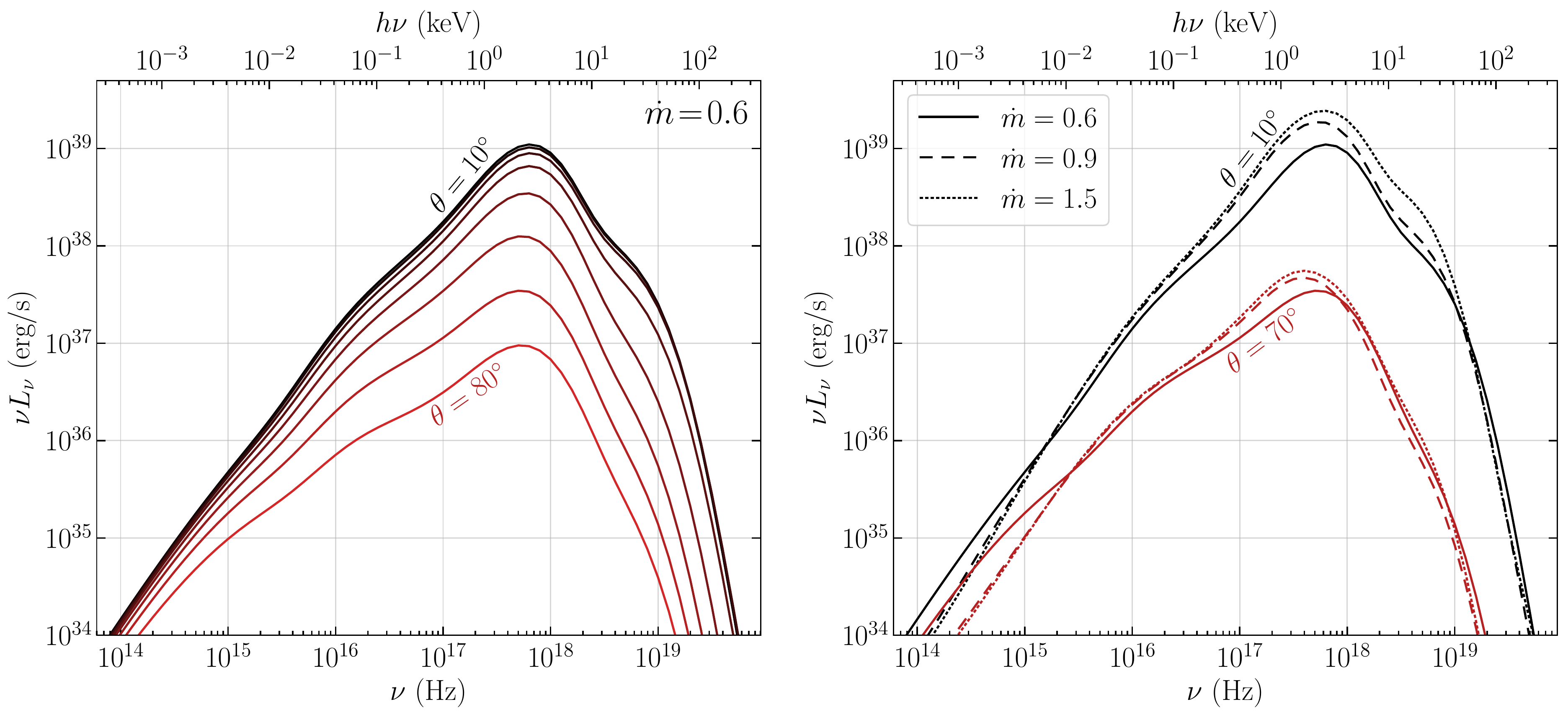}
    \caption{\textit{Left}: isotropic radiative power per logarithmic frequency interval, as a~function of the observer’s inclination, from $10^\circ$ (nearly face-on) to $80^\circ$ (nearly edge-on).
    \textit{Right}: isotropic radiative power per logarithmic frequency interval for three puffy disks simulations, corresponding to different mass accretion rates. }
    \label{fig:4:spectra}
\end{figure}

The shape of the spectra emitted from the puffy disk is similar to analytical thin or slim disk models, with a notable additional power-law component corresponding to thermal Comptonization. However, when comparing the spectra from the three simulation runs at different inclination angles, deviations from the analytical models become larger.  While the observer's inclination angle can influence the position of the maximum of luminosity by up to two orders of magnitude, fixing the inclination angle shows no significant change in the SED for different $\mdot$ values, as shown in Figure \ref{fig:4:spectra}. Regardless of the inclination angle, the frequency of photons associated with the peak luminosity remains constant.

Synthetic spectra of the puffy disk were generated for two inclinations of the \texttt{S06} simulation to test the performance of commonly used spectral fitting models. These spectra were then fitted using the \texttt{XSPEC} package \citep{1996ASPC..101...17A}, simulating the observation of a real astrophysical source. Two models, a fully relativistic thin disk model \citep[\texttt{kerrbb},][]{kerrbb} and a fully relativistic advective slim disk model \citep[\texttt{slimbh},][]{2009ApJS..183..171S,Straub2011}, were used, each accompanied by a thermal Comptonization component \citep[\texttt{nthcomp,}][]{Zdziarski1996,Zycki1999}, which models  seed photons from the disk that are upscattered to higher energies by a homogeneous corona. The same analysis was later repeated for the \texttt{kynbb} \citep{Dovciak04} model.

The BH mass, distance, viscous $\alpha$, and inclination angle were fixed for the fitting, and total luminosity, spin, temperature of the seed and up-scattered photons, and photon index are free parameters. The results showed that the puffy disk could be fitted with standard tools and provide a reasonable set of parameters.

However, the results of the fitting process did not align with the simulation parameters. Notably, the BH spin is overestimated, with obtained values of $a \sim 0.8$ for low inclination and $a \sim 0.5$ for high inclination angles. This discrepancy highlights the problems of fitting spectral states with strong Comptonization, leading to unreliable spin measurements using continuum fitting methods \citep{2004MNRAS.347..885G}.

The puffy disk simulation (\texttt{S06}) can be compared to the ultra-luminous or intermediate state of a microquasar during an outburst, characterized by high luminosity and a soft spectrum with a prominent Compton component. In this state, QPOs are often observed.  However, due to the limitations of the simulation duration and the capabilities of the radiative transfer code, a detailed analysis of the oscillatory behaviour and fast variability is not feasible within the scope of this model.

\chapter{Modelling the rapid time variability of XRBs}
\label{chap:QPOs}

Observations of bright objects in the X-ray sky reveal distinct properties of these sources, such as the quasi-periodic temporal variability across a~wide range of frequencies, as described in Section \ref{sec:1:timing}. Since the discovery of the QPOs in the late 1980s \citep{1986ApJ...306..230M}, they have been brought to the attention of many research groups and have been the target of observations by X-ray telescopes \citep{vdKlis2006,Remillard2006}. With the launch of the Rossi X-Ray Timing Explorer (RXTE) in 1995, high-frequency variability was discovered \citep{1999PhDT........10W,1993A&AS...97..355B}.

Initially observed in NS sources, where the HF QPOs are very strong,  they were also identified in BH sources. This chapter is dedicated to modelling the HF QPOs and their application in extracting information about the geometry and properties of the central object in these sources. 

The frequencies of HF QPOs are apparently linked to the fundamental frequencies of motion in the innermost regions of accretion disks \citep{1999PhRvL..82...17S,tor-etal:2005,1999ApJ...526..953K,1999PASJ...51..317K}. However, accurately modelling the origin of these oscillations is a~complex task that requires long-time GR simulations of the behaviour of matter and radiation in the close vicinity of a~compact object.

\begin{figure}[b]
    \centering
    \includegraphics[width=1\linewidth]{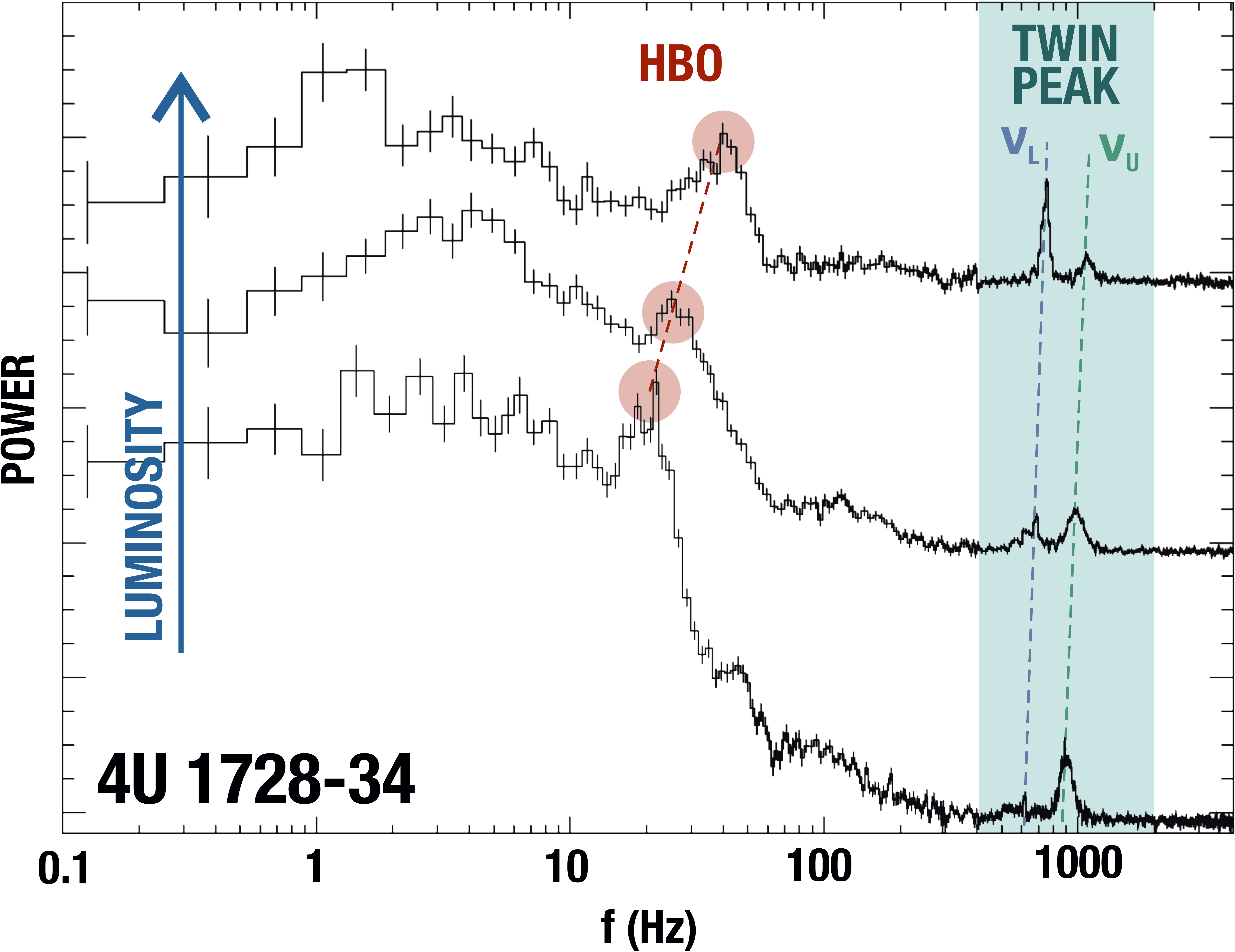}
    \caption{PSD diagram from three observations of NS LMXB \texttt{4U 1728-34}, each corresponding to an increase in luminosity. Both the LF QPOs and thin-peak QPOs are clearly visible, along with changes in the significance of the lower $\nu_L$ and upper $\nu_U$ peak. Data from \cite{Strohmayer1996}.}
    \label{fig:5:qposPSD}
\end{figure}

This chapter describes HF QPO modelling, as presented in the following published papers. Despite the strong correlation between LF QPOs frequencies and those of the HF QPOs, the latter is more complicated to model and explain. In fact, up until now, no widely accepted model comprehensively explains these phenomena.  For convenience,
the HF QPOs are henceforth referred to as QPOs. 

I have contributed to several published results in this field throughout my studies, complementing those described in the previous chapters. Additionally, I have actively participated in several ongoing studies, that have already been published or are currently submitted for peer review.

\section{Fast time variability of black holes and neutron stars in XRBs}

Over the years following their discovery, several models have been proposed to explain the QPOs. These models are founded on the premise that the QPOs originate in the innermost regions of an accretion disk near the central compact object. While each model can explain specific observed properties, they often have limitations when addressing other aspects. Furthermore, it remains to be seen whether the mechanism behind the twin-peak QPOs in NS systems is the same as that governing the QPOs in BH systems \citep{vdKlis2006}. 

In the case of NS sources, the observed variability is often tied to processes occurring on its surface and interactions between the disk, the star, and a~luminous boundary (spreading) layer on the NS surface. As discussed in Section \ref{sec:1:timing}, NS systems display LF QPOs corresponding to the horizontal, normal, and flaring branches on the HID, as well as the twin-peak QPOs (sometimes referred to as kHz QPOs) with very high frequencies comparable to the orbital frequencies in the innermost regions.

An interesting observation is the strong correlation between the HBO frequency and the lower peak of the twin peak QPOs \citep{vdKlis2006}, shown in Figure \ref{fig:5:qposPSD}. This correlation was explained as a~superposition of the Keplerian frequency at the inner edge of an accretion disk (associated with the magnetospheric radius in NS systems) and the rotational frequency of the NS itself. This model is known as the magnetospheric-beat frequency model \citep{1985Natur.316..239A,1998ApJ...509..793M}.

However,  this scenario is limited to specific QPO observations and a~narrow range of NS's surface magnetic field strengths. Eventually, with more twin-peak QPO data being observed, a~decrease in frequency difference with increasing frequency was discovered, contradicting this model.

Another phenomenon often categorized as QPOs is the NS burst oscillations observed during type I X-ray bursts. These bursts occur due to the sudden ignition of thermonuclear burning of accreted material on the NS surface. The frequency of burst oscillations falls within the range of  300 to 600 Hz, which is close to a~typical NS spin frequency. However, frequency shifts during the burst have been observed \citep[see, for example,][]{1997ApJ...486..355S}, consistent with a~small expansion of the NS envelope. Models involving unstable inhomogeneous hydrogen burning on the NS surface and non-radial oscillation modes of the envelope have been proposed, providing a~reasonably accurate representation of the observed data \citep{1998ApJ...506..842B}.

\section{Foundations of the orbital models of QPOs}
\label{sec:5:models}

\begin{figure}[b]
    \centering
    \makebox[0pt]{\includegraphics[width=1.1\linewidth]{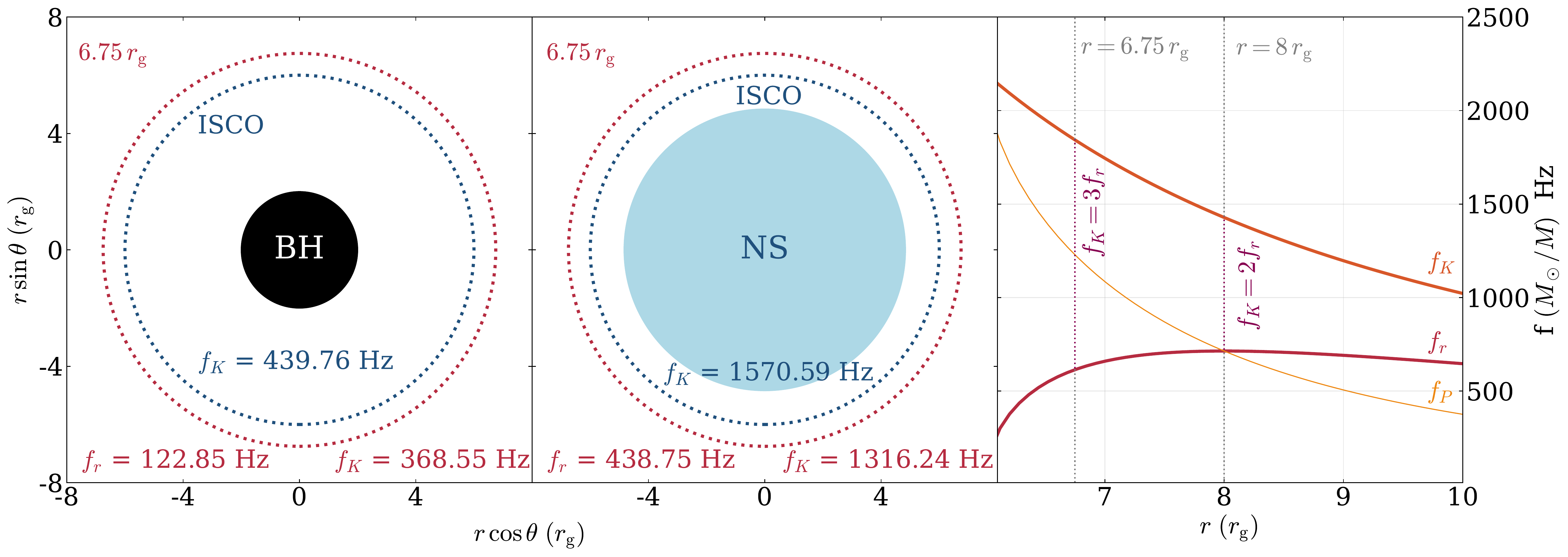}}
    \caption{\textit{Left and Middle}: Illustration of a~non-rotating BH and NS, accompanied by the  corresponding values of Keplerian ($f_K$) and radial epicyclic ($f_r$) frequencies at $\risco$ and $r=6.75\,\rg$, where $f_K = 3f_r$. Values for the BH are calculated for $M = 5\,\msol$, while those for the NS for $M=1.4\,\msol$ and radius $R_* = 10$ km. \textit{Right}: Values of the $f_K$, $f_r$ and $f_P$ in the \schw{} case. Note that in the non-rotating spacetime $f_\theta = f_K$.}
    \label{fig:5:NSBHfreqs}
\end{figure}

QPOs are observed in both NS and BH systems. However, in NS systems, the amplitude of the oscillations is much higher compared to the BH systems, see Figure \ref{fig:1:QPOs}. The observed frequencies of these oscillations closely align with the fundamental frequencies of orbital motion in the vicinity of a~compact object. This  alignment is particularly evident for geodesic circular motion with Keplerian orbital frequency
\begin{equation}
    f_K = \frac{c^3}{2\pi GM}\frac{1}{a+r^{3/2}},
\end{equation}
\noindent and the radial $f_r$ and vertical $f_\theta$ epicyclic frequencies
\begin{align}
    f_r^2 &= f_K^2\left[ \left(1-\frac{6\rg}{r}\right) + 8a\left(\frac{\rg}{r}\right)^{3/2} - 3a^2 \left(\frac{\rg}{r}\right)^2 \right], \label{eq:5:radialepicyclic}\\
    f_\theta^2 &= f_K^2 \left[1-4a\left(\frac{\rg}{r}\right)^{3/2} + 3a^2 \left(\frac{\rg}{r}\right)^2 \right] \label{eq:5:verticalepicyclic}
\end{align}

\noindent in the Kerr metric, where $a$ is the non-dimensional spin parameter of the central object and $r$ is the radius. The radial dependence of these frequencies and their typical values for a~stel\-lar-mass BH and a~NS are illustrated in Figure \ref{fig:5:NSBHfreqs}.

Quasi-periodic modulation in the X-ray band, with frequencies similar to those of the orbital motion, has been observed not only in XRBs but also in galactic centres containing a~supper-massive BH \citep{2018MNRAS.477.3178C}. Observations across a~wide range of masses  demonstrated that the observed QPOs are inversely proportional to the mass of the central object and fall within the range defined by the Keplerian frequency in the innermost regions \citep{Remillard2006,2004ApJ...609L..63A,Gol-etal:2019,IBWS}.

\subsection{Models of orbiting blobs}

Various mechanisms implemented within distinct QPO models can translate the orbital motion into the observable signal. One such mechanism involves the motion of clumps or blobs of matter in the innermost region of accretion disks. These clumps, having higher temperature than the surrounding disk, can modulate the observed light curves as they orbit. This concept was recognized prior to the first QPO observations and was considered a~potential method to measure the NS mass \citep{1990ApJ...358..538K}. 

A specific model based on the kinematics of orbital motion in proximity to a~compact object  is the relativistic precession (RP) model. This model explains the oscillations by considering the geodesic motion of a~test particle within the framework of GR  \citep{1999PhRvL..82...17S}. The RP model connects the frequency of the upper peak of QPOs to the Keplerian frequency ($\nu_U = f_K$), while the lower peak corresponds to the periastron precession frequency ($\nu_L = f_P = f_K - f_r$), as illustrated in the right panel of Figure \ref{fig:5:NSBHfreqs}. Additionally, the model interprets the LF QPOs as the nodal precession frequency. However, the coherence times observed in the QPO signals contradict the predictions of the RP model \citep{2005MNRAS.357.1288B}.

\subsection{Diskoseismology}

The distinct profile of the radial epicyclic frequency motivated the development of a~family of diskoseismic models. The presence of a~maximum in $f_r$ relatively close to the $\risco$ can lead to the trapping of oscillatory modes, particularly the g-mode, within an accretion disk. This trapping can yield observational signatures \citep{1991ApJ...378..656N,1992ApJ...393..697N}. 

One of these models is the disk warp model, where the oscillations are generated by perturbing an initially axisymmetric state of the disk, leading to the trapping of specific modes \citep{2004PASJ...56..905K}.  These modes have also been observed in (GR)MHD simulations \citep{Mishra2019,2020MNRAS.496.3808B,2020PASJ...72...38K,2019MNRAS.483.4811M,2020MNRAS.497..451D}.

\subsection{Oscillations of accretion tori}

An extensive collection of papers has been published, proposing a~connection between the QPOs and  oscillations of accretion tori,   \citep[e.g.,][]{AbramowiczKluzniak2001,2003MNRAS.344..978R,2004ApJ...617L..45B,2005AN....326..849B,2006CQGra..23.1689A,Blaes2006,2010MNRAS.405.2447I,2016MNRAS.461.1356F,2017arXiv170907706D,2017MNRAS.467.4036M}. 

The accretion tori are often approximated using the Polish doughnut model, which assumes a~stationary perfect fluid in purely azimuthal motion. It forms within the surfaces of constant potential, in which the gravitational forces acting on the torus and the internal pressures of the torus are combined \citep{Jarosz1980,Abramowicz+1978}. Although the Polish doughnut model is known to be unstable to the Papaloizou–Pringle instability \citep{1984MNRAS.208..721P}, it, in general, provides a good approximation for a time-averaged realistic turbulent accretion flow.

The specific frequency ratios of QPOs observed in both BH and NS sources have inspired the development of the family of resonance models \citep{KluzniakAbramowicz2001,AbramowiczKluzniak2001}. These models connect the QPO to oscillations at distinct radii, where the ratios of Keplerian and epicyclic frequencies are   small natural numbers. In the simplest case of the \schw{} metric, these radii are $r=6.75\,\rg$, where $f_K = 3f_r$, or $r = 8\, \rg$, where $f_K = 2f_r$ \citep[which also corresponds to the maximum of the radial epicyclic frequency,][]{1980PASJ...32..377K}. 

Oscillations of fluid accretion tori situated very close to the central compact object have been explored within the resonance models, \citep{2004ApJ...603L..89K,2004ApJ...603L..93L,2004AIPC..714...21A}. In this case, the two QPO peaks are associated with the fastest-growing resonant modes in the accretion disk, which correspond to the 3:2 epicyclic frequency ratio \citep{2002astro.ph..3314K,2003MNRAS.344..978R}.

Possible resonances between frequencies of relativistic orbital motion were extensively studied in a large collection of papers \citep{KluzniakAbramowicz2001,AbramowiczKluzniak2001,horak2009,2008NewAR..51..841K,2005A&A...443..777P,2006A&A...451..377H,2013A&A...552A..10S,2015A&A...581A..35B,2021ASSL..461..263M,2004PASJ...56..819H,2013MNRAS.434.2761H}. Currently, the role of resonances in the generation of QPOs is subject to debate  \citep{2003A&A...404L..21A,2008AcA....58....1T,2010MNRAS.401.1290B,2019NewAR..8501524I}. However, the concept of oscillating accretion tori persists as a~robust and plausible explanation for this phenomenon.

\begin{figure}
    \centering
    \includegraphics[width=1\linewidth]{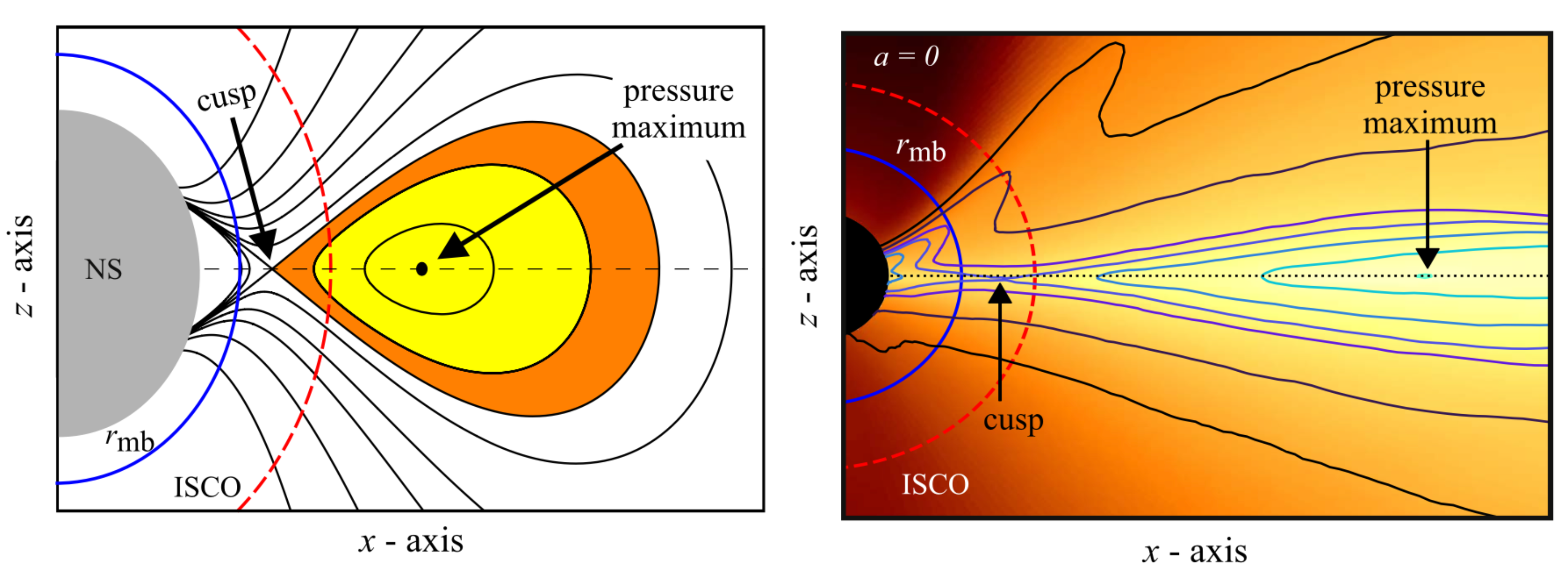}
    \caption{Comparison of the cusp torus geometry based on analytical solution of the Polish doughnut and the equipressure surface in complex GRRMHD simulation of the puffy disk. From \cite{tor-etal:2022}.}
    \label{fig:5:cuspy}
\end{figure}

\section{Marginally overflowing tori and the CT model}

A thick accretion torus can be described using the Polish doughnut model recalled in Sections \ref{sec:2:overwiev} and \ref{sec:5:models}. A torus can fit within any of the equipotential surfaces defined by the model, spanning from an infinitesimally slender torus with an elliptic cross-section (circular in Newtonian gravity or the \schw{} case) to a~maximally thick torus that fills the critical largest closed equipotential, forming a~cusp-like shape. A~typical shape of the equipotential surfaces  is illustrated in the left panel of Figure \ref{fig:5:cuspy}.

\begin{figure}
    \centering
    \includegraphics[width=1\linewidth]{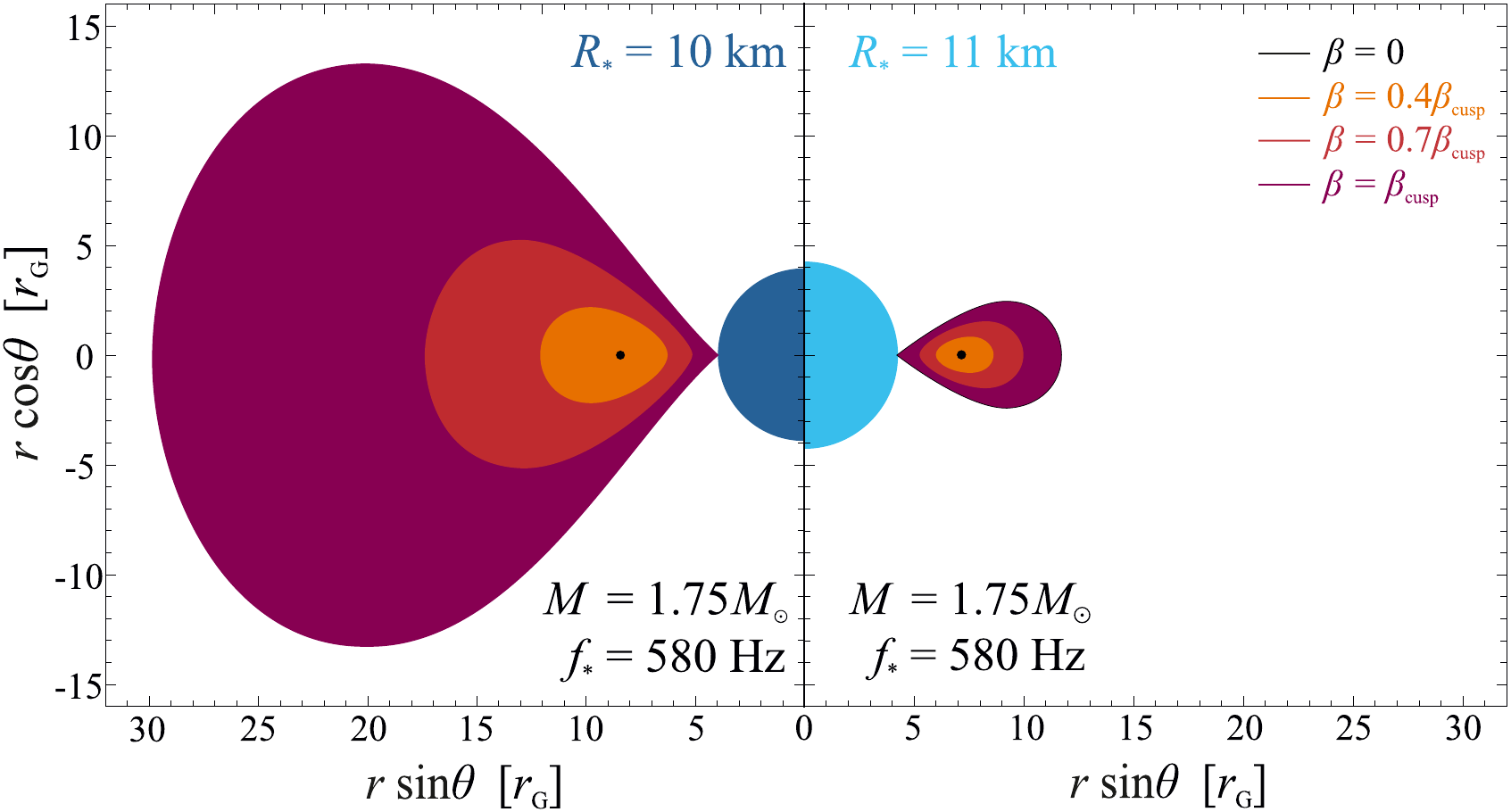}
    \caption{Theoretical maximal size of a~cusp torus in the Hartle-Thorne metric limited by the radius of the NS $R_*$ for fixed mass and rotational frequency $f_*$. This limitation arises from the universal relations linking the metric parameter $q$ to the mass, spin, and radius of the NS \citep{2013MNRAS.433.1903U}. Based on \cite{Matuszkova}.}
    \label{fig:5:maxtorus}
\end{figure}

The torus thickness is quantified by parameter $\beta$, where $\beta=0$ corresponds to a~slender torus and $\beta=1$ to a~theoretical torus of infinite size. The specific value of $\beta$ for the cusp torus configuration ($\beta = \beta_{cusp}$) depends on the spacetime geometry \citep{2009CQGra..26e5011S}, yet it is also constrained by the accreting system geometry, see Figure~\ref{fig:5:maxtorus}.

Analytical formulas for the radial and vertical eigenfrequencies of torus oscillations in the Paczy\'{n}ski-Wiita potential were derived by \cite{2007ApJ...665..642B} and \cite{2007A&A...467..641S}, and a~full-GR  version of these formulas was introduced by \cite{2009CQGra..26e5011S} and \cite{2016MNRAS.461.1356F} for slightly non-slender tori with a~constant angular momentum.

The cusp torus configuration is promising for modelling oscillation modes in real accretion disks, as it corresponds to a~marginally overflowing torus. Oscillations of such torus, situated in the innermost region of the accretion flow, can modulate the amount of matter transferred onto the NS surface, which can be reflected in the luminosity of the boundary layer \citep{1987Natur.327..303P,2005AN....326..845H}. Furthermore, a~cusp-shaped structure is also expected to exist in real accretion disks  \citep{tor-etal:2022,Kotrlova2020}, see Figure \ref{fig:5:cuspy}.

The cusp-torus (CT) model, initially introduced and preliminary tested on observed data from NS sources in \cite{2016MNRAS.457L..19T}, is based on the solution of \cite{2009CQGra..26e5011S}. This model assumes oscillations and Keplerian motion related to cusp tori. In \cite{Kotrlova2020}, the model was applied to observed QPOs in galactic microquasars and compared to other models.

More recently, the CT model has been extended into the Hartle-Thorne metric \citep{2023AN....34420114S,2022arXiv220310653M}, incorporating the influence of the quadrupole moment of a~NS, $q = Q/M^3$. This parameter significantly affects the frequencies of orbital motion \citep{2013MNRAS.433.1903U,Urbancova2019}. The Hartle-Thorne metric is accurate to the second order in $q$ and therefore provides a~sound approximation for the spacetime surrounding an oblate NS, with deviations mainly occurring close to the ISCO. It was shown that this model  matches frequencies of twin-peak QPOs \citep{tor-etal:2022,torok2016mnras}.

\section{Oscillations of accretion tori and estimations of parameters of compact objects}

The frequencies of observed QPOs can be used to derive parameters of the compact object, such as its mass or spin, due to their linkage to the accretion flow properties. Values deduced from such fitting can be directly compared with outcomes obtained from alternative parameter estimation methods, including those based on the spectral properties of the signal. This comparison helps to validate both the results and methodologies.

\subsection{Estimation of black holes spin}

\begin{figure}
    \centering
    \includegraphics[width=1\linewidth]{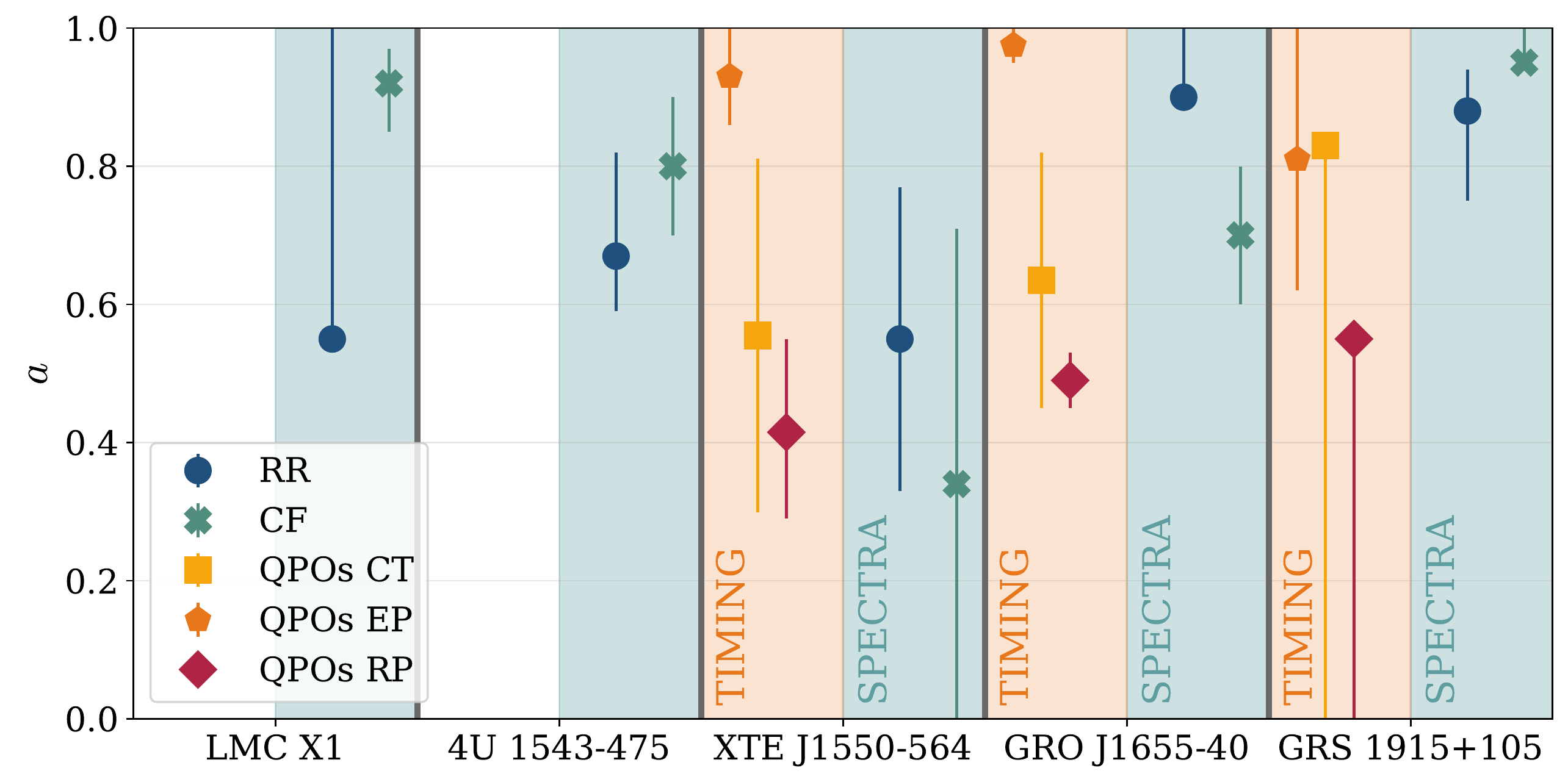}
    \caption{Comparison of spin value estimations for selected BH XRBs. The values obtained via the CF and RR methods, based on the spectral properties, are in the blue segments of the plots. Values derived through the timing-based methods are in the orange segments of the plots. The timing methods employ the CT and RP models and axisymmetric radially and vertically oscillating torus model (EP). Data for \cite{Kotrlova2020} and \cite{Reynolds2021}.}
    \label{fig:5:fits}
\end{figure}

Figure \ref{fig:5:fits} presents a~comparison of multiple spin estimation methods for five BH sources. The first two, \texttt{LMC X1} and \texttt{4U 1543-475}, are included for reference as the QPOs were not observed in them. Several methods were applied to the remaining three sources, some  reproducing results within comparable limits.

The relativistic reflection (RR) method, based on the reflection of X-ray radiation from a~corona on the accretion disk \citep{reflection}, is one of the chosen methods. The relativistic effects broaden the typically narrow iron K$\alpha$ line after reflection, which can reveal the size of the disk and, subsequently, the parameters of the central object. The continuous fitting (CF) method (also discussed in Chapter \ref{chap:AD}) involves fitting observed spectra with a~thermal multi-black body spectrum expected from a~thin accretion disk. Since the spectral shape and total flux strongly depend on the inner edge position, it can reveal important information about the system \citep{CF}. 

In contrast, the timing-based methods often involves fitting the frequencies of QPOs with the fundamental frequencies of motion near a~BH. A~detailed description of these models and the spin estimation process can be found in \cite{Kotrlova2020}.

Figure \ref{fig:5:fits} shows that specific spectral and timing-based methods often yield consistent results, pointing to relatively high spin values in the BH XRBs. However, it is evident that no universally preferred method exists, and the applied methodology remains premature. 
Despite variation in spin estimates from X-ray observations, a~preliminary conclusion is that the CT model tends to recover spin values similar to those obtained via spectral fitting methods. 


\subsection{Estimation of neutron stars mass}

The Hartle-Thorne metric within the CT model provides a sufficient approximation for its application on sources with rotating NS. Furthermore, an approximate analytical formula was derived in \cite{tor-etal:2022}, reproducing the model's predictions for QPO frequencies. Applying this formula to data from several NS sources provides better fits compared to the RP model. This outcome confirms the compatibility of the CT model with the observed data.

\begin{figure}
    \centering
    \includegraphics[width=1\linewidth]{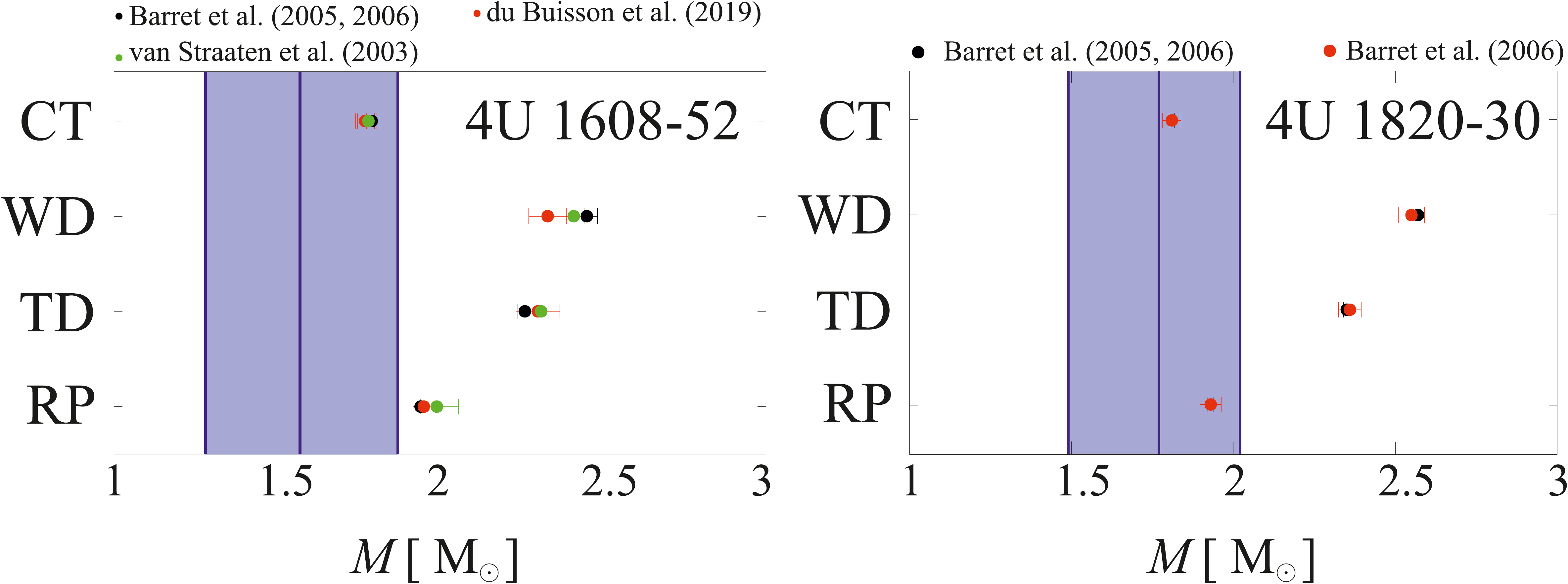}
    \caption{Estimation of the mass for two NS XRBs, comparing the mass derived from type I X-ray bursts (blue rectangles) against results from several QPO models. The WD and TD models are described in \cite{2019MNRAS.488.3896T}. 
    Kateřina Klimovičová (private communication), and \cite{tor-etal:2022,2019MNRAS.488.3896T}.} 
    \label{fig:5:fitsNS}
\end{figure}

The mass of several atoll sources obtained using the CT model in \cite{tor-etal:2022} aligns with estimates obtained from X-ray burst measurements, which is illustrated in Figure~\ref{fig:5:fitsNS} for two chosen sources. The same method was applied to several other atoll sources, yielding mass estimates between 1.5 and 1.9 $\msol$ \citep{tor-etal:2022}.

\section{Oscillations of accretion tori and QPO modulation}

The oscillations of an accretion torus in the vicinity of a~compact object should manifest as observable variations in the total flux. As a~result, the frequencies of these oscillations can be identified in the Fourier image of the observed light curve. 

However, the propagation of light close to a~compact object is governed by the effects of GR. To achieve a~realistic representation, it is necessary to numerically solve the null geodesic equations determining the trajectories of photons. This can be accomplished using a~relativistic ray-tracing code, which accurately simulates the paths of photons in curved spacetimes \citep{Bakala+2015b,1973ApJ...183..237C,2006ApJ...637L.113S,2005MNRAS.359.1217B,2011CQGra..28v5011V,2009ApJ...696.1616D,2013ApJ...777...13C,2018A&A...613A...2B,2023ApJ...950...35P,1992MNRAS.259..569K}. 

\begin{figure}[t]
    \centering
    \includegraphics[width=1\linewidth]{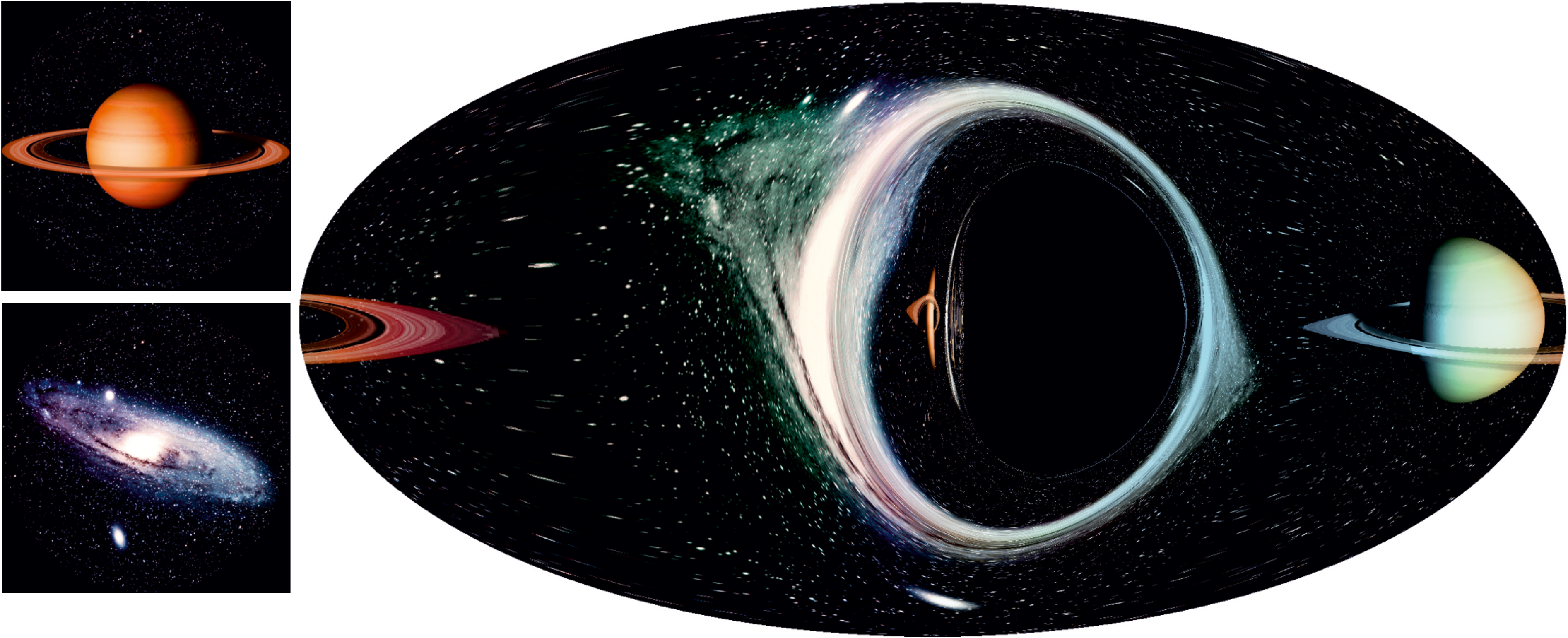}
    \caption{A ray-traced full-sky image for an observer in front of an extreme Kerr BH, a~planet behind, and a~galaxy in front of them. This visualization includes gravitational lensing, higher-order images, and Doppler and gravitational shifts in the photons' energy. Created by Pavel Bakala, \citep{Bakala+2015b}.}
    \label{fig:5:galaxy}
\end{figure}

\subsection{Relativistic ray tracing with the \texttt{LSD} code}

Several ray-tracing codes implementing various numerical methods are available, including the \texttt{HEROIC} code mentioned in Chapter \ref{chap:puffy}. For a~comprehensive overview of relativistic radiative transfer codes, see \cite{2023ApJ...950...35P}. Within the context of QPOs, ray tracing simulations were pioneered by \cite{2004ApJ...606.1098S,2005ApJ...621..940S,2004ApJ...617L..45B, 2005AN....326..849B,2006ApJ...651.1031S,2006ApJ...642..420S}. A~novel ray-tracing code has also been developed in our research group -- the Lensing Simulation Device (\texttt{LSD}) \citep{bak-etal:2007:,Bakala+2015b}.

The \texttt{LSD} code directly integrates the null geodesics in the direction from the observer, allowing for the computation of individual photon trajectories. This approach uniquely considers time delays between different photons, including those from second or higher-order images, which orbit the BH several times before escaping toward the observer. While this time difference may not be significant in most scenarios, its incorporation leads to higher accuracy of results when simulating rapid time variability.

The \texttt{LSD} code consists of two separate components. The first one performs ray tracing, following the photon paths in a~given (curved, axially symmetric) space-time from the observer at a~given location to the point of origin of such photon. This \say{backward} ray tracing  strategy significantly enhances computational efficiency. 

The second part calculates the intersection of these photon paths with objects of general geometry, accounting for obscuration by multiple bodies. These sources are defined by their radiating surface (photosphere) and 4-velocity. The initial step involves generating geodesics for a~specified space-time and the observer's position and resolution. Subsequently, the radiating object is inserted into the photon paths to calculate its lensed picture, light curve, and spectra. The calculated photon geodesics can be reused for different objects, further enhancing computational efficiency.

\begin{figure}[t]
\centering
\includegraphics[width=1\linewidth]{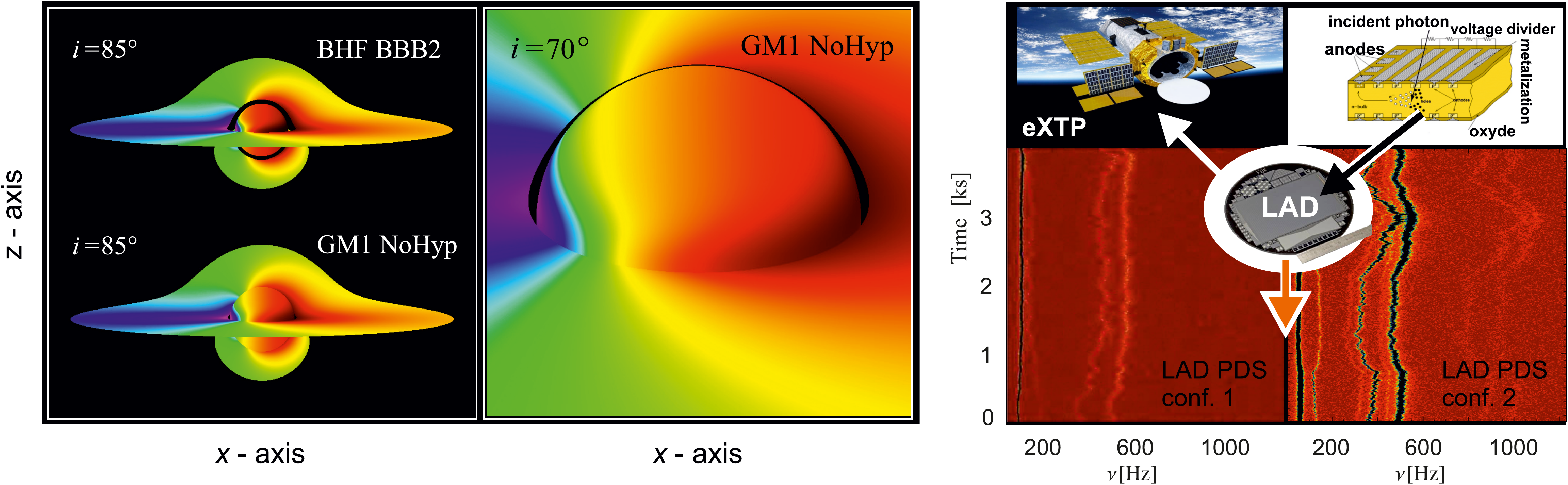}
\caption{\textit{Left:} G-factor maps illustrating the Doppler effect for a~thin disk around a~rotating NS for two different equations of state while maintaining the same mass and rotational frequency. \textit{Middle}: Detail of the spreading boundary layer on the NS surface. Both illustrations are adopted from \cite{Rene}. \textit{Right}: The LAD detector on board the eXTP satellite and examples of simulated spectrograms of an accreting system with a~NS. Illustration from \citep{IBWS}.}
\label{fig:5:LAD}
\end{figure}

\begin{figure}
    \centering
    \includegraphics[width=1\linewidth]{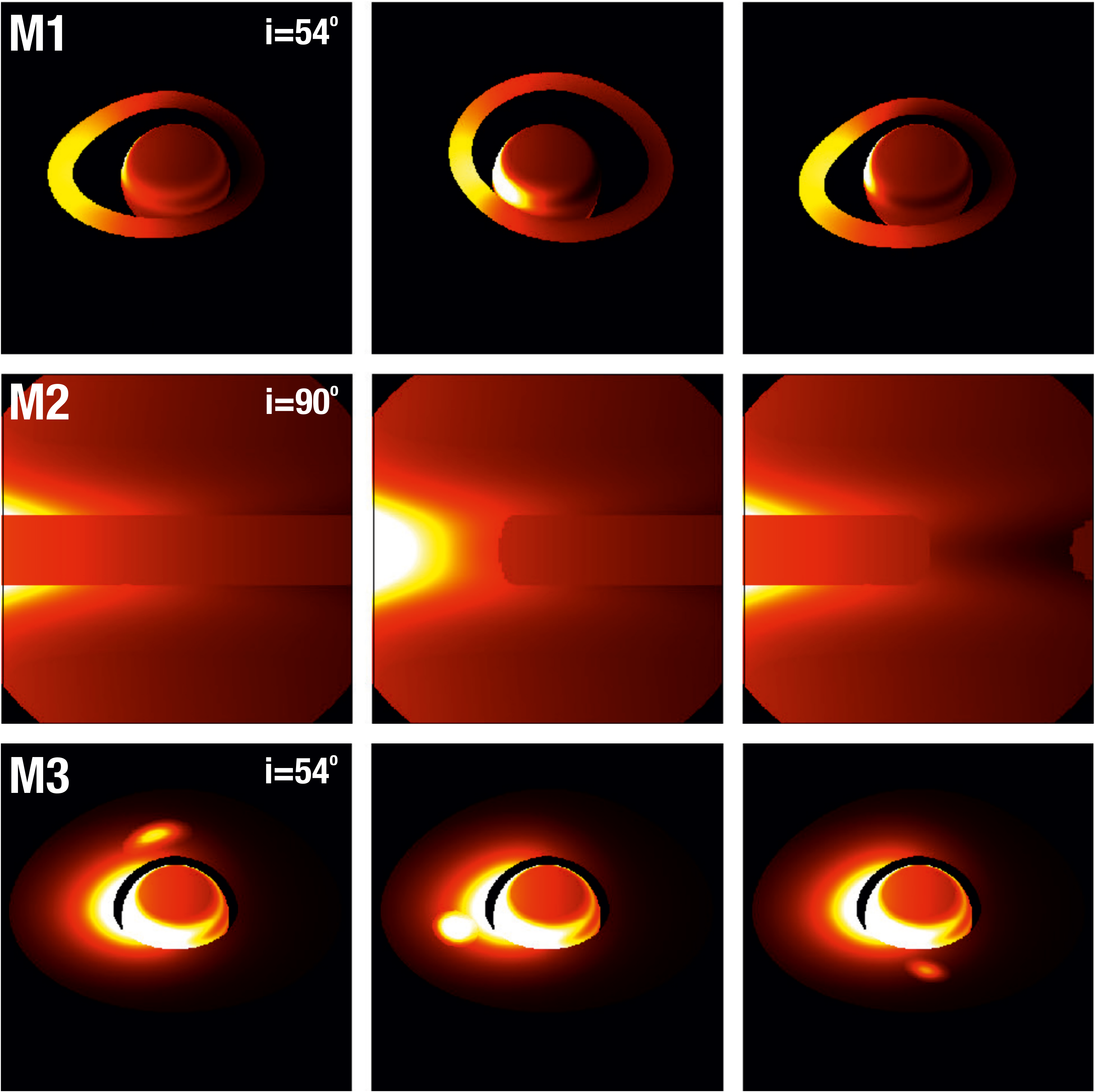}
    \caption{Colourmaps illustrating the relative bolometric intensity of the observable radiation for configurations corresponding to the \textbf{M1-M3} families of models, simulated using the \texttt{LSD} code. Three arbitrarily selected snapshots from the temporal evolution of each configuration are shown. Adopted from \cite{IBWS}.
    \textit{Top}: A~radially oscillating torus around a~NS, where torus oscillations modulate the boundary layer luminosity. \textit{Middle}: A~detailed equatorial view of the NS boundary layer on its surface and a torus fragment orbiting it. \textit{Bottom}: A~hot spot orbiting within a~thin accretion disk around a~NS with a~luminous boundary layer.}
    \label{fig:5:LSD}
\end{figure}

The \texttt{LSD} code has a~wide range of applications, from simulating observable spectra and light curves to studying the subtle, hidden spectral features and immediate surroundings of compact objects, to illustrating the impacts of strong gravity for science popularization and immersive media (see Figure \ref{fig:5:LSD} for the former, and \ref{fig:5:galaxy} for the latter). The \texttt{LSD} code can be applied to precisely model the signal from an X-ray source for a~distant observer in the context of X-ray instrument development,  enabling testing and verifying the capabilities of operating and planned X-ray telescopes. An example of such utilization is shown in Figure \ref{fig:5:LAD}, in the case of the Large Area Detector (LAD) aboard the proposed enhanced X-ray Timing and Polarimetry (eXTP) mission.

Moreover, the \texttt{LSD} code is valuable in modelling the interactions between multiple objects in the strong gravitational fields, accounting for the mutual obscuration effects. Furthermore, it can simulate the observable spectra, e.g., from a~system consisting of a~thin disk around a~NS and a~boundary layer on its surface \citep{Rene}, as illustrated in the left part of Figure \ref{fig:5:LAD} \citep{2014MNRAS.439.1933B,2016SPIE.9905E..1QZ,2019SCPMA..6229504D,2022SPIE12181E..1XF}.

In the context of the QPO modelling, the \texttt{LSD} code can simulate an object in the vicinity of a~compact object and its time evolution, producing individual spectra and spatially resolved pictures. However, the code does not solve the radiative transfer through the material, and thus the effects on the spectral shape and distribution arise solely from the GR effects. The capabilities of the \texttt{LSD} code are illustrated in Figure \ref{fig:5:LSD} for various scenarios of a~BH or a~NS accretion and its application for demonstrating light behaviour in strong gravity. 

\subsection{Ray-traced models}

To demonstrate the capabilities of the \texttt{LSD} code and to compare three distinct approaches to QPO modelling, we analyze both the temporal and frequency domain outputs obtained from simulations of three families of QPO models for both BH and NS sources. These simulations are carried out using the \texttt{LSD} code, incorporating relativistic effects on photon propagation and mutual obscuration of all components through ray tracing. This work is based on \cite{tor-etal:2022} and \cite{IBWS} alongside an ongoing project. Therefore, exhaustive details regarding simulation settings, parameter space exploration, and a~thorough analysis of results are beyond the scope of this thesis. The current discussion should be considered illustrative, demonstrating the differences and similarities between the models. 

The three families of models (\textbf{M1-M3}) for both the NS and BH cases are considered as follows:

\begin{itemize}
    \item \textbf{M1}: Models assuming a~torus oscillating closely to the central object in both vertical and radial directions, based on the work of \cite{KluzniakAbramowicz2001,AbramowiczKluzniak2001,2004AIPC..714...21A}.
    \item \textbf{M2}: Models considering orbital motion and oscillations of torus or products of torus instabilities.  These models are based on the scenario suggested in \cite{tor-etal:2022}, which involve epochs of stable torus oscillations alternating with periods where instability emerges, leading to the formation of one or more torus fragments \citep{1987MNRAS.225..695G}.
    \item \textbf{M3}: Models assuming the presence of one or more hot spots orbiting in proximity to the compact object within an accretion disk. These models are based on work of \cite{1990ApJ...358..538K} and \cite{1999PhRvL..82...17S}.
\end{itemize}

To explore these families of models, we adopt the methodology introduced by \cite{2004ApJ...617L..45B,2005AN....326..849B}. Our approach assumes optically thick tori and employs a~simplified torus thermodynamics \citep{IBWS}. Additionally, for the NS case, a~luminous NS boundary (spreading) layer is included in the simulations \citep{1999AstL...25..269I}. Based on works of \cite{1987Natur.327..303P,2005AN....326..845H,2006CQGra..23.1689A} and findings of \cite{2017MNRAS.470L..34P}, our simulations include the modulation of the NS boundary layer luminosity due to radial oscillations of the torus. 

The {\textbf{M1}} and {\textbf{M2}} families are intrinsically connected, both revolving around the concept of accretion tori, although in very different configurations. On the contrary, the {\textbf{M3}} family is built on different assumptions and is listed for comparison. Illustrations of the observational pictures and evolution of particular configurations of the \textbf{M1}-\textbf{M3} families of models are shown in Figure~\ref{fig:5:LSD}.

\subsection{Simulations setup}

The families of models can each accommodate a~variety of models with diverse parameters, including the radial coordinate of the torus centre, oscillations modes and symmetry, number of spots or torus fragments, and more. The simulations presented in this section are selected as representative examples from these families. For the sake of simplicity, a~non-rotating NS and BH is assumed, and in each case, the torus (or spot) is located at radius $r_0 = 6.75\,\rg$, where, as illustrated in Figure~\ref{fig:5:NSBHfreqs}, $f_K = 3f_r$. The mass of the BH is $M=5\,\msol$, and in the NS case, the mass is $M=1.4\,\msol$ and radius $R_* = 10\,$km (which is the same as in Figure \ref{fig:5:NSBHfreqs}.

The specific parameters for each simulation are:
\begin{itemize}
    \item \textbf{M1}: Torus exhibiting axisymmetric radial and vertical oscillations. 
    \item \textbf{M2}: Torus fragment as the products of instabilities exhibiting axisymmetric radial oscillations. 
    \item \textbf{M3}: A~single orbiting spot with radial precession. 
\end{itemize}

\begin{figure}
    \centering
    \includegraphics[width=1\linewidth]{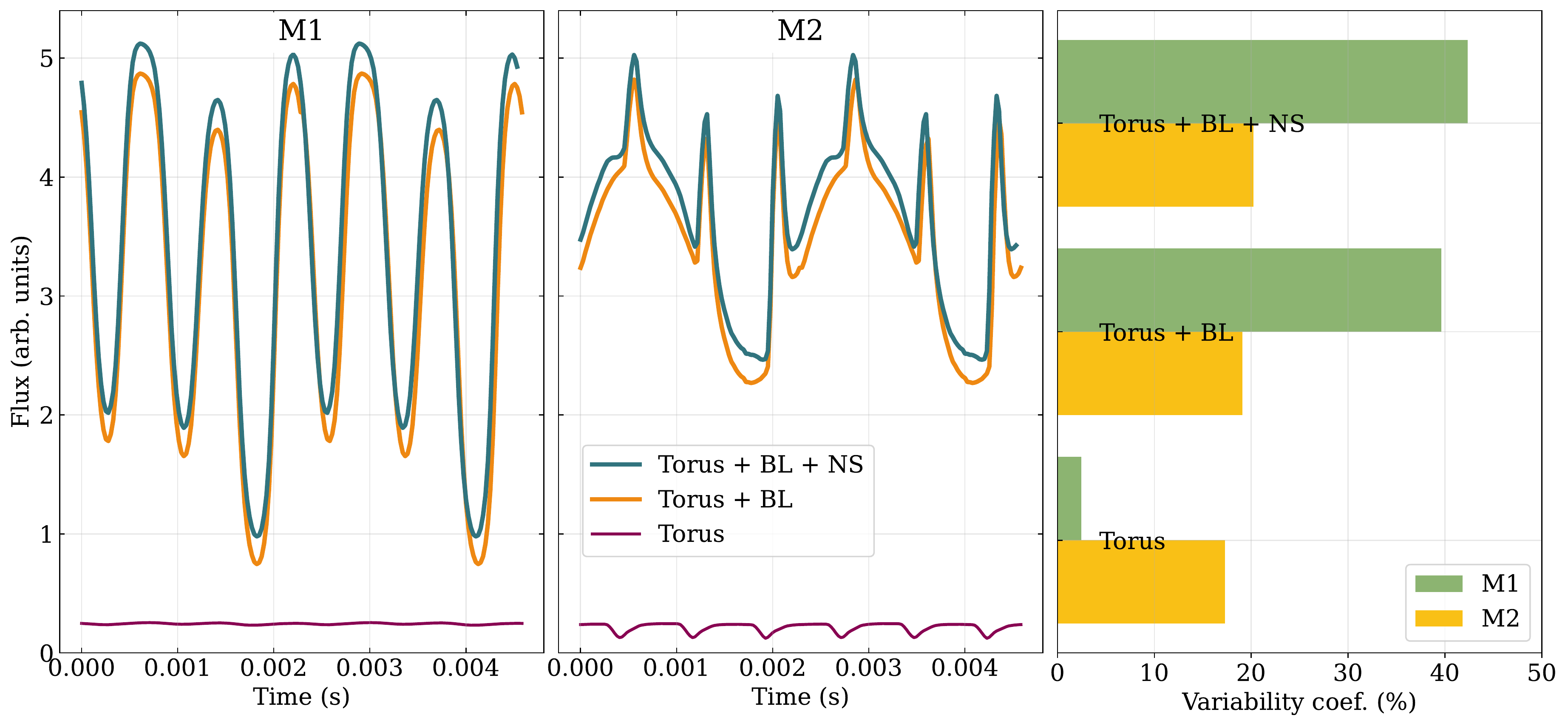}
    \caption{Light curves obtained for models \textbf{M1} (\textit{left}) and \textbf{M2} (\textit{middle}) in the NS case. The extension corresponds to two Keplerian orbits at $r_0$. The individual components of the models are disassembled, and separate light curves are shown while maintaining the effects of mutual obscuration (the objects are included in the simulations, but they do not radiate). The green line corresponds to the full model, the yellow for the torus and boundary layer, and the red one solely for the torus. In the \textit{right} panel, the variability coefficients for both models and all the components are shown.   
    }
    \label{fig:5:lightcurves}
\end{figure}

\subsection{Simulations results}

Figure~\ref{fig:5:lightcurves} presents the light curves and variability for models representing families {\textbf{M1}} and {\textbf{M2}}. Due to the substantial luminosity of the NS boundary layer, which corresponds to releasing a~significant fraction of accreted energy, the torus contributes relatively less to the overall variability than the pulsating, periodically obscured boundary layer. In this context, the torus acts as \say{clock} that determines the observed NS variability, while the NS acts as an \say{amplifier} enhancing the signal. This sets produced signals apart from that of the BH systems.

\begin{figure}
    \centering
    \includegraphics[width=1\linewidth]{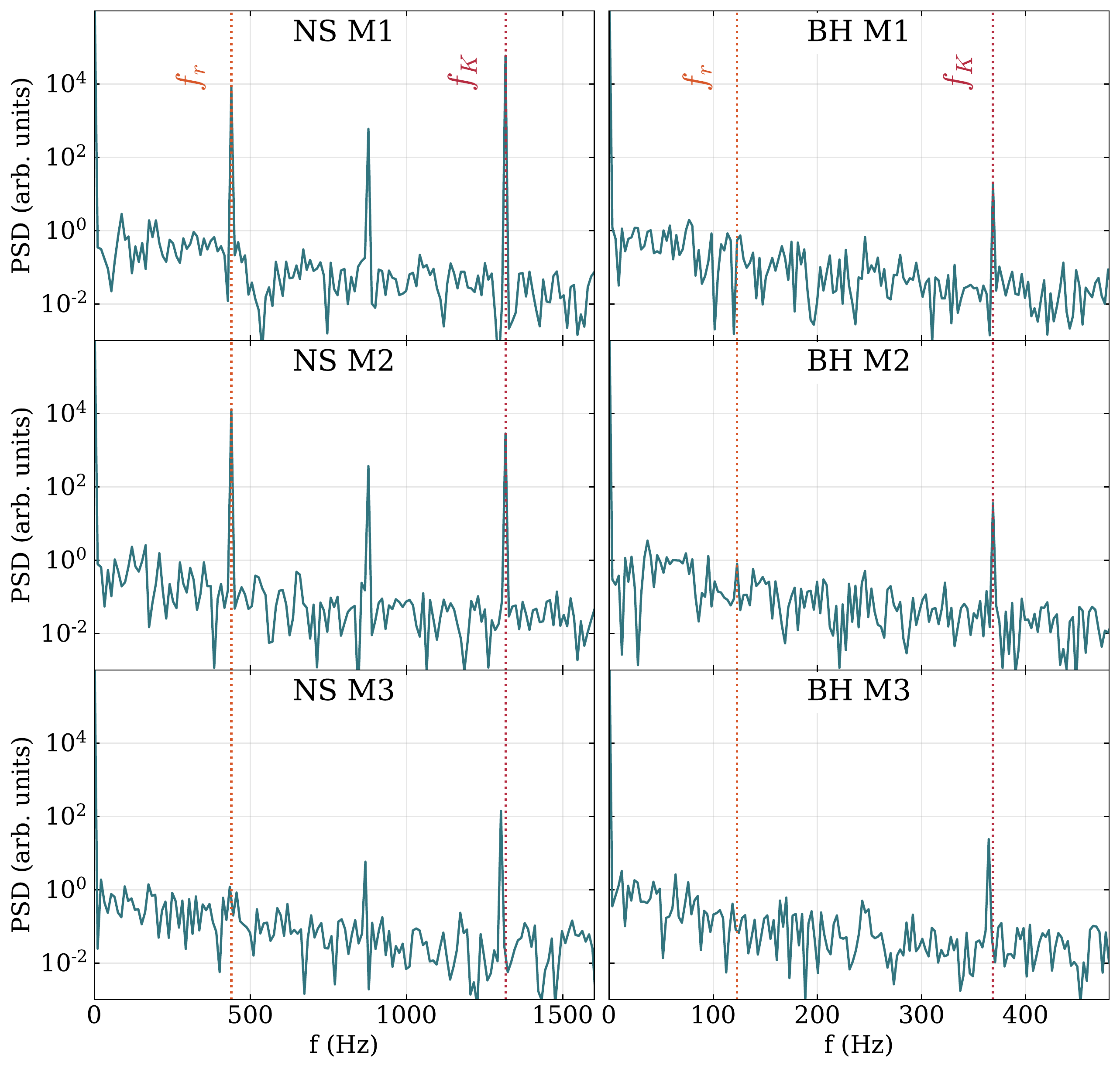}
    \caption{Simulated PSD for selected QPO models in the case of the NS (\textit{left}) and BH (\textit{right}) systems for the inclination angle of the observer $i=70^\circ$.}
    \label{fig:5:thebigone}
\end{figure}

Possible observable PSDs of these simulations are presented in Figure \ref{fig:5:thebigone}. 
Following the approach of \cite{2014MNRAS.439.1933B}, the pure signal obtained from simulations {\textbf{M1-M3}}  is combined with a~typical X-ray background associated with the hard spectral state of an XRB, along with white noise. 


The accretion rate modulation due to the radial oscillations of cusp tori was confirmed in MHD simulations \citep{2017MNRAS.470L..34P}. However, they identified an apparent absence of  modulation caused by vertical oscillations.
The obscuration of  the boundary layer by the torus and the accretion rate modulation, recovers the observed frequency peak corresponding to the vertical oscillations.

The results presented in this section regarding  the combined modulation of QPOs are part of ongoing work and are relevant to a~wide range of models. A~preliminary discussion of these results was presented at a~conference and is included in a~recently submitted paper \citep{IBWS}. These findings have implications for models based on the resonance of the fundamental frequencies of orbital oscillations \citep{AbramowiczKluzniak2001,KluzniakAbramowicz2001}, the CT model \citep{torok2016mnras,tor-etal:2022}, models by Rezzolla and collaborators \citep{2003MNRAS.344..978R}, or the model by Ingram and collaborators \citep{2010MNRAS.405.2447I}. One can also expect that for the last model, strong constraints on the physical properties of the \say{toroidal structure} assumed within the model will be implied.


\chapter*{Summary}

My work aimed to model the accretion disks in XRBs featuring either a~BH or a~NS. The main objective was to employ advanced numerical techniques to uncover the underlying processes governing the observed spectral and timing properties.

One notable outcome of this modelling was the recognition of the puffy accretion disk, a~novel type of accretion disk based purely on the results of numerical simulations. This innovative model covers the previously not fully explored gap between the thin and slim disk analytical models. Moreover, these results have been used to test the simplified analytical models to verify their validity, stability, and whether they can correctly interpret the observed data. 

Modelling accreting systems presents a~challenge due to the broad spectrum of timescales. These timescales span from resolving the turbulent processes that govern viscosity and angular momentum transport within microseconds to accurately modelling the complex oscillations responsible for  observed temporal variability on time\-scales of tens of seconds. Two approaches can be taken to address these challenges. The first is to model the accretion flow within the disk self-consistently during different states. The second approach involves using simplified analytical models of the disk and focusing on its oscillatory behaviour in the context of GR. I have pursued both paths during my studies. 

\section*{Overview of the collection of papers}


\subsection*{Paper 1: Puffy Accretion Disks: Sub-Eddington, Optically Thick, and Stable}

In this Letter, \cite{Lancova2019}, we report on a~new class of solutions for BH accretion disks derived from the results of global 3D GRRMHD simulations. This solution combines features of the canonical thin, slim, and thick disk models but differs from each in crucial ways; thus, we named it the \textit{puffy disk}. The puffy disk provides a~more realistic description of BH accretion. The solution is presented for a~non-spinning BH with a~sub-Eddington mass accretion rate. The disk appears thin based on the density scale height, but most inﬂow occurs through a~highly turbulent, optically thick, Keplerian region. The magnetic field supports the accreting fluid, making the disk thermally stable. The implications for understanding optically thick accretion disks and modelling BH sources are also discussed. 

\subsection*{Paper 2: Observational properties of puffy discs: radiative GRMHD spectra of mildly sub-Eddington accretion}

The puffy disk observational picture is described in this paper, \cite{Wielgus2022}. It explores the observational properties of puffy disks, including their geometrical obscuration at higher inclinations of the observer, and the collimation of radiation along the spin axis, including the discussion on similarities of the output of the simulations with various states of microquasars or ULXs. The paper also examines the fitting of puffy disk spectra using standard spectral models in \texttt{XSPEC} and discusses the limitations of these models in correctly recovering the BH spin and mass accretion rate. Although the synthetic spectra of puffy disks are qualitatively similar to those of a~Comptonized thin disk, other important properties, such as the dependency on the inclination angle of the observer, or the hardness ratio, are distinct from the analytical picture. 

\subsection*{Paper 3: Spectra of puffy accretion discs: the \texttt{kynbb} fit}

This paper, \cite{Lancova2023}, published in the proceedings of 2022 \texttt{XMM\-Newton} meeting in Madrid, we investigated the accuracy of spectral models in interpreting synthetic spectra of the puffy accretion disk in XRBs. The \texttt{kerrbb} and \texttt{kynbb} models in \texttt{XSPEC} were applied to the puffy disk simulated spectra to verify and extend the results of fitting in \cite{Wielgus2022} (Paper 2). The results show that neither of the models accurately recovers the parameters of the accretion disk, the BH spin, or the mass accretion rate. This suggests that new, more accurate spectral models for BH XRBs' luminous regimes are needed. The paper proposes that these new models should be based on the results of numerical simulations. The findings suggest that the traditional thin accretion disk models may not be suitable for accurately interpreting the spectra of XRBs with a~sub-Eddington luminosity. 

\subsection*{Paper 4: Models of high-frequency quasi-periodic oscillations and black hole spin estimates in Galactic microquasars}

This paper, \cite{Kotrlova2020}, examined the influence of pressure forces on the frequencies of oscillations of accretion disks around BHs. The effect of the pressure forces was examined in the framework of various QPOs models and its consequences for BH spin estimations. The study focuses on HF QPOs observed in the X-ray flux of Galactic microquasars. Several QPOs models based on oscillations of accretion tori were considered, and we examined the consequences for estimates of the BH spin. The paper concludes that, for some of the models, the pressure significantly impacts the spin estimates.

\subsection*{Paper 5: Simple Analytic Formula Relating the Mass and Spin of Accreting Compact Objects to Their Rapid X-Ray Variability}

In this paper, \cite{tor-etal:2022}, the CT model was extended and applied to explain observed HF QPOs in NS XRBs. The CT model is based on oscillations of marginally overflowing accretion tori with cusps. The paper derives an analytic formula for rotating compact objects in Kerr and Hartle-Thorne spacetimes. The formula accurately reproduces the predictions of the numerical model for QPOs frequencies. The formula is well applicable to rotating oblate NSs. Its application on data from several NS sources provides better fits of observed data compared to the RP model, confirming the compatibility of the CT model with the data.

\subsection*{Paper 6: Oscillations of non-slender tori in the external Hartle-Thorne geometry}

This paper, \cite{2022arXiv220310653M}, published in the Proceedings of RAGtime 20-22\footnote{\url{https://proceedings.physics.cz/proceedings-of-ragtime-20-22/}} studies the influence of the quadrupole moment of a~slowly rotating NS on the oscillations of non-slender accretion tori. The analytical calculations are applied to determine the frequencies of the radial epicyclic mode of a~torus in the Hartle-Thorne geometry. These frequencies are then compared to those obtained, assuming standard and linearized Kerr metrics. The paper states that while the shape of tori is not significantly affected by the quadrupole moment, the frequencies of oscillations are affected significantly. The paper suggests that considering the quadrupole moment of a~NS is important for HF QPOs modelling.

\newpage

\chapter*{List of Symbols}

\begin{longtable}{ p{.20\textwidth} | p{.80\textwidth} } 
         
         Symbol & Meaning \\
  \hline \hline

$\mathbf{A}_{{dyn}}$ & dynamo-generated vector potential \\
$A_{\widehat{\phi}}$ & azimuthal vector potential \\
$B^i$, $\mathbf{B}$ & magnetic field 3-vector \\
$\mathbf{B}_{{dyn}}$ & dynamo-generated magnetic field \\
$B_{\widehat{p}}$ & orthonormal poloidal component of the magnetic field \\
$B_{\widehat{\phi}}$ & orthonormal  azimuthal component of the magnetic field \\
$B_*$ & magnetic field strength on the surface of star \\
$\widehat{B}$ & frequency-integrated Planck function \\
$E$ & radiative energy density  \\
$\mathbf{E}$ & electric field 3-vector \\
$F$ & local radiative flux  \\
$F^i$ & radiative flux \\
$F_g$ & gravitational force \\
$\mathbf{F}^{\mathrm{L}}$ & left flux through cell face \\
$\mathbf{F}^{\mathrm{R}}$ & right flux through cell face \\
$F_r$ & radiation force \\
$F^{*\mu\nu}$ & dual tensor of the electromagnetic field \\
$G$ & gravitational constant \\
$G_\mu$ & radiative 4-force density \\
$H$ & half-thickness of a disk      \\
$I$ & frequency-integrated specific intensity \\
$I_\nu$ & specific intensity     \\
$J$ & BH angular momentum \\
$\mathbf{J}_{\widehat{p}}$ & poloidal current \\
$L$ &  luminosity \\
$L_{\mathrm{Edd}}$ & Eddington luminosity   \\
$M$ & mass \\
$M_A$ & mass of accretor \\
$M_D$ & mass of donor star \\
$\msol$ & solar mass  \\
$\dot{M}$ & mass accretion rate \\
$\dot{M}_{\mathrm{Edd}}$ & critical mass accretion rate \\
$N^i$ & unit vector in direction $x^i$ \\
$P$ & rotation period \\
$P(\nu)$ & Function of a steep power-law state of XRB\\
$\mathbf{P}$ & vectors of primitive quantities \\
${P}^{ij}$ & radiation pressure tensor \\
$Q$ & BH charge \\
$\mathbf{Q}^L$ & left state \\
$\mathbf{Q}^R$ & right state \\
$Q_\phi$ & quality parameter in $\phi$ direction \\
$Q_\theta$ & quality parameter in $\theta$ direction \\
$Q^+_\mathrm{visc}$ & viscous heating   \\
$Q^-_\mathrm{rad}$ & radiative cooling \\
$R_\mathrm{circ}$ & circularization radius \\
$R^\mu\phantom{}_\nu$ & radiation stress-energy tensor \\
$R^{\mu\nu}_{\mathrm{visc}}$ & radiation viscous term \\
$R_*$ & radius of a star \\
$S$ & gas entropy per unit volume \\
$\mathbf{S}^{\mathrm{L}}$ & left maximal speed in cell face \\
$S^{\mathrm{max}}$ & maximal signal speed in cell \\
$\mathbf{S}^{\mathrm{R}}$ & right maximal speed in cell face \\
$T$ & temperature \\
$T_g$ & gas temperature \\
$T_R$ & radiation temperature \\
$T^{\mu\nu}$ & gas stress-energy tensor \\
$\mathbf{U}$ & vectors of conserved quantities          \\

\hline
$a$ & BH spin \\
$a$ & radiation constant  \\
$b^\mu$ & 4-vector of magnetic field  \\
$c$ & speed of light  \\
$\der s$ & spacetime interval \\
$f_K$ & Keplerian orbital frequency \\
$f_P$ & periastron precession frequency \\
$f_r$ & radial epicyclic frequency \\
$f_\theta$ & vertical epicyclic frequency \\
$f_*$ & rotational frequency of star \\
$g$ & metric determinant \\
$g_{\mu\nu}$ & metric tensor  \\
$h_\rho$ & density scale-height \\
$h_\tau$ & photosphere height  \\
$h^{\mu\nu}_R$ & projection tensor \\
$k_B$ & Boltzmann constant \\
$m$ & mass in units of $\msol$ \\
$m_e$ & electron mass \\
$m_{\mathrm{p^+}}$ & proton mass \\
$\mdot$ & mass accretion rate \\
$p$ & pressure \\
$\pgas$ & gas pressure \\
$p_\mathrm{mag}$ & magnetic pressure \\
$q$ & mass ratio \\
$q$ & non-dimensional quadrupole moment of a NS \\
$r_0$ & position of center of torus \\
$r_{\mathrm{co}}$ & corotation radius \\
$r_{e}^\pm$ & ergosphere in Kerr metric   \\
$\rg$ & gravitational radius \\
$r_h^\pm$ & horizons in Kerr metric \\
$\risco$ & radius of ISCO \\
$r_{\mathrm{lc}}$ & light cylinder radius  \\
$r_{\mathrm{mag}}$ & magnetospheric radius  \\
$r_{\mathrm{mb}}$ & marginally bound orbit \\
$r_{\mathrm{ph}}$ & photon orbit  \\
$r_{Schw}$ & \schw{} radius \\
$r_\mathrm{trap}$ & photon trapping radius  \\
$r_*$ & radius in cm \\
$s$ & binary separation\\
$t_{acc}$ & accretion time  \\
$t_{diff}$ & photon diffusion time \\
$t_{dyn}$ & dynamical time-scale   \\
$\tg$ & gravitational radius crossing time  \\
$t_{r\phi}$ &  viscous stress \\
$t_{th}$ & thermal time-scale   \\
$t^{\mu\nu}$ & stress-energy tensor  \\
$\uint$ & internal energy density \\
$u^\mu$ & lab frame 4-velocity  \\
$\tilde{u}^\mu_R$ & radiation rest frame 4-velocity \\
$v^i = \left(v^r,v^\theta\,v^\phi\right)$ & vector of 3-velocity \\
\hline 
$\Delta t$ & time step  \\
$\Delta x$ & cell size \\
$\Gamma$ & adiabatic index \\
$\Gamma^{\alpha}\phantom{}_{\beta\gamma}$ & Christoffel symbols \\
$\Omega$ & angular velocity \\
$\Omega$ & solid angle \\
$\Sigma$ & surface density \\
$\Omega_K$ & Keplerian angular velocity \\
\hline 
$\alpha$ & viscous coefficient  \\
$\alpha_{dyn}$ & dynamo coefficient \\
$\alpha_{\mathrm{rad}}$ & radiative viscous coefficient \\
$\beta$ & magnetic-to-total pressure ratio  \\
$\beta$ & Thickness parameter of torus \\
$\beta_{cusp}$ & Thickness parameter of a cusp torus \\
$\eta$ & energy conversion efficiency \\
$\eta_{md}$ & magnetic diffusivity coefficient \\
$\eta^{\mu\nu}$ & Minkowski metric \\
$\theta_{MM}$ & minmoddiffusivity parameter  \\
$\kappa_a$ & absorption opacity  \\
$\kappa_{es}$ &  electron scattering opacity \\
$\lambda$ & mean free path of photons \\
$\lambda_\mathrm{MRI}$ & characteristic MRI wavelength  \\
$\mu$ & magnetic moment  \\
$\mu_{30}$ & magnetic moment  in the units of $10^{30}\,\mathrm{G\cdot cm^3}$ \\
$\nu$ & frequency \\
$\nu^A$ & Alfvén speed  \\
$\nu_L$ & Observed lower peak of HF QPOs \\
$\nu_U$ & Observed upper peak of HF QPOs \\
$\nu_{\mathrm{rad}}$ & radiative viscosity coefficient \\
$\xi$ & magnetic tilt angle \\
$\rho$ & Gas density  \\
$\sigma_R^{\mu\nu}$ & radiation shear tensor \\
$\sigma_T$ & Thomson cross-section  \\
$\tau$ & optical depth \\
$\chi $ & total opacity coefficient \\
\hline 
$\left(-,+,+,+\right)$  & metric signature \\
$\left(r,\theta,\phi\right)$  & spacial coordinates \\
$\left(t,r,\theta,\phi\right)$ & spacetime coordinates \\
Latin indices & $\left(i,j,k,...\right)$ (1,2,3) \\
Greek indices & $\left(\alpha,\beta,\gamma,...\right)$  (0,1,2,3) \\

         \hline \hline      
\end{longtable}

\newpage

\chapter*{List of Abbreviations}

\begin{longtable}{ p{.20\textwidth} | p{.80\textwidth} } 

         \hline \hline
         Abbreviations & Meaning \\ \hline
         ADAF & Advection Dominated Accretion Flow \\
         AGN & Active Galactic Nuclei \\
         BH & Black Hole \\
         BL & Boyer–Lindquist (coordinates) \\
         CCD & Colour-Colour Diagram \\
         CFL & Courant–Friedrichs–Lewy (condition) \\
         CT & Constrained Transport \\
         CT & Cusp Torus \\
         EHT & Event Horizon Telescope \\
         EM & Electromagnetic \\
         eXTP & enhanced X-ray Timing and Polarimetry (mission)\\
         FBO & Flaring Branch Oscillations \\
         GR & General Relativity \\
         GRRMHD & General Relativistic Radiative Magnetohydrodynamic \\
         GW & Gravitational Waves \\
         HBO & Horizontal Branch Oscillations \\ 
         HD & Hydrodynamics \\
         HF QPOs & High Frequency QPOs \\
         HID & Hardness Intensity Diagram \\
         HLL & Harten–Lax–van Leer (Riemann solver) \\
         HMXB & High-Mass X-ray Binary \\
         IMEX & Implicit-Explicit \\
         ISCO & Innermost Stable Circular Orbit \\
         KS & Kerr-Schild \\
         LAD & Large Area Detector \\
         LF QPOs & Low Frequency QPOs \\
         LMXB & Low-Mass X-ray Binary \\
         LSD& Lensing Simulation Device \\
         LTE & Local Thermal Equilibrium \\
         MHD & Magnetohydrodynamics \\
         MKS & Modified Kerr-Schild (coordinates) \\
         MPI & Message Parsing Interface \\
         MRI & Magnetorotational Instability \\
         NBO  & Normal Branch Oscillations \\
         NS & Neutron Star \\
         PSD & Power Spectral Density \\
         RP & Relativistic Precession \\
         QPO & Quasi-periodic Oscillations \\
         RR & Relativistic Reflection \\
         RXTE & Rossi X-Ray Timing Explorer \\
         SED & Spectral Energy Distribution \\
         TDE & Tidal Disruption Event \\
         ULX & Ultraluminous X-ray source \\
         UV & Ultraviolet \\
         XRB & X-ray Binary \\
         ZAMO & Zero Angular Momentum Observer \\
         \hline \hline
         
\end{longtable}

\newpage

\renewcommand\bibname{References}
{\small\sloppy
\bibliography{bibliography.bib}

\begin{thebibliography}{253}
\expandafter\ifx\csname natexlab\endcsname\relax\def\natexlab#1{#1}\fi

\bibitem[{{Abbott} {et~al.}(2016){Abbott}, {Abbott}, {Abbott}, {Abernathy},
  {Acernese}, {Ackley}, {Adams}, {Adams}, {Addesso}, {Adhikari}, {Adya},
  {Affeldt}, {Agathos}, {Agatsuma}, {Aggarwal}, {Aguiar}, {Aiello}, {Ain},
  {Ajith}, {Allen}, {Allocca}, {Altin}, {Anderson}, {Anderson}, {Arai},
  {Arain}, {Araya}, {Arceneaux}, {Areeda}, {Arnaud}, {Arun}, {Ascenzi},
  {Ashton}, {Ast}, {Aston}, {Astone}, {Aufmuth}, {Aulbert}, {Babak}, {Bacon},
  {Bader}, {Baker}, {Baldaccini}, {Ballardin}, {Ballmer}, {Barayoga},
  {Barclay}, {Barish}, {Barker}, {Barone}, {Barr}, {Barsotti}, {Barsuglia},
  {Barta}, {Bartlett}, {Barton}, {Bartos}, {Bassiri}, {Basti}, {Batch},
  {Baune}, {Bavigadda}, {Bazzan}, {Behnke}, {Bejger}, {Belczynski}, {Bell},
  {Bell}, {Berger}, {Bergman}, {Bergmann}, {Berry}, {Bersanetti}, {Bertolini},
  {Betzwieser}, {Bhagwat}, {Bhandare}, {Bilenko}, {Billingsley}, {Birch},
  {Birney}, {Birnholtz}, {Biscans}, {Bisht}, {Bitossi}, {Biwer}, {Bizouard},
  {Blackburn}, {Blair}, {Blair}, {Blair}, {Bloemen}, {Bock}, {Bodiya}, {Boer},
  {Bogaert}, {Bogan}, {Bohe}, {Bojtos}, {Bond}, {Bondu}, {Bonnand}, {Boom},
  {Bork}, {Boschi}, {Bose}, {Bouffanais}, {Bozzi}, {Bradaschia}, {Brady},
  {Braginsky}, {Branchesi}, {Brau}, {Briant}, {Brillet}, {Brinkmann},
  {Brisson}, {Brockill}, {Brooks}, {Brown}, {Brown}, {Brown}, {Buchanan},
  {Buikema}, {Bulik}, {Bulten}, {Buonanno}, {Buskulic}, {Buy}, {Byer},
  {Cabero}, {Cadonati}, {Cagnoli}, {Cahillane}, {Bustillo}, {Callister},
  {Calloni}, {Camp}, {Cannon}, {Cao}, {Capano}, {Capocasa}, {Carbognani},
  {Caride}, {Casanueva Diaz}, {Casentini}, {Caudill}, {Cavagli{\`a}},
  {Cavalier}, {Cavalieri}, {Cella}, {Cepeda}, {Baiardi}, {Cerretani},
  {Cesarini}, {Chakraborty}, {Chalermsongsak}, {Chamberlin}, {Chan}, {Chao},
  {Charlton}, {Chassande-Mottin}, {Chen}, {Chen}, {Cheng}, {Chincarini},
  {Chiummo}, {Cho}, {Cho}, {Chow}, {Christensen}, {Chu}, {Chua}, {Chung},
  {Ciani}, {Clara}, {Clark}, {Cleva}, {Coccia}, {Cohadon}, {Colla}, {Collette},
  {Cominsky}, {Constancio}, {Conte}, {Conti}, {Cook}, {Corbitt}, {Cornish},
  {Corsi}, {Cortese}, {Costa}, {Coughlin}, {Coughlin}, {Coulon}, {Countryman},
  {Couvares}, {Cowan}, {Coward}, {Cowart}, {Coyne}, {Coyne}, {Craig},
  {Creighton}, {Creighton}, {Cripe}, {Crowder}, {Cruise}, {Cumming},
  {Cunningham}, {Cuoco}, {Dal Canton}, {Danilishin}, {D'Antonio}, {Danzmann},
  {Darman}, {Da Silva Costa}, {Dattilo}, {Dave}, {Daveloza}, {Davier},
  {Davies}, {Daw}, {Day}, {De}, {DeBra}, {Debreczeni}, {Degallaix}, {De
  Laurentis}, {Del{\'e}glise}, {Del Pozzo}, {Denker}, {Dent}, {Dereli},
  {Dergachev}, {DeRosa}, {De Rosa}, {DeSalvo}, {Dhurandhar}, {D{\'\i}az}, {Di
  Fiore}, {Di Giovanni}, {Di Lieto}, {Di Pace}, {Di Palma}, {Di Virgilio},
  {Dojcinoski}, {Dolique}, {Donovan}, {Dooley}, {Doravari}, {Douglas},
  {Downes}, {Drago}, {Drever}, {Driggers}, {Du}, {Ducrot}, {Dwyer}, {Edo},
  {Edwards}, {Effler}, {Eggenstein}, {Ehrens}, {Eichholz}, {Eikenberry},
  {Engels}, {Essick}, {Etzel}, {Evans}, {Evans}, {Everett}, {Factourovich},
  {Fafone}, {Fair}, {Fairhurst}, {Fan}, {Fang}, {Farinon}, {Farr}, {Farr},
  {Favata}, {Fays}, {Fehrmann}, {Fejer}, {Feldbaum}, {Ferrante}, {Ferreira},
  {Ferrini}, {Fidecaro}, {Finn}, {Fiori}, {Fiorucci}, {Fisher}, {Flaminio},
  {Fletcher}, {Fong}, {Fournier}, {Franco}, {Frasca}, {Frasconi}, {Frede},
  {Frei}, {Freise}, {Frey}, {Frey}, {Fricke}, {Fritschel}, {Frolov}, {Fulda},
  {Fyffe}, {Gabbard}, {Gair}, {Gammaitoni}, {Gaonkar}, {Garufi}, {Gatto},
  {Gaur}, {Gehrels}, {Gemme}, {Gendre}, {Genin}, {Gennai}, {George}, {Gergely},
  {Germain}, {Ghosh}, {Ghosh}, {Ghosh}, {Giaime}, {Giardina}, {Giazotto},
  {Gill}, {Glaefke}, {Gleason}, {Goetz}, {Goetz}, {Gondan}, {Gonz{\'a}lez},
  {Castro}, {Gopakumar}, {Gordon}, {Gorodetsky}, {Gossan}, {Gosselin},
  {Gouaty}, {Graef}, {Graff}, {Granata}, {Grant}, {Gras}, {Gray}, {Greco},
  {Green}, {Greenhalgh}, {Groot}, {Grote}, {Grunewald}, {Guidi}, {Guo},
  {Gupta}, {Gupta}, {Gushwa}, {Gustafson}, {Gustafson}, {Hacker}, {Hall},
  {Hall}, {Hammond}, {Haney}, {Hanke}, {Hanks}, {Hanna}, {Hannam}, {Hanson},
  {Hardwick}, {Harms}, {Harry}, {Harry}, {Hart}, {Hartman}, {Haster},
  {Haughian}, {Healy}, {Heefner}, {Heidmann}, {Heintze}, {Heinzel}, {Heitmann},
  {Hello}, {Hemming}, {Hendry}, {Heng}, {Hennig}, {Heptonstall}, {Heurs},
  {Hild}, {Hoak}, {Hodge}, {Hofman}, {Hollitt}, {Holt}, {Holz}, {Hopkins},
  {Hosken}, {Hough}, {Houston}, {Howell}, {Hu}, {Huang}, {Huerta}, {Huet},
  {Hughey}, {Husa}, {Huttner}, {Huynh-Dinh}, {Idrisy}, {Indik}, {Ingram},
  {Inta}, {Isa}, {Isac}, {Isi}, {Islas}, {Isogai}, {Iyer}, {Izumi}, {Jacobson},
  {Jacqmin}, {Jang}, {Jani}, {Jaranowski}, {Jawahar}, {Jim{\'e}nez-Forteza},
  {Johnson}, {Johnson-McDaniel}, {Jones}, {Jones}, {Jonker}, {Ju}, {Haris},
  {Kalaghatgi}, {Kalogera}, {Kandhasamy}, {Kang}, {Kanner}, {Karki},
  {Kasprzack}, {Katsavounidis}, {Katzman}, {Kaufer}, {Kaur}, {Kawabe},
  {Kawazoe}, {K{\'e}f{\'e}lian}, {Kehl}, {Keitel}, {Kelley}, {Kells},
  {Kennedy}, {Keppel}, {Key}, {Khalaidovski}, {Khalili}, {Khan}, {Khan},
  {Khan}, {Khazanov}, {Kijbunchoo}, {Kim}, {Kim}, {Kim}, {Kim}, {Kim}, {Kim},
  {King}, {King}, {Kinzel}, {Kissel}, {Kleybolte}, {Klimenko}, {Koehlenbeck},
  {Kokeyama}, {Koley}, {Kondrashov}, {Kontos}, {Koranda}, {Korobko}, {Korth},
  {Kowalska}, {Kozak}, {Kringel}, {Krishnan}, {Kr{\'o}lak}, {Krueger}, {Kuehn},
  {Kumar}, {Kumar}, {Kuo}, {Kutynia}, {Kwee}, {Lackey}, {Landry}, {Lange},
  {Lantz}, {Lasky}, {Lazzarini}, {Lazzaro}, {Leaci}, {Leavey}, {Lebigot},
  {Lee}, {Lee}, {Lee}, {Lee}, {Lenon}, {Leonardi}, {Leong}, {Leroy},
  {Letendre}, {Levin}, {Levine}, {Li}, {Libson}, {Littenberg}, {Lockerbie},
  {Logue}, {Lombardi}, {London}, {Lord}, {Lorenzini}, {Loriette}, {Lormand},
  {Losurdo}, {Lough}, {Lousto}, {Lovelace}, {L{\"u}ck}, {Lundgren}, {Luo},
  {Lynch}, {Ma}, {MacDonald}, {Machenschalk}, {MacInnis}, {Macleod},
  {Maga{\~n}a-Sandoval}, {Magee}, {Mageswaran}, {Majorana}, {Maksimovic},
  {Malvezzi}, {Man}, {Mandel}, {Mandic}, {Mangano}, {Mansell}, {Manske},
  {Mantovani}, {Marchesoni}, {Marion}, {M{\'a}rka}, {M{\'a}rka}, {Markosyan},
  {Maros}, {Martelli}, {Martellini}, {Martin}, {Martin}, {Martynov}, {Marx},
  {Mason}, {Masserot}, {Massinger}, {Masso-Reid}, {Matichard}, {Matone},
  {Mavalvala}, {Mazumder}, {Mazzolo}, {McCarthy}, {McClelland}, {McCormick},
  {McGuire}, {McIntyre}, {McIver}, {McManus}, {McWilliams}, {Meacher},
  {Meadors}, {Meidam}, {Melatos}, {Mendell}, {Mendoza-Gandara}, {Mercer},
  {Merilh}, {Merzougui}, {Meshkov}, {Messenger}, {Messick}, {Meyers},
  {Mezzani}, {Miao}, {Michel}, {Middleton}, {Mikhailov}, {Milano}, {Miller},
  {Millhouse}, {Minenkov}, {Ming}, {Mirshekari}, {Mishra}, {Mitra},
  {Mitrofanov}, {Mitselmakher}, {Mittleman}, {Moggi}, {Mohan}, {Mohapatra},
  {Montani}, {Moore}, {Moore}, {Moraru}, {Moreno}, {Morriss}, {Mossavi},
  {Mours}, {Mow-Lowry}, {Mueller}, {Mueller}, {Muir}, {Mukherjee}, {Mukherjee},
  {Mukherjee}, {Mukund}, {Mullavey}, {Munch}, {Murphy}, {Murray}, {Mytidis},
  {Nardecchia}, {Naticchioni}, {Nayak}, {Necula}, {Nedkova}, {Nelemans},
  {Neri}, {Neunzert}, {Newton}, {Nguyen}, {Nielsen}, {Nissanke}, {Nitz},
  {Nocera}, {Nolting}, {Normandin}, {Nuttall}, {Oberling}, {Ochsner}, {O'Dell},
  {Oelker}, {Ogin}, {Oh}, {Oh}, {Ohme}, {Oliver}, {Oppermann}, {Oram},
  {O'Reilly}, {O'Shaughnessy}, {Ott}, {Ottaway}, {Ottens}, {Overmier}, {Owen},
  {Pai}, {Pai}, {Palamos}, {Palashov}, {Palomba}, {Pal-Singh}, {Pan}, {Pan},
  {Pankow}, {Pannarale}, {Pant}, {Paoletti}, {Paoli}, {Papa}, {Paris},
  {Parker}, {Pascucci}, {Pasqualetti}, {Passaquieti}, {Passuello},
  {Patricelli}, {Patrick}, {Pearlstone}, {Pedraza}, {Pedurand}, {Pekowsky},
  {Pele}, {Penn}, {Perreca}, {Pfeiffer}, {Phelps}, {Piccinni}, {Pichot},
  {Pickenpack}, {Piergiovanni}, {Pierro}, {Pillant}, {Pinard}, {Pinto},
  {Pitkin}, {Poeld}, {Poggiani}, {Popolizio}, {Post}, {Powell}, {Prasad},
  {Predoi}, {Premachandra}, {Prestegard}, {Price}, {Prijatelj}, {Principe},
  {Privitera}, {Prix}, {Prodi}, {Prokhorov}, {Puncken}, {Punturo}, {Puppo},
  {P{\"u}rrer}, {Qi}, {Qin}, {Quetschke}, {Quintero}, {Quitzow-James}, {Raab},
  {Rabeling}, {Radkins}, {Raffai}, {Raja}, {Rakhmanov}, {Ramet}, {Rapagnani},
  {Raymond}, {Razzano}, {Re}, {Read}, {Reed}, {Regimbau}, {Rei}, {Reid},
  {Reitze}, {Rew}, {Reyes}, {Ricci}, {Riles}, {Robertson}, {Robie}, {Robinet},
  {Rocchi}, {Rolland}, {Rollins}, {Roma}, {Romano}, {Romano}, {Romanov},
  {Romie}, {Rosi{\'n}ska}, {Rowan}, {R{\"u}diger}, {Ruggi}, {Ryan}, {Sachdev},
  {Sadecki}, {Sadeghian}, {Salconi}, {Saleem}, {Salemi}, {Samajdar}, {Sammut},
  {Sampson}, {Sanchez}, {Sandberg}, {Sandeen}, {Sanders}, {Sanders},
  {Sassolas}, {Sathyaprakash}, {Saulson}, {Sauter}, {Savage}, {Sawadsky},
  {Schale}, {Schilling}, {Schmidt}, {Schmidt}, {Schnabel}, {Schofield},
  {Sch{\"o}nbeck}, {Schreiber}, {Schuette}, {Schutz}, {Scott}, {Scott},
  {Sellers}, {Sengupta}, {Sentenac}, {Sequino}, {Sergeev}, {Serna},
  {Setyawati}, {Sevigny}, {Shaddock}, {Shaffer}, {Shah}, {Shahriar}, {Shaltev},
  {Shao}, {Shapiro}, {Shawhan}, {Sheperd}, {Shoemaker}, {Shoemaker}, {Siellez},
  {Siemens}, {Sigg}, {Silva}, {Simakov}, {Singer}, {Singer}, {Singh}, {Singh},
  {Singhal}, {Sintes}, {Slagmolen}, {Smith}, {Smith}, {Smith}, {Smith}, {Son},
  {Sorazu}, {Sorrentino}, {Souradeep}, {Srivastava}, {Staley}, {Steinke},
  {Steinlechner}, {Steinlechner}, {Steinmeyer}, {Stephens}, {Stevenson},
  {Stone}, {Strain}, {Straniero}, {Stratta}, {Strauss}, {Strigin}, {Sturani},
  {Stuver}, {Summerscales}, {Sun}, {Sutton}, {Swinkels}, {Szczepa{\'n}czyk},
  {Tacca}, {Talukder}, {Tanner}, {T{\'a}pai}, {Tarabrin}, {Taracchini},
  {Taylor}, {Theeg}, {Thirugnanasambandam}, {Thomas}, {Thomas}, {Thomas},
  {Thorne}, {Thorne}, {Thrane}, {Tiwari}, {Tiwari}, {Tokmakov}, {Tomlinson},
  {Tonelli}, {Torres}, {Torrie}, {T{\"o}yr{\"a}}, {Travasso}, {Traylor},
  {Trifir{\`o}}, {Tringali}, {Trozzo}, {Tse}, {Turconi}, {Tuyenbayev},
  {Ugolini}, {Unnikrishnan}, {Urban}, {Usman}, {Vahlbruch}, {Vajente},
  {Valdes}, {Vallisneri}, {van Bakel}, {van Beuzekom}, {van den Brand}, {Van
  Den Broeck}, {Vander-Hyde}, {van der Schaaf}, {van Heijningen}, {van Veggel},
  {Vardaro}, {Vass}, {Vas{\'u}th}, {Vaulin}, {Vecchio}, {Vedovato}, {Veitch},
  {Veitch}, {Venkateswara}, {Verkindt}, {Vetrano}, {Vicer{\'e}}, {Vinciguerra},
  {Vine}, {Vinet}, {Vitale}, {Vo}, {Vocca}, {Vorvick}, {Voss}, {Vousden},
  {Vyatchanin}, {Wade}, {Wade}, {Wade}, {Waldman}, {Walker}, {Wallace},
  {Walsh}, {Wang}, {Wang}, {Wang}, {Wang}, {Wang}, {Ward}, {Ward}, {Warner},
  {Was}, {Weaver}, {Wei}, {Weinert}, {Weinstein}, {Weiss}, {Welborn}, {Wen},
  {We{\ss}els}, {Westphal}, {Wette}, {Whelan}, {Whitcomb}, {White}, {Whiting},
  {Wiesner}, {Wilkinson}, {Willems}, {Williams}, {Williams}, {Williamson},
  {Willis}, {Willke}, {Wimmer}, {Winkelmann}, {Winkler}, {Wipf}, {Wiseman},
  {Wittel}, {Woan}, {Worden}, {Wright}, {Wu}, {Yablon}, {Yakushin}, {Yam},
  {Yamamoto}, {Yancey}, {Yap}, {Yu}, {Yvert}, {Zadro{\.Z}ny}, {Zangrando},
  {Zanolin}, {Zendri}, {Zevin}, {Zhang}, {Zhang}, {Zhang}, {Zhang}, {Zhao},
  {Zhou}, {Zhou}, {Zhu}, {Zucker}, {Zuraw}, {Zweizig}, {LIGO Scientific
  Collaboration}, \& {Virgo Collaboration}}]{Abbott2016}
{Abbott}, B.~P., {Abbott}, R., {Abbott}, T.~D., {et~al.}, {Observation of
  Gravitational Waves from a Binary Black Hole Merger}. 2016, {\it \prl}, {\bf
  116}, 061102, DOI: 10.1103/PhysRevLett.116.061102

\bibitem[{{Abramowicz} {et~al.}(1978){Abramowicz}, {Jaroszy{\'n}ski}, \&
  {Sikora}}]{Abramowicz+1978}
{Abramowicz}, M., {Jaroszy{\'n}ski}, M., \& {Sikora}, M., {Relativistic,
  accreting disks.} 1978, {\it A\&A}, {\bf 63}, 221

\bibitem[{{Abramowicz} {et~al.}(2006){Abramowicz}, {Blaes}, {Hor{\'a}k},
  {Klu{\'z}niak}, \& {Rebusco}}]{2006CQGra..23.1689A}
{Abramowicz}, M.~A., {Blaes}, O.~M., {Hor{\'a}k}, J., {Klu{\'z}niak}, W., \&
  {Rebusco}, P., {Epicyclic oscillations of fluid bodies: II. Strong gravity}.
  2006, {\it Classical and Quantum Gravity}, {\bf 23}, 1689, DOI:
  10.1088/0264-9381/23/5/014

\bibitem[{{Abramowicz} {et~al.}(2003){Abramowicz}, {Bulik}, {Bursa}, \&
  {Klu{\'z}niak}}]{2003A&A...404L..21A}
{Abramowicz}, M.~A., {Bulik}, T., {Bursa}, M., \& {Klu{\'z}niak}, W., {Evidence
  for a 2:3 resonance in Sco X-1 kHz QPOs}. 2003, {\it \aap}, {\bf 404}, L21,
  DOI: 10.1051/0004-6361:20030737

\bibitem[{{Abramowicz} {et~al.}(1988){Abramowicz}, {Czerny}, {Lasota}, \&
  {Szuszkiewicz}}]{AbramSlim}
{Abramowicz}, M.~A., {Czerny}, B., {Lasota}, J.~P., \& {Szuszkiewicz}, E.,
  {Slim Accretion Disks}. 1988, {\it ApJ}, {\bf 332}, 646, DOI: 10.1086/166683

\bibitem[{{Abramowicz} \& {Klu{\'z}niak}(2001)}]{AbramowiczKluzniak2001}
{Abramowicz}, M.~A. \& {Klu{\'z}niak}, W., {A precise determination of black
  hole spin in GRO J1655-40}. 2001, {\it \aap}, {\bf 374}, L19, DOI:
  10.1051/0004-6361:20010791

\bibitem[{{Abramowicz} \& {Klu{\'z}niak}(2004)}]{2004AIPC..714...21A}
{Abramowicz}, M.~A. \& {Klu{\'z}niak}, W., {Interpreting black hole QPOs}.
  2004, in American Institute of Physics Conference Series, Vol. {\bf  714},
  {\it X-ray Timing 2003: Rossi and Beyond}, ed. P.~{Kaaret}, F.~K. {Lamb}, \&
  J.~H. {Swank}, 21--28

\bibitem[{{Abramowicz} {et~al.}(2004){Abramowicz}, {Klu{\'z}niak},
  {McClintock}, \& {Remillard}}]{2004ApJ...609L..63A}
{Abramowicz}, M.~A., {Klu{\'z}niak}, W., {McClintock}, J.~E., \& {Remillard},
  R.~A., {The Importance of Discovering a 3:2 Twin-Peak Quasi-periodic
  Oscillation in an Ultraluminous X-Ray Source, or How to Solve the Puzzle of
  Intermediate-Mass Black Holes}. 2004, {\it \apjl}, {\bf 609}, L63, DOI:
  10.1086/422810

\bibitem[{{Alabarta} {et~al.}(2021){Alabarta}, {Altamirano}, {M{\'e}ndez},
  {C{\'u}neo}, {Vincentelli}, {Castro-Segura}, {Garc{\'\i}a}, {Luff}, \&
  {Veledina}}]{Alabarta2021}
{Alabarta}, K., {Altamirano}, D., {M{\'e}ndez}, M., {et~al.},
  {Failed-transition outbursts in black hole low-mass X-ray binaries}. 2021,
  {\it \mnras}, {\bf 507}, 5507, DOI: 10.1093/mnras/stab2241

\bibitem[{{Alpar} \& {Shaham}(1985)}]{1985Natur.316..239A}
{Alpar}, M.~A. \& {Shaham}, J., {Is GX5 - 1 a millisecond pulsar?} 1985, {\it
  \nat}, {\bf 316}, 239, DOI: 10.1038/316239a0

\bibitem[{Ambartsumian(1958)}]{Ambartsumian:1958qna}
Ambartsumian, V.~A., {On the evolution of galaxies}. 1958, in {\it {11\`eme
  Conseil de Physique de l'Institut International de Physique Solvay}: {La
  structure et l'\'evolution de l'univers : rapports et discussions}}, 241--280

\bibitem[{{Anile}(1989)}]{Anile1989}
{Anile}, A.~M. 1989, {\it {Relativistic fluids and magneto-fluids: With
  applications in astrophysics and plasma physics}} (Cambridge University
  Press)

\bibitem[{{Anninos} {et~al.}(2005){Anninos}, {Fragile}, \&
  {Salmonson}}]{Cosmos}
{Anninos}, P., {Fragile}, P.~C., \& {Salmonson}, J.~D., {Cosmos++: Relativistic
  Magnetohydrodynamics on Unstructured Grids with Local Adaptive Refinement}.
  2005, {\it \apj}, {\bf 635}, 723, DOI: 10.1086/497294

\bibitem[{{Arnaud}(1996)}]{1996ASPC..101...17A}
{Arnaud}, K.~A., {XSPEC: The First Ten Years}. 1996, in Astronomical Society of
  the Pacific Conference Series, Vol. {\bf  101}, {\it Astronomical Data
  Analysis Software and Systems V}, ed. G.~H. {Jacoby} \& J.~{Barnes}, 17

\bibitem[{{Avakyan} {et~al.}(2023){Avakyan}, {Neumann}, {Zainab}, {Doroshenko},
  {Wilms}, \& {Santangelo}}]{Avakyan2023}
{Avakyan}, A., {Neumann}, M., {Zainab}, A., {et~al.}, {XRBcats: Galactic Low
  Mass X-ray Binary Catalogue}. 2023, {\it arXiv e-prints}, arXiv:2303.16168,
  DOI: 10.48550/arXiv.2303.16168

\bibitem[{{Bakala} {et~al.}(2019){Bakala}, {De Falco}, {Battista},
  {Goluchov{\'a}}, {Lan{\v{c}}ov{\'a}}, {Falanga}, \&
  {Stella}}]{2019PhRvD.100j4053B}
{Bakala}, P., {De Falco}, V., {Battista}, E., {et~al.}, {Three-dimensional
  general relativistic Poynting-Robertson effect. II. Radiation field from a
  rigidly rotating spherical source}. 2019, {\it \prd}, {\bf 100}, 104053, DOI:
  10.1103/PhysRevD.100.104053

\bibitem[{{Bakala} {et~al.}(2015{\natexlab{a}}){Bakala}, {Goluchov{\'a}},
  {T{\"o}r{\"o}k}, {{\v{S}}r{\'a}mkov{\'a}}, {Abramowicz}, {Vincent}, \&
  {Mazur}}]{2015A&A...581A..35B}
{Bakala}, P., {Goluchov{\'a}}, K., {T{\"o}r{\"o}k}, G., {et~al.}, {Twin peak
  high-frequency quasi-periodic oscillations as a spectral imprint of dual
  oscillation modes of accretion tori}. 2015{\natexlab{a}}, {\it \aap}, {\bf
  581}, A35, DOI: 10.1051/0004-6361/201525867

\bibitem[{{Bakala} {et~al.}(2015{\natexlab{b}}){Bakala}, {Goluchov{\'a}},
  {T{\"o}r{\"o}k}, {{\v{S}}r{\'a}mkov{\'a}}, {Abramowicz}, {Vincent}, \&
  {Mazur}}]{Bakala+2015b}
{Bakala}, P., {Goluchov{\'a}}, K., {T{\"o}r{\"o}k}, G., {et~al.}, {Twin peak
  high-frequency quasi-periodic oscillations as a spectral imprint of dual
  oscillation modes of accretion tori}. 2015{\natexlab{b}}, {\it A\&A}, {\bf
  581}, A35, DOI: 10.1051/0004-6361/201525867

\bibitem[{{Bakala} {et~al.}(2014){Bakala}, {T{\"o}r{\"o}k}, {Karas},
  {Dov{\v{c}}iak}, {Wildner}, {Wzientek}, {{\v{S}}r{\'a}mkov{\'a}},
  {Abramowicz}, {Goluchov{\'a}}, {Mazur}, \& {Vincent}}]{2014MNRAS.439.1933B}
{Bakala}, P., {T{\"o}r{\"o}k}, G., {Karas}, V., {et~al.}, {Power density
  spectra of modes of orbital motion in strongly curved space-time: obtaining
  the observable signal}. 2014, {\it \mnras}, {\bf 439}, 1933, DOI:
  10.1093/mnras/stu076

\bibitem[{{Bakala} {et~al.}(2007){Bakala}, {{\v{C}}erm{\'a}k}, {Hled{\'\i}k},
  {Stuchl{\'\i}k}, \& {Truparov{\'a}}}]{bak-etal:2007:}
{Bakala}, P., {{\v{C}}erm{\'a}k}, P., {Hled{\'\i}k}, S., {Stuchl{\'\i}k}, Z.,
  \& {Truparov{\'a}}, K., {Extreme gravitational lensing in vicinity of
  Schwarzschild-de Sitter black holes}. 2007, {\it Central European Journal of
  Physics}, {\bf 5}, 599, DOI: 10.2478/s11534-007-0033-6

\bibitem[{{Balbus} \& {Hawley}(1991)}]{Balbus1991}
{Balbus}, S.~A. \& {Hawley}, J.~F., {A Powerful Local Shear Instability in
  Weakly Magnetized Disks. I. Linear Analysis}. 1991, {\it The Astrophysical
  Journal}, {\bf 376}, 214, DOI: 10.1086/170270

\bibitem[{{Balbus} \& {Hawley}(1998)}]{Balbus1998}
{Balbus}, S.~A. \& {Hawley}, J.~F., {Turbulent transport in accretion disks}.
  1998, in American Institute of Physics Conference Series, Vol. {\bf  431},
  {\it Accretion processes in Astrophysical Systems: Some like it hot! - eigth
  AstroPhysics Conference}, ed. S.~S. {Holt} \& T.~R. {Kallman}, 79--88

\bibitem[{{Bambi}(2017)}]{2017bhlt.book.....B}
{Bambi}, C. 2017, {\it {Black Holes: A Laboratory for Testing Strong Gravity}}
  (Springer)

\bibitem[{{Bardeen} {et~al.}(1972){Bardeen}, {Press}, \&
  {Teukolsky}}]{Bardeen1972}
{Bardeen}, J.~M., {Press}, W.~H., \& {Teukolsky}, S.~A., {Rotating Black Holes:
  Locally Nonrotating Frames, Energy Extraction, and Scalar Synchrotron
  Radiation}. 1972, {\it The Astrophysical Journal}, {\bf 178}, 347, DOI:
  10.1086/151796

\bibitem[{{Barret} {et~al.}(2005){Barret}, {Klu{\'z}niak}, {Olive}, {Paltani},
  \& {Skinner}}]{2005MNRAS.357.1288B}
{Barret}, D., {Klu{\'z}niak}, W., {Olive}, J.~F., {Paltani}, S., \& {Skinner},
  G.~K., {On the high coherence of kHz quasi-periodic oscillations}. 2005, {\it
  \mnras}, {\bf 357}, 1288, DOI: 10.1111/j.1365-2966.2005.08734.x

\bibitem[{{Beckwith} \& {Done}(2005)}]{2005MNRAS.359.1217B}
{Beckwith}, K. \& {Done}, C., {Extreme gravitational lensing near rotating
  black holes}. 2005, {\it \mnras}, {\bf 359}, 1217, DOI:
  10.1111/j.1365-2966.2005.08980.x

\bibitem[{{Begelman} \& {Pringle}(2007)}]{Begelman2007}
{Begelman}, M.~C. \& {Pringle}, J.~E., {Accretion discs with strong toroidal
  magnetic fields}. 2007, {\it \mnras}, {\bf 375}, 1070, DOI:
  10.1111/j.1365-2966.2006.11372.x

\bibitem[{{Belczynski} {et~al.}(2010){Belczynski}, {Bulik}, {Fryer}, {Ruiter},
  {Valsecchi}, {Vink}, \& {Hurley}}]{Belczynski2010}
{Belczynski}, K., {Bulik}, T., {Fryer}, C.~L., {et~al.}, {On the Maximum Mass
  of Stellar Black Holes}. 2010, {\it \apj}, {\bf 714}, 1217, DOI:
  10.1088/0004-637X/714/2/1217

\bibitem[{{Bildsten} \& {Cumming}(1998)}]{1998ApJ...506..842B}
{Bildsten}, L. \& {Cumming}, A., {Hydrogen Electron Capture in Accreting
  Neutron Stars and the Resulting g-Mode Oscillation Spectrum}. 1998, {\it
  \apj}, {\bf 506}, 842, DOI: 10.1086/306279

\bibitem[{{Blaes} {et~al.}(2006){Blaes}, {Arras}, \& {Fragile}}]{Blaes2006}
{Blaes}, O.~M., {Arras}, P., \& {Fragile}, P.~C., {Oscillation modes of
  relativistic slender tori}. 2006, {\it \mnras}, {\bf 369}, 1235, DOI:
  10.1111/j.1365-2966.2006.10370.x

\bibitem[{{Blaes} {et~al.}(2007){Blaes}, {{\v{S}}r{\'a}mkov{\'a}},
  {Abramowicz}, {Klu{\'z}niak}, \& {Torkelsson}}]{2007ApJ...665..642B}
{Blaes}, O.~M., {{\v{S}}r{\'a}mkov{\'a}}, E., {Abramowicz}, M.~A.,
  {Klu{\'z}niak}, W., \& {Torkelsson}, U., {Epicyclic Oscillations of Fluid
  Bodies: Newtonian Nonslender Torus}. 2007, {\it \apj}, {\bf 665}, 642, DOI:
  10.1086/519782

\bibitem[{{Blandford} \& {Znajek}(1977)}]{Blandford1977}
{Blandford}, R.~D. \& {Znajek}, R.~L., {Electromagnetic extraction of energy
  from Kerr black holes.} 1977, {\it \mnras}, {\bf 179}, 433, DOI:
  10.1093/mnras/179.3.433

\bibitem[{{Bollimpalli} {et~al.}(2020){Bollimpalli}, {Mahmoud}, {Done},
  {Fragile}, {Klu{\'z}niak}, {Narayan}, \& {White}}]{2020MNRAS.496.3808B}
{Bollimpalli}, D.~A., {Mahmoud}, R., {Done}, C., {et~al.}, {Looking for the
  underlying cause of black hole X-ray variability in GRMHD simulations}. 2020,
  {\it \mnras}, {\bf 496}, 3808, DOI: 10.1093/mnras/staa1808

\bibitem[{{Bombaci}(1996)}]{maxNSmass}
{Bombaci}, I., {The maximum mass of a neutron star.} 1996, {\it \aap}, {\bf
  305}, 871

\bibitem[{{Bondi}(1952)}]{Bondi}
{Bondi}, H., {On spherically symmetrical accretion}. 1952, {\it \mnras}, {\bf
  112}, 195, DOI: 10.1093/mnras/112.2.195

\bibitem[{{Boutelier} {et~al.}(2010){Boutelier}, {Barret}, {Lin}, \&
  {T{\"o}r{\"o}k}}]{2010MNRAS.401.1290B}
{Boutelier}, M., {Barret}, D., {Lin}, Y., \& {T{\"o}r{\"o}k}, G., {On the
  distribution of frequency ratios of kHz quasi-periodic oscillations}. 2010,
  {\it \mnras}, {\bf 401}, 1290, DOI: 10.1111/j.1365-2966.2009.15724.x

\bibitem[{{Boyer} \& {Lindquist}(1967)}]{Boyer1967}
{Boyer}, R.~H. \& {Lindquist}, R.~W., {Maximal Analytic Extension of the Kerr
  Metric}. 1967, {\it Journal of Mathematical Physics}, {\bf 8}, 265, DOI:
  10.1063/1.1705193

\bibitem[{{Bradt} {et~al.}(1993){Bradt}, {Rothschild}, \&
  {Swank}}]{1993A&AS...97..355B}
{Bradt}, H.~V., {Rothschild}, R.~E., \& {Swank}, J.~H., {X-ray timing explorer
  mission}. 1993, {\it \aaps}, {\bf 97}, 355

\bibitem[{{Bradt} \& {McClintock}(1983)}]{Bradt1983}
{Bradt}, H.~V.~D. \& {McClintock}, J.~E., {The optical Counterparts of Compact
  discrete galactic X-Ray sources.} 1983, {\it \araa}, {\bf 21}, 13, DOI:
  10.1146/annurev.aa.21.090183.000305

\bibitem[{{Brandenburg}(2001)}]{Brandenburg2001}
{Brandenburg}, A., {The Inverse Cascade and Nonlinear Alpha-Effect in
  Simulations of Isotropic Helical Hydromagnetic Turbulence}. 2001, {\it \apj},
  {\bf 550}, 824, DOI: 10.1086/319783

\bibitem[{{Brandenburg} {et~al.}(1995){Brandenburg}, {Nordlund}, {Stein}, \&
  {Torkelsson}}]{Brandenburg1995}
{Brandenburg}, A., {Nordlund}, A., {Stein}, R.~F., \& {Torkelsson}, U.,
  {Dynamo-generated Turbulence and Large-Scale Magnetic Fields in a Keplerian
  Shear Flow}. 1995, {\it \apj}, {\bf 446}, 741, DOI: 10.1086/175831

\bibitem[{{Bronzwaer} {et~al.}(2018){Bronzwaer}, {Davelaar}, {Younsi},
  {Mo{\'s}cibrodzka}, {Falcke}, {Kramer}, \& {Rezzolla}}]{2018A&A...613A...2B}
{Bronzwaer}, T., {Davelaar}, J., {Younsi}, Z., {et~al.}, {RAPTOR. I.
  Time-dependent radiative transfer in arbitrary spacetimes}. 2018, {\it \aap},
  {\bf 613}, A2, DOI: 10.1051/0004-6361/201732149

\bibitem[{{Bursa}(2005)}]{2005AN....326..849B}
{Bursa}, M., {Global oscillations of a fluid torus as a modulation mechanism
  for black-hole high-frequency QPOs}. 2005, {\it Astronomische Nachrichten},
  {\bf 326}, 849, DOI: 10.1002/asna.200510426

\bibitem[{{Bursa} {et~al.}(2004){Bursa}, {Abramowicz}, {Karas}, \&
  {Klu{\'z}niak}}]{2004ApJ...617L..45B}
{Bursa}, M., {Abramowicz}, M.~A., {Karas}, V., \& {Klu{\'z}niak}, W., {The
  Upper Kilohertz Quasi-periodic Oscillation: A Gravitationally Lensed Vertical
  Oscillation}. 2004, {\it \apjl}, {\bf 617}, L45, DOI: 10.1086/427167

\bibitem[{{Carpano} \& {Jin}(2018)}]{2018MNRAS.477.3178C}
{Carpano}, S. \& {Jin}, C., {Discovery of a 23.8 h QPO in the Swift light curve
  of XMMU J134736.6+173403}. 2018, {\it \mnras}, {\bf 477}, 3178, DOI:
  10.1093/mnras/sty841

\bibitem[{{Casella} {et~al.}(2005){Casella}, {Belloni}, \&
  {Stella}}]{Casella2005}
{Casella}, P., {Belloni}, T., \& {Stella}, L., {The ABC of Low-Frequency
  Quasi-periodic Oscillations in Black Hole Candidates: Analogies with Z
  Sources}. 2005, {\it \apj}, {\bf 629}, 403, DOI: 10.1086/431174

\bibitem[{{Chael} {et~al.}(2017){Chael}, {Narayan}, \&
  {S\k{a}dowski}}]{Chael2017}
{Chael}, A.~A., {Narayan}, R., \& {S\k{a}dowski}, A., {Evolving non-thermal
  electrons in simulations of black hole accretion}. 2017, {\it \mnras}, {\bf
  470}, 2367, DOI: 10.1093/mnras/stx1345

\bibitem[{{Chan} {et~al.}(2013){Chan}, {Psaltis}, \&
  {{\"O}zel}}]{2013ApJ...777...13C}
{Chan}, C.-k., {Psaltis}, D., \& {{\"O}zel}, F., {GRay: A Massively Parallel
  GPU-based Code for Ray Tracing in Relativistic Spacetimes}. 2013, {\it \apj},
  {\bf 777}, 13, DOI: 10.1088/0004-637X/777/1/13

\bibitem[{{Chandrasekhar}(1998)}]{chandra}
{Chandrasekhar}, S. 1998, {\it {The Mathematical Theory of Black Holes}}
  (Clarendon Press)

\bibitem[{Childs {et~al.}(2012)Childs, Brugger, Whitlock, Meredith, Ahern,
  Pugmire, Biagas, Miller, Harrison, Weber, Krishnan, Fogal, Sanderson, Garth,
  Bethel, Camp, R\"{u}bel, Durant, Favre, \& Navr\'{a}til}]{HPV:VisIt}
Childs, H., Brugger, E., Whitlock, B., {et~al.}, VisIt: An End-User Tool For
  Visualizing and Analyzing Very Large Data. 2012, in {\it High Performance
  Visualization--Enabling Extreme-Scale Scientific Insight} (Taylor \& Francis
  Inc), 357--372

\bibitem[{{Ciesielski} {et~al.}(2012){Ciesielski}, {Wielgus}, {Klu{\'z}niak},
  {S{\k{a}}dowski}, {Abramowicz}, {Lasota}, \& {Rebusco}}]{Ciesielski2012}
{Ciesielski}, A., {Wielgus}, M., {Klu{\'z}niak}, W., {et~al.}, {Stability of
  radiation-pressure dominated disks. I. The dispersion relation for a delayed
  heating {\ensuremath{\alpha}}-viscosity prescription}. 2012, {\it \aap}, {\bf
  538}, A148, DOI: 10.1051/0004-6361/201117478

\bibitem[{{Coughlin} \& {Begelman}(2014)}]{Coughlin2014}
{Coughlin}, E.~R. \& {Begelman}, M.~C., {The General Relativistic Equations of
  Radiation Hydrodynamics in the Viscous Limit}. 2014, {\it \apj}, {\bf 797},
  103, DOI: 10.1088/0004-637X/797/2/103

\bibitem[{{Cowling}(1933)}]{Cowling1933}
{Cowling}, T.~G., {The magnetic field of sunspots}. 1933, {\it \mnras}, {\bf
  94}, 39, DOI: 10.1093/mnras/94.1.39

\bibitem[{{Cunningham} \& {Bardeen}(1973)}]{1973ApJ...183..237C}
{Cunningham}, C.~T. \& {Bardeen}, J.~M., {The Optical Appearance of a Star
  Orbiting an Extreme Kerr Black Hole}. 1973, {\it \apj}, {\bf 183}, 237, DOI:
  10.1086/152223

\bibitem[{{Dauser} {et~al.}(2016){Dauser}, {Garc{\'\i}a}, \&
  {Wilms}}]{reflection}
{Dauser}, T., {Garc{\'\i}a}, J., \& {Wilms}, J., {Relativistic reflection:
  Review and recent developments in modeling}. 2016, {\it Astronomische
  Nachrichten}, {\bf 337}, 362, DOI: 10.1002/asna.201612314

\bibitem[{{de Avellar} {et~al.}(2017){de Avellar}, {Porth}, {Younsi}, \&
  {Rezzolla}}]{2017arXiv170907706D}
{de Avellar}, M. G.~B., {Porth}, O., {Younsi}, Z., \& {Rezzolla}, L., {The kilo
  Hertz quasi-periodic oscillations in neutron star low-mass X-ray binaries as
  tori oscillation modes. I}. 2017, {\it arXiv e-prints}, arXiv:1709.07706,
  DOI: 10.48550/arXiv.1709.07706

\bibitem[{{De Falco} {et~al.}(2019){De Falco}, {Bakala}, {Battista},
  {Lan{\v{c}}ov{\'a}}, {Falanga}, \& {Stella}}]{2019PhRvD..99b3014D}
{De Falco}, V., {Bakala}, P., {Battista}, E., {et~al.}, {Three-dimensional
  general relativistic Poynting-Robertson effect: Radial radiation field}.
  2019, {\it \prd}, {\bf 99}, 023014, DOI: 10.1103/PhysRevD.99.023014

\bibitem[{{De Rosa} {et~al.}(2019){De Rosa}, {Uttley}, {Gou}, {Liu}, {Bambi},
  {Barret}, {Belloni}, {Berti}, {Bianchi}, {Caiazzo}, {Casella}, {Feroci},
  {Ferrari}, {Gualtieri}, {Heyl}, {Ingram}, {Karas}, {Lu}, {Luo}, {Matt},
  {Motta}, {Neilsen}, {Pani}, {Santangelo}, {Shu}, {Wang}, {Wang}, {Xue}, {Xu},
  {Yuan}, {Yuan}, {Zhang}, {Zhang}, {Agudo}, {Amati}, {Andersson}, {Baglio},
  {Bakala}, {Baykal}, {Bhattacharyya}, {Bombaci}, {Bucciantini}, {Capitanio},
  {Ciolfi}, {Cui}, {D'Ammando}, {Dauser}, {Del Santo}, {De Marco}, {Di Salvo},
  {Done}, {Dov{\v{c}}iak}, {Fabian}, {Falanga}, {Gambino}, {Gendre},
  {Grinberg}, {Heger}, {Homan}, {Iaria}, {Jiang}, {Jin}, {Koerding}, {Linares},
  {Liu}, {Maccarone}, {Malzac}, {Manousakis}, {Marin}, {Marinucci},
  {Mehdipour}, {M{\'e}ndez}, {Migliari}, {Miller}, {Miniutti}, {Nardini},
  {O'Brien}, {Osborne}, {Petrucci}, {Possenti}, {Riggio}, {Rodriguez}, {Sanna},
  {Shao}, {Sobolewska}, {Sramkova}, {Stevens}, {Stiele}, {Stratta}, {Stuchlik},
  {Svoboda}, {Tamburini}, {Tauris}, {Tombesi}, {Torok}, {Urbanec}, {Vincent},
  {Wu}, {Yuan}, {in't Zand}, {Zdziarski}, \& {Zhou}}]{2019SCPMA..6229504D}
{De Rosa}, A., {Uttley}, P., {Gou}, L., {et~al.}, {Accretion in strong field
  gravity with eXTP}. 2019, {\it Science China Physics, Mechanics, and
  Astronomy}, {\bf 62}, 29504, DOI: 10.1007/s11433-018-9297-0

\bibitem[{{Dewberry} {et~al.}(2020){Dewberry}, {Latter}, {Ogilvie}, \&
  {Fromang}}]{2020MNRAS.497..451D}
{Dewberry}, J.~W., {Latter}, H.~N., {Ogilvie}, G.~I., \& {Fromang}, S., {HFQPOs
  and discoseismic mode excitation in eccentric, relativistic discs. II.
  Magnetohydrodynamic simulations}. 2020, {\it \mnras}, {\bf 497}, 451, DOI:
  10.1093/mnras/staa1898

\bibitem[{{Dexter} \& {Agol}(2009)}]{2009ApJ...696.1616D}
{Dexter}, J. \& {Agol}, E., {A Fast New Public Code for Computing Photon Orbits
  in a Kerr Spacetime}. 2009, {\it \apj}, {\bf 696}, 1616, DOI:
  10.1088/0004-637X/696/2/1616

\bibitem[{{Done} {et~al.}(2007){Done}, {Gierli{\'n}ski}, \&
  {Kubota}}]{Done2007}
{Done}, C., {Gierli{\'n}ski}, M., \& {Kubota}, A., {Modelling the behaviour of
  accretion flows in X-ray binaries. Everything you always wanted to know about
  accretion but were afraid to ask}. 2007, {\it Astronomy and Astrophysics
  Review}, {\bf 15}, 1, DOI: 10.1007/s00159-007-0006-1

\bibitem[{{Done} {et~al.}(2004){Done}, {Wardzi{\'n}ski}, \&
  {Gierli{\'n}ski}}]{Done2004}
{Done}, C., {Wardzi{\'n}ski}, G., \& {Gierli{\'n}ski}, M., {GRS 1915+105: the
  brightest Galactic black hole}. 2004, {\it Monthly Notices of the Royal
  Astronomical Society}, {\bf 349}, 393, DOI: 10.1111/j.1365-2966.2004.07545.x

\bibitem[{{Dov{\v{c}}iak} {et~al.}(2004){Dov{\v{c}}iak}, {Karas}, \&
  {Yaqoob}}]{Dovciak04}
{Dov{\v{c}}iak}, M., {Karas}, V., \& {Yaqoob}, T., {An Extended Scheme for
  Fitting X-Ray Data with Accretion Disk Spectra in the Strong Gravity Regime}.
  2004, {\it \apjs}, {\bf 153}, 205, DOI: 10.1086/421115

\bibitem[{{Dullo} {et~al.}(2021){Dullo}, {Gil de Paz}, \& {Knapen}}]{Dullo2021}
{Dullo}, B.~T., {Gil de Paz}, A., \& {Knapen}, J.~H., {Ultramassive Black Holes
  in the Most Massive Galaxies: M$_{BH}$-{\ensuremath{\sigma}} versus
  M$_{BH}$-R$_{b}$}. 2021, {\it \apj}, {\bf 908}, 134, DOI:
  10.3847/1538-4357/abceae

\bibitem[{{Etienne} {et~al.}(2015){Etienne}, {Paschalidis}, {Haas},
  {M{\"o}sta}, \& {Shapiro}}]{Etienne2015}
{Etienne}, Z.~B., {Paschalidis}, V., {Haas}, R., {M{\"o}sta}, P., \& {Shapiro},
  S.~L., {IllinoisGRMHD: an open-source, user-friendly GRMHD code for dynamical
  spacetimes}. 2015, {\it Classical and Quantum Gravity}, {\bf 32}, 175009,
  DOI: 10.1088/0264-9381/32/17/175009

\bibitem[{{Fath}(1909)}]{StrangeSpectra}
{Fath}, E.~A., {The spectra of some spiral nebulae and globular star clusters}.
  1909, {\it Lick Observatory Bulletin}, {\bf 149}, 71, DOI:
  10.5479/ADS/bib/1909LicOB.5.71F

\bibitem[{{Feroci} {et~al.}(2022){Feroci}, {Ambrosi}, {Ambrosino}, {Antonelli},
  {Argan}, {Babinec}, {Barbera}, {Bayer}, {Bellutti}, {Bertucci}, {Bertuccio},
  {Bi}, {Boezio}, {Bonvicini}, {Borghi}, {Bozzo}, {Baudin}, {Bouyjou},
  {Brienza}, {Cadoux}, {Campana}, {Cao}, {Cavazzuti}, {Ceraudo}, {Chen},
  {Chen}, {Cirrincione}, {De Angelis}, {De Rosa}, {Del Monte}, {Di Cosimo},
  {Dilillo}, {Dohnal}, {Donnarumma}, {Evangelista}, {Fan}, {Favre},
  {Fiandrini}, {Ficorella}, {Fuschino}, {Gao}, {Gevin}, {Grassi}, {Guedel},
  {Han}, {He}, {Hedderman}, {den Herder}, {Hynek}, {Hong}, {Jin}, {Kole},
  {Karas}, {Komarek}, {Labanti}, {Li}, {Li}, {Liang}, {Limousin}, {Liu}, {Lo
  Cicero}, {Lohering}, {Lombardi}, {Lu}, {Luo}, {Malcovati}, {Mao},
  {Marinucci}, {Mele}, {Mendes}, {Merkl}, {Meuris}, {Michalska}, {Morbidini},
  {Morgante}, {Muleri}, {Munini}, {Mussolin}, {Negri}, {Nov{\'a}k},
  {Nowosielski}, {Nuti}, {Orleanski}, {Ottensamer}, {Pacciani}, {Paltani},
  {Pan}, {Pepponi}, {Perinati}, {Piazzolla}, {Picciotto}, {Pliego},
  {Rachevski}, {Rashevskaia}, {Santangelo}, {Schanne}, {Serafinelli}, {Skup},
  {Sveda}, {Svoboda}, {Tenzer}, {Todaro}, {Torok}, {Trois}, {Vacchi},
  {Virgilli}, {Xiong}, {Wang}, {Wang}, {Winter}, {Wu}, {Xu}, {Zampa}, {Zampa},
  {Zdziarski}, {Zhang}, {Zhang}, {Zhang}, {Zhang}, {Zhang}, {Zhang}, {Zhou}, \&
  {Zorzi}}]{2022SPIE12181E..1XF}
{Feroci}, M., {Ambrosi}, G., {Ambrosino}, F., {et~al.}, {The large area
  detector onboard the eXTP mission}. 2022, in Society of Photo-Optical
  Instrumentation Engineers (SPIE) Conference Series, Vol. {\bf  12181}, {\it
  Space Telescopes and Instrumentation 2022: Ultraviolet to Gamma Ray}, ed.
  J.-W.~A. {den Herder}, S.~{Nikzad}, \& K.~{Nakazawa}, 121811X

\bibitem[{{Foschini} {et~al.}(2006){Foschini}, {Ebisawa}, {Kawaguchi},
  {Cappelluti}, {Grandi}, {Malaguti}, {Rodriguez}, {Courvoisier}, {Di Cocco},
  {Ho}, \& {Palumbo}}]{Foschini2006}
{Foschini}, L., {Ebisawa}, K., {Kawaguchi}, T., {et~al.}, {The application of
  slim disk models to ULX: The case of M33 X-8}. 2006, {\it Advances in Space
  Research}, {\bf 38}, 1378, DOI: 10.1016/j.asr.2005.06.024

\bibitem[{{Fragile} {et~al.}(2018){Fragile}, {Etheridge}, {Anninos}, {Mishra},
  \& {Klu{\'z}niak}}]{Fragile2018}
{Fragile}, P.~C., {Etheridge}, S.~M., {Anninos}, P., {Mishra}, B., \&
  {Klu{\'z}niak}, W., {Relativistic, Viscous, Radiation Hydrodynamic
  Simulations of Geometrically Thin Disks. I. Thermal and Other Instabilities}.
  2018, {\it ApJ}, {\bf 857}, 1, DOI: 10.3847/1538-4357/aab788

\bibitem[{{Fragile} {et~al.}(2016){Fragile}, {Straub}, \&
  {Blaes}}]{2016MNRAS.461.1356F}
{Fragile}, P.~C., {Straub}, O., \& {Blaes}, O., {High-frequency and type-C QPOs
  from oscillating, precessing hot, thick flow}. 2016, {\it \mnras}, {\bf 461},
  1356, DOI: 10.1093/mnras/stw1428

\bibitem[{{Frank} {et~al.}(2002){Frank}, {King}, \& {Raine}}]{AccretionPower}
{Frank}, J., {King}, A., \& {Raine}, D.~J. 2002, {\it {Accretion Power in
  Astrophysics: Third Edition}} (Cambridge University Press)

\bibitem[{Galassi \& Gough(2009)}]{galassi_gnu_2009}
Galassi, M. \& Gough, B. 2009, {\it {GNU} {Scientific} {Library}: {Reference}
  {Manual}}, {GNU} manual (Network Theory)

\bibitem[{{Gammie} {et~al.}(2003){Gammie}, {McKinney}, \&
  {T{\'o}th}}]{Gammie2003}
{Gammie}, C.~F., {McKinney}, J.~C., \& {T{\'o}th}, G., {HARM: A Numerical
  Scheme for General Relativistic Magnetohydrodynamics}. 2003, {\it \apj}, {\bf
  589}, 444, DOI: 10.1086/374594

\bibitem[{{Gierli{\'n}ski} \& {Done}(2004)}]{2004MNRAS.347..885G}
{Gierli{\'n}ski}, M. \& {Done}, C., {Black hole accretion discs: reality
  confronts theory}. 2004, {\it \mnras}, {\bf 347}, 885, DOI:
  10.1111/j.1365-2966.2004.07266.x

\bibitem[{{Gierli{\'n}ski} {et~al.}(1999){Gierli{\'n}ski}, {Zdziarski},
  {Poutanen}, {Coppi}, {Ebisawa}, \& {Johnson}}]{Gierlinski1999}
{Gierli{\'n}ski}, M., {Zdziarski}, A.~A., {Poutanen}, J., {et~al.}, {Radiation
  mechanisms and geometry of Cygnus X-1 in the soft state}. 1999, {\it \mnras},
  {\bf 309}, 496, DOI: 10.1046/j.1365-8711.1999.02875.x

\bibitem[{{Goluchov{\'a}} {et~al.}(2019){Goluchov{\'a}}, {T{\"o}r{\"o}k},
  {{\v{S}}r{\'a}mkov{\'a}}, {Abramowicz}, {Stuchl{\'\i}k}, \&
  {Hor{\'a}k}}]{Gol-etal:2019}
{Goluchov{\'a}}, K., {T{\"o}r{\"o}k}, G., {{\v{S}}r{\'a}mkov{\'a}}, E.,
  {et~al.}, {Mass of the active galactic nucleus black hole
  XMMUJ134736.6+173403}. 2019, {\it A\&A}, {\bf 622}, L8, DOI:
  10.1051/0004-6361/201834774

\bibitem[{{Goodman} {et~al.}(1987){Goodman}, {Narayan}, \&
  {Goldreich}}]{1987MNRAS.225..695G}
{Goodman}, J., {Narayan}, R., \& {Goldreich}, P., {The stability of accretion
  tori. II - Non-linear evolution to discrete planets}. 1987, {\it \mnras},
  {\bf 225}, 695, DOI: 10.1093/mnras/225.3.695

\bibitem[{{Gou} {et~al.}(2011){Gou}, {McClintock}, {Reid}, {Orosz}, {Steiner},
  {Narayan}, {Xiang}, {Remillard}, {Arnaud}, \& {Davis}}]{2011ApJ...742...85G}
{Gou}, L., {McClintock}, J.~E., {Reid}, M.~J., {et~al.}, {The Extreme Spin of
  the Black Hole in Cygnus X-1}. 2011, {\it \apj}, {\bf 742}, 85, DOI:
  10.1088/0004-637X/742/2/85

\bibitem[{{Gronkiewicz} \& {R{\'o}{\.z}a{\'n}ska}(2020)}]{Gronki2020}
{Gronkiewicz}, D. \& {R{\'o}{\.z}a{\'n}ska}, A., {Warm and thick corona for a
  magnetically supported disk in galactic black hole binaries}. 2020, {\it
  \aap}, {\bf 633}, A35, DOI: 10.1051/0004-6361/201935033

\bibitem[{Harten {et~al.}(1983)Harten, Lax, \& van Leer}]{Harten1983}
Harten, A., Lax, P., \& van Leer, B., On Upstream Differencing and Godunov-Type
  Schemes for Hyperbolic Conservation Laws. 1983, {\it SIAM Rev}, {\bf 25}, 35

\bibitem[{{Hartle} \& {Thorne}(1968)}]{HartleThorne1968}
{Hartle}, J.~B. \& {Thorne}, K.~S., {Slowly Rotating Relativistic Stars. II.
  Models for Neutron Stars and Supermassive Stars}. 1968, {\it APJ}, {\bf 153},
  807, DOI: 10.1086/149707

\bibitem[{{Hasinger} \& {van der Klis}(1989)}]{1989A&A...225...79H}
{Hasinger}, G. \& {van der Klis}, M., {Two patterns of correlated X-ray timing
  and spectral behaviour in low-mass X-ray binaries.} 1989, {\it \aap}, {\bf
  225}, 79

\bibitem[{{Hawley} {et~al.}(2011){Hawley}, {Guan}, \& {Krolik}}]{Hawley2011}
{Hawley}, J.~F., {Guan}, X., \& {Krolik}, J.~H., {Assessing Quantitative
  Results in Accretion Simulations: From Local to Global}. 2011, {\it \apj},
  {\bf 738}, 84, DOI: 10.1088/0004-637X/738/1/84

\bibitem[{{Hawley} {et~al.}(2013){Hawley}, {Richers}, {Guan}, \&
  {Krolik}}]{Hawley2013}
{Hawley}, J.~F., {Richers}, S.~A., {Guan}, X., \& {Krolik}, J.~H., {Testing
  Convergence for Global Accretion Disks}. 2013, {\it \apj}, {\bf 772}, 102,
  DOI: 10.1088/0004-637X/772/2/102

\bibitem[{{Hills}(1975)}]{Hills1975}
{Hills}, J.~G., {Possible power source of Seyfert galaxies and QSOs}. 1975,
  {\it \nat}, {\bf 254}, 295, DOI: 10.1038/254295a0

\bibitem[{{Hor{\'a}k}(2005)}]{2005AN....326..845H}
{Hor{\'a}k}, J., {A possible mechanism for QPOs modulation in neutron star
  sources}. 2005, {\it Astronomische Nachrichten}, {\bf 326}, 845, DOI:
  10.1002/asna.200510425

\bibitem[{{Hor{\'a}k} {et~al.}(2004){Hor{\'a}k}, {Abramowicz}, {Karas}, \&
  {Kluzniak}}]{2004PASJ...56..819H}
{Hor{\'a}k}, J., {Abramowicz}, M.~A., {Karas}, V., \& {Kluzniak}, W., {Of NBOs
  and kHz QPOs: a Low-Frequency Modulation in Resonant Oscillations of
  Relativistic Accretion Disks}. 2004, {\it \pasj}, {\bf 56}, 819, DOI:
  10.1093/pasj/56.5.819

\bibitem[{{Hor{\'a}k} {et~al.}(2009){Hor{\'a}k}, {Abramowicz}, {Klu{\'z}niak},
  {Rebusco}, \& {T{\"o}r{\"o}k}}]{horak2009}
{Hor{\'a}k}, J., {Abramowicz}, M.~A., {Klu{\'z}niak}, W., {Rebusco}, P., \&
  {T{\"o}r{\"o}k}, G., {Internal resonance in nonlinear disk oscillations and
  the amplitude evolution of neutron-star kilohertz QPOs}. 2009, {\it A\&A},
  {\bf 499}, 535, DOI: 10.1051/0004-6361/200810740

\bibitem[{{Hor{\'a}k} \& {Karas}(2006)}]{2006A&A...451..377H}
{Hor{\'a}k}, J. \& {Karas}, V., {Twin-peak quasiperiodic oscillations as an
  internal resonance}. 2006, {\it \aap}, {\bf 451}, 377, DOI:
  10.1051/0004-6361:20054039

\bibitem[{{Hor{\'a}k} \& {Lai}(2013)}]{2013MNRAS.434.2761H}
{Hor{\'a}k}, J. \& {Lai}, D., {Corotation resonance and overstable oscillations
  in black hole accretion discs: general relativistic calculations}. 2013, {\it
  \mnras}, {\bf 434}, 2761, DOI: 10.1093/mnras/stt1120

\bibitem[{{Huang} {et~al.}(2023){Huang}, {Jiang}, {Feng}, {Davis}, {Stone}, \&
  {Middleton}}]{Huang2023}
{Huang}, J., {Jiang}, Y.-F., {Feng}, H., {et~al.}, {Global 3D Radiation
  Magnetohydrodynamic Simulations of Accretion onto a Stellar-mass Black Hole
  at Sub- and Near-critical Accretion Rates}. 2023, {\it The Astrophysical
  Journal}, {\bf 945}, 57, DOI: 10.3847/1538-4357/acb6fc

\bibitem[{{Ingram} \& {Done}(2010)}]{2010MNRAS.405.2447I}
{Ingram}, A. \& {Done}, C., {A physical interpretation of the variability power
  spectral components in accreting neutron stars}. 2010, {\it \mnras}, {\bf
  405}, 2447, DOI: 10.1111/j.1365-2966.2010.16614.x

\bibitem[{{Ingram} \& {Motta}(2019)}]{2019NewAR..8501524I}
{Ingram}, A.~R. \& {Motta}, S.~E., {A review of quasi-periodic oscillations
  from black hole X-ray binaries: Observation and theory}. 2019, {\it \nar},
  {\bf 85}, 101524, DOI: 10.1016/j.newar.2020.101524

\bibitem[{{Inogamov} \& {Sunyaev}(1999)}]{1999AstL...25..269I}
{Inogamov}, N.~A. \& {Sunyaev}, R.~A., {Spread of matter over a neutron-star
  surface during disk accretion}. 1999, {\it Astronomy Letters}, {\bf 25}, 269,
  DOI: 10.48550/arXiv.astro-ph/9904333

\bibitem[{{Janiuk}(2023)}]{Janiuk23}
{Janiuk}, A., {Neutrino cooled disk in post-merger system studied via numerical
  GR MHD simulation with a composition-dependent equation of state}. 2023, {\it
  arXiv e-prints}, arXiv:2303.18129, DOI: 10.48550/arXiv.2303.18129

\bibitem[{{Jaroszy{\'n}ski} {et~al.}(1980){Jaroszy{\'n}ski}, {Abramowicz}, \&
  {Paczy{\'n}ski}}]{Jarosz1980}
{Jaroszy{\'n}ski}, M., {Abramowicz}, M.~A., \& {Paczy{\'n}ski}, B.,
  {Supercritical accretion disks around black holes}. 1980, {\it Acta Astron.},
  {\bf 30}, 1

\bibitem[{{Jiang} {et~al.}(2013){Jiang}, {Stone}, \& {Davis}}]{Jiang2013}
{Jiang}, Y.-F., {Stone}, J.~M., \& {Davis}, S.~W., {On the Thermal Stability of
  Radiation-dominated Accretion Disks}. 2013, {\it \apj}, {\bf 778}, 65, DOI:
  10.1088/0004-637X/778/1/65

\bibitem[{{Jiang} {et~al.}(2019){Jiang}, {Stone}, \& {Davis}}]{Jiang2019}
{Jiang}, Y.-F., {Stone}, J.~M., \& {Davis}, S.~W., {Super-Eddington Accretion
  Disks around Supermassive Black Holes}. 2019, {\it ApJ}, {\bf 880}, 67, DOI:
  10.3847/1538-4357/ab29ff

\bibitem[{{Kaaret} {et~al.}(2017){Kaaret}, {Feng}, \& {Roberts}}]{ULXs}
{Kaaret}, P., {Feng}, H., \& {Roberts}, T.~P., {Ultraluminous X-Ray Sources}.
  2017, {\it \araa}, {\bf 55}, 303, DOI: 10.1146/annurev-astro-091916-055259

\bibitem[{{Karas}(1999{\natexlab{a}})}]{1999PASJ...51..317K}
{Karas}, V., {Quasi-Periodic Features Due to Clumps Orbiting around a Black
  Hole}. 1999{\natexlab{a}}, {\it \pasj}, {\bf 51}, 317, DOI:
  10.1093/pasj/51.3.317

\bibitem[{{Karas}(1999{\natexlab{b}})}]{1999ApJ...526..953K}
{Karas}, V., {Twin Peak Separation in Sources with Kilohertz Quasi-periodic
  Oscillations Caused by Orbital Motion}. 1999{\natexlab{b}}, {\it \apj}, {\bf
  526}, 953, DOI: 10.1086/308015

\bibitem[{{Karas} {et~al.}(2023){Karas}, {Klimovi{\v{c}}ov{\'a}},
  {Lan{\v{c}}ov{\'a}}, {Svoboda}, {T{\"o}r{\"o}k}, {Matuszkov{\'a}},
  {{\v{S}}r{\'a}mkov{\'a}}, {{\v{S}}pr\v{n}a}, \& {Urbanec}}]{IBWS}
{Karas}, V., {Klimovi{\v{c}}ov{\'a}}, K., {Lan{\v{c}}ov{\'a}}, D., {et~al.}
  2023, manuscript submitted for publication to Contrib. Astron. Obs.
  Skalnat\'{e} Pleso

\bibitem[{{Karas} {et~al.}(1992){Karas}, {Vokrouhlick\'y}, \&
  {Polnarev}}]{1992MNRAS.259..569K}
{Karas}, V., {Vokrouhlick\'y}, D., \& {Polnarev}, A.~G., {In the vicinity of a
  rotating black hole: a fast numerical code for computing observational
  effects.} 1992, {\it \mnras}, {\bf 259}, 569, DOI: 10.1093/mnras/259.3.569

\bibitem[{{Kato}(2004)}]{2004PASJ...56..905K}
{Kato}, S., {Resonant Excitation of Disk Oscillations by Warps: A Model of kHz
  QPOs}. 2004, {\it \pasj}, {\bf 56}, 905, DOI: 10.1093/pasj/56.5.905

\bibitem[{{Kato} \& {Fukue}(1980)}]{1980PASJ...32..377K}
{Kato}, S. \& {Fukue}, J., {Trapped Radial Oscillations of Gaseous Disks around
  a Black Hole}. 1980, {\it \pasj}, {\bf 32}, 377

\bibitem[{{Kato} {et~al.}(2008){Kato}, {Fukue}, \& {Mineshige}}]{Kato2008}
{Kato}, S., {Fukue}, J., \& {Mineshige}, S. 2008, {\it {Black-Hole Accretion
  Disks: Towards a New Paradigm}} (Kyoto University Press)

\bibitem[{{Kato} \& {Machida}(2020)}]{2020PASJ...72...38K}
{Kato}, S. \& {Machida}, M., {A possible origin of kilohertz quasi-periodic
  oscillations in low-mass X-ray binaries}. 2020, {\it \pasj}, {\bf 72}, 38,
  DOI: 10.1093/pasj/psaa019

\bibitem[{{Kerr}(1963)}]{kerr:1963}
{Kerr}, R.~P., {Gravitational Field of a Spinning Mass as an Example of
  Algebraically Special Metrics}. 1963, {\it Phys.~Rev.~Lett}, {\bf 11}, 237,
  DOI: 10.1103/PhysRevLett.11.237

\bibitem[{{Kerr} \& {Schild}(2009)}]{Kerr2009}
{Kerr}, R.~P. \& {Schild}, A., {Republication of: A new class of vacuum
  solutions of the Einstein field equations}. 2009, {\it General Relativity and
  Gravitation}, {\bf 41}, 2485, DOI: 10.1007/s10714-009-0857-z

\bibitem[{{Klu{\'z}niak}(2008)}]{2008NewAR..51..841K}
{Klu{\'z}niak}, W., {Resonance model for high-frequency QPOs in white dwarfs,
  neutron stars and black holes}. 2008, {\it \nar}, {\bf 51}, 841, DOI:
  10.1016/j.newar.2008.03.014

\bibitem[{{Klu{\'z}niak} \& {Abramowicz}(2001)}]{KluzniakAbramowicz2001}
{Klu{\'z}niak}, W. \& {Abramowicz}, M.~A., {Strong-Field Gravity and Orbital
  Resonance in Black Holes and Neutron Stars --- kHz Quasi-Periodic
  Oscillations (QPO)}. 2001, {\it Acta Physica Polonica B}, {\bf 32}, 3605

\bibitem[{{Klu{\'z}niak} \& {Abramowicz}(2002)}]{2002astro.ph..3314K}
{Klu{\'z}niak}, W. \& {Abramowicz}, M.~A., {Parametric epicyclic resonance in
  black hole disks: QPOs in micro-quasars}. 2002, {\it arXiv e-prints}, astro,
  DOI: 10.48550/arXiv.astro-ph/0203314

\bibitem[{{Klu{\'z}niak} {et~al.}(2004){Klu{\'z}niak}, {Abramowicz}, {Kato},
  {Lee}, \& {Stergioulas}}]{2004ApJ...603L..89K}
{Klu{\'z}niak}, W., {Abramowicz}, M.~A., {Kato}, S., {Lee}, W.~H., \&
  {Stergioulas}, N., {Nonlinear Resonance in the Accretion Disk of a
  Millisecond Pulsar}. 2004, {\it \apjl}, {\bf 603}, L89, DOI: 10.1086/383143

\bibitem[{{Klu{\'z}niak} \& {Kita}(2000)}]{Kluzniak2000}
{Klu{\'z}niak}, W. \& {Kita}, D., {Three-dimensional structure of an alpha
  accretion disk}. 2000, {\it arXiv e-prints}, astro, DOI:
  10.48550/arXiv.astro-ph/0006266

\bibitem[{{Klu{\'z}niak} {et~al.}(1990){Klu{\'z}niak}, {Michelson}, \&
  {Wagoner}}]{1990ApJ...358..538K}
{Klu{\'z}niak}, W., {Michelson}, P., \& {Wagoner}, R.~V., {Determining the
  Properties of Accretion-Gap Neutron Stars}. 1990, {\it \apj}, {\bf 358}, 538,
  DOI: 10.1086/169006

\bibitem[{{Komissarov}(1999)}]{Komissarov1999}
{Komissarov}, S.~S., {A Godunov-type scheme for relativistic
  magnetohydrodynamics}. 1999, {\it \mnras}, {\bf 303}, 343, DOI:
  10.1046/j.1365-8711.1999.02244.x

\bibitem[{{Kotrlov{\'a}} {et~al.}(2020){Kotrlov{\'a}},
  {{\v{S}}r{\'a}mkov{\'a}}, {T{\"o}r{\"o}k}, {Goluchov{\'a}}, {Hor{\'a}k},
  {Straub}, {Lan{\v{c}}ov{\'a}}, {Stuchl{\'\i}k}, \&
  {Abramowicz}}]{Kotrlova2020}
{Kotrlov{\'a}}, A., {{\v{S}}r{\'a}mkov{\'a}}, E., {T{\"o}r{\"o}k}, G.,
  {et~al.}, {Models of high-frequency quasi-periodic oscillations and black
  hole spin estimates in Galactic microquasars}. 2020, {\it \aap}, {\bf 643},
  A31, DOI: 10.1051/0004-6361/201937097

\bibitem[{{Lan{\v{c}}ov{\'a}} {et~al.}(2019){Lan{\v{c}}ov{\'a}}, {Abarca},
  {Klu{\'z}niak}, {Wielgus}, {S{\k a}dowski}, {Narayan}, {Schee},
  {T{\"o}r{\"o}k}, \& {Abramowicz}}]{Lancova2019}
{Lan{\v{c}}ov{\'a}}, D., {Abarca}, D., {Klu{\'z}niak}, W., {et~al.}, {Puffy
  Accretion Disks: Sub-Eddington, Optically Thick, and Stable}. 2019, {\it ApJ
  Letters}, {\bf 884}, L37, DOI: 10.3847/2041-8213/ab48f5

\bibitem[{{Lan{\v{c}}ov{\'a}} {et~al.}(2023){Lan{\v{c}}ov{\'a}}, {Yilmaz},
  {Wielgus}, {Dov{\v{c}}iak}, {Straub}, \& {T{\"o}r{\"o}k}}]{Lancova2023}
{Lan{\v{c}}ov{\'a}}, D., {Yilmaz}, A., {Wielgus}, M., {et~al.}, Spectra of
  puffy accretion discs: the kynbb fit. 2023, {\it Astronomische Nachrichten},
  DOI: 10.1002/asna.20230023

\bibitem[{{Lee} {et~al.}(2004){Lee}, {Abramowicz}, \&
  {Klu{\'z}niak}}]{2004ApJ...603L..93L}
{Lee}, W.~H., {Abramowicz}, M.~A., \& {Klu{\'z}niak}, W., {Resonance in Forced
  Oscillations of an Accretion Disk and Kilohertz Quasi-periodic Oscillations}.
  2004, {\it \apjl}, {\bf 603}, L93, DOI: 10.1086/383245

\bibitem[{{Levermore}(1984)}]{Levermore1984}
{Levermore}, C., Relating Eddington factors to flux limiters. 1984, {\it
  Journal of Quantitative Spectroscopy and Radiative Transfer}, {\bf 31}, 149 ,
  DOI: https://doi.org/10.1016/0022-4073(84)90112-2

\bibitem[{{Li} {et~al.}(2005{\natexlab{a}}){Li}, {Zimmerman}, {Narayan}, \&
  {McClintock}}]{kerrbb}
{Li}, L.-X., {Zimmerman}, E.~R., {Narayan}, R., \& {McClintock}, J.~E.,
  {Multitemperature Blackbody Spectrum of a Thin Accretion Disk around a Kerr
  Black Hole: Model Computations and Comparison with Observations}.
  2005{\natexlab{a}}, {\it \apjs}, {\bf 157}, 335, DOI: 10.1086/428089

\bibitem[{{Li} {et~al.}(2005{\natexlab{b}}){Li}, {Zimmerman}, {Narayan}, \&
  {McClintock}}]{Li2005}
{Li}, L.-X., {Zimmerman}, E.~R., {Narayan}, R., \& {McClintock}, J.~E.,
  {Multitemperature Blackbody Spectrum of a Thin Accretion Disk around a Kerr
  Black Hole: Model Computations and Comparison with Observations}.
  2005{\natexlab{b}}, {\it \apjs}, {\bf 157}, 335, DOI: 10.1086/428089

\bibitem[{{Li} \& {Begelman}(2014)}]{Li2014}
{Li}, S.-L. \& {Begelman}, M.~C., {Thermal Stability of a Thin Disk with
  Magnetically Driven Winds}. 2014, {\it \apj}, {\bf 786}, 6, DOI:
  10.1088/0004-637X/786/1/6

\bibitem[{{Lightman} \& {Eardley}(1974)}]{Lightman1974}
{Lightman}, A.~P. \& {Eardley}, D.~M., {Black Holes in Binary Systems:
  Instability of Disk Accretion}. 1974, {\it The Astrophysical Journall}, {\bf
  187}, L1, DOI: 10.1086/181377

\bibitem[{{Lin} {et~al.}(2009){Lin}, {Remillard}, \&
  {Homan}}]{2009ApJ...696.1257L}
{Lin}, D., {Remillard}, R.~A., \& {Homan}, J., {Spectral States of XTE J1701 -
  462: Link Between Z and Atoll Sources}. 2009, {\it \apj}, {\bf 696}, 1257,
  DOI: 10.1088/0004-637X/696/2/1257

\bibitem[{{Liska} {et~al.}(2019){Liska}, {Tchekhovskoy}, {Ingram}, \& {van der
  Klis}}]{Liska2019}
{Liska}, M., {Tchekhovskoy}, A., {Ingram}, A., \& {van der Klis}, M.,
  {Bardeen-Petterson alignment, jets, and magnetic truncation in GRMHD
  simulations of tilted thin accretion discs}. 2019, {\it Monthly Notices of
  the Royal Astronomical Society}, {\bf 487}, 550, DOI: 10.1093/mnras/stz834

\bibitem[{{Liska} {et~al.}(2022){Liska}, {Chatterjee}, {Issa}, {Yoon}, {Kaaz},
  {Tchekhovskoy}, {van Eijnatten}, {Musoke}, {Hesp}, {Rohoza}, {Markoff},
  {Ingram}, \& {van der Klis}}]{HAMR}
{Liska}, M.~T.~P., {Chatterjee}, K., {Issa}, D., {et~al.}, {H-AMR: A New
  GPU-accelerated GRMHD Code for Exascale Computing with 3D Adaptive Mesh
  Refinement and Local Adaptive Time Stepping}. 2022, {\it \apjs}, {\bf 263},
  26, DOI: 10.3847/1538-4365/ac9966

\bibitem[{{L{\"o}ffler} {et~al.}(2012){L{\"o}ffler}, {Faber}, {Bentivegna},
  {Bode}, {Diener}, {Haas}, {Hinder}, {Mundim}, {Ott}, {Schnetter}, {Allen},
  {Campanelli}, \& {Laguna}}]{EinsteinToolkit}
{L{\"o}ffler}, F., {Faber}, J., {Bentivegna}, E., {et~al.}, {The Einstein
  Toolkit: a community computational infrastructure for relativistic
  astrophysics}. 2012, {\it Classical and Quantum Gravity}, {\bf 29}, 115001,
  DOI: 10.1088/0264-9381/29/11/115001

\bibitem[{{Lynden-Bell}(1969)}]{Lynden-Bell1969}
{Lynden-Bell}, D., {Galactic Nuclei as Collapsed Old Quasars}. 1969, {\it
  Nature}, {\bf 223}, 690, DOI: 10.1038/223690a0

\bibitem[{{Matuszkov{\'a}} {et~al.}(2022){Matuszkov{\'a}},
  {Klimovi{\v{c}}ov{\'a}}, {Urbancov{\'a}}, {Lan{\v{c}}ov{\'a}},
  {{\v{S}}r{\'a}mkov{\'a}}, \& {T{\"o}r{\"o}k}}]{2022arXiv220310653M}
{Matuszkov{\'a}}, M., {Klimovi{\v{c}}ov{\'a}}, K., {Urbancov{\'a}}, G.,
  {et~al.}, {Oscillations of non-slender tori in the external Hartle-Thorne
  geometry}. 2022, {\it arXiv e-prints}, arXiv:2203.10653, DOI:
  10.48550/arXiv.2203.10653

\bibitem[{{Matuszkov\'a} {et~al.}(in preparation){Matuszkov\'a},
  {T{\"o}r{\"o}k}, {Klimovi{\v{c}}ov{\'a}}, Ji\v{r}\'{\i}, Odele,
  {\v{S}r\'{a}mkov\'{a}}, {Urbancov\'a}, {Lan{\v{c}}ov{\'a}}, {Urbanec}, \&
  {Karas}}]{Matuszkova}
{Matuszkov\'a}, M., {T{\"o}r{\"o}k}, G., {Klimovi{\v{c}}ov{\'a}}, K., {et~al.}
  in preparation, in preparation

\bibitem[{{McClintock} {et~al.}(2014){McClintock}, {Narayan}, \&
  {Steiner}}]{McClintock2014}
{McClintock}, J.~E., {Narayan}, R., \& {Steiner}, J.~F., {Black Hole Spin via
  Continuum Fitting and the Role of Spin in Powering Transient Jets}. 2014,
  {\it \ssr}, {\bf 183}, 295, DOI: 10.1007/s11214-013-0003-9

\bibitem[{{McKinney} \& {Gammie}(2004)}]{McKinney2004}
{McKinney}, J.~C. \& {Gammie}, C.~F., {A Measurement of the Electromagnetic
  Luminosity of a Kerr Black Hole}. 2004, {\it \apj}, {\bf 611}, 977, DOI:
  10.1086/422244

\bibitem[{{McKinney} {et~al.}(2014){McKinney}, {Tchekhovskoy}, {Sadowski}, \&
  {Narayan}}]{McKinney2014}
{McKinney}, J.~C., {Tchekhovskoy}, A., {Sadowski}, A., \& {Narayan}, R.,
  {Three-dimensional general relativistic radiation magnetohydrodynamical
  simulation of super-Eddington accretion, using a new code HARMRAD with M1
  closure}. 2014, {\it \mnras}, {\bf 441}, 3177, DOI: 10.1093/mnras/stu762

\bibitem[{{M{\'e}ndez} \& {Belloni}(2021)}]{2021ASSL..461..263M}
{M{\'e}ndez}, M. \& {Belloni}, T.~M., {High-Frequency Variability in
  Neutron-Star Low-Mass X-ray Binaries}. 2021, in Astrophysics and Space
  Science Library, Vol. {\bf  461}, {\it Timing Neutron Stars: Pulsations,
  Oscillations and Explosions}, ed. T.~M. {Belloni}, M.~{M{\'e}ndez}, \&
  C.~{Zhang}, 263--331

\bibitem[{{M{\'e}ndez} {et~al.}(2022){M{\'e}ndez}, {Karpouzas}, {Garc{\'\i}a},
  {Zhang}, {Zhang}, {Belloni}, \& {Altamirano}}]{vsechnosecsim}
{M{\'e}ndez}, M., {Karpouzas}, K., {Garc{\'\i}a}, F., {et~al.}, {Coupling
  between the accreting corona and the relativistic jet in the microquasar GRS
  1915+105}. 2022, {\it Nature Astronomy}, {\bf 6}, 577, DOI:
  10.1038/s41550-022-01617-y

\bibitem[{{Middleditch} \& {Priedhorsky}(1986)}]{1986ApJ...306..230M}
{Middleditch}, J. \& {Priedhorsky}, W.~C., {Discovery of Rapid Quasi-periodic
  Oscillations in Scorpius X-1}. 1986, {\it \apj}, {\bf 306}, 230, DOI:
  10.1086/164335

\bibitem[{{Mignone} {et~al.}(2007){Mignone}, {Bodo}, {Massaglia}, {Matsakos},
  {Tesileanu}, {Zanni}, \& {Ferrari}}]{Mignone2007}
{Mignone}, A., {Bodo}, G., {Massaglia}, S., {et~al.}, {PLUTO: A Numerical Code
  for Computational Astrophysics}. 2007, {\it \apjs}, {\bf 170}, 228, DOI:
  10.1086/513316

\bibitem[{{Mihalas} \& {Mihalas}(1984)}]{MihalasMihalas}
{Mihalas}, D. \& {Mihalas}, B.~W. 1984, {\it {Foundations of radiation
  hydrodynamics}} (Dover Books on Physics)

\bibitem[{{Miller} {et~al.}(1998){Miller}, {Lamb}, \&
  {Cook}}]{1998ApJ...509..793M}
{Miller}, M.~C., {Lamb}, F.~K., \& {Cook}, G.~B., {Effects of Rapid Stellar
  Rotation on Equation-of-State Constraints Derived from Quasi-periodic
  Brightness Oscillations}. 1998, {\it \apj}, {\bf 509}, 793, DOI:
  10.1086/306533

\bibitem[{{Mirabel} {et~al.}(1992){Mirabel}, {Rodriguez}, {Cordier}, {Paul}, \&
  {Lebrun}}]{MIrabel1992}
{Mirabel}, I.~F., {Rodriguez}, L.~F., {Cordier}, B., {Paul}, J., \& {Lebrun},
  F., {A double-sided radio jet from the compact Galactic Centre annihilator
  1E1740.7-2942}. 1992, {\it Nature}, {\bf 358}, 215, DOI: 10.1038/358215a0

\bibitem[{{Mishra} {et~al.}(2022){Mishra}, {Fragile}, {Anderson},
  {Blankenship}, {Li}, \& {Nalewajko}}]{Mishra2022}
{Mishra}, B., {Fragile}, P.~C., {Anderson}, J., {et~al.}, {The Role of Strong
  Magnetic Fields in Stabilizing Highly Luminous Thin Disks}. 2022, {\it The
  Astrophysical Journal}, {\bf 939}, 31, DOI: 10.3847/1538-4357/ac938b

\bibitem[{{Mishra} {et~al.}(2016){Mishra}, {Fragile}, {Johnson}, \&
  {Klu{\'z}niak}}]{Mishra2016}
{Mishra}, B., {Fragile}, P.~C., {Johnson}, L.~C., \& {Klu{\'z}niak}, W.,
  {Three-dimensional, global, radiative GRMHD simulations of a thermally
  unstable disc}. 2016, {\it MNRAS}, {\bf 463}, 3437, DOI:
  10.1093/mnras/stw2245

\bibitem[{{Mishra} {et~al.}(2019){Mishra}, {Klu{\'z}niak}, \&
  {Fragile}}]{2019MNRAS.483.4811M}
{Mishra}, B., {Klu{\'z}niak}, W., \& {Fragile}, P.~C., {Breathing Oscillations
  in a Global Simulation of a Thin Accretion Disk}. 2019, {\it \mnras}, {\bf
  483}, 4811, DOI: 10.1093/mnras/sty3124

\bibitem[{{Mishra} {et~al.}(2020){Mishra}, {Klu{\'z}niak}, \&
  {Fragile}}]{Mishra2019}
{Mishra}, B., {Klu{\'z}niak}, W., \& {Fragile}, P.~C., {Relativistic,
  axisymmetric, viscous, radiation hydrodynamic simulations of geometrically
  thin discs. II. Disc variability}. 2020, {\it Monthly Notices of the Royal
  Astronomical Society}, {\bf 497}, 1066, DOI: 10.1093/mnras/staa1848

\bibitem[{{Mishra} {et~al.}(2017){Mishra}, {Vincent}, {Manousakis}, {Fragile},
  {Paumard}, \& {Klu{\'z}niak}}]{2017MNRAS.467.4036M}
{Mishra}, B., {Vincent}, F.~H., {Manousakis}, A., {et~al.}, {Quasi-periodic
  oscillations from relativistic ray-traced hydrodynamical tori}. 2017, {\it
  \mnras}, {\bf 467}, 4036, DOI: 10.1093/mnras/stx299

\bibitem[{{Misner} {et~al.}(1973){Misner}, {Thorne}, \&
  {Wheeler}}]{1973grav.book.....M}
{Misner}, C.~W., {Thorne}, K.~S., \& {Wheeler}, J.~A. 1973, {\it {Gravitation}}
  (Princeton University Press)

\bibitem[{{Miyamoto} {et~al.}(1995){Miyamoto}, {Kitamoto}, {Hayashida}, \&
  {Egoshi}}]{1995ApJ...442L..13M}
{Miyamoto}, S., {Kitamoto}, S., {Hayashida}, K., \& {Egoshi}, W., {Large
  Hysteretic Behavior of Stellar Black Hole Candidate X-Ray Binaries}. 1995,
  {\it \apjl}, {\bf 442}, L13, DOI: 10.1086/187804

\bibitem[{{Motta} {et~al.}(2022){Motta}, {Belloni}, {Stella}, {Pappas},
  {Casares}, {Mu{\~n}oz-Darias}, {Torres}, \& {Yanes-Rizo}}]{Motta2022}
{Motta}, S.~E., {Belloni}, T., {Stella}, L., {et~al.}, {Black hole mass and
  spin measurements through the relativistic precession model: XTE J1859+226}.
  2022, {\it \mnras}, {\bf 517}, 1469, DOI: 10.1093/mnras/stac2142

\bibitem[{{Motta} {et~al.}(2021){Motta}, {Kajava}, {Giustini}, {Williams}, {Del
  Santo}, {Fender}, {Green}, {Heywood}, {Rhodes}, {Segreto}, {Sivakoff}, \&
  {Woudt}}]{2021MNRAS.503..152M}
{Motta}, S.~E., {Kajava}, J.~J.~E., {Giustini}, M., {et~al.}, {Observations of
  a radio-bright, X-ray obscured GRS 1915+105}. 2021, {\it \mnras}, {\bf 503},
  152, DOI: 10.1093/mnras/stab511

\bibitem[{{Motta} {et~al.}(2014){Motta}, {Munoz-Darias}, {Sanna}, {Fender},
  {Belloni}, \& {Stella}}]{Motta2014}
{Motta}, S.~E., {Munoz-Darias}, T., {Sanna}, A., {et~al.}, {Black hole spin
  measurements through the relativistic precession model: XTE J1550-564.} 2014,
  {\it \mnras}, {\bf 439}, L65, DOI: 10.1093/mnrasl/slt181

\bibitem[{{Mu{\~n}oz-Darias} {et~al.}(2014){Mu{\~n}oz-Darias}, {Fender},
  {Motta}, \& {Belloni}}]{2014MNRAS.443.3270M}
{Mu{\~n}oz-Darias}, T., {Fender}, R.~P., {Motta}, S.~E., \& {Belloni}, T.~M.,
  {Black hole-like hysteresis and accretion states in neutron star low-mass
  X-ray binaries}. 2014, {\it \mnras}, {\bf 443}, 3270, DOI:
  10.1093/mnras/stu1334

\bibitem[{{Narayan} \& {McClintock}(2008)}]{NarayanMcClintock2008}
{Narayan}, R. \& {McClintock}, J.~E., {Advection-dominated accretion and the
  black hole event horizon}. 2008, {\it \nar}, {\bf 51}, 733, DOI:
  10.1016/j.newar.2008.03.002

\bibitem[{{Narayan} {et~al.}(2016){Narayan}, {Zhu}, {Psaltis}, \&
  {S\k{a}dowski}}]{2016MNRAS.457..608N}
{Narayan}, R., {Zhu}, Y., {Psaltis}, D., \& {S\k{a}dowski}, A., {HEROIC: 3D
  general relativistic radiative post-processor with comptonization for black
  hole accretion discs}. 2016, {\it \mnras}, {\bf 457}, 608, DOI:
  10.1093/mnras/stv2979

\bibitem[{{Neumann} {et~al.}(2023){Neumann}, {Avakyan}, {Doroshenko}, \&
  {Santangelo}}]{Neumann2023}
{Neumann}, M., {Avakyan}, A., {Doroshenko}, V., \& {Santangelo}, A., {XRBcats:
  Galactic High Mass X-ray Binary Catalogue}. 2023, {\it arXiv e-prints},
  arXiv:2303.16137, DOI: 10.48550/arXiv.2303.16137

\bibitem[{{Noble} {et~al.}(2006){Noble}, {Gammie}, {McKinney}, \& {Del
  Zanna}}]{Noble2006}
{Noble}, S.~C., {Gammie}, C.~F., {McKinney}, J.~C., \& {Del Zanna}, L.,
  {Primitive Variable Solvers for Conservative General Relativistic
  Magnetohydrodynamics}. 2006, {\it \apj}, {\bf 641}, 626, DOI: 10.1086/500349

\bibitem[{{Novikov} \& {Thorne}(1973)}]{Novikov+Thorne1973}
{Novikov}, I.~D. \& {Thorne}, K.~S., {Astrophysics of black holes.} 1973, in
  {\it Black Holes (Les Astres Occlus)}, 343--450

\bibitem[{{Nowak} \& {Wagoner}(1991)}]{1991ApJ...378..656N}
{Nowak}, M.~A. \& {Wagoner}, R.~V., {Diskoseismology: Probing Accretion Disks.
  I. Trapped Adiabatic Oscillations}. 1991, {\it \apj}, {\bf 378}, 656, DOI:
  10.1086/170465

\bibitem[{{Nowak} \& {Wagoner}(1992)}]{1992ApJ...393..697N}
{Nowak}, M.~A. \& {Wagoner}, R.~V., {Diskoseismology: Probing Accretion Disks.
  II. G-Modes, Gravitational Radiation Reaction, and Viscosity}. 1992, {\it
  \apj}, {\bf 393}, 697, DOI: 10.1086/171538

\bibitem[{{Oda} {et~al.}(2009){Oda}, {Machida}, {Nakamura}, \&
  {Matsumoto}}]{Oda2009}
{Oda}, H., {Machida}, M., {Nakamura}, K.~E., \& {Matsumoto}, R., {Thermal
  Equilibria of Magnetically Supported Black Hole Accretion Disks}. 2009, {\it
  \apj}, {\bf 697}, 16, DOI: 10.1088/0004-637X/697/1/16

\bibitem[{{Paczynski}(1987)}]{1987Natur.327..303P}
{Paczynski}, B., {Possible relation between the X-ray QPO phenomenon and
  general relativity}. 1987, {\it \nat}, {\bf 327}, 303, DOI: 10.1038/327303a0

\bibitem[{{Page} \& {Thorne}(1974)}]{Page1974}
{Page}, D.~N. \& {Thorne}, K.~S., {Disk-Accretion onto a Black Hole.
  Time-Averaged Structure of Accretion Disk}. 1974, {\it \apj}, {\bf 191}, 499,
  DOI: 10.1086/152990

\bibitem[{{Papaloizou} \& {Pringle}(1984)}]{1984MNRAS.208..721P}
{Papaloizou}, J.~C.~B. \& {Pringle}, J.~E., {The dynamical stability of
  differentially rotating discs with constant specific angular momentum}. 1984,
  {\it \mnras}, {\bf 208}, 721, DOI: 10.1093/mnras/208.4.721

\bibitem[{{Pareschi} {et~al.}(2005){Pareschi}, {Puppo}, \&
  {Russo}}]{Pareschi2005}
{Pareschi}, L., {Puppo}, G., \& {Russo}, G., {Central Runge-Kutta Schemes for
  Conservation Laws}. 2005, {\it SIAM Journal on Scientific Computing}, {\bf
  26}, 979, DOI: 10.1137/S1064827503420696

\bibitem[{{Parker}(1958)}]{Parker1958}
{Parker}, E.~N., {Dynamics of the Interplanetary Gas and Magnetic Fields.}
  1958, {\it \apj}, {\bf 128}, 664, DOI: 10.1086/146579

\bibitem[{{Parthasarathy} {et~al.}(2017){Parthasarathy}, {Klu{\'z}niak}, \&
  {{\v{C}}emelji{\'c}}}]{2017MNRAS.470L..34P}
{Parthasarathy}, V., {Klu{\'z}niak}, W., \& {{\v{C}}emelji{\'c}}, M., {MHD
  simulations of oscillating cusp-filling tori around neutron stars - missing
  upper kHz QPO}. 2017, {\it \mnras}, {\bf 470}, L34, DOI:
  10.1093/mnrasl/slx070

\bibitem[{{Paugnat} {et~al.}(2022){Paugnat}, {Lupsasca}, {Vincent}, \&
  {Wielgus}}]{Paugnat2022}
{Paugnat}, H., {Lupsasca}, A., {Vincent}, F.~H., \& {Wielgus}, M., {Photon ring
  test of the Kerr hypothesis: Variation in the ring shape}. 2022, {\it \aap},
  {\bf 668}, A11, DOI: 10.1051/0004-6361/202244216

\bibitem[{{Penna} {et~al.}(2013{\natexlab{a}}){Penna}, {Kulkarni}, \&
  {Narayan}}]{Penna2013_torus}
{Penna}, R.~F., {Kulkarni}, A., \& {Narayan}, R., {A new equilibrium torus
  solution and GRMHD initial conditions}. 2013{\natexlab{a}}, {\it \aap}, {\bf
  559}, A116, DOI: 10.1051/0004-6361/201219666

\bibitem[{{Penna} {et~al.}(2010){Penna}, {McKinney}, {Narayan}, {Tchekhovskoy},
  {Shafee}, \& {McClintock}}]{2010MNRAS.408..752P}
{Penna}, R.~F., {McKinney}, J.~C., {Narayan}, R., {et~al.}, {Simulations of
  magnetized discs around black holes: effects of black hole spin, disc
  thickness and magnetic field geometry}. 2010, {\it \mnras}, {\bf 408}, 752,
  DOI: 10.1111/j.1365-2966.2010.17170.x

\bibitem[{{Penna} {et~al.}(2013{\natexlab{b}}){Penna}, {S{\k{a}}dowski},
  {Kulkarni}, \& {Narayan}}]{2013MNRAS.428.2255P}
{Penna}, R.~F., {S{\k{a}}dowski}, A., {Kulkarni}, A.~K., \& {Narayan}, R., {The
  Shakura-Sunyaev viscosity prescription with variable {\ensuremath{\alpha}}
  (r)}. 2013{\natexlab{b}}, {\it \mnras}, {\bf 428}, 2255, DOI:
  10.1093/mnras/sts185

\bibitem[{{Penna} {et~al.}(2013{\natexlab{c}}){Penna}, {S{\k{a}}dowski},
  {Kulkarni}, \& {Narayan}}]{Penna2013}
{Penna}, R.~F., {S{\k{a}}dowski}, A., {Kulkarni}, A.~K., \& {Narayan}, R., {The
  Shakura-Sunyaev viscosity prescription with variable {\ensuremath{\alpha}}
  (r)}. 2013{\natexlab{c}}, {\it MNRAS}, {\bf 428}, 2255, DOI:
  10.1093/mnras/sts185

\bibitem[{{Penna} {et~al.}(2012){Penna}, {S{\k{a}}dowski}, \&
  {McKinney}}]{2012MNRAS.420..684P}
{Penna}, R.~F., {S{\k{a}}dowski}, A., \& {McKinney}, J.~C., {Thin-disc theory
  with a non-zero-torque boundary condition and comparisons with simulations}.
  2012, {\it \mnras}, {\bf 420}, 684, DOI: 10.1111/j.1365-2966.2011.20084.x

\bibitem[{{Penrose} \& {Floyd}(1971)}]{Penrose1971}
{Penrose}, R. \& {Floyd}, R.~M., {Extraction of Rotational Energy from a Black
  Hole}. 1971, {\it Nature Physical Science}, {\bf 229}, 177, DOI:
  10.1038/physci229177a0

\bibitem[{{P{\'e}tri}(2005)}]{2005A&A...443..777P}
{P{\'e}tri}, J., {A toy model for coupling accretion disk oscillations to the
  neutron star spin}. 2005, {\it \aap}, {\bf 443}, 777, DOI:
  10.1051/0004-6361:20054119

\bibitem[{{Podgorny} {et~al.}(2023){Podgorny}, {Marra}, {Muleri}, {Rodriguez
  Cavero}, {Ratheesh}, {Dovciak}, {Mikusincova}, {Brigitte}, {Steiner},
  {Veledina}, {Bianchi}, {Krawczynski}, {Svoboda}, {Kaaret}, {Matt}, {Garcia},
  {Petrucci}, {Lutovinov}, {Semena}, {Di Marco}, {Negro}, {Weisskopf},
  {Ingram}, {Poutanen}, {Beheshtipour}, {Chun}, {Hu}, {Mizuno}, {Sixuan},
  {Tombesi}, {Zane}, {Agudo}, {Antonelli}, {Bachetti}, {Baldini},
  {Baumgartner}, {Bellazzini}, {Bongiorno}, {Bonino}, {Brez}, {Bucciantini},
  {Capitanio}, {Castellano}, {Cavazzuti}, {Chen}, {Ciprini}, {Costa}, {De
  Rosa}, {Del Monte}, {Di Gesu}, {Di Lalla}, {Donnarumma}, {Doroshenko},
  {Ehlert}, {Enoto}, {Evangelista}, {Fabiani}, {Ferrazzoli}, {Gunji},
  {Hayashida}, {Heyl}, {Iwakiri}, {Jorstad}, {Karas}, {Kislat}, {Kitaguchi},
  {Kolodziejczak}, {La Monaca}, {Latronico}, {Liodakis}, {Maldera}, {Manfreda},
  {Marin}, {Marinucci}, {Marscher}, {Marshall}, {Massaro}, {Mitsuishi}, {Ng},
  {O'Dell}, {Omodei}, {Oppedisano}, {Papitto}, {Pavlov}, {Peirson}, {Perri},
  {Pesce-Rollins}, {Pilia}, {Possenti}, {Puccetti}, {Ramsey}, {Rankin},
  {Roberts}, {Romani}, {Sgro}, {Slane}, {Soffitta}, {Spandre}, {Swartz},
  {Tamagawa}, {Tavecchio}, {Taverna}, {Tawara}, {Tennant}, {Thomas}, {Trois},
  {Tsygankov}, {Turolla}, {Vink}, {Wu}, \& {Xie}}]{Podgorny2023}
{Podgorny}, J., {Marra}, L., {Muleri}, F., {et~al.}, {The first X-ray
  polarimetric observation of the black hole binary LMC X-1}. 2023, {\it arXiv
  e-prints}, arXiv:2303.12034, DOI: 10.48550/arXiv.2303.12034

\bibitem[{{Porth} {et~al.}(2019){Porth}, {Chatterjee}, {Narayan}, {Gammie},
  {Mizuno}, {Anninos}, {Baker}, {Bugli}, {Chan}, {Davelaar}, {Del Zanna},
  {Etienne}, {Fragile}, {Kelly}, {Liska}, {Markoff}, {McKinney}, {Mishra},
  {Noble}, {Olivares}, {Prather}, {Rezzolla}, {Ryan}, {Stone}, {Tomei},
  {White}, {Younsi}, {Akiyama}, {Alberdi}, {Alef}, {Asada}, {Azulay}, {Baczko},
  {Ball}, {Balokovi{\'c}}, {Barrett}, {Bintley}, {Blackburn}, {Boland},
  {Bouman}, {Bower}, {Bremer}, {Brinkerink}, {Brissenden}, {Britzen},
  {Broderick}, {Broguiere}, {Bronzwaer}, {Byun}, {Carlstrom}, {Chael},
  {Chatterjee}, {Chen}, {Chen}, {Cho}, {Christian}, {Conway}, {Cordes},
  {Geoffrey}, {Crew}, {Cui}, {De Laurentis}, {Deane}, {Dempsey}, {Desvignes},
  {Doeleman}, {Eatough}, {Falcke}, {Fish}, {Fomalont}, {Fraga-Encinas},
  {Freeman}, {Friberg}, {Fromm}, {G{\'o}mez}, {Galison}, {Garc{\'\i}a},
  {Gentaz}, {Georgiev}, {Goddi}, {Gold}, {Gu}, {Gurwell}, {Hada}, {Hecht},
  {Hesper}, {Ho}, {Ho}, {Honma}, {Huang}, {Huang}, {Hughes}, {Ikeda}, {Inoue},
  {Issaoun}, {James}, {Jannuzi}, {Janssen}, {Jeter}, {Jiang}, {Johnson},
  {Jorstad}, {Jung}, {Karami}, {Karuppusamy}, {Kawashima}, {Keating},
  {Kettenis}, {Kim}, {Kim}, {Kim}, {Kino}, {Koay}, {Patrick}, {Koch}, {Koyama},
  {Kramer}, {Kramer}, {Krichbaum}, {Kuo}, {Lauer}, {Lee}, {Li}, {Li},
  {Lindqvist}, {Liu}, {Liuzzo}, {Lo}, {Lobanov}, {Loinard}, {Lonsdale}, {Lu},
  {MacDonald}, {Mao}, {Marrone}, {Marscher}, {Mart{\'\i}-Vidal}, {Matsushita},
  {Matthews}, {Medeiros}, {Menten}, {Mizuno}, {Moran}, {Moriyama},
  {Moscibrodzka}, {M{\"u}ller}, {Nagai}, {Nagar}, {Nakamura}, {Narayanan},
  {Natarajan}, {Neri}, {Ni}, {Noutsos}, {Okino}, {Oyama}, {{\"O}zel},
  {Palumbo}, {Patel}, {Pen}, {Pesce}, {Pi{\'e}tu}, {Plambeck}, {PopStefanija},
  {Preciado-L{\'o}pez}, {Psaltis}, {Pu}, {Ramakrishnan}, {Rao}, {Rawlings},
  {Raymond}, {Ripperda}, {Roelofs}, {Rogers}, {Ros}, {Rose}, {Roshanineshat},
  {Rottmann}, {Roy}, {Ruszczyk}, {Rygl}, {S{\'a}nchez},
  {S{\'a}nchez-Arguelles}, {Sasada}, {Savolainen}, {Schloerb}, {Schuster},
  {Shao}, {Shen}, {Small}, {Sohn}, {SooHoo}, {Tazaki}, {Tiede}, {Tilanus},
  {Titus}, {Toma}, {Torne}, {Trent}, {Trippe}, {Tsuda}, {van Bemmel}, {van
  Langevelde}, {van Rossum}, {Wagner}, {Wardle}, {Weintroub}, {Wex}, {Wharton},
  {Wielgus}, {Wong}, {Wu}, {Young}, {Young}, {Yuan}, {Yuan}, {Zensus}, {Zhao},
  {Zhao}, {Zhu}, \& {Event Horizon Telescope Collaboration}}]{Porth2019}
{Porth}, O., {Chatterjee}, K., {Narayan}, R., {et~al.}, {The Event Horizon
  General Relativistic Magnetohydrodynamic Code Comparison Project}. 2019, {\it
  \apjs}, {\bf 243}, 26, DOI: 10.3847/1538-4365/ab29fd

\bibitem[{{Porth} {et~al.}(2017){Porth}, {Olivares}, {Mizuno}, {Younsi},
  {Rezzolla}, {Moscibrodzka}, {Falcke}, \& {Kramer}}]{BHAC}
{Porth}, O., {Olivares}, H., {Mizuno}, Y., {et~al.}, {The black hole accretion
  code}. 2017, {\it Computational Astrophysics and Cosmology}, {\bf 4}, 1, DOI:
  10.1186/s40668-017-0020-2

\bibitem[{{Prather} {et~al.}(2023){Prather}, {Dexter}, {Moscibrodzka}, {Pu},
  {Bronzwaer}, {Davelaar}, {Younsi}, {Gammie}, {Gold}, {Wong}, {Akiyama},
  {Alberdi}, {Alef}, {Algaba}, {Anantua}, {Asada}, {Azulay}, {Bach}, {Baczko},
  {Ball}, {Balokovi{\'c}}, {Barrett}, {Baub{\"o}ck}, {Benson}, {Bintley},
  {Blackburn}, {Blundell}, {Bouman}, {Bower}, {Boyce}, {Bremer}, {Brinkerink},
  {Brissenden}, {Britzen}, {Broderick}, {Broguiere}, {Bustamante}, {Byun},
  {Carlstrom}, {Ceccobello}, {Chael}, {Chan}, {Chang}, {Chatterjee},
  {Chatterjee}, {Chen}, {Chen}, {Cheng}, {Cho}, {Christian}, {Conroy},
  {Conway}, {Cordes}, {Crawford}, {Crew}, {Cruz-Osorio}, {Cui}, {De Laurentis},
  {Deane}, {Dempsey}, {Desvignes}, {Dhruv}, {Doeleman}, {Dougal}, {Dzib},
  {Eatough}, {Emami}, {Falcke}, {Farah}, {Fish}, {Fomalont}, {Ford},
  {Fraga-Encinas}, {Freeman}, {Friberg}, {Fromm}, {Fuentes}, {Galison},
  {Garc{\'\i}a}, {Gentaz}, {Georgiev}, {Goddi}, {G{\'o}mez-Ruiz}, {G{\'o}mez},
  {Gu}, {Gurwell}, {Hada}, {Haggard}, {Haworth}, {Hecht}, {Hesper}, {Heumann},
  {Ho}, {Ho}, {Honma}, {Huang}, {Huang}, {Hughes}, {Ikeda}, {Impellizzeri},
  {Inoue}, {Issaoun}, {James}, {Jannuzi}, {Janssen}, {Jeter}, {Jiang},
  {Jim{\'e}nez-Rosales}, {Johnson}, {Jorstad}, {Joshi}, {Jung}, {Karami},
  {Karuppusamy}, {Kawashima}, {Keating}, {Kettenis}, {Kim}, {Kim}, {Kim},
  {Kim}, {Kino}, {Koay}, {Kocherlakota}, {Kofuji}, {Koyama}, {Kramer},
  {Kramer}, {Krichbaum}, {Kuo}, {La Bella}, {Lauer}, {Lee}, {Lee}, {Leung},
  {Levis}, {Li}, {Lico}, {Lindahl}, {Lindqvist}, {Lisakov}, {Liu}, {Liu},
  {Liuzzo}, {Lo}, {Lobanov}, {Loinard}, {Lonsdale}, {Lu}, {MacDonald}, {Mao},
  {Marchili}, {Markoff}, {Marrone}, {Marscher}, {Mart{\'\i}-Vidal},
  {Matsushita}, {Matthews}, {Medeiros}, {Menten}, {Michalik}, {Mizuno},
  {Mizuno}, {Moran}, {Moriyama}, {M{\"u}ller}, {Mus}, {Musoke}, {Myserlis},
  {Nadolski}, {Nagai}, {Nagar}, {Nakamura}, {Narayan}, {Narayanan},
  {Natarajan}, {Nathanail}, {Fuentes}, {Neilsen}, {Neri}, {Ni}, {Noutsos},
  {Nowak}, {Oh}, {Okino}, {Olivares}, {Ortiz-Le{\'o}n}, {Oyama}, {{\"O}zel},
  {Palumbo}, {Paraschos}, {Park}, {Parsons}, {Patel}, {Pen}, {Pesce},
  {Pi{\'e}tu}, {Plambeck}, {PopStefanija}, {Porth}, {P{\"o}tzl},
  {Preciado-L{\'o}pez}, {Psaltis}, {Ramakrishnan}, {Rao}, {Rawlings},
  {Raymond}, {Rezzolla}, {Ricarte}, {Ripperda}, {Roelofs}, {Rogers}, {Ros},
  {Romero-Ca{\~n}izales}, {Roshanineshat}, {Rottmann}, {Roy}, {Ruiz},
  {Ruszczyk}, {Rygl}, {S{\'a}nchez}, {S{\'a}nchez-Arg{\"u}elles},
  {S{\'a}nchez-Portal}, {Sasada}, {Satapathy}, {Savolainen}, {Schloerb},
  {Schonfeld}, {Schuster}, {Shao}, {Shen}, {Small}, {Sohn}, {SooHoo},
  {Souccar}, {Sun}, {Tazaki}, {Tetarenko}, {Tiede}, {Tilanus}, {Titus},
  {Torne}, {Traianou}, {Trent}, {Trippe}, {Turk}, {van Bemmel}, {van
  Langevelde}, {van Rossum}, {Vos}, {Wagner}, {Ward-Thompson}, {Wardle},
  {Weintroub}, {Wex}, {Wharton}, {Wielgus}, {Wiik}, {Witzel}, {Wondrak}, {Wu},
  {Yamaguchi}, {Yfantis}, {Yoon}, {Young}, {Young}, {Yu}, {Yuan}, {Yuan},
  {Zensus}, {Zhang}, {Zhao}, {Zhao}, \& {Event Horizon Telescope
  Collaboration}}]{2023ApJ...950...35P}
{Prather}, B.~S., {Dexter}, J., {Moscibrodzka}, M., {et~al.}, {Comparison of
  Polarized Radiative Transfer Codes Used by the EHT Collaboration}. 2023, {\it
  \apj}, {\bf 950}, 35, DOI: 10.3847/1538-4357/acc586

\bibitem[{{Remillard} \& {McClintock}(2006)}]{Remillard2006}
{Remillard}, R.~A. \& {McClintock}, J.~E., {X-Ray Properties of Black-Hole
  Binaries}. 2006, {\it \araa}, {\bf 44}, 49, DOI:
  10.1146/annurev.astro.44.051905.092532

\bibitem[{{Ressler} {et~al.}(2017){Ressler}, {Tchekhovskoy}, {Quataert}, \&
  {Gammie}}]{Ressler2017}
{Ressler}, S.~M., {Tchekhovskoy}, A., {Quataert}, E., \& {Gammie}, C.~F., {The
  disc-jet symbiosis emerges: modelling the emission of Sagittarius A* with
  electron thermodynamics}. 2017, {\it \mnras}, {\bf 467}, 3604, DOI:
  10.1093/mnras/stx364

\bibitem[{{Revnivtsev} {et~al.}(2001){Revnivtsev}, {Churazov}, {Gilfanov}, \&
  {Sunyaev}}]{Revnivtsev2001}
{Revnivtsev}, M., {Churazov}, E., {Gilfanov}, M., \& {Sunyaev}, R., {New class
  of low frequency QPOs: Signature of nuclear burning or accretion disk
  instabilities?} 2001, {\it \aap}, {\bf 372}, 138, DOI:
  10.1051/0004-6361:20010434

\bibitem[{{Reynolds}(2021)}]{Reynolds2021}
{Reynolds}, C.~S., {Observational Constraints on Black Hole Spin}. 2021, {\it
  \araa}, {\bf 59}, 117, DOI: 10.1146/annurev-astro-112420-035022

\bibitem[{{Rezzolla} {et~al.}(2003){Rezzolla}, {Yoshida}, \&
  {Zanotti}}]{2003MNRAS.344..978R}
{Rezzolla}, L., {Yoshida}, S., \& {Zanotti}, O., {Oscillations of vertically
  integrated relativistic tori - I. Axisymmetric modes in a Schwarzschild
  space-time}. 2003, {\it \mnras}, {\bf 344}, 978, DOI:
  10.1046/j.1365-8711.2003.07023.x

\bibitem[{{Rhoades} \& {Ruffini}(1974)}]{Rhoades1974}
{Rhoades}, C.~E. \& {Ruffini}, R., {Maximum Mass of a Neutron Star}. 1974, {\it
  \prl}, {\bf 32}, 324, DOI: 10.1103/PhysRevLett.32.324

\bibitem[{{R{\'o}{\.z}a{\'n}ska} {et~al.}(1999){R{\'o}{\.z}a{\'n}ska},
  {Czerny}, {{\.Z}ycki}, \& {Pojma{\'n}ski}}]{Rozanska1999}
{R{\'o}{\.z}a{\'n}ska}, A., {Czerny}, B., {{\.Z}ycki}, P.~T., \&
  {Pojma{\'n}ski}, G., {Vertical structure of accretion discs with hot coronae
  in active galactic nuclei}. 1999, {\it \mnras}, {\bf 305}, 481, DOI:
  10.1046/j.1365-8711.1999.02425.x

\bibitem[{{R{\'o}{\.z}a{\'n}ska} {et~al.}(2015){R{\'o}{\.z}a{\'n}ska},
  {Malzac}, {Belmont}, {Czerny}, \& {Petrucci}}]{Rozanska2015}
{R{\'o}{\.z}a{\'n}ska}, A., {Malzac}, J., {Belmont}, R., {Czerny}, B., \&
  {Petrucci}, P.~O., {Warm and optically thick dissipative coronae above
  accretion disks}. 2015, {\it \aap}, {\bf 580}, A77, DOI:
  10.1051/0004-6361/201526288

\bibitem[{{Salpeter}(1964)}]{Salpeter1964}
{Salpeter}, E.~E., {Accretion of Interstellar Matter by Massive Objects.} 1964,
  {\it The Astrophysical Journal}, {\bf 140}, 796, DOI: 10.1086/147973

\bibitem[{{Schmidt}(1963)}]{1963Natur.197.1040S}
{Schmidt}, M., {3C 273 : A Star-Like Object with Large Red-Shift}. 1963, {\it
  \nat}, {\bf 197}, 1040, DOI: 10.1038/1971040a0

\bibitem[{{Schnittman}(2005)}]{2005ApJ...621..940S}
{Schnittman}, J.~D., {Interpreting the High-Frequency Quasi-periodic
  Oscillation Power Spectra of Accreting Black Holes}. 2005, {\it \apj}, {\bf
  621}, 940, DOI: 10.1086/427646

\bibitem[{{Schnittman} \& {Bertschinger}(2004)}]{2004ApJ...606.1098S}
{Schnittman}, J.~D. \& {Bertschinger}, E., {The Harmonic Structure of
  High-Frequency Quasi-periodic Oscillations in Accreting Black Holes}. 2004,
  {\it \apj}, {\bf 606}, 1098, DOI: 10.1086/383180

\bibitem[{{Schnittman} {et~al.}(2006{\natexlab{a}}){Schnittman}, {Homan}, \&
  {Miller}}]{2006ApJ...642..420S}
{Schnittman}, J.~D., {Homan}, J., \& {Miller}, J.~M., {A Precessing Ring Model
  for Low-Frequency Quasi-periodic Oscillations}. 2006{\natexlab{a}}, {\it
  \apj}, {\bf 642}, 420, DOI: 10.1086/500923

\bibitem[{{Schnittman} {et~al.}(2006{\natexlab{b}}){Schnittman}, {Krolik}, \&
  {Hawley}}]{2006ApJ...651.1031S}
{Schnittman}, J.~D., {Krolik}, J.~H., \& {Hawley}, J.~F., {Light Curves from an
  MHD Simulation of a Black Hole Accretion Disk}. 2006{\natexlab{b}}, {\it
  \apj}, {\bf 651}, 1031, DOI: 10.1086/507421

\bibitem[{{Schnittman} \& {Rezzolla}(2006)}]{2006ApJ...637L.113S}
{Schnittman}, J.~D. \& {Rezzolla}, L., {Quasi-periodic Oscillations in the
  X-Ray Light Curves from Relativistic Tori}. 2006, {\it \apjl}, {\bf 637},
  L113, DOI: 10.1086/500545

\bibitem[{{Schwarzschild}(1916)}]{1916AbhKP1916..189S}
{Schwarzschild}, K., {On the Gravitational Field of a Mass Point According to
  Einstein's Theory}. 1916, {\it Abh. Konigl. Preuss. Akad. Wissenschaften
  Jahre 1906,92, Berlin,1907}, {\bf 1916}, 189

\bibitem[{{Seguin}(1975)}]{1975ApJ...197..745S}
{Seguin}, F.~H., {The stability of nonuniform rotation in relativistic stars.}
  1975, {\it \apj}, {\bf 197}, 745, DOI: 10.1086/153563

\bibitem[{{Seward} \& {Charles}(2010)}]{XrayUniverse}
{Seward}, F.~D. \& {Charles}, P.~A. 2010, {\it {Exploring the X-ray Universe}}
  (Cambridge University Press)

\bibitem[{{Shafee} {et~al.}(2008){Shafee}, {McKinney}, {Narayan},
  {Tchekhovskoy}, {Gammie}, \& {McClintock}}]{Shafee2008}
{Shafee}, R., {McKinney}, J.~C., {Narayan}, R., {et~al.}, {Three-Dimensional
  Simulations of Magnetized Thin Accretion Disks around Black Holes: Stress in
  the Plunging Region}. 2008, {\it \apjl}, {\bf 687}, L25, DOI: 10.1086/593148

\bibitem[{{Shakura} \& {Sunyaev}(1973)}]{Shakura+Sunyaev1973}
{Shakura}, N.~I. \& {Sunyaev}, R.~A., {Black Holes in Binary Systems:
  Observational Appearances}. 1973, in IAU Symposium, Vol. {\bf ~55}, {\it X-
  and Gamma-Ray Astronomy}, 155

\bibitem[{{Shakura} \& {Sunyaev}(1976)}]{Shakura1976}
{Shakura}, N.~I. \& {Sunyaev}, R.~A., {A theory of the instability of disk
  accretion on to black holes and the variability of binary X-ray sources,
  galactic nuclei and quasars.} 1976, {\it MNRAS}, {\bf 175}, 613, DOI:
  10.1093/mnras/175.3.613

\bibitem[{{S{\k a}dowski}(2016)}]{Sadowski2016}
{S{\k a}dowski}, A., {Thin accretion discs are stabilized by a strong magnetic
  field}. 2016, {\it Monthly Notices of the Royal Astronomical Society}, {\bf
  459}, 4397, DOI: 10.1093/mnras/stw913

\bibitem[{{S{\k{a}}dowski}(2009)}]{2009ApJS..183..171S}
{S{\k{a}}dowski}, A., {Slim Disks Around Kerr Black Holes Revisited}. 2009,
  {\it \apjs}, {\bf 183}, 171, DOI: 10.1088/0067-0049/183/2/171

\bibitem[{{S{\k{a}}dowski} \& {Narayan}(2015)}]{Sadowski2015_compton}
{S{\k{a}}dowski}, A. \& {Narayan}, R., {Photon-conserving Comptonization in
  simulations of accretion discs around black holes}. 2015, {\it Monthly
  Notices of the Royal Astronomical Society}, {\bf 454}, 2372, DOI:
  10.1093/mnras/stv2022

\bibitem[{{S{\k{a}}dowski} \& {Narayan}(2016)}]{SadowskiNarayan2016}
{S{\k{a}}dowski}, A. \& {Narayan}, R., {Three-dimensional simulations of
  supercritical black hole accretion discs - luminosities, photon trapping and
  variability}. 2016, {\it \mnras}, {\bf 456}, 3929, DOI: 10.1093/mnras/stv2941

\bibitem[{{S{\k{a}}dowski} {et~al.}(2014){S{\k{a}}dowski}, {Narayan},
  {McKinney}, \& {Tchekhovskoy}}]{Sadowski2014}
{S{\k{a}}dowski}, A., {Narayan}, R., {McKinney}, J.~C., \& {Tchekhovskoy}, A.,
  {Numerical simulations of super-critical black hole accretion flows in
  general relativity}. 2014, {\it MNRAS}, {\bf 439}, 503, DOI:
  10.1093/mnras/stt2479

\bibitem[{{S{\k{a}}dowski} {et~al.}(2015){S{\k{a}}dowski}, {Narayan},
  {Tchekhovskoy}, {Abarca}, {Zhu}, \& {McKinney}}]{Sadowski2015_dynamo}
{S{\k{a}}dowski}, A., {Narayan}, R., {Tchekhovskoy}, A., {et~al.}, {Global
  simulations of axisymmetric radiative black hole accretion discs in general
  relativity with a mean-field magnetic dynamo}. 2015, {\it \mnras}, {\bf 447},
  49, DOI: 10.1093/mnras/stu2387

\bibitem[{{S{\k{a}}dowski} {et~al.}(2013){S{\k{a}}dowski}, {Narayan},
  {Tchekhovskoy}, \& {Zhu}}]{Sadowski2013}
{S{\k{a}}dowski}, A., {Narayan}, R., {Tchekhovskoy}, A., \& {Zhu}, Y.,
  {Semi-implicit scheme for treating radiation under M1 closure in general
  relativistic conservative fluid dynamics codes}. 2013, {\it MNRAS}, {\bf
  429}, 3533, DOI: 10.1093/mnras/sts632

\bibitem[{{S{\k{a}}dowski} {et~al.}(2017){S{\k{a}}dowski}, {Wielgus},
  {Narayan}, {Abarca}, {McKinney}, \& {Chael}}]{Sadowski2017}
{S{\k{a}}dowski}, A., {Wielgus}, M., {Narayan}, R., {et~al.}, {Radiative,
  two-temperature simulations of low-luminosity black hole accretion flows in
  general relativity}. 2017, {\it MNRAS}, {\bf 466}, 705, DOI:
  10.1093/mnras/stw3116

\bibitem[{{Sod}(1978)}]{1978JCoPh..27....1S}
{Sod}, G.~A., {Review. A Survey of Several Finite Difference Methods for
  Systems of Nonlinear Hyperbolic Conservation Laws}. 1978, {\it Journal of
  Computational Physics}, {\bf 27}, 1, DOI: 10.1016/0021-9991(78)90023-2

\bibitem[{{Sorathia} {et~al.}(2012){Sorathia}, {Reynolds}, {Stone}, \&
  {Beckwith}}]{Sorathia2012}
{Sorathia}, K.~A., {Reynolds}, C.~S., {Stone}, J.~M., \& {Beckwith}, K.,
  {Global Simulations of Accretion Disks. I. Convergence and Comparisons with
  Local Models}. 2012, {\it \apj}, {\bf 749}, 189, DOI:
  10.1088/0004-637X/749/2/189

\bibitem[{{\v{S}pr\v{n}a} {et~al.}(in preparation){\v{S}pr\v{n}a}, {Falanga},
  {T{\"o}r{\"o}k}, {Klimovi{\v{c}}ov{\'a}}, {Urbanec}, {Lan{\v{c}}ov{\'a}},
  {Malacaria}, \& {Marmat}}]{Rene}
{\v{S}pr\v{n}a}, R., {Falanga}, M., {T{\"o}r{\"o}k}, G., {et~al.} in
  preparation, in preparation

\bibitem[{{\v{S}r{\'a}mkov{\'a}} {et~al.}(2007){\v{S}r{\'a}mkov{\'a}},
  {Torkelsson}, \& {Abramowicz}}]{2007A&A...467..641S}
{\v{S}r{\'a}mkov{\'a}}, E., {Torkelsson}, U., \& {Abramowicz}, M.~A.,
  {Oscillations of tori in the pseudo-Newtonian potential}. 2007, {\it \aap},
  {\bf 467}, 641, DOI: 10.1051/0004-6361:20065979

\bibitem[{{Stahl} {et~al.}(2012){Stahl}, {Wielgus}, {Abramowicz},
  {Klu{\'z}niak}, \& {Yu}}]{2012A&A...546A..54S}
{Stahl}, A., {Wielgus}, M., {Abramowicz}, M., {Klu{\'z}niak}, W., \& {Yu}, W.,
  {Eddington capture sphere around luminous stars}. 2012, {\it \aap}, {\bf
  546}, A54, DOI: 10.1051/0004-6361/201220187

\bibitem[{{Stella} \& {Vietri}(1999)}]{1999PhRvL..82...17S}
{Stella}, L. \& {Vietri}, M., {kHz Quasiperiodic Oscillations in Low-Mass X-Ray
  Binaries as Probes of General Relativity in the Strong-Field Regime}. 1999,
  {\it \prl}, {\bf 82}, 17, DOI: 10.1103/PhysRevLett.82.17

\bibitem[{{Straub} {et~al.}(2011){Straub}, {Bursa}, {S{\k{a}}dowski},
  {Steiner}, {Abramowicz}, {Klu{\'z}niak}, {McClintock}, {Narayan}, \&
  {Remillard}}]{Straub2011}
{Straub}, O., {Bursa}, M., {S{\k{a}}dowski}, A., {et~al.}, {Testing slim-disk
  models on the thermal spectra of LMC X-3}. 2011, {\it \aap}, {\bf 533}, A67,
  DOI: 10.1051/0004-6361/201117385

\bibitem[{{Straub} \& {{\v{S}}r{\'a}mkov{\'a}}(2009)}]{2009CQGra..26e5011S}
{Straub}, O. \& {{\v{S}}r{\'a}mkov{\'a}}, E., {Epicyclic oscillations of
  non-slender fluid tori around Kerr black holes}. 2009, {\it Classical and
  Quantum Gravity}, {\bf 26}, 055011, DOI: 10.1088/0264-9381/26/5/055011

\bibitem[{{Strohmayer} {et~al.}(1997){Strohmayer}, {Jahoda}, {Giles}, \&
  {Lee}}]{1997ApJ...486..355S}
{Strohmayer}, T.~E., {Jahoda}, K., {Giles}, A.~B., \& {Lee}, U., {Millisecond
  Pulsations from a Low-Mass X-Ray Binary in the Galactic Center Region}. 1997,
  {\it \apj}, {\bf 486}, 355, DOI: 10.1086/304522

\bibitem[{{Strohmayer} {et~al.}(1996){Strohmayer}, {Zhang}, {Swank}, {Smale},
  {Titarchuk}, {Day}, \& {Lee}}]{Strohmayer1996}
{Strohmayer}, T.~E., {Zhang}, W., {Swank}, J.~H., {et~al.}, {Millisecond X-Ray
  Variability from an Accreting Neutron Star System}. 1996, {\it \apjl}, {\bf
  469}, L9, DOI: 10.1086/310261

\bibitem[{{Stuchl{\'\i}k} {et~al.}(2013){Stuchl{\'\i}k}, {Kotrlov{\'a}}, \&
  {T{\"o}r{\"o}k}}]{2013A&A...552A..10S}
{Stuchl{\'\i}k}, Z., {Kotrlov{\'a}}, A., \& {T{\"o}r{\"o}k}, G.,
  {Multi-resonance orbital model of high-frequency quasi-periodic oscillations:
  possible high-precision determination of black hole and neutron star spin}.
  2013, {\it \aap}, {\bf 552}, A10, DOI: 10.1051/0004-6361/201219724

\bibitem[{{Tchekhovskoy}(2019)}]{HARMPI}
{Tchekhovskoy}, A. 2019, {HARMPI: 3D massively parallel general relativictic
  MHD code}, Astrophysics Source Code Library, record ascl:1912.014

\bibitem[{{Tchekhovskoy} {et~al.}(2011){Tchekhovskoy}, {Narayan}, \&
  {McKinney}}]{Tchekhovskoy2011}
{Tchekhovskoy}, A., {Narayan}, R., \& {McKinney}, J.~C., {Efficient generation
  of jets from magnetically arrested accretion on a rapidly spinning black
  hole}. 2011, {\it \mnras}, {\bf 418}, L79, DOI:
  10.1111/j.1745-3933.2011.01147.x

\bibitem[{{Tetarenko} {et~al.}(2016){Tetarenko}, {Sivakoff}, {Heinke}, \&
  {Gladstone}}]{Tetarenko2016}
{Tetarenko}, B.~E., {Sivakoff}, G.~R., {Heinke}, C.~O., \& {Gladstone}, J.~C.,
  {WATCHDOG: A Comprehensive All-sky Database of Galactic Black Hole X-ray
  Binaries}. 2016, {\it \apjs}, {\bf 222}, 15, DOI: 10.3847/0067-0049/222/2/15

\bibitem[{Toro(2009)}]{Toro}
Toro, E. 2009, {\it Riemann Solvers and Numerical Methods for Fluid Dynamics: A
  Practical Introduction} (Springer)

\bibitem[{{T{\"o}r{\"o}k} {et~al.}(2005){T{\"o}r{\"o}k}, {Abramowicz},
  {Klu{\'z}niak}, \& {Stuchl{\'\i}k}}]{tor-etal:2005}
{T{\"o}r{\"o}k}, G., {Abramowicz}, M.~A., {Klu{\'z}niak}, W., \&
  {Stuchl{\'\i}k}, Z., {The orbital resonance model for twin peak kHz quasi
  periodic oscillations in microquasars}. 2005, {\it \aap}, {\bf 436}, 1, DOI:
  10.1051/0004-6361:20047115

\bibitem[{{T{\"o}r{\"o}k} {et~al.}(2008){T{\"o}r{\"o}k}, {Bakala},
  {Stuchl\'{i}k}, \& {\v{C}ech}}]{2008AcA....58....1T}
{T{\"o}r{\"o}k}, G., {Bakala}, P., {Stuchl\'{i}k}, Z., \& {\v{C}ech}, P.,
  {Modeling the Twin Peak QPO Distribution in the Atoll Source 4U 1636-53}.
  2008, {\it \actaa}, {\bf 58}, 1

\bibitem[{{T{\"o}r{\"o}k} {et~al.}(2016{\natexlab{a}}){T{\"o}r{\"o}k},
  {Goluchov{\'a}}, {Hor{\'a}k}, {{\v{S}}r{\'a}mkov{\'a}}, {Urbanec},
  {Pech{\'a}{\v{c}}ek}, \& {Bakala}}]{torok2016mnras}
{T{\"o}r{\"o}k}, G., {Goluchov{\'a}}, K., {Hor{\'a}k}, J., {et~al.}, {Twin peak
  quasi-periodic oscillations as signature of oscillating cusp torus}.
  2016{\natexlab{a}}, {\it \mnras}, {\bf 457}, L19, DOI: 10.1093/mnrasl/slv196

\bibitem[{{T{\"o}r{\"o}k} {et~al.}(2016{\natexlab{b}}){T{\"o}r{\"o}k},
  {Goluchov{\'a}}, {Hor{\'a}k}, {{\v{S}}r{\'a}mkov{\'a}}, {Urbanec},
  {Pech{\'a}{\v{c}}ek}, \& {Bakala}}]{2016MNRAS.457L..19T}
{T{\"o}r{\"o}k}, G., {Goluchov{\'a}}, K., {Hor{\'a}k}, J., {et~al.}, {Twin peak
  quasi-periodic oscillations as signature of oscillating cusp torus}.
  2016{\natexlab{b}}, {\it \mnras}, {\bf 457}, L19, DOI: 10.1093/mnrasl/slv196

\bibitem[{{T{\"o}r{\"o}k} {et~al.}(2019){T{\"o}r{\"o}k}, {Goluchov{\'a}},
  {{\v{S}}r{\'a}mkov{\'a}}, {Urbanec}, \& {Straub}}]{2019MNRAS.488.3896T}
{T{\"o}r{\"o}k}, G., {Goluchov{\'a}}, K., {{\v{S}}r{\'a}mkov{\'a}}, E.,
  {Urbanec}, M., \& {Straub}, O., {Time-scale of twin-peak quasi-periodic
  oscillations and mass of accreting neutron stars}. 2019, {\it \mnras}, {\bf
  488}, 3896, DOI: 10.1093/mnras/stz1929

\bibitem[{{T{\"o}r{\"o}k} {et~al.}(2022){T{\"o}r{\"o}k}, {Kotrlov{\'a}},
  {Matuszkov{\'a}}, {Klimovi{\v{c}}ov{\'a}}, {Lan{\v{c}}ov{\'a}},
  {Urbancov{\'a}}, \& {{\v{S}}r{\'a}mkov{\'a}}}]{tor-etal:2022}
{T{\"o}r{\"o}k}, G., {Kotrlov{\'a}}, A., {Matuszkov{\'a}}, M., {et~al.},
  {Simple analytic formula relating the mass and spin of accreting compact
  objects to their rapid X-ray variability}. 2022, {\it arXiv e-prints},
  arXiv:2203.04787

\bibitem[{{T{\'o}th}(2000)}]{Toth2000}
{T{\'o}th}, G., {The {\ensuremath{\nabla}}{\textperiodcentered} B=0 Constraint
  in Shock-Capturing Magnetohydrodynamics Codes}. 2000, {\it Journal of
  Computational Physics}, {\bf 161}, 605, DOI: 10.1006/jcph.2000.6519

\bibitem[{{Treves} {et~al.}(2000){Treves}, {Turolla}, {Zane}, \&
  {Colpi}}]{Treves2000}
{Treves}, A., {Turolla}, R., {Zane}, S., \& {Colpi}, M., {Isolated Neutron
  Stars: Accretors and Coolers}. 2000, {\it \pasp}, {\bf 112}, 297, DOI:
  10.1086/316529

\bibitem[{{Urbancov{\'a}} {et~al.}(2019){Urbancov{\'a}}, {Urbanec},
  {T{\"o}r{\"o}k}, {Stuchl{\'\i}k}, {Blaschke}, \& {Miller}}]{Urbancova2019}
{Urbancov{\'a}}, G., {Urbanec}, M., {T{\"o}r{\"o}k}, G., {et~al.}, {Epicyclic
  Oscillations in the Hartle-Thorne External Geometry}. 2019, {\it \apj}, {\bf
  877}, 66, DOI: 10.3847/1538-4357/ab1b4c

\bibitem[{{Urbanec} {et~al.}(2013){Urbanec}, {Miller}, \&
  {Stuchl{\'\i}k}}]{2013MNRAS.433.1903U}
{Urbanec}, M., {Miller}, J.~C., \& {Stuchl{\'\i}k}, Z., {Quadrupole moments of
  rotating neutron stars and strange stars}. 2013, {\it \mnras}, {\bf 433},
  1903, DOI: 10.1093/mnras/stt858

\bibitem[{{Urry} \& {Padovani}(1995)}]{1995PASP..107..803U}
{Urry}, C.~M. \& {Padovani}, P., {Unified Schemes for Radio-Loud Active
  Galactic Nuclei}. 1995, {\it \pasp}, {\bf 107}, 803, DOI: 10.1086/133630

\bibitem[{{van der Klis}(2004)}]{vdKlis2004}
{van der Klis}, M., {A review of rapid X-ray variability in X-ray binaries}.
  2004, {\it arXiv e-prints}, astro, DOI: 10.48550/arXiv.astro-ph/0410551

\bibitem[{Van~der Klis(2006)}]{vdKlis2006}
Van~der Klis, M., Rapid X-ray variability. 2006, {\it in Compact stellar X-ray
  sources}, {\bf 39}, 39

\bibitem[{{van Leer}(1979)}]{vLeer1979}
{van Leer}, B., {Towards the Ultimate Conservative Difference Scheme. V. A
  Second-Order Sequel to Godunov's Method}. 1979, {\it Journal of Computational
  Physics}, {\bf 32}, 101, DOI: 10.1016/0021-9991(79)90145-1

\bibitem[{{Vincent} {et~al.}(2011){Vincent}, {Paumard}, {Gourgoulhon}, \&
  {Perrin}}]{2011CQGra..28v5011V}
{Vincent}, F.~H., {Paumard}, T., {Gourgoulhon}, E., \& {Perrin}, G., {GYOTO: a
  new general relativistic ray-tracing code}. 2011, {\it Classical and Quantum
  Gravity}, {\bf 28}, 225011, DOI: 10.1088/0264-9381/28/22/225011

\bibitem[{{Vincentelli} {et~al.}(2023){Vincentelli}, {Neilsen}, {Tetarenko},
  {Cavecchi}, {Castro Segura}, {del Palacio}, {van den Eijnden},
  {Vasilopoulos}, {Altamirano}, {Armas Padilla}, {Bailyn}, {Belloni},
  {Buisson}, {C{\'u}neo}, {Degenaar}, {Knigge}, {Long}, {Jim{\'e}nez-Ibarra},
  {Milburn}, {Mu{\~n}oz Darias}, {{\"O}zbey Arabac{\i}}, {Remillard}, \&
  {Russell}}]{Vincentelli2023}
{Vincentelli}, F.~M., {Neilsen}, J., {Tetarenko}, A.~J., {et~al.}, {A shared
  accretion instability for black holes and neutron stars}. 2023, {\it Nature},
  {\bf 615}, 45, DOI: 10.1038/s41586-022-05648-3

\bibitem[{{von Zeipel}(1924)}]{1924MNRAS..84..665V}
{von Zeipel}, H., {The radiative equilibrium of a rotating system of gaseous
  masses}. 1924, {\it \mnras}, {\bf 84}, 665, DOI: 10.1093/mnras/84.9.665

\bibitem[{{{\v{S}}r{\'a}mkov{\'a}} {et~al.}(2023){{\v{S}}r{\'a}mkov{\'a}},
  {Matuszkov{\'a}}, {Klimovi{\v{c}}ov{\'a}}, {Hor{\'a}k}, {Straub},
  {Urbancov{\'a}}, {Urbanec}, {Karas}, {T{\"o}r{\"o}k}, \&
  {Lan{\v{c}}ov{\'a}}}]{2023AN....34420114S}
{{\v{S}}r{\'a}mkov{\'a}}, E., {Matuszkov{\'a}}, M., {Klimovi{\v{c}}ov{\'a}},
  K., {et~al.}, {Oscillations of fluid tori around neutron stars}. 2023, {\it
  Astronomische Nachrichten}, {\bf 344}, e20220114, DOI: 10.1002/asna.20220114

\bibitem[{{Wang} {et~al.}(2013){Wang}, {Nowak}, {Markoff}, {Baganoff},
  {Nayakshin}, {Yuan}, {Cuadra}, {Davis}, {Dexter}, {Fabian}, {Grosso},
  {Haggard}, {Houck}, {Ji}, {Li}, {Neilsen}, {Porquet}, {Ripple}, \&
  {Shcherbakov}}]{Wang}
{Wang}, Q.~D., {Nowak}, M.~A., {Markoff}, S.~B., {et~al.}, {Dissecting
  X-ray-Emitting Gas Around the Center of Our Galaxy}. 2013, {\it Science},
  {\bf 341}, 981, DOI: 10.1126/science.1240755

\bibitem[{{Webster} \& {Murdin}(1972)}]{Webster1972}
{Webster}, B.~L. \& {Murdin}, P., {Cygnus X-1-a Spectroscopic Binary with a
  Heavy Companion ?} 1972, {\it \nat}, {\bf 235}, 37, DOI: 10.1038/235037a0

\bibitem[{{White} {et~al.}(2016){White}, {Stone}, \& {Gammie}}]{Athena}
{White}, C.~J., {Stone}, J.~M., \& {Gammie}, C.~F., {An Extension of the
  Athena++ Code Framework for GRMHD Based on Advanced Riemann Solvers and
  Staggered-mesh Constrained Transport}. 2016, {\it \apjs}, {\bf 225}, 22, DOI:
  10.3847/0067-0049/225/2/22

\bibitem[{{Wielgus} {et~al.}(2020){Wielgus}, {Hor{\'a}k}, {Vincent}, \&
  {Abramowicz}}]{Wielgus2020}
{Wielgus}, M., {Hor{\'a}k}, J., {Vincent}, F., \& {Abramowicz}, M.,
  {Reflection-asymmetric wormholes and their double shadows}. 2020, {\it \prd},
  {\bf 102}, 084044, DOI: 10.1103/PhysRevD.102.084044

\bibitem[{{Wielgus} {et~al.}(2022){Wielgus}, {Lan{\v{c}}ov{\'a}}, {Straub},
  {Klu{\'z}niak}, {Narayan}, {Abarca}, {R{\'o}{\.z}a{\'n}ska}, {Vincent},
  {T{\"o}r{\"o}k}, \& {Abramowicz}}]{Wielgus2022}
{Wielgus}, M., {Lan{\v{c}}ov{\'a}}, D., {Straub}, O., {et~al.}, {Observational
  properties of puffy discs: radiative GRMHD spectra of mildly sub-Eddington
  accretion}. 2022, {\it Monthly Notices of the Royal Astronomical Society},
  {\bf 514}, 780, DOI: 10.1093/mnras/stac1317

\bibitem[{{Wijnands}(1999)}]{1999PhDT........10W}
{Wijnands}, R. 1999, {Millisecond phenomena in X-ray binaries}, PhD thesis, -

\bibitem[{{Zdziarski} {et~al.}(1996){Zdziarski}, {Johnson}, \&
  {Magdziarz}}]{Zdziarski1996}
{Zdziarski}, A.~A., {Johnson}, W.~N., \& {Magdziarz}, P., {Broad-band
  {\ensuremath{\gamma}}-ray and X-ray spectra of NGC 4151 and their
  implications for physical processes and geometry.} 1996, {\it \mnras}, {\bf
  283}, 193, DOI: 10.1093/mnras/283.1.193

\bibitem[{{Zhang} {et~al.}(1997){Zhang}, {Cui}, \& {Chen}}]{CF}
{Zhang}, S.~N., {Cui}, W., \& {Chen}, W., {Black Hole Spin in X-Ray Binaries:
  Observational Consequences}. 1997, {\it \apjl}, {\bf 482}, L155, DOI:
  10.1086/310705

\bibitem[{{Zhang} {et~al.}(2016){Zhang}, {Feroci}, {Santangelo}, {Dong},
  {Feng}, {Lu}, {Nandra}, {Wang}, {Zhang}, {Bozzo}, {Brandt}, {De Rosa}, {Gou},
  {Hernanz}, {van der Klis}, {Li}, {Liu}, {Orleanski}, {Pareschi}, {Pohl},
  {Poutanen}, {Qu}, {Schanne}, {Stella}, {Uttley}, {Watts}, {Xu}, {Yu}, {in 't
  Zand}, {Zane}, {Alvarez}, {Amati}, {Baldini}, {Bambi}, {Basso},
  {Bhattacharyya S.}, {}, {Belloni}, {Bellutti}, {Bianchi}, {Brez}, {Bursa},
  {Burwitz}, {Budtz-J{\o}rgensen}, {Caiazzo}, {Campana}, {Cao}, {Casella},
  {Chen}, {Chen}, {Chen}, {Chen}, {Chen}, {Chen}, {Civitani}, {Coti Zelati},
  {Cui}, {Cui}, {Dai}, {Del Monte}, {de Martino}, {Di Cosimo}, {Diebold},
  {Dovciak}, {Donnarumma}, {Doroshenko}, {Esposito}, {Evangelista}, {Favre},
  {Friedrich}, {Fuschino}, {Galvez}, {Gao}, {Ge}, {Gevin}, {Goetz}, {Han},
  {Heyl}, {Horak}, {Hu}, {Huang}, {Huang}, {Hudec}, {Huppenkothen}, {Israel},
  {Ingram}, {Karas}, {Karelin}, {Jenke}, {Ji}, {Korpela}, {Kunneriath},
  {Labanti}, {Li}, {Li}, {Li}, {Liang}, {Limousin}, {Lin}, {Ling}, {Liu},
  {Liu}, {Liu}, {Lu}, {Lund}, {Lai}, {Luo}, {Luo}, {Ma}, {Mahmoodifar},
  {Marisaldi}, {Martindale}, {Meidinger}, {Men}, {Michalska}, {Mignani},
  {Minuti}, {Motta}, {Muleri}, {Neilsen}, {Orlandini}, {Pan}, {Patruno},
  {Perinati}, {Picciotto}, {Piemonte}, {Pinchera}, {Rachevski A.}, {Rapisarda},
  {Rea}, {Rossi}, {Rubini}, {Sala}, {Shu}, {Sgro}, {Shen}, {Soffitta}, {Song},
  {Spandre}, {Stratta}, {Strohmayer}, {Sun}, {Svoboda}, {Tagliaferri},
  {Tenzer}, {Hong}, {Taverna}, {Torok}, {Turolla}, {Vacchi}, {Wang}, {Walton},
  {Wang}, {Wang}, {Wang}, {Wang}, {Weng}, {Wilms}, {Winter}, {Wu}, {Wu},
  {Xiong}, {Xu}, {Xue}, {Yan}, {Yang}, {Yang}, {Yang}, {Yuan}, {Yuan}, {Yuan},
  {Zampa}, {Zampa}, {Zdziarski}, {Zhang}, {Zhang}, {Zhang}, {Zhang}, {Zhang},
  {Zhang}, {Zheng}, {Zhou}, \& {Zhou X.~L.}}]{2016SPIE.9905E..1QZ}
{Zhang}, S.~N., {Feroci}, M., {Santangelo}, A., {et~al.}, {eXTP: Enhanced X-ray
  Timing and Polarization mission}. 2016, in Society of Photo-Optical
  Instrumentation Engineers (SPIE) Conference Series, Vol. {\bf  9905}, {\it
  Space Telescopes and Instrumentation 2016: Ultraviolet to Gamma Ray}, ed.
  J.-W.~A. {den Herder}, T.~{Takahashi}, \& M.~{Bautz}, 99051Q

\bibitem[{{Zhu} \& {Narayan}(2013)}]{Zhu2013}
{Zhu}, Y. \& {Narayan}, R., {Thermal stability in turbulent accretion discs}.
  2013, {\it \mnras}, {\bf 434}, 2262, DOI: 10.1093/mnras/stt1161

\bibitem[{{Zhu} {et~al.}(2015){Zhu}, {Narayan}, {S\k{a}dowski}, \&
  {Psaltis}}]{2015MNRAS.451.1661Z}
{Zhu}, Y., {Narayan}, R., {S\k{a}dowski}, A., \& {Psaltis}, D., {HERO - A 3D
  general relativistic radiative post-processor for accretion discs around
  black holes}. 2015, {\it \mnras}, {\bf 451}, 1661, DOI: 10.1093/mnras/stv1046

\bibitem[{{{\.Z}ycki} {et~al.}(1999){{\.Z}ycki}, {Done}, \&
  {Smith}}]{Zycki1999}
{{\.Z}ycki}, P.~T., {Done}, C., \& {Smith}, D.~A., {The 1989 May outburst of
  the soft X-ray transient GS 2023+338 (V404 Cyg)}. 1999, {\it \mnras}, {\bf
  309}, 561, DOI: 10.1046/j.1365-8711.1999.02885.x

\end{thebibliography}
\par}

\part{Collection of papers}

\noindent \textit{Note: The published papers are replaced by the abstracts and DOIs in the arXiv version of this dissertation.}

\chapter*{Paper 1: Puffy Accretion Disks: Sub-Eddington, Optically Thick, and Stable}

\noindent DOI: \href{https://ui.adsabs.harvard.edu/link_gateway/2019ApJ...884L..37L/doi:10.3847/2041-8213/ab48f5}{10.3847/2041-8213/ab48f5}

\vspace{1cm}

\noindent \Large \textbf{Abstract}

\vspace{0.5cm}

\small We report on a new class of solutions of black hole accretion disks that we found through radiative magnetohydrodynamic global simulations. The presented solution has a near-Eddington luminosity $0.6\,L_{\rm Edd}$. It combines features of the canonical thin, slim and thick disk models but differs in crucial respects from each of them. By the density scale-height measure it appears to be thin, having a high density core near the equatorial plane of height $\sim 0.1 r$, but it is sandwiched by a highly advective, turbulent, optically thick, Keplerian inflow of substantial geometrical thickness $H \sim r$. It is partially supported by an advected poloidal magnetic field with $\beta \sim 0.5$, which makes the disk thermally stable.

\newpage

\chapter*{Paper 2: Observational properties of puffy discs: radiative GRMHD spectra of mildly sub-Eddington accretion}
\setcounter{page}{135}

\noindent DOI: \href{https://ui.adsabs.harvard.edu/link_gateway/2022MNRAS.514..780W/doi:10.1093/mnras/stac1317}{10.1093/mnras/stac1317}

\vspace{1cm}

\noindent \Large \textbf{Abstract}

\vspace{0.5cm}

\small Numerical general relativistic radiative magnetohydrodynamic simulations of accretion disks around a stellar mass black hole with a luminosity above 0.5 of the Eddington value reveal their stratified, elevated vertical structure. We refer to these thermally stable numerical solutions as puffy disks. Above a dense and geometrically thin core of dimensionless thickness $h/r \sim 0.1$, crudely resembling a classic thin accretion disk, a puffed-up, geometrically thick layer of lower density is formed. This puffy layer corresponds to $h/r \sim 1.0$, with a very limited dependence of the dimensionless thickness on the mass accretion rate. We discuss the observational properties of puffy disks, in particular the geometrical obscuration of the inner disk by the elevated puffy region at higher observing inclinations, and collimation of the radiation along the accretion disk spin axis, which may explain the apparent super-Eddington luminosity of some X-ray objects. We also present synthetic spectra of puffy disks, and show that they are qualitatively similar to those of a Comptonized thin disk. We demonstrate that the existing \textsc{xspec} spectral fitting models provide good fits to synthetic observations of puffy disks, but cannot correctly recover the input black hole spin. The puffy region remains optically thick to scattering; in its spectral properties the puffy disk roughly resembles that of a warm corona sandwiching the disk core. We suggest that puffy disks may correspond to X-ray binary systems of luminosities above 0.3 of the Eddington luminosity in the intermediate spectral states.

\newpage

\chapter*{Paper 3: Spectra of puffy accretion discs: the \texttt{kynbb} fit}
\setcounter{page}{147}
\noindent DOI: \href{https://ui.adsabs.harvard.edu/link_gateway/2023AN....34430023L/doi:10.1002/asna.20230023}{10.1002/asna.20230023}

\vspace{1cm}

\noindent \Large \textbf{Abstract}

\vspace{0.5cm}

\small Puffy disc is a~numerical model, expected to capture the properties of the accretion flow in X-ray black hole binaries in the luminous, mildly sub-Eddington state. We fit the \texttt{kerrbb} and \texttt{kynbb} spectral models in \textsc{xspec} to synthetic spectra of puffy accretion discs, obtained in general relativistic radiative magnetohydrodynamic simulations, to see if they correctly recover the black hole spin and mass accretion rate assumed in the numerical simulation. We conclude that neither of the two models is capable of correctly recovering the puffy disc parameters, which highlights a~necessity to develop new, more accurate spectral models for the luminous regime of accretion in X-ray black hole binaries. We propose that such spectral models should be based on the results of numerical simulations of accretion.

\newpage

\chapter*{Paper 4: Models of high-frequency quasi-periodic oscillations and black hole spin estimates in Galactic microquasars}
\setcounter{page}{155}
\noindent DOI: \href{https://ui.adsabs.harvard.edu/link_gateway/2020A&A...643A..31K/doi:10.1051/0004-6361/201937097}{10.1051/0004-6361/201937097}

\vspace{1cm}

\noindent \Large \textbf{Abstract}

\vspace{0.5cm}

\small  We explore the influence of non-geodesic pressure forces that are present in an accretion disk on the frequencies of its axisymmetric and non-axisymmetric epicyclic oscillation modes. {We discuss its implications for models of high frequency quasi-periodic oscillations (QPOs) that have been observed in the X-ray flux of accreting black holes (BHs) in the three Galactic microquasars, GRS 1915+105, GRO J1655$-$40 and XTE J1550$-$564. We focus on previously considered QPO models that deal with low azimuthal number epicyclic modes, $\lvert m \rvert \leq 2$, and outline the consequences for the estimations of BH spin, $a\in[0,1]$.} For four out of six examined models, we find only small, rather insignificant changes compared to the geodesic case. For the other two models, on the other hand, there is a fair increase of the estimated upper limit on the spin. Regarding the QPO model's falsifiability, we find that one particular model from the examined set is incompatible with the data. If the microquasar's spectral spin estimates that point to $a>0.65$ were fully confirmed, two more QPO models would be ruled out. Moreover, if two very different values of the spin, such as $a\approx 0.65$ in GRO~J1655$-$40 vs. $a\approx 1$ in GRS~1915+105, were confirmed, all the models except one would remain unsupported by our results. Finally, we discuss the implications for a model recently proposed in the context of neutron star (NS) QPOs as a disk-oscillation-based modification of the relativistic precession model. This model provides overall better fits of the NS data and predicts more realistic values of the NS mass compared to the relativistic precession model. We conclude that it also implies a significantly higher upper limit on the microquasar's BH spin ($a\sim 0.75$ vs. $a\sim 0.55$).

\newpage

\chapter*{Paper 5: Simple Analytic Formula Relating the Mass and Spin of Accreting Compact Objects to Their Rapid X-Ray Variability}
\setcounter{page}{165}
\noindent DOI: \href{https://ui.adsabs.harvard.edu/link_gateway/2022ApJ...929...28T/doi:10.3847/1538-4357/ac5ab6}{10.3847/1538-4357/ac5ab6}

\vspace{1cm}

\noindent \Large \textbf{Abstract}

\vspace{0.5cm}

\small Following the previous  research on epicyclic oscillations of accretion disks around black holes (BHs) and neutron stars (NSs), a new  model of high-frequency quasi-periodic oscillations (QPOs) has been proposed (CT model), which deals with oscillations of fluid in marginally overflowing accretion tori (i.e., tori terminated by cusps). According to preliminary investigations, the model provides better fits of the NS QPO data compared to the relativistic precession (RP) model. It also implies a significantly higher upper limit on the Galactic microquasars BH spin. A short analytic formula has been noticed to well reproduce the model's predictions on the QPO frequencies in Schwarzschild spacetimes. Here we derive an extended version of this formula that applies to rotating compact objects. We start with the consideration of Kerr spacetimes and derive a formula that is not restricted to a particular specific angular momentum distribution of the inner accretion flow, such as Keplerian or constant. Finally, we consider Hartle-Thorne spacetimes and include corrections implied by the NS oblateness. For a particular choice of a single parameter, our relation  provides frequencies predicted by the CT model. For another value, it provides frequencies predicted by the RP model. We conclude that the formula is well applicable for rotating oblateness NSs and both models. We briefly illustrate application of our simple formula on several NS sources and confirm the expectation that the CT model is compatible with realistic values of the NS mass and provides better fits of data than the RP model.

\newpage

\chapter*{Paper 6: Oscillations of non-slender tori in the external Hartle-Thorne geometry}
\setcounter{page}{175}

\noindent DOI: \href{https://ui.adsabs.harvard.edu/link_gateway/2022arXiv220310653M/doi:10.48550/arXiv.2203.10653}{10.48550/arXiv.2203.10653}

\vspace{1cm}

\noindent \Large \textbf{Abstract}

\vspace{0.5cm}

\small We examine the influence of the quadrupole moment of a slowly rotating neutron star on the oscillations of non-slender accretion tori. We apply previously developed methods to perform analytical calculations of frequencies of the radial epicyclic mode of a torus in the specific case of the Hartle-Thorne geometry. We present here our preliminary results and provide a brief comparison between the calculated frequencies and the frequencies previously obtained assuming both standard and linearized Kerr geometry. Finally, we shortly discuss the consequences for models of high-frequency quasi-periodic oscillations observed in low-mass X-ray binaries.

\newpage

\end{document}